\documentclass[10pt,twoside]{article}
\usepackage{times}
\usepackage{cite}
\usepackage{afterpage}
\usepackage{latexsym}
\usepackage{amssymb,amsbsy}
\usepackage{nonfloat}
\usepackage[section]{placeins}
\usepackage{multirow}
 \usepackage[dvips]{graphicx}
                \DeclareGraphicsExtensions{.eps, .jpg}
\def\pbar      {\ensuremath{\bar{\mathrm{p}}}}
\def\nbar      {\ensuremath{\bar{\mathrm{n}}}}
\def\p            {\ensuremath{\mathrm{p}}}

\def\n            {\ensuremath{\mathrm{n}}}
\def\Nb          {\ensuremath{\overline{\mathrm{N}}}}
\def\N           {\ensuremath{\mathrm{N}}}

\def\Pe     {\ensuremath{\mathrm{e}}}
\def\piz      {\ensuremath{\pi^0}}
\def\pip      {\ensuremath{\pi^+}}
\def\pim      {\ensuremath{\pi^-}}
\def\pipm      {\ensuremath{\pi^{\pm}}}
\def\pimp      {\ensuremath{\pi^{\mp}}}
\def\ftmass           {\ensuremath{{\mathrm f}_2(1270)   }}
\def\at                {\ensuremath{{\mathrm{a}_2}}}
\def\atmass           {\ensuremath{{\mathrm a}_2(1320)   }}
\def\atpmmass           {\ensuremath{{\mathrm a}_2^\pm(1320)   }}
\def\atzmass           {\ensuremath{{\mathrm a}_2^0(1320)   }}
\def\ft                {\ensuremath{{\mathrm{f}_2}}}

\def\K     {\ensuremath{\mathrm{K}}}                 
\let\PK=\K
\def\Kp    {\ensuremath{\mathrm{K}^+}}
\def\Km    {\ensuremath{\mathrm{K}^-}}
\def\Kpm   {\ensuremath{\mathrm{K}^{\pm}}}

\def\Kn    {\ensuremath{\mathrm{K}^0}}

\let\Kbn=\Knb
\def\Kb    {\ensuremath{\overline{\mathrm{K}}}{}}
\def\Ks    {\ensuremath{\mathrm{K_s}}}
\def\Kl    {\ensuremath{\mathrm{K_l}}}

\def\ppb              {\ensuremath{\bar{\mathrm{p}}\mathrm{p}}}
\let\pbp=\ppb

\let\pbarp=\ppb
\def\pbard            {\ensuremath{\bar{\mathrm{p}}\mathrm{d}}}
\let\pbd=\pbard
\def\pbn              {\ensuremath{\bar{\mathrm{p}}\mathrm{n}}}

\def\nbp       {\ensuremath{\bar{\mathrm{n}}\mathrm{p}}}
\def\pbN              {\ensuremath{\bar{\mathrm{p}}\mathrm{N}}}

\def\NNb              {\ensuremath{\overline{\mathrm{N}}\mathrm{N}}}

\def\HT      {\ensuremath{\mathrm{H}_2}}
\def\DT      {\ensuremath{\mathrm{D}_2}}

\def\qqb        {\ensuremath{\bar{q}q}}
\let\qqbar=\qqb
\def\ssb        {\ensuremath{\bar{{s}}{s}}} 

\def\uub        {\ensuremath{\bar{{u}}{u}}} 
\def\ddb        {\ensuremath{\bar{{d}}{d}}} 
\def\nnb        {\ensuremath{\bar{{n}}{n}}} 
\def\udb        {\ensuremath{\bar{{d}}{u}}} 
\def\dub        {\ensuremath{\bar{{u}}{d}}} 

\newcommand{\SLJ}[3]    {\relax\ensuremath{{}^{#1}{{\mathrm #2}}_{#3}}}
\newcommand{\ISLJ}[4]   {\relax\ensuremath{{}^{#1,#2}{{\mathrm #3}}_{#4}}}

\newcommand{\islj}{\relax\ensuremath{{}^{2I+1,2S+1}L_J}}
\newcommand{\slj}{\relax\ensuremath{{}^{2S+1}L_J}}

\newcommand{\jpc}{\relax\ensuremath{J^{PC}}}
\newcommand{\nslj}{\relax\ensuremath{n\,{}^{2S+1}L_J}}

\newcommand{\stso}{\ISLJ13S1}
\newcommand{\ttso}{\ISLJ33S1}
\newcommand{\tssz}{\ISLJ31S0}
\newcommand{\sssz}{\ISLJ11S0}
\newcommand{\stpz}{\ISLJ13P0}
\newcommand{\ttpo}{\ISLJ33P1}
\newcommand{\ttpt}{\ISLJ33P2}
\newcommand{\tspo}{\ISLJ31P1}
\newcommand{\sspo}{\ISLJ11P1}
\newcommand{\tso}{\SLJ3S1}
\newcommand{\ssz}{\SLJ1S0}
\newcommand{\tpz}{\SLJ3P0}
\newcommand{\tpo}{\SLJ3P1}
\newcommand{\tpt}{\SLJ3P2}
\newcommand{\spo}{\SLJ1P1}

\newcommand{\pann}  {\relax\ensuremath{\mathrm{P}_{\mathrm{ann}}  }}
\newcommand{\er}           {\ensuremath{\pm}}
\newcommand{\rstp}         {\ensuremath{\rho_{\rm{STP}}}}
\newcommand{\fp}          {\ensuremath{\mathrm{f}_{\rm{P}}(\rho)}} 
\newcommand{\fpl}          {\ensuremath{\mathrm{f}_{\rm{P}}(\rm liq.)}} 
\newcommand{\fps}         {\ensuremath{\mathrm{f}_{\rm{P}}(\rstp)}}
\newcommand{\fpa}          {\ensuremath{\mathrm{f}_{\rm{P}_{\rm{ann}}}(\rho)}}
\newcommand{\fpal}         {\mbox{f$_{\rm{P}_{\rm{ann}}}({\rm liq.})$}}
\newcommand{\fpas}         {\mbox{f$_{\rm{P}_{\rm{ann}}}(\rstp)$}}

\newcommand{\vi}[2]{\ensuremath{{\vec{#1}}_{#2}}} 
\newcommand{\vis}[2]{\ensuremath{{\vec{#1}}_{#2}^{\, 2}}}
\newcommand{\vs}[1]{\ensuremath{{\vec{#1}}{}^{\,2}}}
\newcommand{\fracd}[2]{{\displaystyle #1\over \displaystyle #2}}

\def\be{\begin{equation}}                \def\ee{\end{equation}}
\def\ba{\begin{eqnarray}}                \def\ea{\end{eqnarray}}

\catcode`\@=11
\def\eqalign#1{\null\,\vcenter{\openup\jot\m@th
\ialign{\strut\hfil$\displaystyle{##}$&$\displaystyle{{}##}$\hfil
     \crcr#1\crcr}}\,}
\catcode`\@=12
\catcode`\@=11
\def\eqalignt#1{\null\,\vcenter{\openup\jot\m@th
\ialign{\strut\hfil$\displaystyle{##}$&$\displaystyle{{}##}$\hfil
&&%
\strut\hfil$\displaystyle{##}$&$\displaystyle{{}##}$\hfil
     \crcr#1\crcr}}\,}
\catcode`\@=12
\def\sectionb#1{\section{\boldmath #1\unboldmath}}
\def\subsectionb#1{\boldmath\subsection{#1}\unboldmath}
\def\subsubsectionb#1{\boldmath\subsubsection{#1}\unboldmath}
\def\etal{{et al.}}

\def\mevc {\ensuremath{\,{\mathrm{Me\hskip -1pt V}\hskip -2pt/\hskip -1pt c}}}
\def\gevc {\ensuremath{\,{\mathrm{Ge\hskip -1pt V}\hskip -2pt/\hskip -1pt c}}}

\def\text#1{{\mathrm{#1}}}
\newcommand{\tfrac}[2]{\ensuremath{\textstyle\frac{#1}{#2}}}

\newcommand{\ket}[1]{\ensuremath{\vert\,#1\,\rangle}}

\newcommand{\y}{$\surd$}
\newcommand{\bs}{\hspace*{-4pt}}

\newcommand{\BR}{\ensuremath{\mathrm{BR}}}
\newcommand{\AF}{\ensuremath{\mathrm{AF}}}

 {\everymath{\displaystyle\everymath{}}\array}%
 {\endarray}
 \def\DR{{\rm DR}}
\def\Rso{R_\mathrm{so}}
\def\Rsi{R_\mathrm{si}}
\newcommand{\dsr}{dynamical\ selection\ rule}
\newcommand{\dsrs}{dynamical\ selection\ rules}
\newcommand{\rpp}{$\rho\pi$ puzzle}

\newcommand{\captionfonts}{\small\sf}

\makeatletter  
\long\def\@makecaption#1#2{%
  \vskip\abovecaptionskip
  \sbox\@tempboxa{{\captionfonts #1: #2}}%
  \ifdim \wd\@tempboxa >\hsize
    {\captionfonts #1: #2\par}
  \else
    \hbox to\hsize{\hfil\box\@tempboxa\hfil}%
  \fi
  \vskip\belowcaptionskip}
\makeatother   

\begin{document}

\begin{titlepage}
\title{
{\Large\bf THE ANTINUCLEON--NUCLEON INTERACTION}
\\[3pt]
 {\Large\bf AT LOW ENERGY: ANNIHILATION DYNAMICS}   
\\[20pt]}
\author{
Eberhard Klempt\\
{\small Helmholtz-Institut f\"ur Strahlen- und Kernphysik}\\[-2pt]
{\small der Rheinische Friedrich-Wilhelms Universit\"at}\\[-2pt]
{\small Nu\ss allee 14-16, D--53115 Bonn, Germany}\\[10pt]
Chris Batty\\
{\small Rutherford Appleton Laboratory}\\[-2pt]
{\small Chilton, Didcot}\\[-2pt]
{\small Oxon,   OX11 OQX, United Kingdom}\\[10pt]
{Jean-Marc Richard}\\
{\small Laboratoire de Physique Subatomique et Cosmologie}\\[-2pt]
{\small Universit\'e Joseph Fourier--CNRS--IN2P3}\\[-2pt]
{\small 53, avenue des Martyrs, F--38026 Grenoble Cedex, France}\\[10pt]}
\date{\today \\[12pt]}
\vfil
\maketitle
\begin{abstract}
The general properties of antiproton--proton annihilation at rest are presented,
with special focus on the two-meson final states. The data exhibit 
remarkable dynamical selection rules: some allowed annihilation modes are
suppressed by one order of magnitude with respect to modes of comparable phase-space. 
Various phenomenological analyses are reviewed, based on microscopic quark
dynamics or symmetry considerations. The role of initial- and final-state interaction is
also examined.
\end{abstract}
\vspace*{2cm}
\begin{flushleft}
\end{flushleft}
\end{titlepage}
\clearpage\markboth{\sl Annihilation dynamics} {\sl Table of contents}
\tableofcontents
\clearpage\markboth{\sl Annihilation dynamics} {\sl Introduction}
\setcounter{equation}{0}
%
\section{Introduction}\label{se:intro}
\subsection{Annihilation in hadron physics}
\label{intro:sub:anni}
Annihilation is a fascinating process, one of the most interesting in
low-energy hadron physics, in which matter undergoes a transition from
its baryon structure to one consisting solely of mesons. In the early
days of antiproton physics, antinucleon--nucleon (\NNb) annihilation
was considered by analogy with positronium annihilation in QED, and
described as a short-range process mediated by baryon
exchange. Nowadays the quark model offers a drastic alternative, where
the so-called ``annihilation'' does not imply actual annihilation of
all incoming quarks and antiquarks, but simply results from their
rearrangement into quark--antiquark pairs. Were quark rearrangement to
be the leading mechanism, \NNb\ annihilation would be better
considered by analogy with rearrangement collisions in atomic or
molecular physics. Intermediate scenarios are however conceivable,
where some of the incoming quarks and antiquarks annihilate, and new
quark--antiquark pairs are created.

This review is part of a project devoted to strong
interaction physics with low-energy antiprotons, as measured at the
LEAR facility of CERN. A first part \cite{Klempt:2002ap} was devoted
to \NNb\ scattering and to antiprotonic hydrogen and deuterium.  The
present review covers the general properties of annihilation, the
results on two-meson final states and their phenomenological
analysis. A third article will concentrate on meson spectroscopy, as
studied from multimeson final states of annihilation.
\subsection{Historical considerations}\label{intro:sub:hist}
Detailed studies of antiproton--proton annihilation at rest were
carried out in the 1960's, and the results are still significant for
studies of annihilation dynamics and meson spectroscopy. These
experiments were performed at the Brookhaven National Laboratory (BNL)
and at CERN in Geneva by stopping antiprotons in bubble
chambers. Analysis methods and early results were reviewed in detail
by Armenteros and French \cite{Armenteros69}, but many important
results were not included.  Later reviews
\cite{Walcher:1989qp,Amsler:1998up,Dover:1992vj} focused primarily on
new concepts and developments and did not aim at a comprehensive
description of all experimental data.  The physics results
on \pbar N\ annihilation obtained from bubble chambers filled with
\HT\ or \DT\ are included in our report.

In the 70's, \NNb\ physics was dominated by claims for narrow
baryonium states, which were not confirmed by more systematic
searches. The motivation for quasi-nuclear \NNb\ states and for
multiquark states preferentially coupled to \NNb\, and the
experimental results have been reviewed extensively in several
articles \cite{Shapiro:1978wi,Buck:1979rt,Dover:1979zj,Montanet:1980te,Amsler:1991cf}.  

Research on \pbp\ annihilation was resumed in 1983 when LEAR came into
operation. The Asterix collaboration investigated annihilation from
P-states of the \pbp\ atom formed in H$_2$ gas with a 2$\pi$
electronic detector. The focus of the research was {\em dynamical
selection rules} which will be discussed in some detail in
Sec.~\ref{se:dsr}. A broad resonance, called AX(1565), possibly a
quasi-nuclear state, was discovered. The search for narrow states
produced in annihilation at rest continued both at LEAR and KEK,
eventually yielding negative answers.

In more recent years, two $4\pi$ spectrometers, Crystal Barrel (PS197) and
Obelix (PS201), took data on \pbp\ annihilation at LEAR. The
Crystal-Barrel research activity was directed towards annihilation at
rest and in flight.  Obelix investigated antiproton and antineutron \cite{Bressani:2003pv}
interactions at rest and with very low momenta. Nuclear physics was
also an important part of the Obelix program. 

The experimental progress was accompanied by active groups of
theoreticians trying to understand the basic mechanisms responsible
for annihilation.  From a theoretical point of view annihilation is a
very complicated process which is likely driven by both the underlying
quark dynamics and by conventional hadronic interactions.  If for
instance, the \NNb\ potential is attractive in one partial wave, and
repulsive in another, one expects annihilation from the former to be
enhanced, and annihilation from the latter to be suppressed. Similar
remarks hold for final state interactions with, in addition, the
possibility of interferences between, for instance, primary $\rho$
mesons formed by \qqb\ pairs and $\rho$ mesons built from $\pi\pi$
final-state interactions. An accurate description for all annihilation
rates seems therefore to be unlikely. It is hoped, however, that the
leading mechanisms of annihilation will be identified, in particular
to explain the so-called {\em dynamical selection rules}, the
observation that some annihilation modes are strongly suppressed in
certain partial waves, while still being allowed by energy and
quantum-number conservation.
\subsection{Outline}\label{intro:sub:outl}
This review begins, in Sec.~\ref{se:exp}, with a presentation of the
beams and detector facilities used to measure annihilation
properties. In Sec.~\ref{se:mes}, we briefly summarise the properties
of the mesons seen in annihilation experiments.  Kinematics and
conservation laws are reviewed in Sec.~\ref{se:kin}.  The main
features of annihilation, as seen in various experiments, are
presented in Sec.~\ref{se:glob}, while Sec.~\ref{se:tm} is devoted to
a thorough review of the rates into various two-meson final
states. The dynamical selection rules are presented and discussed in
Sec.~\ref{se:dsr}. Section~\ref{se:phe} contains a critical survey
of various approaches to annihilation mechanisms, and an analysis of
what is learned from the systematics of two-body branching ratios.
Some conclusions are presented in Sec.~\ref{se:conc}.

\subsection{A guide to the literature}\label{intro:sub:guide}
The physics mediated by antiprotons has been presented at many
Conferences, in particular the NAN conferences, the LEAR Workshops,
and the LEAP conferences resulting from the merging of these two
series, as well as at some Schools
\cite{NAN72,NAN74,NAN76,NAN78,NAN80,NAN82,NAN84,NAN86,NAN91,NAN93,NAN95,LEAP90,LEAP92,LEAP94,LEAP96,LEAP98,LEAP00,LEAP03,%
LEAR79,LEAR82,LEAR85,LEAR87,ERICE86,ERICE87,ERICE88,ERICE90,%
SUPERLEAR87, SUPERLEAR91, Turin89}.

The early review by Armenteros and French \cite{Armenteros69} remains
a reference for early annihilation data. Before the completion of LEAR
measurements and the final analysis of the data, important review
articles became available; some concerning general aspects of LEAR
physics \cite{Walcher:1989qp,Amsler:1991cf}, whilst others specialised
more on the annihilation process \cite{Amsler:1998up,Dover:1992vj}. Antineutron physics, including antineutron annihilation, is reviewed in~\cite{Bressani:2003pv}. A review devoted to annihilation in flight appeared recently~\cite{Bugg:2004xu}.

\clearpage\markboth{\sl Annihilation dynamics} {\sl Beams and experiments}
\setcounter{equation}{0}
 

\section{Beams and experiments}\label{se:exp}
\FloatBarrier\subsection{Early experiments at CERN and BNL}\label{exp:sub:early}
Following the discovery of the antiproton in 1955, \pbar\ beams were
rapidly developed and a first survey of antiproton annihilation on
protons or neutrons was possible, as early as in the 1960's, by stopping
antiprotons in hydrogen- and deuterium-filled bubble chambers. These
experiments demonstrated that \pbp\ annihilation is a powerful tool to
discover meson resonances, even though  only
limited statistics were achieved. Some of the early results are still
important, and it seems appropriate to include a short discussion
as to how the data were obtained.

Two main experiments were carried out at that time: the first one at Brookhaven
by a group from Columbia University and the other at CERN by a
CERN--Coll\`ege de France collaboration. The experimental conditions
were closely similar and it is sufficient to discuss just one of
them.

The CERN bubble chamber,  built at Saclay,  had an illuminated
volume of 80\,cm in length, and of 30\,cm in height and
depth. Antiprotons from a separated antiproton beam of momentum
700\mevc\ were moderated in a Cu degrader and stopped in the
target. The chamber was situated in a magnetic field of $2.1\,$T.  Due
to the momentum spread in the incident beam and multiple scattering in
the degrader and target, the stopping distribution was rather wide.  A
cut on a minimum track length of 5\,cm guaranteed a minimum momentum
resolution; the average track length was 16\,cm. The intensity of the
antiproton beam was adjusted to allow for several (3 or 4)
annihilation events for each bubble chamber expansion. Three stereoscopic
pictures were taken of each expansion to enable a three-dimensional
reconstruction of the tracks.

Scanning  the films and reconstructing events was a major enterprise.
The spatial coordinates of four points for each track were measured
from the films, with a precision of 80$\,\mu$m. From the coordinates
the charged--particle momenta were determined. We estimate the
momentum resolution for 928\mevc\ pions from the reaction $\pbp
\to\pi^+\pi^-$ to be 25\mevc.

A total of $1.6 \times 10^6$ events were recorded at CERN, $7.5
\times10^5$ at BNL. These numbers exceeded the scanning capabilities
available at that time, and only a fraction of the data was analysed:
about 80,000 events at CERN and 45,000 events at BNL. From the
momentum of the incoming antiprotons their range was estimated and
compared to the true range; thus contamination due to in flight
annihilation could be avoided, or at least reduced.

The Brookhaven results are documented in  
Refs.~\cite{Baltay65a,Baltay65b,Barash65a,Baltay65c,Baltay65d,Barash65b,%
Baltay66,Barash66,Baltay67}.             
At CERN, more aspects of the annihilation process were investigated,
leading to a larger number of publications.  An incomplete list of
publications in refereed journals includes
\cite{Chadwick:1963ab,Armenteros:1964ac,Boserup:1964ad,%
Bettini:1965ae,%
Armenteros:1965af,Armenteros:1965ag,Andlau:1965ah,Ndili:1965ai,%
Conforto:1967aj,Astier:1967ak,Baillon:1967al,%
Foster:1968am,Foster:1968di,James:1968an,Fowler:1968ao,%
Armenteros69,Astier:1969ap,Bizzarri:1968aq,Bizzarri:1970ta,%
Aguilar:1969ar,Diaz:1970as,Bloch:1970at,Bizzarri:1971ax,%
Chung:1971ay,Chung:1971az,Espigat:1972aa,Frenkiel:1972ab}
for H$_2$-filled and
\cite{Bettini:1966ba,Bettini:1967bb,Bettini:1969bc,Bettini69a,%
Bizzarri68,Bizzarri:1974,Bizzarri:1974xf,Bizzarri:1974hr}
D$_2$-filled chambers.  Further publications discussed the
interpretation of these results.
\par
In the 1970's, a first set of counter experiments were performed at BNL to study 
$\gamma$-rays from antiproton annihilation. Antiprotons from a 
separated beam were stopped in a liquid H$_2$ or D$_2$ target.
Photons were detected by their conversion in Cu(Pb) plates sandwiched between 
scintillation counters. In some experiments a NaI detector, surrounded by scintillation 
counters, was used to measure $\gamma$-rays with better resolution. 
Data from the Rome--Syracuse collaboration, taken with the D$_2$-filled BNL bubble chamber,
were analysed in parallel. Results can be found 
in~\cite{Bizzarri:Anninos:1968pa,Bizzarri:1970yw,Kalogeropoulos:1970pc,%
Gray:1971pf,Devons:1971rn,Papadopoulou:1973kc,Gray:1973dh,%
Kalogeropoulos:1975hv,Kalogeropoulos:1976ws,%
Kalogeropoulos:1981ua,Kalogeropoulos:1982qv,Gray:1983cw,Brando:1984mi,Bridges:1986kw,Bridges:1986kx,Bridges:1986vc,Daftari:1987td}.

\FloatBarrier\subsection{Experiments at KEK}\label{exp:sub:KEK}
This experiment was designed to search for narrow lines in the
momentum distributions of $\pi^0$ and $\eta$ from
\pbp\ annihilation. The initial aim 
was to find narrow multiquark or quasi-nuclear bound states
\cite{Chiba:1986cf,Chiba:1987ge,Chiba:1991ag,Chiba:1999cy}. 
Later, frequencies for annihilation into
two narrow mesons were determined with both 
\HT\ \cite{Chiba:1988cg,Chiba:1989yw} and \DT\ \cite{Chiba:2000ev} targets. 

\begin{figure}[!h]
\includegraphics[width=.8\textwidth]{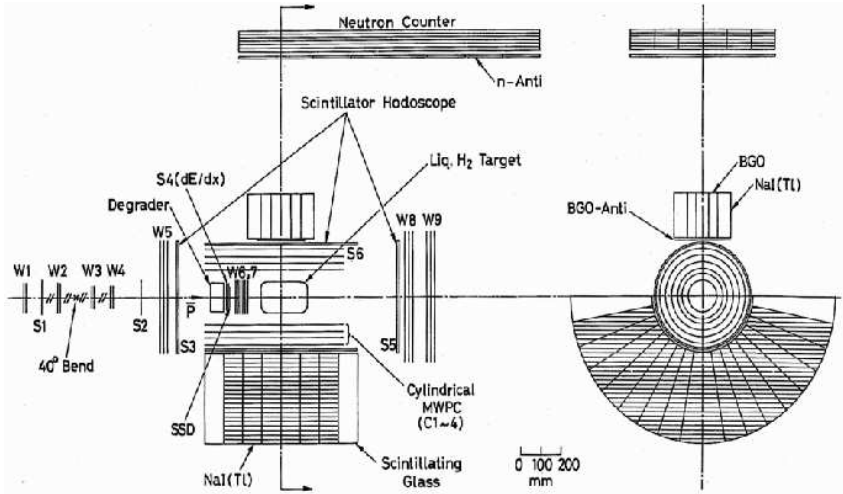}
 \caption{\label{exp:fig:KEK}%
Side and end view of the KEK detector.}
\end{figure}

The layout of the experimental set-up is shown in
Fig. \ref{exp:fig:KEK}.  A full description of the detector can be
found in
\cite{Chiba:1986cf,Chiba:1987ge}. Antiprotons at 580\mevc, produced at 
the KEK 12\,GeV proton synchrotron, were degraded in a graphite slab 
and stopped in a liquid \HT\ target of 14\,cm diameter and 
23\,cm in length. The \pbar\ beam used double-stage mass separation 
and a contamination ratio of $\Pe\mu\pi/\pbar\simeq8$ 
was obtained with a typical 
stopping intensity of 270\,\pbar/synchrotron pulse. The charged 
particles produced in \pbp\ annihilation were detected with 
scintillation counter hodoscopes and tracked with cylindrical and 
planar multiwire proportional chambers whose total coverage was 
$93\% \times 4\pi$ sr. Photons were measured with a calorimeter 
consisting of 96 NaI(Tl) crystals surrounded by 48 scintillating 
glass modules and assembled into a half barrel \cite{Kobayashi:1986vh}, see Fig.\ref{exp:fig:KEK}.
The geometrical acceptance for \piz\ increased from 10.5\% at a 
\piz\ energy of 500\,MeV to 14.5\% at 900\,MeV. The overall energy 
resolution at FWHM for photons was approximately 
$\Delta E_{\gamma}/E_{\gamma} = 6.2\%/(E_{\gamma} \mbox{ in \,GeV})^{1/4}$,
for energies above 80\,MeV.

Events were recorded when the detector signalled that a slow antiproton
was incident on the liquid \HT\ target and one or two photons were
measured in the NaI detector. A fast cluster counting logic counted
the multiplicities of charged and neutral clusters separately, and if
they satisfied preselected criteria the data was recorded. In some of
the later experiments \cite{Chiba:1989yw} an additional small $(1.3\%
\times 4\pi$ sr.) BGO detector surrounded by NaI modules was used. No
cluster counting logic was used in this case and the energy resolution
(FWHM) was estimated to be $\Delta E_{\gamma}/E_{\gamma} =
6.8\%/(E_{\gamma}\ \mbox{in \,GeV})^{1/4}$.  As the NaI photon
spectrometer has less than $2\pi$ acceptance,  for measurements of
two-body branching ratios it is not possible to detect both 
mesons. In this case the existence of the
second meson and its mass are deduced from the
inclusive energy spectrum recorded for a single \piz\ or $\eta$.
\FloatBarrier\subsection{Cooled antiproton beams}\label{exp:sub:cool}
Early experiments with electronic detection techniques, including
those carried out at KEK, used partially separated secondary
antiproton beams produced from an external target. These beams were
characterised by a relatively low rate of stopped antiprotons over a
large volume and with a contamination of unwanted particles
$\Pe\mu\pi/\pbar\simeq 100$ in the earliest experiments to
$\Pe\mu\pi/\pbar\simeq 8$ in the more recent ones. This situation was
transformed by the availability of cooled antiproton beams at CERN,
together with the construction and operation of the LEAR facility.

A description of the cooled antiproton beams used at CERN has been
given in a previous review article \cite{Klempt:2002ap}.  For a proton
beam of 23\gevc\ incident on a Be target, antiprotons are produced with a
broad maximum in momentum at 3.5\gevc. The use of cooling allows these
antiprotons to be decelerated to low momenta whilst keeping the same
flux. Additionally, cooling gives the antiproton beams a small size
and a reduced momentum.

The LEAR facility was constructed at CERN to handle pure antiproton
beams in the momentum range from 105\mevc\ to 2000\mevc\ with small
physical size and a typical momentum spread of $\Delta
p/p\sim0.1\%$. This small momentum spread for low momentum protons
gave a very small stopping region in liquid \HT\ and \DT\ targets and
also enabled gas targets to be used. The use of gas targets is
particularly important since the fraction of P-state annihilation is
considerably increased in gaseous \HT\ targets due to the reduced
effect of Stark mixing.  An ultra-slow extraction system enabled
essentially DC beams to be produced, with spills typically lasting
several hours. Typical beam intensities were in the range $10^4$ to
$10^6$ \pbar /sec. The beam purity was 100\%.

The LEAR project was approved by CERN in 1980, and in July 1983 the
first antiproton beams were delivered to users. After a break in
1987, to construct a new Antiproton Collector (ACOL) which resulted in
a flux gain of a factor of 10, the facility was operated until the end of
1996, when it was closed for financial reasons.  The Asterix, Obelix
and Crystal Barrel (CBAR) experiments were all carried out at LEAR;
Asterix in the first, Obelix and Crystal Barrel in the second phase.

\FloatBarrier\subsection{Detectors at LEAR}\label{exp:sub:LEAR}

\subsubsection{PS 171: The Asterix experiment}\label{exp:sub:AST}
Liquid targets were used in both the bubble chamber and counter
experiments described earlier (Sec. \ref{exp:sub:early} and
\ref{exp:sub:KEK}).  In liquid H$_2$ or D$_2$, annihilation occurs at
rest and is preceded by capture of an antiproton by a hydrogen or
deuterium atom. Collisions between the protonium atom and H$_2$
molecules induce transitions from high orbital angular momentum states
via Stark mixing; and this mixing is fast enough to ensure dominant
capture from S-wave orbitals.  In H$_2$ gas, the collision frequency
is reduced and P-wave annihilation makes significantly larger
contributions. In particular at very low target pressures the P-wave
fractional contribution is very large. Alternatively, rather pure
samples of P-wave annihilation can also be studied by coincident
detection of X-rays emitted in the atomic cascade of the \ppb\ system
(which feed mostly the 2P level).

\begin{figure*}[!htpc]
\includegraphics[width=.9\textwidth]{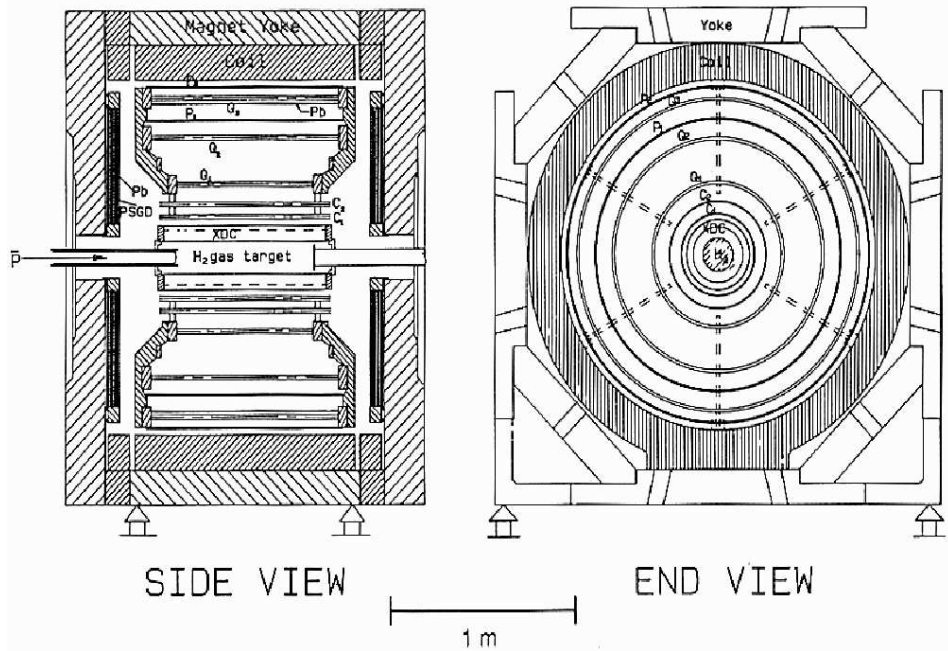}
 \caption{ \label{exp:fig:AST}%
Side and front view of the PS172 Asterix detector.} \end{figure*}

The Asterix experiment was designed to study \ppb\ annihilation from
P-wave orbitals by stopping antiprotons in H$_2$ gas at room
temperature and pressure and observing the coincident X-ray
spectrum. The detector, shown in Fig.~\ref{exp:fig:AST}, consisted of
the following main components:
\begin{enumerate}\itemsep -1mm
\item A gas target of 45\,cm length and 14\,cm in diameter contained
the full \pbar\ stop distribution for an antiproton beams at 105\mevc.  
\item The target was surrounded by a X-ray drift chamber (also used to
improve the tracking capability and for particle identification via 
dE/dx). The energy resolution of the detector for 8 keV X-rays was
about 20\%. Pions and kaons could be separated up to 400\mevc. 
The target and X-ray drift chamber were separated by a 6 $\mu$m 
aluminised mylar foil to guarantee gas tightness and good
X-ray transmission even at low energies.
\item Charged particles were tracked in a set of seven 
multi-wire proportional chambers, partly with cathode readout to 
provide spatial resolution along the wires. The momentum resolution  for $\pbp\to\pip\pim$ events 
at 928\mevc\ was 3\%. 
\item A one-radiation-length lead foil in front of the outer 
chambers permitted reconstruction of the impact points of photons. 
\item Two end-cap detectors with three wire planes and cathode readout 
on both sides gave large solid-angle coverage. A lead foil
was mounted behind the first chamber. The end cap detectors were used 
to identify $\gamma$'s but not for reconstruction of charged tracks. 
\item The assembly was situated in a homogeneous magnetic field of 
0.8\,Tesla.
\end{enumerate}

With the experimental resolution of the detector, there was nearly no
background for fully-constrained final states and up to 14\% for final
states with one missing \piz .

The main data sets taken with the Asterix detector consisted of
$1.38\times 10^6$ events with two long tracks (passing at least the
first five chambers) without triggering on X-rays, $2.13\times 10^6$
events with two long tracks with a trigger on X-rays, and $1.89\times
10^6$ events with four long tracks and with the X-ray trigger. The
``long-track'' criterion guaranteed that the particles reached the
outermost chambers and gave optimum momentum resolution. The X-ray
enhancing trigger had an efficiency of 25\%; one quarter of the
triggered events had --- after all cuts --- an identified low-energy
X-ray. There was a contamination from Bremsstrahlung X-rays of about
15\% in the X-ray data sample.

The detector is fully described in \cite{Ahmad:1990gx}. Physics results 
were published in 
\cite{Ahmad:1984bp,Ahmad:1985nx,Doser:1988yi,Doser:1988fw,%
Ziegler:1988bp,Reifenrother:1988uc,Schafer:1989wq,Duch:1989sx,%
May:1989sw,Riedlberger:1989kn,May:1990ju,May:1990jv,%
Klempt:1990vc,Reifenroether:1991ik,Weidenauer:1990fy,Weidenauer:1993mv}.
\subsubsection{PS 201: The Obelix experiment}\label{exp:sub:OBX}
\begin{figure*}[!htpc]
\includegraphics[width=.9\textwidth]{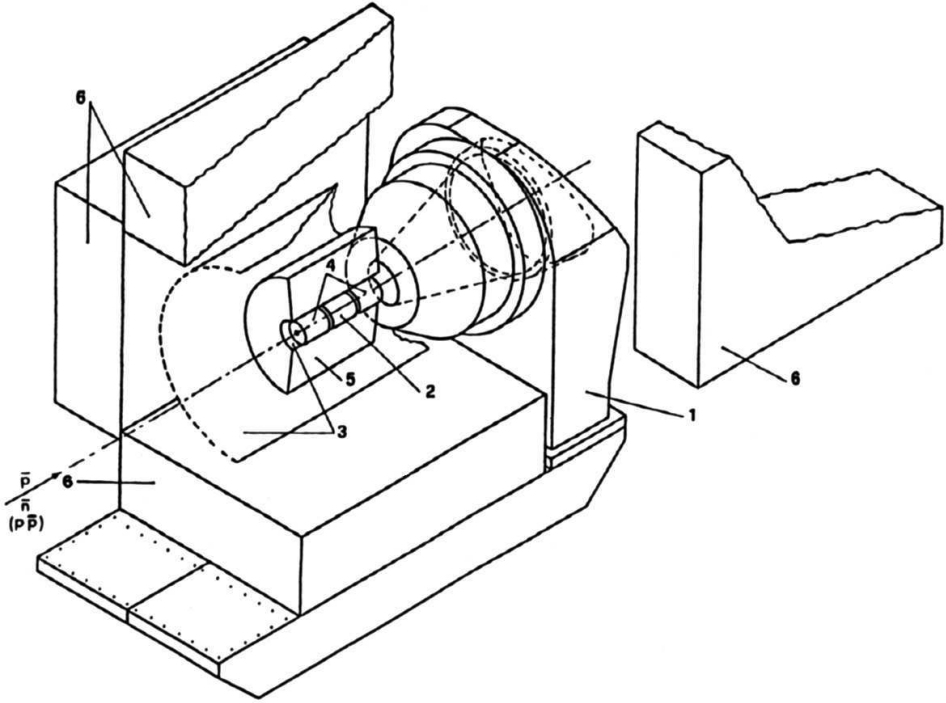}
\caption{ \label{exp:fig:OBX}%
Schematic view of the Obelix experiment set-up.
The numbers indicate the main components of the
apparatus: the Open Axial Field magnet (1),
the SPC (2, 4), the TOF (3), the JDC (5), the HARGD (6).}
\end{figure*}

The layout of the Obelix spectrometer is shown in
Fig.~\ref{exp:fig:OBX}.  A full description of the detector can be
found in \cite{Affatato:1993mw,Adamo:1992bb}.
It consists of four sub-detectors
arranged inside and around the open-axial field magnet which had
previously been used for experiments at the ISR. The magnet provides a
field of 0.5\,T in an open volume of about 3 m$^3$. The subdetectors
are:
\begin{enumerate}\itemsep -1mm
\item 
A spiral projection chamber (SPC): an imaging vertex detector with 
three-dimensional readout for charged tracks and X-ray detection.
This detector allowed data to be taken with a large fraction of P-wave
annihilation, and to measure angular correlations between X-rays from
the \pbp\ atomic cascade and annihilation products.
\item 
A time-of-flight (TOF) system: two coaxial barrels of plastic 
scintillators consisting of 30 (84) slabs positioned at a distance of
18\,cm (136\,cm) from the beam axis; a time resolution of 800\,ps FWHM
is achieved.
\item 
A jet drift chamber (JDC) for tracking and particle identification 
by dE/dx measurement with 3280 wires and flash-analog-to-digital
readout.  The chamber was split into two half-cylinders (160\,cm in
diameter, 140\,cm long). The intrinsic spatial resolution was
$\sigma_z = 12\,$mm, $\sigma_{r\phi} = 200\,\mu$m; the momentum
resolution for monoenergetic pions (with 928\mevc) from the reaction
$\pbp\to\pi^+\pi^-$ was found to be 3.5\%.
\item 
A high-angular-resolution gamma detector (HARGD)
\cite{Affatato:1993mw}. The calorimeter consisted of four modules
made of layers of $3 \times 4\,\mathrm{m}^2$ lead converter foils with
planes of limited streamer tubes as the active elements. Twenty
converter layers, each 3\,mm thick, were used corresponding to a total
depth of about $10$ radiation lengths. Due to their excellent spatial
resolution, good energy resolution in the reconstruction of final
states is obtained:
\piz\ are reconstructed with a mass resolution of $\sigma_{\piz} =
10$\,MeV and a momentum-dependent efficiency of 15 to 25\%.
\end{enumerate}

The detector system allowed a variety of targets to be used: a liquid
H$_2$ target, a gaseous H$_2$ target at room temperature and pressure,
also a target at low pressures (down to 30 mbar). The wide range of
target densities could be used to study in detail the influence of the
atomic cascade on the annihilation process. The H$_2$ could also be
replaced by D$_2$. A further special feature of the detector was the
possibility to study antineutron interactions. The $\nbar$ beam was
produced by charge exchange in a liquid H$_2$ target (positioned 2\,m
upstream of centre of the main detector).  The intensity of the
collimated beam was about  $40\,\nbar/10^6 \pbar$ of which 
about 30\% interact in the central target. The $\nbar$ beam
intensity was monitored by a downstream $\nbar$ detector.

The Obelix Collaboration had a broad program of experiments covering
atomic, nuclear and particle physics \cite{Adamo:1992bb}.  The main
results can be found in
\cite{Bressani:2003pv,Adamo:1992bb,Adamo:1992ci,Ableev:1993uz,Agnello:1994kg,Ableev:1994uy,Ableev:1994vs,%
Ableev:1994bg,Bertin:1995fx,Ableev:1995uu,Ableev:1995aq,Bertin:1996nq,%
Bertin:1996pb,Bertin:1996rr,Bertin:1997dw,%
Bertin:1997jg,Bertin:1997kh,Bertin:1997vf,%
Bertin:1997wc,Bertin:1997zu,Alberico:1998uj,Alberico:1998fr,%
Bertin:1998sb,Bertin:1998hu,%
Cicalo:1999sn,Denisov:1999tr,Filippi:1999xk,%
Filippi:1999ym,Filippi:2000is,%
Bendiscioli:2001sv,Gorchakov:2002dm,Bargiotti:2002mu,
Nichitiu:2002cj,Bargiotti:2003ev,Bargiotti:2003bv,Salvini:2004gz}.
\subsubsection{PS 197: The Crystal Barrel experiment}\label{exp:sub:CBAR}
The main objective of the Crystal Barrel experiment was the study of
meson spectroscopy and in particular the search for glueballs (gg) and
hybrid (g\qqb) mesons produced in
\pbp\ and \pbd\ annihilation at rest and in flight. Other objectives 
were the study of \pbp\ and \pbd\ annihilation dynamics and 
the study of radiative and rare meson decays. A particular feature 
of the experiment was its photon detection over a large solid angle 
with good energy resolution.
 Physics results are published in 
\cite{Amsler:1998up,Bugg:2004xu,%
Aker:1991bk,Amsler:1992rx,Amsler:1992wm,Amsler:1992ku,%
Amsler:1993kg,%
Amsler:1993pr,Amsler:1993jz,Amsler:1993xd,%
Amsler:1994gt,Amsler:1994aa,Amsler:1994ah,Amsler:1995aa,Amsler:1995sy,%
Amsler:1995up,Amsler:1995nw,Amsler:1995bf,Amsler:1995bz,%
Amsler:1995ct,Amsler:1995wz,Amsler:1996hb,%
Adomeit:1996nr,Abele:1996fr,Abele:1996nn,%
Abele:1997wh,Abele:1997qy,Abele:1997yi,Abele:1997dz,%
Abele:1997wg,Abele:1997vu,Abele:1997vv,%
Abele:1998gn,Abele:1998yi,Abele:1998yj,Abele:1998qd,McCrady:1998th,%
Abele:1998kv,Baker:1999ac,Abele:1999tf,Abele:1999fw,Anisovich:1999jw,%
Anisovich:1999jx,Abele:1999en,Abele:1999ac,Abele:1999uv,Abele:1999pf,%
Abele:2000qq,Abele:2000xt,Abele:2001ek,%
Abele:2001js,Amsler:2001fh,Amsler:2002qq,Amsler:2002ua,Amsler:2003bq,%
Amsler:2004rd,Amsler:2004kn}

\begin{figure*}[!h]
\includegraphics{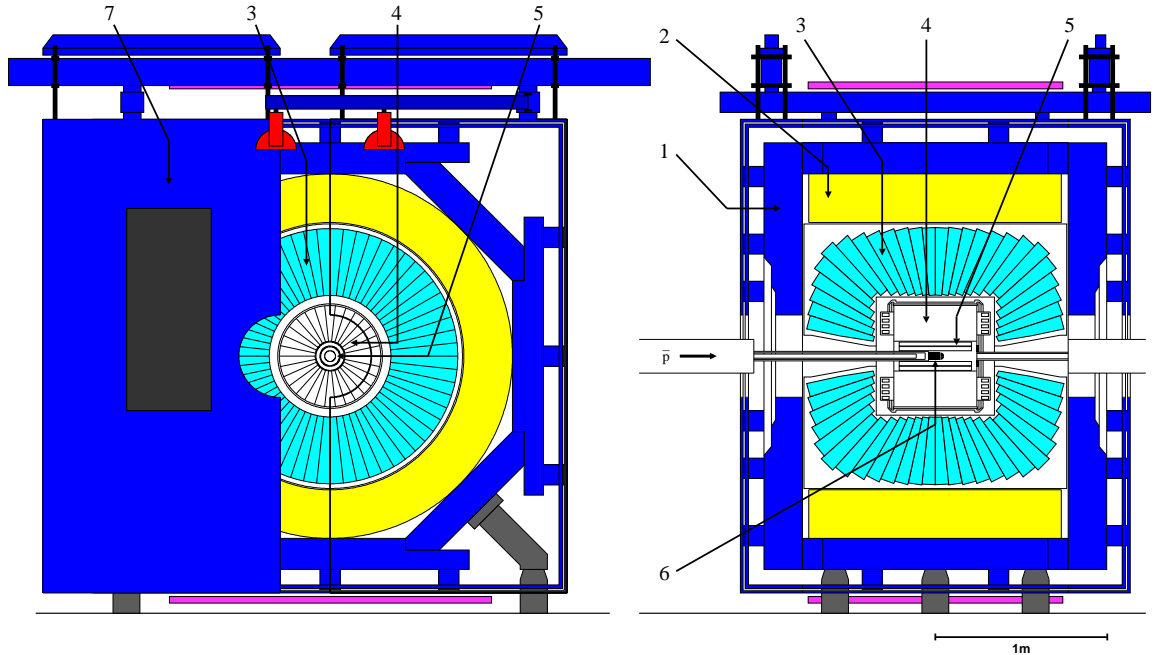}
\caption{\label{exp:fig:CBAR}%
Overall layout of the Crystal Barrel detector showing 
(1) magnet yoke, (2) magnet coils, (3) CsI barrel, 
(4) jet drift chamber, (5) proportional chamber, (6) liquid hydrogen 
target, (7) one half of endplate. Left - longitudinal cross section; 
Right- transverse view.}
\end{figure*}

The layout of the Crystal Barrel spectrometer is shown in
Fig.~\ref{exp:fig:CBAR}.  A detailed description of the apparatus, as
used for early data-taking (1989 onwards), is given in
\cite{Aker:1992ny}.  To study annihilation at rest, a beam of
200\mevc\ antiprotons, extracted from LEAR, was stopped in a 4\,cm
long liquid hydrogen target at the centre of the detector. The whole
detector was situated in a 1.5\,T solenoidal magnet with the incident
antiproton beam direction along its axis.  The target was surrounded
by a pair of multiwire proportional chambers (PWC's) and a cylindrical
jet drift chamber (JDC). The JDC had 30 sectors with each sector
having 23 sense wires at radial distances between 63\,mm and
239\,mm. The position resolution in the plane transverse to the beam
axis was $\sigma =125\,\mu$m. The coordinate along the wire was
determined by charge division with a resolution of $\sigma=8$\,mm. This
gave a momentum resolution for pions of $\sigma/p \simeq 2\%$ at
200\mevc, rising to 7\% at 1\gevc\ for those tracks that tracked all
layers of the JDC. The JDC also provided $\pi$/K separation below
500\mevc\ by ionisation sampling.

The JDC was surrounded by a barrel shaped calorimeter consisting of
1380 CsI(Tl) crystals in a pointing geometry. The CsI calorimeter
covered the polar angles between $12^\circ$ and $168^\circ$ with full
coverage in azimuth. The overall acceptance for shower detection was
0.95 $\times\ 4\pi$ sr.  Typical photon energy resolutions for energy
$E$ (in GeV) were $\sigma_{E}/E = 2.5\%/E^{1/4}$, and
$\sigma_{\phi,\theta} = 1.2^\circ$ in both polar and azimuthal angles.
The mass resolution was $\sigma\ = 10$\,MeV for \piz\ and 17\,MeV for
$\eta \rightarrow 2\gamma$.

In 1995 the PWC's were replaced by a microstrip vertex detector (SVTX)
consisting of 15 single-sided silicon detectors, each having 128
strips with a pitch of $50\,\mu$m running parallel to the beam axis
\cite{Regenfus:1997eq,Doser:1998ge}. (See \cite[Fig.~1]{Doser:1998ge} for an overall view of the detector.) As well as giving improved identification of secondary vertices, 
this detector provided better vertex resolution in $r,\phi$ and improved momentum 
determination with a resolution $\Delta p/p$ for charged tracks of 3.4\% at 0.8\gevc\
and 4.2\% at 1.0\gevc. 

To study annihilation in hydrogen gas, the liquid target was replaced
by a 12 \,cm long Mylar vessel with $230\mu$m thick walls and a
$195\mu$m thick entrance window, containing hydrogen gas at room
temperature and 12\,bar pressure.  A $55\mu$m thick Si detector was
used to count the incident 105\mevc\ antiproton beam.

A particular feature of the detector system was a multi-level trigger
\cite{Aker:1992ny} on charged and neutral multiplicities and on 
invariant mass combinations of the neutral secondary particles. 
This allowed the suppression of well-known channels and the 
enhancement of rare channels of specific interest. The PWC/SVTX and the 
inner layers of the jet drift chamber determined the charged 
multiplicity of the final state. Events with long tracks could be 
selected to give optimum momentum resolution by 
counting the charged multiplicity in the outer layers of the JDC. 
A hardwired processor determined the cluster multiplicity in the 
CsI barrel, whilst a software trigger, which was an integral part 
of the calorimeter read out system, allowed a trigger on the 
total deposited energy in the barrel or on the \piz\ or $\eta$ 
multiplicity. 

Typical incident beam intensities were $\rm 10^4\,\pbar/sec$ at 200\mevc\
for stopping in liquid \HT\ or 105\mevc\ when a 12 bar gas target was
used. For experiments to study interactions in flight, larger
intensities in the range from $10^5$ to $\rm 10^6\,\pbar/sec$ were required at
beam momenta in the range 600 to 1940\mevc.

A convenient summary of the data taken by the experiment, both at rest
and in flight, for liquid \HT\, liquid \DT\ and gaseous
\HT\ targets has been given by 
Amsler (see \cite[Table 1]{Amsler:1998up}). Typical data sets contain 
$10^6$ to $2\times 10^7$ events.
\FloatBarrier\subsection{Future experiments}\label{exp:sub:future}
With the closure of the LEAR facility in 1996, an era of intensive
experimental study of low and medium energy antiproton annihilations
came to an end. CERN has continued its involvement in the production
of \pbar\ beams with the construction of the AD (Antiproton
Decelerator) \cite{Hemery:1998ps}.  This provides \pbar\ beams with
momentum from 100 to 300 \mevc\ but without slow extraction. Slow
extraction is not possible without making major modifications to the
AD \cite{Erikson:2000he,Belochitskii04}.  The space for experiments
\cite{Giovannozzi:1998qt} is very limited and the experimental program
is solely devoted to the production of trapped antihydrogen and
studies of the formation and cascade in light antiprotonic atoms.

For many years Fermilab has had the world's most intense antiproton
source. However the opportunities for medium energy \pbar\ physics
have been very limited and experiments have focused on the observation
and measurement of charmonium states. There has been no low-energy
program. The possibility of building a new antiproton
facility which could decelerate \pbar\ to below 2\gevc\ for injection
into a new storage ring has been discussed \cite{Kaplan:2000nx}. The
storage ring would be equipped with RF to decelerate \pbar\ down to
the hundreds of \mevc\ range. There are however, as yet, no firm plans
to build such a facility.

The design of the Japan Proton Accelerator Research Complex (J-PARC,
\cite{jkj2003}) is well under way. The J-PARC project includes studies
of particle and nuclear physics, materials science, life sciences and
nuclear technology. The accelerator complex consists of an injection
linac, a 3\,GeV synchrotron and a 50\,GeV synchrotron
\cite{Nagamiya:2004ei}. At this latter machine 
nuclear and particle physics experiments using neutrinos, antiprotons, 
kaons, hyperons and the primary proton beam are planned. At one stage 
it was hoped that the LEAR facility could be moved to J-PARC. However 
the LEAR ring is now required for the injection of heavy ions into 
the CERN LHC. It also seems likely that neutrino physics and the study 
of rare kaon decays will be topics for the first experiments at JFK
and the construction of a dedicated low/medium energy antiproton facility
is now some way off.


At the GSI laboratory in Darmstadt the construction of a new facility
is planned and conditionally approved~\cite{gsi2003}, the International
Facility for Antiproton and Ion Research, FAIR. 
A conceptual design report~\cite{Gutbrod:2001xg} outlines 
a wide physics program 
and the envisaged accelerator complex. In particular a High Energy
Storage Ring (HESR, \cite{Koch:2004db}) 
will provide stored antiproton 
beams~\cite{Kienle:1999ud,Metag:2001gm,Henning:2004eh}
in the range 3 to 15 \gevc\ with very good momentum resolution 
($\Delta p/p \approx 10^{-5}$). The charm region is thus
accessible with high rates and excellent resolution. 
A target inside the storage ring 
will be used, together with a large multi-purpose detector for
neutral and charged particles with good particle 
identification~\cite{Ritman:2004ef}. 
Production and use of polarised antiprotons  is a
further future option for studying spin aspects of antiproton--proton scattering and annihilation. 
 It is planned to broaden the program by including
a \underline{l}ow--energy component, F\underline{L}AIR
for the study of antimatter and highly-charged ions at low energies or
nearly at rest.

\clearpage\markboth{\sl Annihilation dynamics}{\sl Mesons and their quantum numbers}
\setcounter{equation}{0}
%
\newcommand{\nonett}[9]
{
\setlength{\unitlength}{1mm}
\begin{picture}(150.00,90.00)
\put(10.00,45.00){\vector(1,0){70.00}}
\put(45.00,10.00){\vector(0,1){70.00}}
\put(110.00,45.00){\vector(1,0){30.00}}
\put(125.00,30.00){\vector(0,1){30.00}}
\put(82.50,42.50){\makebox(5.00,5.00){$I_3$}}
\put(142.50,42.50){\makebox(5.00,5.00){$I_3$}}
\put(127.50,60.00){\makebox(5.00,5.00){S\quad\ Singlet }}
\put(47.50,80.00){\makebox(5.00,5.00){S\quad\ Octet }}
\put(45.00,45.00){\circle*{2.00}}
\put(70.00,45.00){\circle*{2.00}}
\put(20.00,45.00){\circle*{2.00}}
\put(32.50,70.00){\circle*{2.00}}
\put(57.50,70.00){\circle*{2.00}}
\put(32.50,20.00){\circle*{2.00}}
\put(57.50,20.00){\circle*{2.00}}
\put(125.00,45.00){\circle*{2.00}}
\put(45.00,45.00){\circle{0.00}}
\put(45.00,45.00){\circle{5.00}}
\put(32.50,70.00){\line(1,0){25.00}}
\put(57.50,70.00){\line(1,-2){12.50}}
\put(70.00,45.00){\line(-1,-2){12.50}}
\put(57.50,20.00){\line(-1,0){25.00}}
\put(32.50,20.00){\line(-1,2){12.50}}
\put(20.00,45.00){\line(1,2){12.50}}
\put(60.00,70.00){\makebox(15.00,5.00)[l]{#2 $(\bar{s} u)$}}
\put(15.00,70.00){\makebox(15.00,5.00)[r]{#1 $(\bar{s}d)$}}
\put(15.00,15.00){\makebox(15.00,5.00)[r]{#3 $(\bar{u}s)$}}
\put(6.75,37.50){\makebox(13.25,5.00)[r]{#5 $(\bar{u}d)$}}
\put(60.00,15.00){\makebox(15.00,5.00)[l]{#4 $(\bar{d}s)$}}
\put(70.00,37.50){\makebox(13.75,5.00)[l]{#6 $(\bar{d}u)$}}
\put(47.50,37.50){\makebox(12.50,5.00)[l]{#7}}
\put(47.50,47.50){\makebox(12.50,5.00)[l]{#8}}
\put(127.50,47.50){\makebox(12.50,5.00)[l]{#9}}
\end{picture}
}

\section{Mesons and their quantum numbers}\label{se:mes}
\subsectionb{\protect$q\bar q$ mesons and beyond}
\label{mes:sub:qqb}
Since the annihilation process leads to production of mesons, it is 
useful to recall some basic definitions and properties of the meson
spectrum.

Mesons are strongly-interacting particles with integer spin.  The well
established mesons have flavour structure and other quantum numbers
which allows us to describe them as bound states of a quark and an
antiquark. These valence quarks which describe the flavour content are
surrounded by many gluons and quark--antiquark pairs. Other forms of
mesons are also predicted to exist: glueballs should have no
valence quarks at all; in hybrids, the hypothetical gluon string
transmitting the colour forces between quark and antiquark is supposed
to be dynamically excited; and multiquark states are predicted,
described either as states of $(qq\bar q\bar q)$ or higher
valence-quark structure, or as meson--meson or baryon--antibaryon
bound states or resonances. These unconventional states are presently
searched for intensively; they are however not the subject of this
review.

Quarks have spin $s=1/2$ and baryon number $B=1/3$, antiquarks $s=1/2$
and $B=-1/3$.  Quark and antiquark combine to $B=0$ and to a spin
triplet ($S =1$) or singlet ($S = 0$).  In conventional mesons, the
total spin $\vec S$ of the quark $q$ and the antiquark $\bar q$, and
the orbital angular momentum $\vec L$ between $q$ and $\bar q$ couple
to the total angular momentum $\vec J$ of the meson: $\vec J = \vec L
+ \vec S$. Light mesons are restricted to $u,d$, and $s$ quarks.

\FloatBarrier\subsection{Quantum numbers}\label{mes:sub:qn}
\paragraph{Parity:} 
The parity $P$ of a meson involves the orbital 
angular momentum $L$ between quark and antiquark and the
product of the intrinsic parities which is 
$P_{q}P_{\bar q}=-1$ for a fermion and its antiparticle: 
\begin{equation}
P=(-1)^{L+1}.
\end{equation}
\paragraph{Charge conjugation:} 
Neutral mesons are eigenstates of the charge conjugation operator 
\begin{equation}
C = (-1)^{S+L}.
\end{equation} 
It turns out convenient to use the same sign convention within a
multiplet. For instance, since $C\piz=\piz$, we choose
$C\pi^\pm=\pi^\mp$, and $C\K^0=\Kb{}^0$, while since
$C\rho^0=-\rho^0$, we adopt $C\rho^\pm=-\rho^\mp$ and
$C\K^{\star\,\pm}=-\Kb{}^{\star\,\mp}$.
\paragraph{Isospin:} 
Proton and neutron form an isospin doublet and so do the {\it up} and
the {\it down} quark. We define antiquarks by $\bar u\,=\,C\,u$ and
$\bar d\,=\,C\,d$, antinucleons by $\rm\bar p\,=\,\it G\,\rm n$ and
$\rm\bar n\,=\,\it G\,\rm p$.  This means we use the ${\bf\bar{2}}$
representation of SU(2) for $\{\bar{u},\,\bar{d}\}$ and the ${\bf 2}$
representation for $\{\pbar,\,\nbar\}$. See, e.g., \cite{Martin70} for
a detailed discussion on phase conventions for isospin states of
antiparticles. We obtain:

\begin{equation}\label{mes:eq:states}
\renewcommand{\arraystretch}{1.3}
\begin{tabular}{ccccc}
$\ket{I=1, I_3=1 }$&=&$-\ket{\udb}~,$\\
$\ket{I=1, I_3=0 }$&=&$\frac{1}{\sqrt{2}}(\ket{\uub} - \ket{\ddb})~,$\\
$\ket{I=1, I_3=-1}$&=&$\ket{\dub}~,$\\
$\ket{I=0, I_3=0 }$&=&$\frac{1}{\sqrt{2}}(\ket{\uub} + \ket{\ddb})$&=&
$\ket{\nnb}~,$\\
$\ket{I=0, I_3=0 }$&=&$\ket{\ssb}~.$\\
\end{tabular} \renewcommand{\arraystretch}{1.0}
\end{equation}
The \ket{\nnb} and \ket{\ssb} states have the same quantum numbers
and mix to form two physical states. With $n$ we denote
the two lightest quarks, $u$ and $d$, while $\rm n$ stands for the 
neutron.
\paragraph{The $G$-parity:}  
The $G$-parity is defined as charge conjugation  followed
by a rotation in isospin space  about the $y$-axis,  
\begin{equation}
G =C e^{i\pi I_y} =  (-1)^I C = (-1)^{L+S+I} ~,
\end{equation}
and is approximately conserved in strong interactions.
It is a useful concept since 
 $G=(-1)^{n_\pi}$ for a system of $n_{\pi}$ pions. This 
generalises the 
selection rule for $\Pe^+\Pe^-\rightarrow n\gamma$ in QED, namely
$C=(-1)^n$. 

The \atmass, for instance, decays into $\rho\pi$ with $\rho\to\pi\pi$,
hence into three pions, and into $\eta\pi$ with one pion in the final
state. The \atmass\ never decays into $\pi\pi$ or $\eta\pi\pi$: 
$G$-parity is conserved. 
The $\eta$ having $G=+1$ nevertheless decays into three pions; the
$\omega$ has a small partial width for decays into two pions. These
decay modes break isospin invariance; they vanish in the limit where
$u$- and $d$-quark have equal masses and electromagnetic interactions are neglected. 
$\K\overline{\K}$ pairs may have $G=-1$ or $+1$.

\FloatBarrier\subsection{\label{mes:sub:nonets}Meson nonets}
Mesons are characterised by their quantum numbers 
$J^{PC}$ and their flavour content. Quark-antiquark states with 
quantum numbers $J^{PC}$ are often referred to by the 
spectroscopic notation \nslj\ borrowed from atomic 
physics. In the light-quark domain, any \nslj\ 
leads to a nonet of states. Based on SU(3) symmetry, we expect an octet and a
singlet. However, the $s$ quark is heavier than the
$u$ and $d$ quark. This results into SU(3) breaking. The actual
mesons can be decomposed either on a basis of SU(3) eigenstates or 
according to their  \uub , \ddb\ and \ssb\ content.

\subsubsection{\label{mes:sub:psmeson}The pseudoscalar mesons}
\begin{figure}[!h]
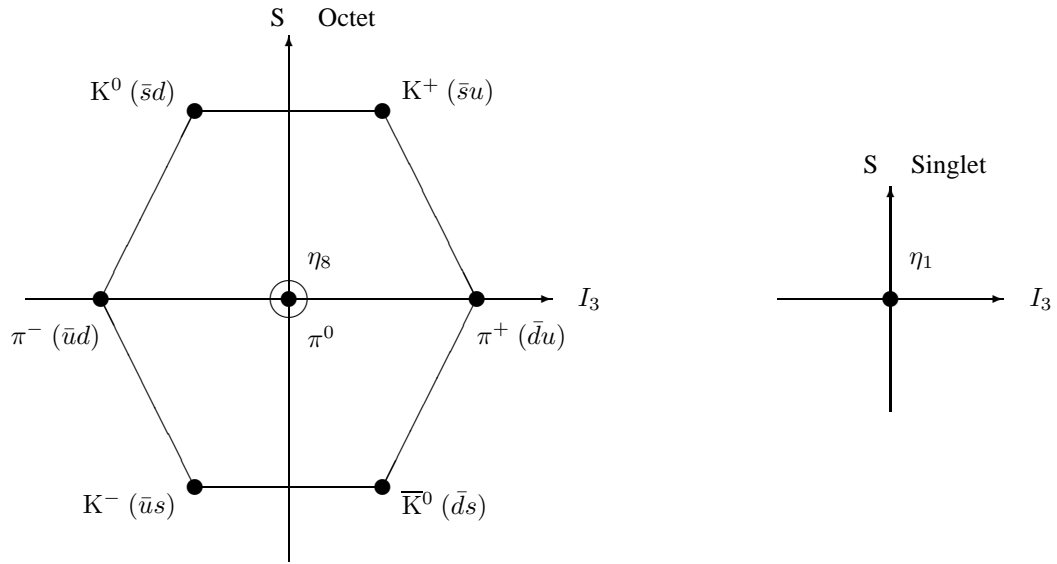

\nonett{$\K^0$}{$\Kp$}{$\Km$}
{$\Kb^0$}{$\pi^-$}{$\pi^+$}{$\pi^0$}
{$\eta_8$}{$\eta_1$}
\vspace*{-15mm}
\caption{The nonet of pseudoscalar mesons.  \label{mes:fig:pseudo}
}
\end{figure}

The pseudoscalar mesons correspond to 
$\jpc =0^{-+}$ and \nslj$ = 1^1S_0$. 
The nine orthogonal SU(3) eigenstates are shown in Fig.~\ref{mes:fig:pseudo}.
The quark representation of
the neutral members is 
\begin{equation}\label{mes:eq:neutrals}
\pi^0 = \frac{1}{\sqrt 2}(\uub - \ddb)~,\quad
\eta_8 = \sqrt{\frac{1}{6}}(\uub +\ddb -2\ssb )~,\quad
\eta_1 = \sqrt{\frac{1}{3}}(\uub +\ddb +\ssb )~.
\end{equation}
The actual mesons $\eta$ and $\eta^{\prime}$ can be written as
\begin{equation}
\eqalign{
&\ket{\eta\phantom{^\prime}} = \cos\Theta_{\rm PS}\ket{\eta_8} - \sin\Theta_{\rm PS}\ket{\eta_1}\cr
&\ket{\eta^{\prime}} = \sin\Theta_{\rm PS}\ket{\eta_8} + 
\cos\Theta_{\rm PS}\ket{\eta_1}}
\end{equation}
with the pseudoscalar mixing angle $\Theta_{\rm PS}$.
A mixing angle $\Theta_{\rm PS}=\Theta_{\rm id}=\arctan(1/\sqrt{2})$,
called the {\em ideal} mixing angle,
would lead to a decoupling  $\eta\propto\ket{\ssb}$ and $\eta'\propto\ket{\nnb}$.
The $\eta$ and $\eta^{\prime}$ wave functions can be decomposed
into the \nnb\ and \ssb\ basis. With $\Theta=\Theta_{\rm PS}-\Theta_{\rm id}+\pi/2$,
\begin{equation}
\label{meson:eq:psmix}
\eqalign{
&\ket{\eta\phantom{^\prime}} = 
\cos\Theta\,\ket{\nnb} - \sin\Theta\,\ket{\ssb} = 
X_{\eta\phantom{^\prime}}\,\ket{\nnb} +
Y_{\eta\phantom{^\prime}}\,\ket{\ssb}
\cr
&\ket{\eta^{\prime}} = 
\sin\Theta\,\ket{\nnb} + \cos\Theta\,\ket{\ssb} = 
X_{\eta^{\prime}}\,\ket{\nnb} +
Y_{\eta^{\prime}}\,\ket{\ssb}}
\end{equation}

The mixing angle can be determined experimentally from 
$\eta$ and $\eta^{\prime}$ production and decay rates. 
It is shown in \cite{Feldmann:2002kz} that the 
quark flavour basis $\nnb$ and $\ssb$ is better
suited to describe the data than the octet--singlet
basis. The latter can describe data only by the
introduction of a second mixing angle. 

The $\eta$ and $\eta^{\prime}$ could also mix with
other states, in particular radial excitations or glueballs.
The $\eta^{\prime}$ is nearly a flavour singlet state and can hence 
couple directly to the gluon field;  this 
has led to speculations that the $\eta^{\prime}$ (and 
to a lesser extend also the $\eta$) 
may contain a large fraction of glue. 
This requires an extension of the mixing
scheme (\ref{meson:eq:psmix}) by introduction of a non-$q\bar q$ or 
{\it inert} component,
with a third state of unknown mass which could, e.g.,  be dominantly
a glueball. 

\begin{eqnarray}
\ket{\eta\phantom{^\prime} } = & 
X_{\eta\phantom{^\prime}}\,\ket{\nnb} \qquad +\qquad 
Y_{\eta\phantom{^\prime}}\,\ket{\ssb} \qquad +\qquad 
Z_{\eta\phantom{^\prime}}\,\ket{\rm glue} \nonumber \\
\ket{\eta^{\prime} } = & 
X_{\eta^{\prime}}\,\ket{\nnb} \qquad +\qquad 
Y_{\eta^{\prime}}\,\ket{\ssb} \qquad +\qquad  
Z_{\eta^{\prime}}\,\ket{\rm glue} \\
& \hspace*{-7mm}\mathrm{light\ quark}\quad\ \qquad \mathrm{strange\ quark} 
\qquad\ \qquad \mathrm{inert} \nonumber
\end{eqnarray}
At present there is no convincing evidence for a
glueball content in the $\eta^{\prime}$ wave function, and we
will assume ${Z_{\eta}}$ and ${Z_{\eta^{\prime}}}$ to vanish,
i.e., $Z_{\eta}=Z_{\eta^{\prime}}\sim 0$.

\subsubsection{Other meson nonets}
\label{mes:sub:others}
A meson nonet consists of five isospin multiplets. The
pseudoscalar nonet, for instance, contains the pion triplet, two kaon doublets, 
the $\eta^{\prime}$ and the
$\eta$. Well known are also the nonet of vector mesons
with quantum numbers $J^{PC} = 1^{--}$ (three $\rho$, four K$^*$,
$\phi$ and $\omega$), and the nonet of tensor mesons
with quantum number $2^{++}$ (\atmass , K$_2^*(1430)$, 
f$_2(1525)$, \ftmass). In a spectroscopic notation, these are
the 1$\tso$ and 1$\tpt$ states. Both nonets have a
nearly {\it ideal mixing angle} $\Theta_{\rm id} = 
35.3^{\circ}$ for which one meson is a purely $\nnb$ and the other
one a purely $\ssb$ state.
These are the $\omega$ and $\phi (1020)$, and the \ftmass\ and
f$_2(1525)$ mesons, respectively.
Note that the mass difference between the $s\bar s$
and the $n\bar n$ state is about 250 MeV. This mass difference
is due to the larger constituent mass of strange quarks.

The mixing angles for these meson nonets (all except the pseudoscalar
nonet) are defined as
\begin{equation}
\label{meson:eq:vmix}
\eqalign{
&\ket{\omega} = 
\cos\Theta_V\,\ket{\nnb} - \sin\Theta_V\,\ket{\ssb} 
\cr
&\ket{\phi} = 
\sin\Theta_V\,\ket{\nnb} + \cos\Theta_V\,\ket{\ssb}}
\end{equation}

In Table \ref{mes:tab:nonets} some meson nonets are collected. 
The assignment shown here is reproduced from the quark-model description
of Amsler and Wohl in \cite{Eidelman:2004wy} and represents one 
possible scenario. In particular the scalar-meson nonet is hotly 
debated \cite{Minkowski:2002nf,Klempt:2000ud}
but there are also open questions in the axial-vector nonet 
\cite{Aston:1988ak,Barberis:1997vf}
and for the radial excitations of vector \cite{Donnachie:2001zn}
and pseudoscalar  \cite{Klempt:2004yz} mesons.

\begin{table}[h!]
\caption{The light mesons. The two mesons K$_{1A}$ and K$_{1B}$ 
mix to form the observed resonances K$_{1}(1280)$ and K$_{1}(1400)$.
The scalar mesons resist an unambiguous classification; the scenario
reproduced here assumes that the $\mathrm{a}_0(980)$ and
$\mathrm{f}_0(980)$ are $\K\Kb$ molecules or generated dynamically but
are not $\qqb$ states, that the $\mathrm{f}_0(1370)$ is a $\qqb$ state
and not generated dynamically, and that the $\mathrm{f}_0(1500)$ is a
glueball. The $\mathrm{f}_1(1510)$ is discarded in the listing
below. There are considerable difficulties for the nonets of
pseudoscalar and vector radial excitations.  }
\label{mes:tab:nonets}
\renewcommand{\arraystretch}{1.3}
\begin{tabular}{cccccccccc}
\hline\hline
\ $L$\qquad &\ $S$\qquad &\ $J$\qquad &\ $n$\qquad & $I=1$\qquad & $I=1/2$
\qquad & $I=0$\qquad & $I=0$\qquad & $J^{PC}$\qquad & $n^{2s+1}L_J$\qquad  \\
\hline
0 & 0 & 0 & 1 & $\pi$  &  \K & $\eta$& $\eta^{\prime}$   & $0^{-+}$ & $1\ssz$\\
0 & 1 & 1 & 1 & $\rho$ & $\K^*$ &$\phi$ & $\omega$ & $1^{--}$ & $1\tso$ \\
\hline
1 & 0 & 1 & 1 &b$_1(1235)$&$\K_{1B}$ &h$_1(1380)$ &h$_1(1170)$ \quad
 & $1^{+-}$ & $1\spo$\\
1 & 1 & 0 & 1 &a$_0(1450)$&$\K_{0}^*(1430)$ &f$_0(1710)$ &f$_0(1370)$ 
 & $0^{++}$ & $1\tpz$\\
1 & 1 & 1 & 1 &a$_1(1260)$&$\K_{1A}$ &f$_1(1420)$ &f$_1(1285)$ 
 & $1^{++}$ & $1\tpo$\\
1 & 1 & 2 & 1 &a$_2(1320)$&$\K_{2}^*(1430)$ &f$_2(1525)$ &f$_2(1270)$ 
 & $2^{++}$ & $1\tpt$\\
\hline
2 & 0 & 2 & 1 &$\pi_2(1670)$&$\K_{2}(1770)$ &$\eta_2(1870)$ &$\eta_2(1645)$ 
 & $2^{-+}$ & $\rm 1^1D_2$\\
2 & 1 & 1 & 1 &$\rho(1700)$&$\K^*(1680)$ &$\phi(????)$ &$\omega(1650)$ 
 & $1^{--}$ & $\rm 1^3D_1$\\
2 & 1 & 2 & 1 &$\rho_2(????)$&$\K_2(1820)$ &$\phi_2(????)$ &$\omega_2(????)$ 
 & $2^{--}$ & $\rm 1^3D_2$\\
2 & 1 & 3 & 1 &$\rho_3(1690)$&$\K^*_3(1780)$ &$\phi_3(1850)$&$\omega_3(1670)$ 
& $3^{--}$ & $\rm 1^3D_3$\\
\hline
0 & 0 & 0 & 2 &$\pi$(1370) &  $\K(1460)$  &$\eta (1440)$  & $\eta (1295)$
 & $0^{-+}$ & $2\ssz$\\
0 & 1 & 1 & 2 &$\rho$(1450)& $\K^*(1410)$ &$\phi (1680)$ &$\omega$(1420)
 & $1^{--}$ & $2\tso$\\
\hline\hline
\end{tabular}
\renewcommand{\arraystretch}{1.0}
\end{table}
\subsubsection{The Gell-Mann--Okubo mass formula}
\label{mes:sub:GMO}
The Gell-Mann--Okubo mass formula relates the masses of a meson nonet
and its mixing angle. It can be derived by
ascribing to mesons a common mass $M_0$
plus the (constituent) masses of the quark and antiquark
it is composed of. The relation is written as
\begin{equation}
\tan^2\Theta = \frac{3M_{\eta} + M_{\pi} - 4M_{K}}
{4M_{K} - 3M_{\eta^{\prime}} - M_{\pi}}\,.
\end{equation}
Often, the linear GMO mass formula is replaced by
the quadratic GMO formula which is given as above but
with $M^2$ values instead of masses $M$. Note that in the limit of chiral
symmetry quark masses are proportional to the mass square of the
meson masses. The quadratic GMO formula reads
\begin{equation}
\tan^2\Theta = \frac{3M^{2}_{\eta} + M_{\pi}^2 - 4M_{K}^2}
{4M_{K}^2 - 3M^{2}_{\eta^{\prime}} - M_{\pi}^2}\,.
\end{equation}
Table \ref{mes:tab:mix} gives the mixing angles derived from the
linear and quadratic GMO formula.
\begin{table}[!h]
\caption{Mixing angles of meson nonets }
\renewcommand{\arraystretch}{1.3}
\begin{tabular}{ccc}
\hline\hline
   Nonet members &\quad $\Theta_{\rm linear}$\quad &\quad $\Theta_{\rm quad} $ \\
$\pi , \K , \eta^{\prime} , \eta $ & $-23^{\circ}$ & $-10^{\circ}$ \\
$\rho , \K^* , \phi\ , \omega $    & $\phantom{-}36^{\circ}$ &$\phantom{-}39^{\circ}$  \\
$\atmass , \K^{*}_{2}(1430) , \ft(1525) , \ft(1270)$ & 
$\phantom{-}26^{\circ}$ & $\phantom{-}29^{\circ}$ \\
$\rho_3(1690) , \K^*_3(1780) , \phi_3(1850) , \omega_3(1670)$\quad & $\phantom{-}29^{\circ}$ &$\phantom{-}28^{\circ}$  \\
\hline\hline
\end{tabular}
\renewcommand{\arraystretch}{1.0}
\label{mes:tab:mix}
\end{table}
\subsubsection{The Zweig rule}\label{mes:sub:Zweig}
The ``quark line rule'', or Zweig rule, is also called the ``OZI rule'',
after Okubo, Zweig and Iizuka, or even the ``A--Z rule'' to account for
all the various contributions to its study. See, e.g.,
Ref.~\cite{Lipkin:1977pi}, for a comprehensive list of
references. This rule has played a crucial role in the development of
the quark model.  

\begin{figure}[!htpc]
\vspace*{.2cm}
\includegraphics*[width=.7\textwidth]{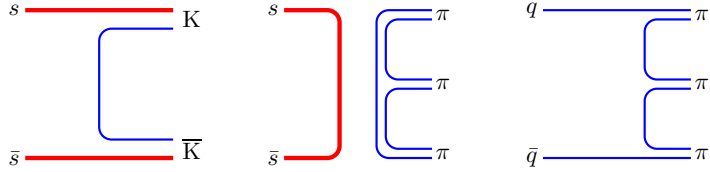}
\vspace*{-.2cm}
\caption{Connected (left) and disconnected (centre)  contribution to 
$\phi(1020)$ decay. The latter contribution to $\phi(1020)\to\pi\pi\pi$
can be described as an allowed decay 
from a small impurity in the wave function (right).\label{mes:fig:OZI1}}
\end{figure}

For instance, the $\phi(1020)$ is a vector meson with isospin $I=0$,
seemingly similar to $\omega(780)$, but much narrower, in spite of the
more favourable phase-space. It decays preferentially into \K\Kb\
pairs, and rarely into three pions. The explanation is that the
$\phi(1020)$ has an almost pure \ssb\ content, and that the decay
proceeds mostly with the strange quarks and antiquarks flowing from
the initial state to one of the final mesons, as per
Fig.~\ref{mes:fig:OZI1}, left, while the process with an internal
\ssb\ annihilation (centre) is suppressed. The decay
$\phi\to \pi\pi\pi$  is attributed to the main $\ssb$ component being
slightly mixed with a $\qqb$ component, which in turn decays into
pions by a perfectly allowed process (right).

The rule is quite strictly observed, as the non-strange width of the
$\phi(1020)$ is less than 1$\;$MeV! Even more remarkably, the rule
works better and better for the charm and beauty analogues, for which
the decay into naked-flavour mesons is energetically forbidden: the
total width is only about 90$\;$keV for the
$\mathrm{J}/\Psi(c\bar{c})$ and 50$\;$keV for
$\Upsilon(b\bar{b})$. 

 The magnitude of OZI violation in mesons is primarily described by the mixing angle, though an OZI violation from the decay amplitude cannot be excluded.
The vector mesons, e.g., have a mixing angle of $\Theta_V = 39^{\circ}$
which deviates from the ideal mixing angle $\Theta_{\rm id}$ by $\delta =3.7^{\circ}$.
The physical $\phi (1020)$ is then written
\begin{equation}
\label{mes:eq:OZI }
\ket{\phi}= \cos\delta\,\ket{\ssb} + \sin\delta\,\ket{nnb}.
\end{equation}
If there are no $s$-quarks in the initial state we expect the production
of $\phi$ mesons to be suppressed compared to 
$\omega$ production by $\sin^2\delta \simeq0.042$.
Indeed, a $\phi/\omega$ ratio of $0.0032\pm 0.0004$ was found  at Argonne
in the reaction $\pi^- \p \to \phi (\omega )\p$ at 6 GeV/\,c 
\cite{Cohen:1977ep}. This has to be compared
with the extremely large values (up to 0.6\,!) found 
in WA56 data on $\pi^- \p \to (\phi (\omega )+\pi ) \p$ at 12 and 20 GeV/\,c
\cite{Ferrer:1997cz}. In \pbp\ annihilation at rest, 
the $\phi/\omega$ production ratios were found to depend strongly
on the \pbp\ initial state and on the recoiling particles. These
results will be discussed further in Sec.~\ref{se:phe}.

There is a wide consensus that ideal mixing and OZI rule can be, if not derived, at 
least justified from QCD, but there are different approaches: the $1/N_c$ 
expansion, where $N_c$ is the number of colour degrees of freedom, 
cancellations of loop diagrams, lattice simulations, instantons effects, etc. 
The generally accepted conclusion is that ideal mixing is nearly achieved for mesons, except in 
the scalar and in the pseudoscalar sectors. See, e.g., Ref.~\cite{Isgur:2000ts} and references there.

\subsubsection{Meson decays}\label{mes:sub:decay}
The decays of mesons belonging to a given nonet are related
by SU(3) symmetry. The coefficients governing these relations are
called SU(3) isoscalar factors and listed by the Particle Data
Group \cite{Eidelman:2004wy}. We show here two simple examples.

A glueball is, by definition, a flavour singlet. It may decay
into two octet mesons, schematically $1 \rightarrow\ 8 \times\ 8$.
The isoscalar factors for this decay are%
\footnote{The parenthesis reads 
$(\sqrt2,\sqrt3,-1,-\sqrt2)$, 
and similarly for Eqs.~(\protect\ref{mes:eq:888a}).}
\begin{equation}
({\rm glueball}) \rightarrow\ (\K\Kb\ ,\ \pi\pi\ ,\ \eta_8\eta_8\ ,\
\Kb\K) =
\frac{1}{\sqrt 8}(2\ ,\ 3\ ,\ -1\ ,\ -2)^{1/2}\,.
\end{equation}
Hence glueballs have squared couplings to $\K\Kb$, $\pi\pi$ , 
$\eta_8\eta_8$ of
4\,:\,3\,:\,1\,. The decay into two isosinglet mesons $\eta_1\eta_1$ has
an independent coupling and is not restricted by these SU(3)
relations. It could be large leading to the notion that the
$\eta^{\prime}$ couples strongly to glueballs, and that 
$\eta^{\prime}$ are \emph{gluish}. The
decay into $\eta_1\eta_8$ is forbidden: a singlet cannot decay into
a singlet and an octet meson. This selection rule holds for
any pseudoscalar mixing angle: the two mesons $\eta$ and $\eta^{\prime}$ have 
orthogonal SU(3) flavour states and a flavour singlet cannot dissociate
into two states which are orthogonal. 

As a second example, we choose decays of vector mesons into two
pseudoscalar mesons. We compare the two decays $\K^*\to\K\pi$ and
$\rho\to\pi\pi$. These are decays of octet particles into 
two octet particles. This  $8 \rightarrow\ 8 \times 8$ coupling is either symmetric ($8_1\to 8\times 8$) or antisymmetric ($8_2\to 8\times 8$) under the exchange of the final-state particles. The two pseudoscalars in $\K^*$ or $\rho$ decay having orbital momentum $\ell=1$,  one should use the antisymmetric flavour coupling, 
$8_2 \rightarrow\ 8 \times 8$, whose isoscalar factors are
\begin{equation}\label{mes:eq:888a}
\eqalign{%
(\K^*) \rightarrow\ (\K\pi\ \ \K\eta\ \ \pi\K\ \ \eta\K) &=
\frac{1}{\sqrt{12}}(3 \ ,\ 3 \ ,\ 3 ,\ -3)^{1/2}\cr
(\rho ) \rightarrow\ (\K\Kb\ \ \pi\pi\ \ \eta\pi\ \ \pi\eta\ \ \Kb\K) &=
\frac{1}{\sqrt{12}}(2 \ ,\ 8 \ ,\ 0 \ ,\ 0 ,\ -2)^{1/2}}
\end{equation}
Hence we derive $\K^*\to\K\pi +\pi\K$ $\propto 6$, \ \ $\rho\to\pi\pi\ 
\propto\ 8$, \ \ or
\begin{equation}
\frac{\Gamma_{\K^*\to\K\pi +\pi\K}}{\Gamma_{\rho\to\pi\pi}} = \frac{6}{8}
\left(\frac{0.291}{0.358}\right)^3 = 0.40
\end{equation}
The latter factor is the ratio of the decay
momenta $q$ to the 3rd power. 
The transition probability is proportional to $q$; for
low momenta (or point-like particles), the centrifugal barrier 
scales with $q^{2\ell}$ where $\ell$ is the orbital angular momentum.

From data we know that the width ratio is 0.34 and so the relations
are fulfilled at the level of about $20\%$,  a typical magnitude for
SU(3) breaking effects. We have neglected many aspects: the transition
rates are proportional to the squared matrix element (given by SU(3))
and the wave function overlap. The latter can be different for
the two decays. Mesons are not point-like; the angular 
barrier factor should hence include Blatt--Weisskopf 
corrections \cite{VonHippel:1972fg}. An
application of SU(3) to vector and tensor mesons can be found
in \cite{Peters:1995jv}.

\clearpage\markboth{\sl Annihilation dynamics}%
{\sl Kinematics and conservation laws}
\setcounter{equation}{0}
%
\section{Kinematics and conservation laws}\label{se:kin}%
In this section, we discuss the kinematics of annihilation into two,
three, or more mesons, and the selection rules due to exactly or
approximately conserved quantum numbers.
\FloatBarrier\subsection{Kinematics}\label{kin:sub:Kin}
We consider annihilation at rest. The formalism below can be used for
annihilation in flight in the centre of mass, if $2m$ is replaced by
$s^{1/2}$, where $m$ is the proton or antiproton mass, and $s$ the
usual Mandelstam variable $s=(\tilde{p}_1+\tilde{p}_2)^2$ built of the
four-momenta of the initial proton and antiproton.

If $m_1$, $m_2$, etc.\ denote the mass of the final mesons,
annihilation is possible if
\begin{equation}
\label{kin:eq:possible }
2m\ge m_1+m_2+\cdots~,
\end{equation}
Up to 13 pions could be produced.
\subsubsection{Two-body annihilation}
\begin{figure}[!h]
\includegraphics[width=.6\textwidth]{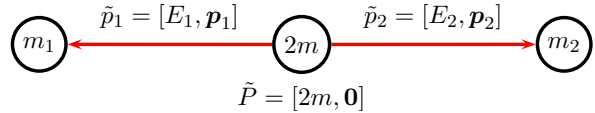}
\caption{\label{kin:fig:twob} Kinematics for two-body annihilation}
\end{figure}
The notation is defined in Fig.~\ref{kin:fig:twob}.  From the
energy--momentum balance rewritten as
$\tilde{p}_2=\tilde{P}-\tilde{p}_1$ and squared, one gets
\begin{equation}\label{kin:eq:tbe1}\eqalign{%
&E_1={4m^2-m_2^2+m_1^2\over 4m}~,\cr
&\vert\vec{p}_1\vert=\vert\vec{p}_2\vert
=\frac{\left[4m^2-(m_1+m_2)^2\right]^{\frac{1}{2}}\left[4m^2-(m_1-m_2)^2\right]^{\frac{1}{2}}}{4m}~.}
\end{equation}
In case of two identical mesons the momentum can be written in the
form
\begin{equation}
\vert\vec{p}_1\vert=
\sqrt{m^2-m_1^2}~.
\label{kin:eq:tbe2}
\end{equation}
The occurrence of two narrow mesons is easily observed due to the
narrow momentum distribution of the two produced particles. Other
annihilation modes such as $\rm p\bar p\to$ $\rho\pi$ or \atmass$\pi$
involve short-lived resonances.  In these cases, the measurement of
the annihilation frequency is more complicated.

Due to its relatively large mass, the antiproton-proton system
possesses a large variety of two-body annihilation modes.  In Table
\ref{kin:tab:momenta} we list the momenta of typical reactions.  For
broad resonances the momenta are calculated for the nominal meson
masses. Some annihilation modes like $\pbp\to\omega\ftmass$
are at the edge of the phase space and only the part of the \ftmass\
below 1094 MeV is produced. They may, nevertheless, make a significant
contribution to the annihilation process.

The mean decay length, $D_l = \gamma\beta c\tau = \hbar
p/(m\Gamma)$, varies over a wide
range. For $\rm p\bar p\to\pi^0\eta$, the mean path of the
$\pi^0$ is $D_l = 1\,$nm within the $\eta$ lifetime. 
Assume the $\eta$ then decays into
$\pi^+\pi^-\pi^0$. Because of the large $D_l$ there will be no
interaction between the $\pi^0$\ recoiling against the $\eta$ and the
pions from $\eta$-decays.  In case of $\pi\rho$ annihilations, the
recoil pion travels 7.5\,fm within the $\rho$ mean live time, and
rescattering of the primarily produced pion and pions from
$\rho$ decays is unlikely. The situation is different for production
of two short-lived high-mass mesons. In annihilation into
$\K^*\Kb^* + \mathrm{c.c.}$, the $\K^*$'s have moved only a mean distance 
of 1.3\,fm when they decay and interactions between the 
kaons and pions from  $\K^*$ decays are likely to occur.

\begin{table}[!h]
\caption{\label{kin:tab:momenta}Momenta for \pbp\  annihilation at rest  into two mesons.}
\renewcommand{\baselinestretch}{1.8}
\begin{tabular}{lcrlcr}
\hline\hline
\multicolumn{2}{c}{Channel} & Momentum & \multicolumn{2}{c}{Channel} &
Momentum \\ \hline
\quad $\pbp\to$\  & $\pi^0\pi^0$\  & 928.5 \mevc & 
\quad $\pbp\to$\ & $\rho\rho$     & 536.3 \mevc \\  
& $\pi^+\pi^-$\ & 927.8 \mevc & & $\omega\rho$ &527.5 \mevc \\ 
& $\pi^0\eta$ & 852.3 \mevc & & $\omega\omega$ & 518.5\mevc \\ 
& \Kp\Km & 797.9 \mevc & & $\eta\phi$ & 499.7 \mevc \\ 
&
$\Kn\Kb^0$ & 795.4 \mevc & & $\pi$\ftmass\ & 491.2 \mevc \\ 
&$\pi\rho$ & 773.2 \mevc & & $\pi$\atmass\ & 459.9 \mevc \\ 
&$\pi^0\omega$ & 768.4 \mevc & & $\eta^{\prime}\rho$ & 364.4 \mevc \\ 
&$\eta\eta$ & 761.0 \mevc & & $\eta^{\prime}\omega$ & 350.5 \mevc \\ 
&$\eta\rho$ & 663.5 \mevc & & $\rm\K^*\Kb^*$ & 285.2 \mevc \\ 
&$\pi^0\eta^{\prime}$ & 658.7 \mevc & & $\rho\phi$ & 280.3 \mevc \\ 
&$\eta\omega$ & 656.4 \mevc & & $\omega\phi$ & 260.8 \mevc \\ 
&$\pi^0\phi$ & 652.4 \mevc & & $\eta$\ftmass & 206.2 \mevc \\ 
& $\rm\K\Kb^*$& 616.2 \mevc & & $\eta$\atmass & 91.2 \mevc \\ 
&$\eta\eta^{\prime}$ & 546.1 \mevc & & $\omega$\ftmass & See text\\
\hline\hline
\end{tabular}
\end{table}
\subsubsection{Three-body annihilation}
\begin{figure}[!h]
\centering{\hspace{-1cm}\includegraphics{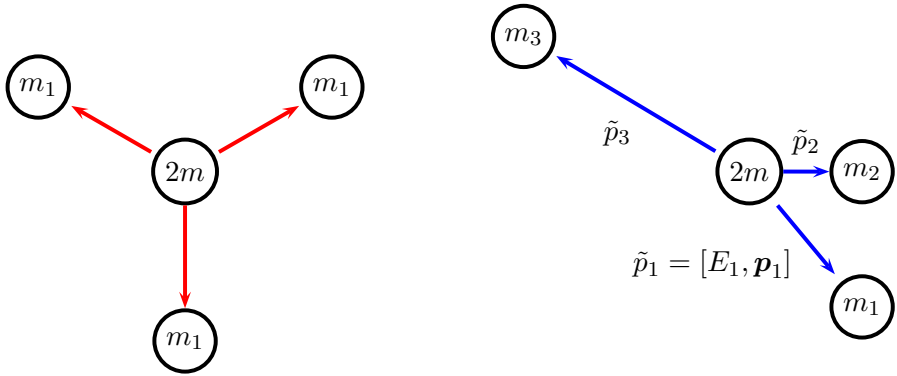}}
\caption{\label{kin:fig:3b}%
Notation for the kinematics of three-body annihilation (right). Even
for equal masses, the symmetric star (left) is just one possibility
among many others.}
\end{figure}

Unlike the two-body case, the energy of a given particle can vary over
a certain range. Even for equal masses, the symmetric star of
Fig.~\ref{kin:fig:3b} (left) is only a very particular case.
The minimal energy of particle~1, for instance, is obviously obtained when it is  produced at rest and particles 2 and 3 share the remaining energy $2m-m_1$, as per Eq.~(\ref{kin:eq:tbe1}),
{\sl mutatis mutandis}. However, the maximal value of $E_1$ does not correspond to either particle 2 or 3 being being produced at rest. Equation~(\ref{kin:eq:tbe1}), if rewritten as
\begin{equation}
\label{kin:eq:3b1a}
2m E_1=4m^2-m_{23}^2+m_1^2~,
\end{equation}
 where $m_{23}$ is the invariant mass of the $\{2,3\}$
 subsystem, indicates that $E_1$ is maximal, with value\begin{equation}
\label{kin:eq:3b2 }
\max E_1={4m^2-(m_2+m_3)^2+m_1^2\over 4m}~,
\end{equation}
when $m_{23}$ is minimal, i.e., $m_{23}=m_2+m_3$, when particles 2 and
3 are at rest relative to each other.
\subsubsection{Dalitz plot}
\begin{figure}[!hb]
\parbox{.71\textwidth}{%
\includegraphics[width=.70\textwidth]{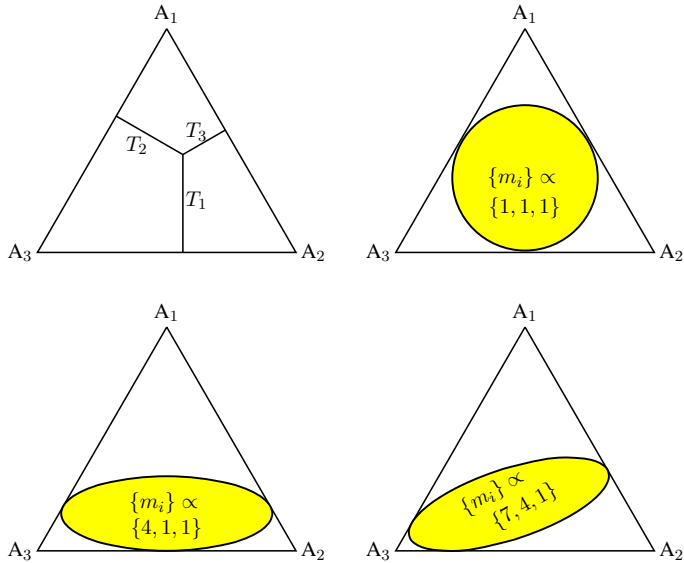}}
\parbox{.28\textwidth}{%
\caption{\label{kin:fig:Da1}%
Definition of the Dalitz plot for kinetic energies (upper left), and
boundary in the non-relativistic limit, for a decay into three
identical particles (upper right) and for particles with masses in
ratio [4:1:1] (lower left) or [7:4:1] (lower right), corresponding to
the mass ratios for $\eta\to3\pi$, $\eta'\to \eta\pi\pi$ and
$\ppb\to\eta'\eta\pi$, respectively.}}
\end{figure}

Energy conservation, rewritten for the kinetic part $T_i=E_i-m_i$
implies that
\begin{equation}
\label{kin:eq:Da1 }
T_1+T_2+T_3=T=2m-m_1-m_2-m_3~,
\end{equation}
remains constant from one event to another. This property is fulfilled
if $T_i$ is represented by the distance of a point to the
$i^\text{th}$ side of an equilateral triangle of height $T$, as in
Fig.~\ref{kin:fig:Da1}. We just saw that while $T_1=0$ is possible,
$T_1=T$ contradicts momentum conservation, which requires
\begin{equation}
\label{ kin:eq:Da2}
\vert p_2-p_3\vert\le p_1\le p_2+p_3~.
\end{equation}
Saturating this inequality, i.e., fixing the three momenta $\vec{p}_i$
to be parallel, gives the boundary of the {\em Dalitz plot}, which
corresponds to all possible sets $\{ T_i\}$ allowed by energy and
momentum conservation.  In the non-relativistic limit, the frontier is
a circle for identical particles, and an ellipsis for unequal masses,
as shown in Fig.~\ref{kin:fig:Da1}.

\begin{figure}[!h]
\parbox{.71\textwidth}{%
\includegraphics*[width=0.70\textwidth]{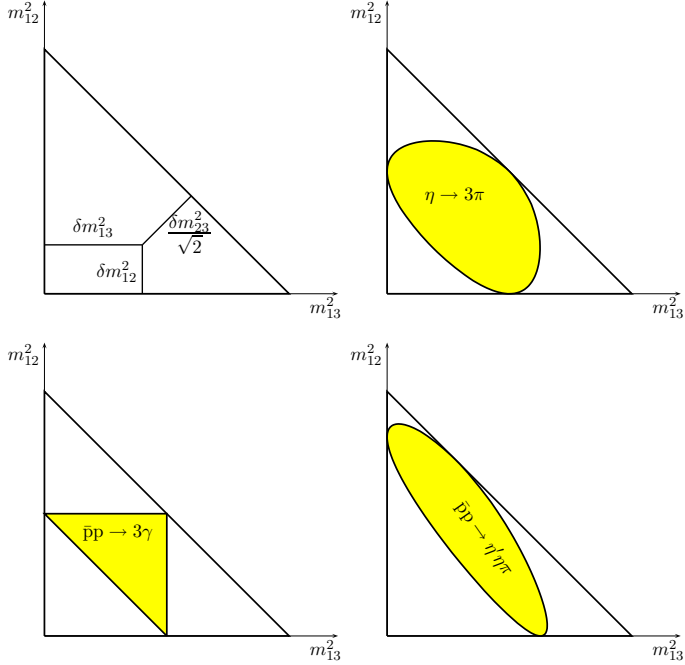}}
\parbox{.28\textwidth}{%
\caption{\label{kin:fig:Da2}%
Dalitz plots with relativistic kinematics:
interpretation of the distances ($\delta m_{ij}^2$ represents the
excess of $m_{ij}^2$ with respect to the minimal value $(m_i+m_j)^2$),
decays $\eta\to3\pi$, $\ppb\to3\gamma$, and $\ppb\to\eta'\eta\pi$.  }}
\end{figure}

In the relativistic case, the shape of the Dalitz plot becomes more
angular. For the (rare) $\pbp\to 3\gamma$ decays, it reduces to
a triangle limiting the middles of the sides.  Using
Eq.~(\ref{kin:eq:3b1a}), the Dalitz plot can be rescaled to substitute
the kinetic energies $T_k$ with the invariant masses $m_{ij}^2$, which
fulfil
\begin{equation}
\label{ kin:eq:Da5}
\sum_{i<j} m_{ij}^2=4m^2+m_1^2+m_2^2+m_3^3~,\quad
(m_1+m_2)^2\le m_{12}^2\le(2m-m_3)^2~.
\end{equation}
Also, the equilateral frame is replaced in recent literature by
rectangular triangles. Some examples are shown in
Fig.~\ref{kin:fig:Da2}.
%

\FloatBarrier\subsubsection{Multiparticle final state}
The results on the energy range are easily generalised to more than
three mesons in the final state.  If $2m\ge m_1+m_2+\cdots +m_n$,
annihilation into the $n$ mesons $\{m_i\}$ is allowed, and the energy
of particle 1, for instance, is bounded by
\begin{equation}
\label{kin:eq:nb1 }
m_1\le E_1\le {4m^2-(m_2+m_3+\cdots +m_n)^2+m_1^2\over 4m}~.
\end{equation}
\FloatBarrier\subsection{Phase space}\label{kin:sub:phsp}
Heavier particles are less easily produced than lighter ones. Removing
this obvious kinematical effect leads to a more meaningful comparison
among various reaction rates. We give below some basic
results. For a more detailed treatment, see, e.g., the reviews
\cite{Eidelman:2004wy,Byckling:1973}.

The typical decay rate of protonium, of mass $2m$, into a set of $n$
mesons $\{m_i\}$ is given by
\begin{equation}
\label{kin:eq:ps1}
\Gamma={ (2\pi)^{4-3n}\over 2^{2+n}m}\int
\vert\mathcal{M}\vert^2\prod_{i=1}^n 
\mathrm{d}^3 {p_i\over E_i}\,
\delta^4(\tilde P-\tilde{p}_1-\cdots\tilde{p}_n)~.
\end{equation}
If $\vert\mathcal{M}\vert^2$ is removed, one obtains the phase-space
integral. For two-body decays ($n=2$), $\vert\mathcal{M}\vert^2$
contains the non-trivial dynamical variations when going from one
channel to another. For $n\ge3$, $\vert\mathcal{M}\vert^2$ also
contains information on the correlation or anticorrelation of momenta,
allowing tests of scenarios for the formation of resonances, etc.

For $n=2$, the phase-space integral is evaluated in the c.o.m.\ frame
\cite{Eidelman:2004wy} as
\begin{equation}
\label{kin:eq:ps2 }
\Gamma={\vert\vec{p}_1\vert\over 32\pi m^2}\,\vert\mathcal{M}\vert^2~.
\end{equation}

For $n=3$, the phase-space integral is conveniently expressed as an
integral over two kinetic energies or, equivalently, two invariant
2-body masses, i.e., the variables used to draw the Dalitz plot. The
result is \cite{Eidelman:2004wy}
\begin{equation}
\label{kin:eq:ps3}
\mathrm{d}\Gamma={\vert\mathcal{M}\vert^2\over (2\pi)^3 16m}\mathrm{d}T_1\mathrm{d}T_2
={\vert\mathcal{M}\vert^2\over (2\pi)^3
256m^3}\mathrm{d}m_{12}^2\mathrm{d}m_{23}^2~,
\end{equation}
where an average over initial spins is implied. Unfortunately,
measurements of annihilation at rest with a polarised protonium state
have not been performed.

Equation (\ref{kin:eq:ps3}) implies that for the fictitious  dynamics where
$\vert{\cal M}\vert$ is constant, the Dalitz plot in coordinates
$\{T_1,T_2\}$ or $\{ m_{12}^2,m_{23}^2\}$ is uniformly
populated. Peaks in the population density indicate the formation of
resonances, or at least constructive interferences of several
dynamical contributions to the amplitude.
\FloatBarrier\subsection{Conservation laws}\label{kin:sub:cons}
Strong interactions obey very strictly conservation laws due to basic
symmetries: energy $E$, momentum $\vec p$, angular momentum $J$,
parity $P$ and charge conjugation $C$ are conserved, as well as
flavours.  Isospin is also a rather good symmetry. $G$-parity (or
isotopic parity) is a combination of charge conjugation and isospin,
$G=C\exp(-i\pi I_2)$.
\subsubsection{Partial waves}
The algebra of \NNb\ quantum numbers is very similar to that of \qqb\
reviewed in the section on mesons.  The partial wave
${}^{2I+1,2S+1}L_J$, where $L$ is the orbital momentum and $S$ the
total spin has $C=(-1)^{L+S}$ (if the system is neutral),
$P=(-1)^{L+1}$, and $G=(-1)^{L+S+I}$.  This is summarised in
Table~\ref{kin:tab:pw}, for S and P waves.
\renewcommand{\bs}{\hspace*{-4pt}}
\begin{table}[!h]
\caption{\label{kin:tab:pw}%
Quantum numbers of the S and P partial waves (PW) of the {\NNb} system. The notation is \islj.}
\vskip 4pt
\renewcommand{\arraystretch}{1.2}
\begin{tabular}{ccccccccccccc}
\hline\hline
\bs PW \bs&\bs\ISLJ11S0\bs&\bs\ISLJ31S0\bs&\bs\ISLJ13S1\bs&\bs\ISLJ33S1\bs&%
\bs\ISLJ11P1\bs&\bs\ISLJ31P1\bs&\bs\ISLJ13P0\bs&\bs\ISLJ33P0\bs&%
\bs\ISLJ13P1\bs&\bs\ISLJ33P1\bs&\bs\ISLJ13P2\bs&\bs\ISLJ33P2\bs\\
\hline
$J^{PC}$  &$ 0^{-+}   $&$ 0^{-+}   $&$ 1^{--}  $&$ 1^{--}    $&$
 1^{+-}   $&$ 1^{+-}   $&$ 0^{++}   $&$  0^{++}  $&$
 1^{++}   $&$  1^{++}  $&$ 2^{++}   $&$  2^{++}  $\\
$I^G$ &$ 0^+   $&$ 1^-  $&$  0^- $&$  1^+   $&$
 0^-   $&$ 1^+   $&$ 0^+  $&$  1^-  $&$
 0^+   $&$ 1^-   $&$ 0^+  $&$   1^- $\\
\hline\hline\end{tabular}     
\end{table}
\subsubsection{Simple rules}
The simplest and most effective selection rule comes from $G$-parity.
Only half of the partial waves, for instance, are candidates for
annihilation into five pions, those with $G=-1$.  Once $G$-parity is obeyed, one can
generally match any set of quantum numbers by cleverly arranging the
spins and internal orbital momenta of the mesons. Important exceptions
are observed in channels involving a small number of spinless mesons
or identical particles. We review these selection rules below.  The
reasoning was elaborated in the 60's (see, for instance,
\cite{Henley:1962,Zemach:1964}) to identify the quantum numbers of
meson resonances from their observed decays into a few mesons. The
large angular acceptance and improved resolution of modern detectors
forces us to consider higher multiplicities.
\subsubsection{Two spinless mesons}
For \NNb$\rightarrow$ two scalars or two pseudoscalars, only natural
parity $P=(-1)^J$ is allowed.  For $\pi^0\pi^0$, $J$ has to be even;
$I+J$ is even for $\pi^-\pi^+$, while there is no correlation between
$I$ and $J$ for $\PK^-\PK^+$.  For $\NNb\rightarrow\K^0\Kb{}^0$, one
detects \Ks\ or \Kl\ and $C\!P$ invariance provides selection rules
for \Ks\Ks\ or \Ks\Kl\ channels.  The results are summarised in
Table~\ref{kin:tab:allowed}.  Similarly, for $\NNb\rightarrow$ scalar
+ pseudoscalar, only states with unnatural parity $P=(-1)^{J+1}$
contribute.
\subsubsection{Identical vector mesons}
For two identical vector mesons, the Pauli principle does not lead to
further restrictions, once $G$ or $C$ conservation is enforced. Total
spin $S=0$ or 2 of mesons implies even $\ell$ and opens the
$J^P=0^+,1^+,2^+,\ldots$ channels, while spin $S=1$ requires an
antisymmetric space configuration and thus $J^P=0^-,1^-,2^-,\ldots$
Note, however, that $\ISLJ13P1\rightarrow\omega\omega$ (or $\phi\phi$,
or $\rho^0\rho^0$) involves a total spin $S=2$ and an angular momentum
$\ell=2$ in the final state.
\renewcommand{\bs}{\hspace*{-3pt}}
\renewcommand{\baselinestretch}{1.3}
\begin{table}[!h]
\caption{\label{kin:tab:allowed}%
Allowed decays from S and P-wave protonium states into selected
two-meson final states (FS).}
\begin{tabular}{ccccccccccccc}
\hline\hline
\bs FS\bs&\bs\ISLJ11S0\bs&\bs\ISLJ31S0\bs&\bs\ISLJ13S1\bs&\bs\ISLJ33S1\bs&%
\bs\ISLJ11P1\bs&\bs\ISLJ31P1\bs&\bs\ISLJ13P0\bs&\bs\ISLJ33P0\bs&%
\bs\ISLJ13P1\bs&\bs\ISLJ33P1\bs&\bs\ISLJ13P2\bs&\bs\ISLJ33P2\bs\\
\hline
$\piz\piz$&     &       &     &     &
    &     & \y    &      &
    &    &   \y &       \\
 $\pim\pip$&     &       &     &   \y  &
    &     & \y    &      &
    &    &   \y &       \\
 $\piz\eta^{(\prime)}$&     &       &     &     &
    &     &    &  \y     &
    &    &   &   \y     \\
 $\eta^{(\prime)}\eta^{(\prime)} $&     &       &     &     &
    &     &  \y  &       &
    &    &   \y &        \\
 $\Km\Kp$&     &       & \y    &   \y  &
    &     & \y   &  \y     &
    &    &\y   &   \y     \\
 $\Ks\Kl$&     &       & \y    &   \y  &
    &     &   &       &
    &    &   &      \\
 $\Ks\Ks$&     &       &     &   &
    &     & \y  &  \y     &
    &    &  \y & \y     \\
$\piz\omega(\phi)$&     &       &     &\y   &
    &  \y   &   &       &
    &    &   &       \\
\bs $\eta^{(\prime)}\omega(\phi)$\bs&     &       & \y    &   &
 \y   &     &    &       &
    &    &   &     \\
$\piz\rho^0$&     &       & \y    &   &
 \y   &     &    &       &
    &    &   &     \\
$\eta^{(\prime)}\rho^0$&     &       &    & \y   &
    &   \y   &    &       &
    &    &   &     \\
 $\pi^\pm\rho^\mp$&     &    \y   & \y   &   &
 \y   &      &    &       &
    &  \y  &   &   \y  \\
\hline\hline
\end{tabular}  
\renewcommand{\baselinestretch}{1.0}
\end{table}
\subsubsectionb{Symmetric multi-$\pi^0$ states}
 More delicate is the case of three or more $\pi^0$. This is no longer
an academic problem, since these channels are seen in modern
detectors. For instance, \ISLJ11S0\ has adequate $C$ and $G$ for
decaying into $4\pi^0$. One can thus say that $\ISLJ11S0
\rightarrow4\pi^0$ is not forbidden by charge conjugation.  To claim
that it is actually allowed, one should exhibit at least one example
of a four-body wave function that is together symmetric and
pseudoscalar. At first sight, this seems impossible. In fact, one can
build such a wave function, with the desired coupling $\vec{\ell}_1+\vec{\ell}_2
+\vec{\ell}_3=0$ of the internal orbital momenta, but with $\ell_1$, $\ell_2$ and $\ell_3$
not vanishing, as it will be explained shortly. As a first step, let us
consider the $\pi^0+\pi^0+\pi^0$ case.

The method adopted below is simple, but a little empirical.  We refer
to Ref.\ \cite{Zupancic:1994bf} for a group-theoretical treatment.  We
tentatively write down minimal polynomials in the Jacobi variables
with the required quantum numbers.  Their angular-momentum content is
then the lowest term in a systematic partial-wave expansion.
\subsubsectionb{Three $\pi^0$}
There is a copious literature on 3-body wave functions and their
permutation properties. For instance, in the simple quark model of
baryons, one should write down spatial wave functions with
well-identified permutation symmetry, to be associated with spin,
isospin and colour wave functions, to form a state with overall
antisymmetry.  Let $p$ be the parity of the orbital wave-function.
In the harmonic-oscillator scheme
\cite{Hey:1983aj}, the symmetric states are labelled as $[56,J^p]$, and the
allowed values of $J^p$ in the lowest multiplets with $N\le3$ quanta
of excitation are $0^+$, $1^-$, $2^+,\ldots$ \cite{Hey:1983aj}. There is no pseudoscalar $(J^p=0^-$),
since to get $J=0$, the two internal orbital momenta $\ell_1$ and $\ell_2$
should be equal, and thus $p=(-1)^{\ell_1+\ell_2}=+1$.  The states $J^p=1^+$
and $2^-$ are allowed \cite{Henley:1962,Zemach:1964}, but with
complicated wave functions, since the coupling of internal
momenta, $\vec{\ell}_+\\vec{\ell}_2=\vec{J}$ is achieved with $\ell_1+\ell_2=8$ and $5$,
respectively.

To show that $J^p=0^+, \, 1^-$ and $2^-$ are allowed for three bosons,
it is sufficient to give explicit examples. The Jacobi variables
\be
\label{kin:eq:Jacobi1}
\vec{\rho} =\vi p 2- \vi p 1~,  \quad
\vec{\lambda} =( 2\vi p 3 -\vi p 1-\vi p 2 )/\sqrt 3 ~,
\ee
are built out of the individual momenta $\vi pi$.  $1\leftrightarrow2$
exchange and circular permutation results in
\be
\label{kin:eq:3-body-perm}
(\vec\lambda,\vec\rho\,)\rightarrow(\vec\lambda,-\vec\rho\,),\
(-\vec\lambda\pm\sqrt3\vec\rho,
\,\mp\sqrt3\vec\lambda-\vec\rho\,)/2~.
\ee
A constant, or $\rho^2+\lambda^2$, is scalar and symmetric, leading to
the existence of $J^p=0^+$ spatial wave functions. The vector
$(J^p=1^-)$
\be\label{kin:eq:3-body-symm-vector}
(\lambda^2-\rho^2)\vec\lambda-(2\vec\lambda\!\cdot\!\vec\rho\,)\vec\rho,
\ee
is also symmetric. The vector and axial vector
\be\label{kin:eq:3-body-antisymm-vector}
\vec{V}=(\lambda^2-\rho^2)\vec\rho+
(2\vec\lambda\!\cdot\!\vec\rho\,)\vec\lambda, \quad
\vec{W}=\vec\rho\times\vec\lambda~,
\ee
are both antisymmetric and thus
\be\label{kin:eq:3-body-symm-pseudo-tensor}
V_+W_+       \ee
has $J=2$, $J_z=2$, and orbital parity $p=-1$, and is symmetric.  In
Eq.\ (\ref{kin:eq:3-body-symm-pseudo-tensor}), we use the standard
notation $V_+=-(V_x+iV_y)/\sqrt2$.

For three pions, the parity is $P=-p$, once the intrinsic parities are taken into account. Hence $J^P=0^-$, $1^+$ and $2^-$ are allowed, i.e., \ISLJ31S0, \ISLJ33P1\ and \ISLJ33P2\ can decay into
$3\pi^0$.
\subsubsectionb{$\NNb\rightarrow4\pi^0$}
One first eliminates all channels but those with $C=G=+1$ and thus is restricted to 
$I=0$. While it seems $J^P=0^+$ or $2^+$ are obviously allowed, it
seems at first rather difficult to enforce all requirements of
permutation symmetry for $0^-$ and $1^+$. For instance, with the
system of Jacobi coordinates consisting of $\vec x\propto\vi p2-\vi
p1$, $\vec y\propto\vi p4-\vi p3$ and $\vec z\propto\vi p4+\vi p3-\vi
p2-\vi p1$, $1\leftrightarrow2$ and $3\leftrightarrow4$ symmetries are
simply translated into an even behaviour in $\vec x$ and $\vec y$, but
the effect of transpositions like $1\leftrightarrow3$ is not easily
written down. The task is simplified by using the variables
\ba
\label{kin:eq:Jacobi2}
\vec u&=&(\vi p4+\vi p1-\vi p2-\vi p3)~,\nonumber\\
\vec v&=&(\vi p4+\vi p2-\vi p3-\vi p1)~,\\
\vec w&=&(\vi p4+\vi p3-\vi p1-\vi p2)~,\nonumber
\ea
since the wave function simply has to survive the changes $\vec
u\leftrightarrow\vec v$, $\vec u\leftrightarrow-\vec v$, $\vec
v\leftrightarrow\vec w$, etc. The following wave functions can be
written down
\ba
J^P=0^-&\quad
&(u^2-v^2)(v^2-w^2)(w^2-u^2)\vec u\cdot(\vec v\times\vec w)~,\nonumber\\
J^P=0^+&\quad         &1~,\nonumber\\ 
J^P=1^+&\quad
&(u^2-v^2)(v^2-w^2)(w^2-u^2)
(\vec u\times\vec v+\vec v\times\vec w+\vec w\times\vec u)~,\nonumber\\ 
{\rm or}&&~\\ 
&&(u^2- v^2)(\vec w\times \vec
u)\times(\vec w\times \vec v)+ (v^2-w^2)(\vec u\times \vec
v)\times(\vec u\times \vec w) ~\nonumber\\
&\quad&\phantom{(u^2-v^2)(\vec w\times \vec u)\times(\vec w\times \vec
u)} {}+(w^2-u^2)(\vec v\times \vec w)\times(\vec v\times \vec
u)~\nonumber\\ 
J^P=2^+&\quad
&u_+^2+v_+^2+w_+^2~.\nonumber
\ea
In summary \ISLJ11S0, \ISLJ13P0, \ISLJ13P1, and \ISLJ13P2\ can decay
into four neutral pions.  One can probably establish these results in
a more physical way, by symmetrising amplitudes written as a product
of terms describing the successive steps of sequential decays,
provided there is no cancellation when the overall Bose--Einstein
symmetry is implemented. For instance, $\ISLJ11S0\to\pi^0\mathrm{a}_2$
is allowed, with a relative angular momentum $\ell=2$. Also
$\mathrm{a}_2\to\mathrm{f}_2\pi^0$, with $\ell'=1$, and in turn,
$\mathrm{f}_2\to\pi^0\pi^0$.

\subsubsectionb{Five or more $\pi^0$}
 For $\ppb\rightarrow5\pi^0$, \ISLJ31S0\ and the triplet states
$\ISLJ33P{0,1,2}$ are allowed. We refer here to the paper by Henley and
Jacobsohn \cite{Henley:1962}. Symmetric states are found in any $J^p$,
sometimes with many internal excitations. In particular, it is
difficult to produce five \piz\ from $J^p=0^-$, corresponding to
\ISLJ33P0.

The pattern is seemingly generalisable for any number $n>5$ of
identical pions. No state is strictly forbidden to decay into
$n\,\piz$, but the transition is sometimes suppressed by the
requirement of having many internal excitations in the final wave
function.
\FloatBarrier\subsection{Isospin considerations}\label{theo:sub:iso}
We summarise here some results on isospin symmetry and its violation
in \ppb\ annihilation.
\subsubsection{Relations for two-body annihilation}
Simple relations can be written down for two-body decays.  Consider,
for instance, two-pion events with a trigger on a X-ray, ensuring that
annihilation takes place from a P-state of protonium. Isospin symmetry
presumably holds for such pions since their energy is large compared
to the $\pi^\pm-\piz$ mass differences, and Coulomb effects are likely
to be negligible in the final state. Then $\pi\pi$ states from a
$J^P=0^+$ or $2^+$ state are pure $I=0$, and
\begin{equation}
\label{theo:eq:iso3}
\AF_{\rm P}(\ppb\to\pip\pim)=2\,\AF_{\rm P}(\ppb\to\piz\piz)~,
\end{equation}
where the subscript P indicates annihilation from P-states only.
\subsubsection{Relations for three-body annihilation}
For more than two particles in the final state, there are usually more
than one isospin amplitude contributing to the transition. Consider
for instance
\begin{equation}
\label{theo:eq:iso4}
\nbar\p\to \Kp\Kb^0\piz~,\ \ \Kp\Km\pip~,\ \ \K^0\Kb^0\pip~,
\end{equation}
for which one can identify two amplitudes, one with $(\K\Kb)$ in a $I=0$
state, i.e.,
\begin{equation}
\label{theo:eq:iso5}
\mathcal{A}_0=\mathcal{A}(\nbar\p\to (\K\Kb)^0\pip)~,
\quad (\K\Kb)^0={\Kp\Km-\K^0\Kb^0\over \sqrt{2}}~,
\end{equation}
and another one where $(\K\Kb)$ has $I=1$
\begin{equation}
\label{theo:eq:iso6}
\mathcal{A}_1=\mathcal{A}\left[\nbar\p\to{1\over\sqrt{2}}\Kp\Kb^0\piz
-{1\over2}(\Kp\Km\pip+\K^0\Kb^0\pip)\right]~.
\end{equation}
One can extract the contribution of $\mathcal{A}_0$ and
$\mathcal{A}_1$ to each of the final states and, after squaring,
deduce (up to a common phase- space factor) that
\begin{equation}
\label{theo:eq:iso7}\eqalign{%
\AF(\Kp\Km\pip)+\AF(\K^0\Kb^0\pip)&=
{1\over2}|\mathcal{A}_1|^2+|\mathcal{A}_0|^2~,\cr
\AF(\Kp\Kb^0\piz)&=
{1\over 2}|\mathcal{A}_1|^2~.}
\end{equation}
Hence
\begin{equation}
\label{theo:eq:iso8}
\AF(\Kp\Km\pip)+\AF(\K^0\Kb^0\pip)\ge \AF(\Kp\Kb^0\piz)~,
\end{equation}
the difference being the $I=0$ contribution. In other words, one gets often
inequalities instead of equalities.
\subsubsection{Isospin equalities for three-body final states}
\label{theo:sub:eq3}
However, one may sometimes get equalities, either by straightforward
Clebsch--Gordan recoupling or by more sophisticated methods.
An example is antinucleon annihilation into two pions on deuterium. Lipkin
and Peskin \cite{Lipkin:1972ju} have shown that
\begin{equation}
\label{theo:eq:eq31}
\sigma(\pbard\to\pim\pip\n)={1\over2}\,\sigma(\pbard\to\pim\piz\p)+2\,
\sigma(\pbard\to\piz\piz\n)~.
\end{equation}
This is based on examination of the dependence of the amplitudes on
the isospin projection $I_z$ of the outgoing pions. Alternatively, one
can write down the amplitudes $\mathcal{A}_I$ corresponding to a
$(\pi\pi)$ pair in isospin $I$, and obtain (again, to an overall
phase-space factor)
\begin{equation}\label{theo:eq:eq32}
\eqalign{%
\sigma(\pbard\to\pim\pip\n)&={2\over3}|\mathcal{A}_0|^2+
{1\over3}|\mathcal{A}_1|^2~,\cr
\sigma(\pbard\to\piz\piz\n)&={1\over3}|\mathcal{A}_0|^2~,\cr
\sigma(\pbard\to\pim\piz\p)&={2\over3}|\mathcal{A}_1|^2~,}
\end{equation}
in the limit of isospin symmetry where, in particular, the initial
state has pure $I=1/2$.

Here, we are dealing with integrated cross-sections. More subtle
effects can be observed if one considers the rate for a given set of
relative angles.
\subsubsection{Charge content of  final states}
As the number of mesons increases, it becomes more difficult to set
limits on the relative abundance of different charge states. Consider
for instance in $n_{\pi}=4$ pions, $\rm p\bar
p\rightarrow\pi^+\pi^+\pi^-\pi^-$, $\pi^+\pi^-\pi^0\pi^0$ and
$\pi^0\pi^0\pi^0\pi^0$. Since intermediate isospin coupling $I\ge2$
are allowed, there is no unique wave-function corresponding to the
total isospin $I=0$ or 1. The relative abundance depends on
the detailed dynamics. Some general results have however been obtained. 
See, e.g., Ref.~\cite{LlewellynSmith:1972pe}.

If $n_+$ is the average number of $\pi^+$, etc., with the obvious
relations
\be\label{kin:eq:reln}
n_++n_0+n_-=n_{\pi}~,\quad\ n_+=n_-~,
\ee
an $I=0$ initial state, which is isotropic in isospin space, will lead
to
\be\label{kin:eq:reln0}
n_+=n_-=n_0=n_{\pi}/3~.
\ee
For the isospin $I=1$ case, it is found \cite{LlewellynSmith:1972pe}
that the ratio
\be\label{kin:eq:reldef}
R(n_\pi)=\frac{n_0}{n_++n_-}~,
\ee
fulfils
\ba\label{kin:eq:rel1}
&&R(2)=0~,\quad \frac{1}{2}\le R(3)\le\frac{11}{4}~,\nonumber\\
&&\frac{n_\pi-2}{4n_\pi+2}\le R(n_\pi) \le \frac{n_\pi-2}{2n_\pi+2}\quad\hbox{if}\ n_\pi\ge4
\ \hbox{and\ is even}~,\\
&&\frac{1}{4}\le R(n_\pi) \le \frac{3n_\pi+2}{2n_\pi-2}\quad\hbox{if}\ n_\pi\ge5 \ 
\hbox{and\ is odd}~.\nonumber
\ea
\subsubsection{Isospin mixing in protonium}\label{kin:sub:imp}
Isospin symmetry is known to be approximate. Isospin breaking
effects are, indeed, observed, such as the masses $m(\piz)$ and $m(\pip)$
being different, or the transition $\Psi'\to\piz+\mathrm{J}/\Psi$ not
vanishing.

The long-range part of the protonium wave function is built by the
electromagnetic interaction and thus corresponds to the simple isospin
combination (using our convention)
\begin{equation}
\label{theo:eq:imp1}\ket{\pbp}={\ket{I=1}-\ket{I=0}\over\sqrt{2}}~.
\end{equation}
If this content had remained the same at short distances, and the strength of
annihilation been  independent of isospin, $\ppb$ annihilation would always
contain $50\%$ $I=0$ and $50\%$ $I=1$.

However, at short distances, the protonium wave function is distorted
by the onset of strong interaction. In particular, charged mesons can
be exchanged. More generally, the $\NNb$ interaction contains an
isoscalar and an isovector part. This latter component induces
$\ppb\leftrightarrow\nbar\n$ transitions.  Potential model
calculations, e.g.\ 
\cite{Kaufmann:1978vp,Richard:1982zr,Carbonell:1989cs,Furui:1990ex},
found that this effect is very large, especially for triplet
P-states. For instance, it is found that $\tpz$ is dominantly $I=0$ at
short distances, i.e., in the region where annihilation takes place.

How reliable are these predictions of potential models? On the one
hand, they are based on the $G$-parity transformation applied to the
part of nuclear forces which is best established, pion exchange. These
long-range forces predict a hierarchy of energy shifts that is well
observed in experiments on protonium spectroscopy, as reviewed in
\cite{Klempt:2002ap}. The hierarchy of widths, 
in particular $\Gamma(\tpz)$ being larger than other $P$-state widths,
is also a prediction of potential models, and this is a crucial
ingredient of cascade calculations used to extract the branching
ratios from measurements done at various values of the density (see
Sec. \ref{se:tm}). On the other hand, when comparing, e.g., $\piz\piz$
and $\eta\piz$ frequencies, one does not observe the hierarchy one
would infer from potential model calculations.  Hence the problem of
the isospin content of annihilation remains open. It will be further
discussed in Sec.~\ref{se:phe}.

The isospin distortion of protonium is described by writing the reduced wave
function of a typical protonium state as
\begin{equation}\Psi={u(r)\over r} \ket{\ppb}+{w(r)\over r}\ket{\nbar\n}~,
\end{equation}
with
\begin{equation}  
\int_0^{+\infty} (|u(r)|^2+ |w(r)|^2)\,\mathrm{d} r=1~.
\end{equation}
In potential models, one can solve the coupled $(\ppb,\,\nbar\n)$ equations,
rearrange $u(r)$ and $w(r)$ into components $u_I$ of given isospin $I$, to separate
the width  into its $I=0$ and $I=1$ components, 
\begin{equation}
\Gamma_I=-\int |u_I(r)|^2\,\Im\mathrm{m} V(r)\,\mathrm{d} r~,
\end{equation}
where $\Im\mathrm{m} V$ could depend on isospin $I$, though, for simplicity,
this was not introduced in early optical models. In this approach, the short-range part
of the potential, including the annihilation component  $\Im\mathrm{m} V$, is tuned to
reproduce the scattering data.

\subsubsection{Isospin content in antiproton-deuterium annihilation}\label{theo:sub:iad}
 The problem of the isospin content in $\pbard$ annihilation has been less frequently
discussed. One can write the wave function as
\begin{equation}
\Psi= u \ket{\pbar\p\n}+ w \ket{\nbar\n\n}= 
a \ket{(\NNb)_{I=0}\,\n}+b\ket{(\NNb)_{I=1}\,\n}~,
\end{equation}
where $u$ and $w$, or $a$ and $b$,  depend on the relative distances of
the particles, through a set of Jacobi variables $\vec{x}$ and $\vec{y}$. If
 $\mathrm{d} \tau$ denotes the integration over these variables, then the isospin distortion can tentatively be expressed in terms of the
probabilities $\alpha$ and $\beta$ being non equal, where
\begin{equation}
\alpha=\int |a|^2\,\mathrm{d}\tau~,\qquad  \beta=\int |b|^2\,\mathrm{d}\tau~.
\end{equation}
Note that one cannot exclude a more complicated scenario where the radial profiles
of $|a|^2$ and $|b|^2$ are rather different, in which case the isospin
distortion would depend of the part of the wave function that is most
explored. For instance, one may argue that a decay with two light
pseudoscalar involves a large momentum and thus the short-range part of the wave
function, while two heavier resonances are produced with small momentum by
the external part of the annihilation region.

\clearpage\markboth{\sl Annihilation dynamics}{\sl Global features}
\setcounter{equation}{0}
%
\sectionb{Global features of $\NNb$  annihilation}\label{se:glob}
Annihilation can be described by a few simple variables of a statistical
nature such as the mean number of pions and their respective
multiplicity distribution, the fraction of events in which an $\eta$
is produced or the fraction with strange mesons in the final state.
The inclusive momentum distribution can be used to argue for a
thermodynamic picture of annihilation. Pion interferometry can give information
about the correlation of pions and hence the size of the fire-ball
formed by the annihilating proton and antiproton.

The frequencies of annihilation modes such as $\rm p\bar
p\to\pi^+\pi^-\pi^+\pi^-\pi^0$ are needed when annihilation
frequencies for intermediate 2-body modes like $\rho\omega$ are to be
determined. In this section we give a survey on these global aspects
of annihilation.
\subsection{Pionic multiplicity distribution}\label{glob:sub:piomult}
In bubble chamber experiments, global annihilation frequencies can be
determined by a thorough scan of events. Special care is needed to
avoid contamination by Dalitz pairs (e.g., from $\pi^0\to\gamma
e^+e^-$) faking charged pions.  A correction needs to be applied for
charge-exchange scattering at the end of the antiproton range
simulating zero-prong annihilation.  Such scans were performed for
  bubble-chamber experiments at Brookhaven and CERN.  The scan at
CERN gave the following results for the charged-particle multiplicity
distribution:
\begin{table}[h!]
\caption{\label{glob:tab:freq}%
Charged-particle multiplicity distribution (from
\protect\cite{Armenteros:1980})}
\renewcommand{\arraystretch}{1.2}
\begin{tabular}{lcr}
\hline\hline
0 prongs           &\qquad\qquad  &$       4.1^{+0.2}_{-0.6}\% $  \\
2 prongs           &  &$      43.2^{+0.9}_{-0.7}\% $  \\
4 prongs           &  &$      48.6^{+0.9}_{-0.7}\% $  \\
6 prongs           &  &$       4.1^{+0.2}_{-0.2}\% $  \\
\hline\hline
\end{tabular}
\renewcommand{\arraystretch}{1.0}
\end{table}

From Table~\ref{glob:tab:freq} we derive the mean number of charged pions
per \pbar p annihilation to be
\begin{equation}
n_{\pi^{\pm}} = 3.054\,{}^{+0.040}_{-0.036}~.
\end{equation}

In events in which the momenta of all particles are measured, a new
set of kinematical variables can be derived that automatically satisfy
energy and momentum conservation. The new momenta are then improved in
accuracy. In events with all particles reconstructed, four constraints
can be used; such fits are called fits with four-constraints or 4C
fits. Due to the nature of bubble chamber experiments, only charged
particles are detected. But kinematical constraints allow the
three-momentum of one unseen neutral particle to be reconstructed in a
1C kinematical fit.

The bubble chamber data were split into classes with defined number of
charged tracks and fitted to different kinematical hypotheses. When no
visible \Ks\ was present, the tracks were assumed to correspond to
pions. A kinematical fit identified with high reliability events
without missing particles (4C events); events with one missing \piz\
(1C events) contained up to 12-14\% contamination of 2\piz\
events. Events not passing the 4C or 1C hypothesis were called {\it
missing-mass} events. The distribution of pionic states to \ppb\
annihilation at rest is shown in Table \ref{glob:tab:pimult}, which
is based on publications of the Columbia group, on a compilation
of (partly unpublished) CERN results
\cite{Armenteros:1980}, and on Crystal Barrel data, also partly
unpublished.
\begin{table}[!h]
\caption{\label{glob:tab:pimult}
Annihilation frequencies of \protect$\pbp$ annihilation at rest
in liquid H$_2$ into pionic final states (in units of $10^{-3}$),
from \cite{Armenteros69,Baltay66,Amsler:2003bq}. Events with more
than one $\pi^0$ cannot be reconstructed without neutral particle
detection; they are listed as {\it missing mass} (MM) events.  In
the Crystal Barrel experiment, events with $\eta\to\gamma\gamma , 
\pi^+\pi^-\gamma$, $\omega\to\pi^0\gamma$ and a fraction
of $\eta^{\prime}$ decays do not lead to multi--pionic final states,
they contribute to the Crystal Barrel {\it missing mass} events.
Their fraction is estimated in the text.  Final states with more
than 5$\pi^0$ are difficult to reconstruct.  Their contribution
is estimated by assuming that one ($^1$) or two ($^2$) $\eta$ mesons
were produced and decayed into 3$\pi^0$. Their rate was measured
from the $\eta\to 2\gamma$ decay mode.}
\renewcommand{\arraystretch}{1.1}
\begin{tabular}{lccc}
\hline\hline
\quad Final state             &   BNL     \qquad   & CERN  \qquad  &
Crystal Barrel \\ 
\hline \quad all neutral             &\quad
$32\pm 5$& $41\,{}^{+2}_{-6}$ &  $35\pm 3 $\\ 
\quad 2$\pi^0$                &                         &  & $0.65\pm 0.03$ \\ 
\quad 3$\pi^0$                & & & $7.0\pm 0.4$\\ 
\quad 4$\pi^0$                & &  & $3.1\pm 0.2$\\ 
\quad 5$\pi^0$                &&  & $9.2\pm 0.4$\\ 
\quad 6$\pi^0$ \ $(^1)$       & &  & $0.12\pm 0.01$\\ 
\quad 7$\pi^0$ \ $(^1)$       & &  & $1.3\pm 0.1$\\ 
\quad 8$\pi^0$ \ $(^2)$       &&  & $0.012\pm 0.001$\\ 
\quad 9$\pi^0$ \ $(^2)$       &&  & $0.025\pm 0.003$\\ 
non-multipion                 &&&$15\pm 5$\\ 
\hline \quad $\pi^+\pi^-$  &$3.2\pm 0.3$   & $  3.33\pm
0.17$ & $3.14\pm 0.12$\\ 
\quad $\pi^+\pi^-\pi^0$ & $ 78\pm 9$&$  69.0\pm 3.5$& $67\pm 10$\\ 
\quad
$\pi^+\pi^-2\pi^0$      &                         &  & $122\pm 18$\\ 
\quad $\pi^+\pi^-3\pi^0$      &              &  &$133\pm20$\\ 
\quad $\pi^+\pi^-4\pi^0$      &&  & $36\pm 5$\\ 
\quad $\pi^+\pi^-5\pi^0$ \ $(^1)$&&  &$13\pm 2$\\ 
\ \quad $\pi^+\pi^-$MM          & $345\pm 12$&      $358\pm 8 $&$65\pm 20^*$\\ 
\hline 
\quad$2\pi^{+}2\pi^-$         &$     58\pm 3 $&$  69\pm 6  $ &$ 56\pm9$\\ 
\quad $2\pi^{+}2\pi^-\pi^0$   &$    187\pm 7 $&$ 196\pm 6  $ &$210\pm 32$\\ 
\quad $2\pi^{+}2\pi^-2\pi^0$  &               &              &$177\pm 27 $\\ 
\quad $2\pi^{+}2\pi^-3\pi^0$  &               &              &$  6\pm 2  $\\ 
\quad $2\pi^{+}2\pi^-$MM      &$    213\pm 11$&$ 208\pm 7  $ &$ 30\pm 15^*$\\
\hline \quad $3\pi^{+}3\pi^-$ &$ 19\pm 2     $&$21.0\pm 2.5$ &  \multirow{3}{1.5cm}{$\Biggl\}\, 40\pm 3^*$}             \\
\quad $3\pi^{+}3\pi^-\pi^0$   &$     16\pm 3  $&$18.5\pm 1.5$&   \\
\quad $3\pi^{+}3\pi^-$MM      &$     3\pm 1   $&$  3 \pm 1$  &  \\
\hline 
Sum    &$954\pm 18$      &    $986\pm 6$ &$970\pm 58$ \\
\hline\hline
\end{tabular}
\renewcommand{\arraystretch}{1.0}
\vskip .05mm
\centerline{\footnotesize{$^*$ Including final states with open strangeness
and other non-multipion events.}}
\end{table}
The three frequency distributions are differently normalised: the CERN
data exclude only events with an detected \Ks . The BNL distribution
is corrected for all annihilation modes containing kaons. Crystal
Barrel data are given as branching fractions determined from exclusive
final states. Hence data containing an $\eta$ are not included for
$\eta\to\gamma\gamma , \pi^+\pi^-\gamma$ decays. With an estimated
inclusive $\eta$ frequency of 7\% (see below), about $3.2$\% of all
annihilation events are lost. The inclusive $\omega$ rate is not
known; if one $\omega$ is produced in 10 annihilations, there is a
fraction of 0.9\% of all annihilation not leading to one of the final
states listed in Table~\ref{glob:tab:pimult}. Similarly, 1.5\% of all
annihilations are missing if the (unknown) inclusive $\eta^{\prime}$
rate is 2\%. Including the 5.4\% kaonic annihilations (see below), we
expect altogether 11\% of all annihilations not to contribute to the
frequencies in Table~\ref{glob:tab:pimult}. We assign (educated guess)
6.5\% to 2-prong, and 3\% to 4-prong and 1.5\% to zero-prong
annihilation, respectively. The total Crystal Barrel 6-prong yield is
taken from the BNL and CERN results.

We notice a very significant even--odd staggering in the frequencies
of multi-$\pi^0$ final states: annihilation into $2n\pi^0$ is
reduced  since their production from S-states is forbidden or
suppressed.  As seen in Sec.~\ref{kin:sub:cons}, $\stso$
and $\ttso$ cannot decay into any number of $\pi^0$ by charge
conjugation invariance, $\tssz$ is
forbidden by $G$-parity for any $2n\piz$ mode;
$\sssz$ cannot decay into $2\piz$, and its decays into
$4\piz$ or into higher even modes are anyhow suppressed by the internal
orbital barrier required to match parity and Bose statistics.

 From Table~\ref{glob:tab:pimult} we derive frequencies into multipion
final states. They are presented in Table~\ref{glob:tab:stat} and
compared to the frequency distribution estimated by Ghesqui\`ere
invoking arguments based on isospin invariance
\cite{Ghesquiere:1974}. The frequencies of \cite{Ghesquiere:1974}
are normalised to unity; in the central column, only multi-pion final
states are included and the expected sum of all contributions
is 88\%.

\begin{table}[h!]
\caption{Pionic multiplicity distribution.\label{glob:tab:stat}}
\renewcommand{\arraystretch}{1.2}
\begin{tabular}{lrrr}
\hline\hline
&& From Table~\protect\ref{glob:tab:pimult}
&From \protect\cite{Ghesquiere:1974}  \\
\hline
\quad 2 pions  &\qquad &$0.38\pm 0.03\%$ \quad &$0.38\pm 0.03\%$ \\
\quad 3 pions  &\qquad &$ 7.4\pm 0.3\% $ \quad &$ 7.8\pm 0.4\% $ \\
\quad 4 pions  &\qquad &$18.1\pm 1.8\% $ \quad &$17.5\pm 3.0\% $ \\
\quad 5 pions  &\qquad &$35.2\pm 3.7\% $ \quad &$45.8\pm 3.0\% $ \\
\quad 6 pions  &\qquad &$23.3\pm 2.8\% $ \quad &$22.1\pm 1.5\% $ \\
\quad 7 pions  &\qquad &$ 3.3\pm 0.3\% $ \quad &$ 6.1\pm 1.0\% $ \\
\quad 8 pions  &\qquad &  \quad &$ 0.3\pm 0.1\% $ \\
\hline\hline
\end{tabular}
\renewcommand{\arraystretch}{1.0}
\end{table}

The mean number of pions per annihilation into
multi-pionic events is estimated to 
\begin{equation}
n_{\pi} = 4.98\pm 0.35~, \qquad n_{\pi^{\pm}} = 3.14\pm 0.28~, \qquad
n_{\pi^{0}} = 1.83\pm 0.21~.
\label{glob:eq:n1neutral}
\end{equation}
Using Table~\ref{glob:tab:freq} and the number of
photons per annihilation $n_{\gamma} = 3.93\pm 0.24$
\cite{Backenstoss:1983gu} (and allowing for
small fractions from $\eta\to\gamma\gamma$ and similar
decays), the numbers in (\ref{glob:eq:n1neutral}) can be refined to
\begin{equation}
n_{\pi} = 4.98\pm 0.13~, \qquad n_{\pi^{\pm}} = 3.05\pm 0.04~,
\qquad n_{\pi^{0}} = 1.93\pm 0.12~.
\label{glob:eq:n2neutral}
\end{equation}
The inclusive $\eta$ rate was determined to be 
\begin{equation}
n_{\eta} = 0.0698\pm 0.0079~,
\label{glob:eq:eta}
\end{equation}
per annihilation \cite{Chiba:1989yw}.

\begin{figure}[tbh!]
\setlength{\unitlength}{.01\textwidth}
\begin{picture}(50,41)(25,6)
\put(0,0){\vbox{\includegraphics[width=50\unitlength]{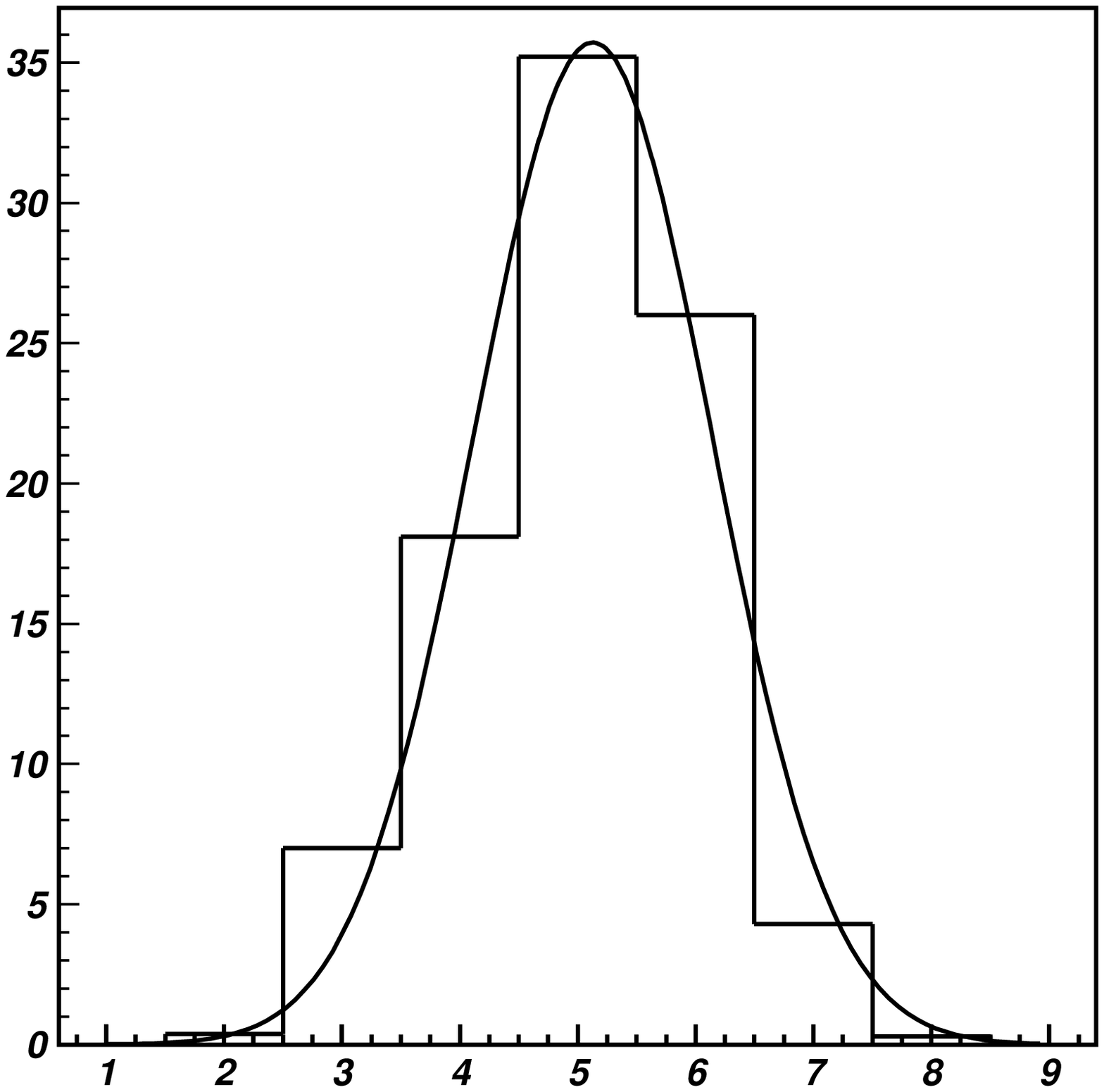}}}
\put(26,40){$\%$}
\put(64,2){$n_\pi$}
\end{picture}
\caption{\label{glob:fig:pimult}The pion multiplicity distribution
(in \%) from Crystal Barrel data, Table~\protect\ref{glob:tab:stat}.}
\end{figure}

Figure~\ref{glob:fig:pimult} shows the pion multiplicity distribution
from $\pbp$ annihilation at rest and a comparison with a Gaussian fit.
The mean number of pions is now $5.03\pm 0.05$, the width is $1.13\pm
0.07$. Amado {\it et al.} \cite{Amado:1994ym} have shown that a fit
with a Poisson distribution constrained by energy and momentum
conservation reproduces both the mean value and the variance.
\subsection{Inclusive spectra\label{glob:sub:incl}}
\begin{figure}[tbh]
\begin{tabular}{ccc}
\hspace*{-5mm}\includegraphics[width=0.32\textwidth]{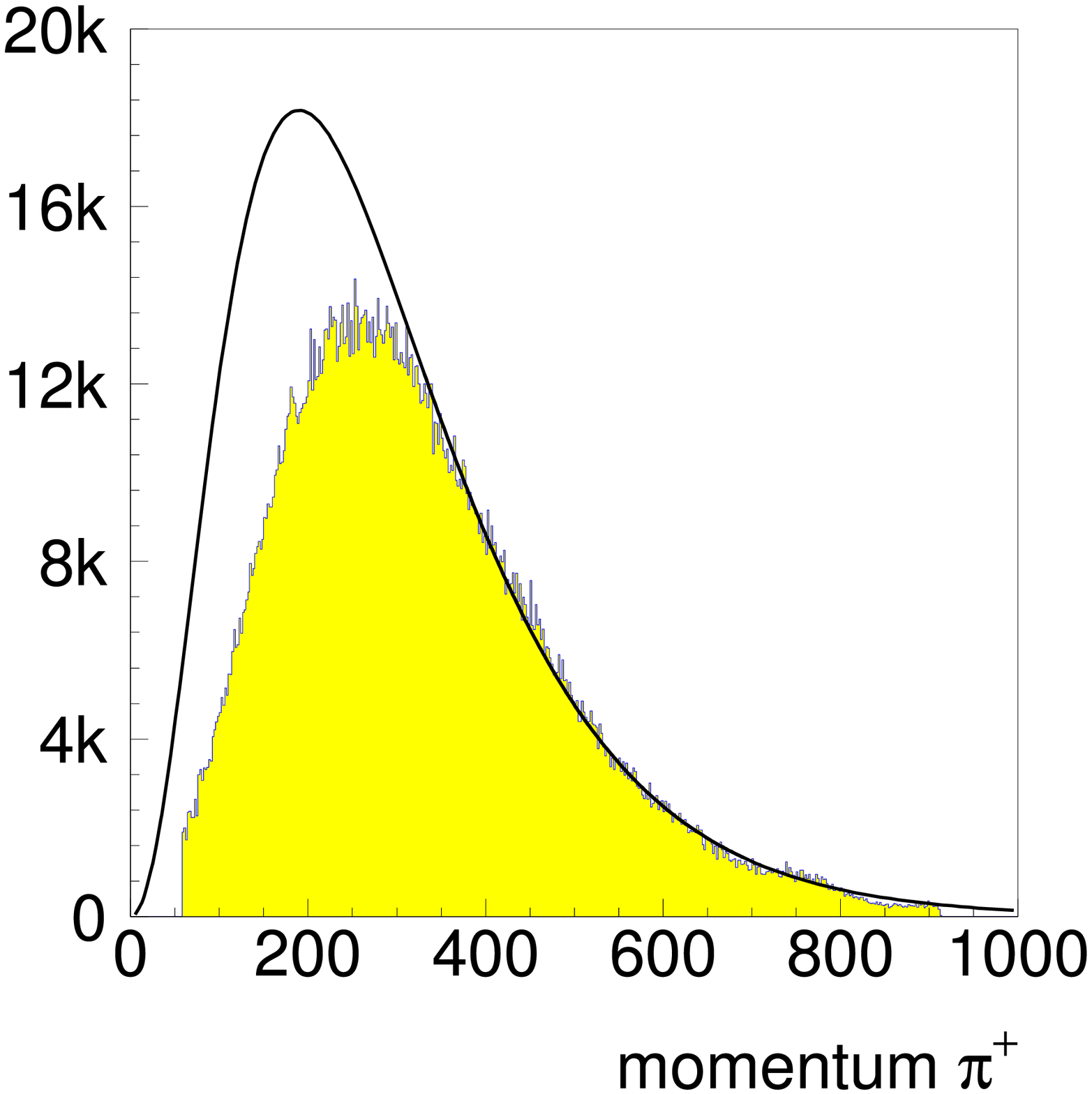}&
\hspace*{-3mm}\includegraphics[width=0.35\textwidth]{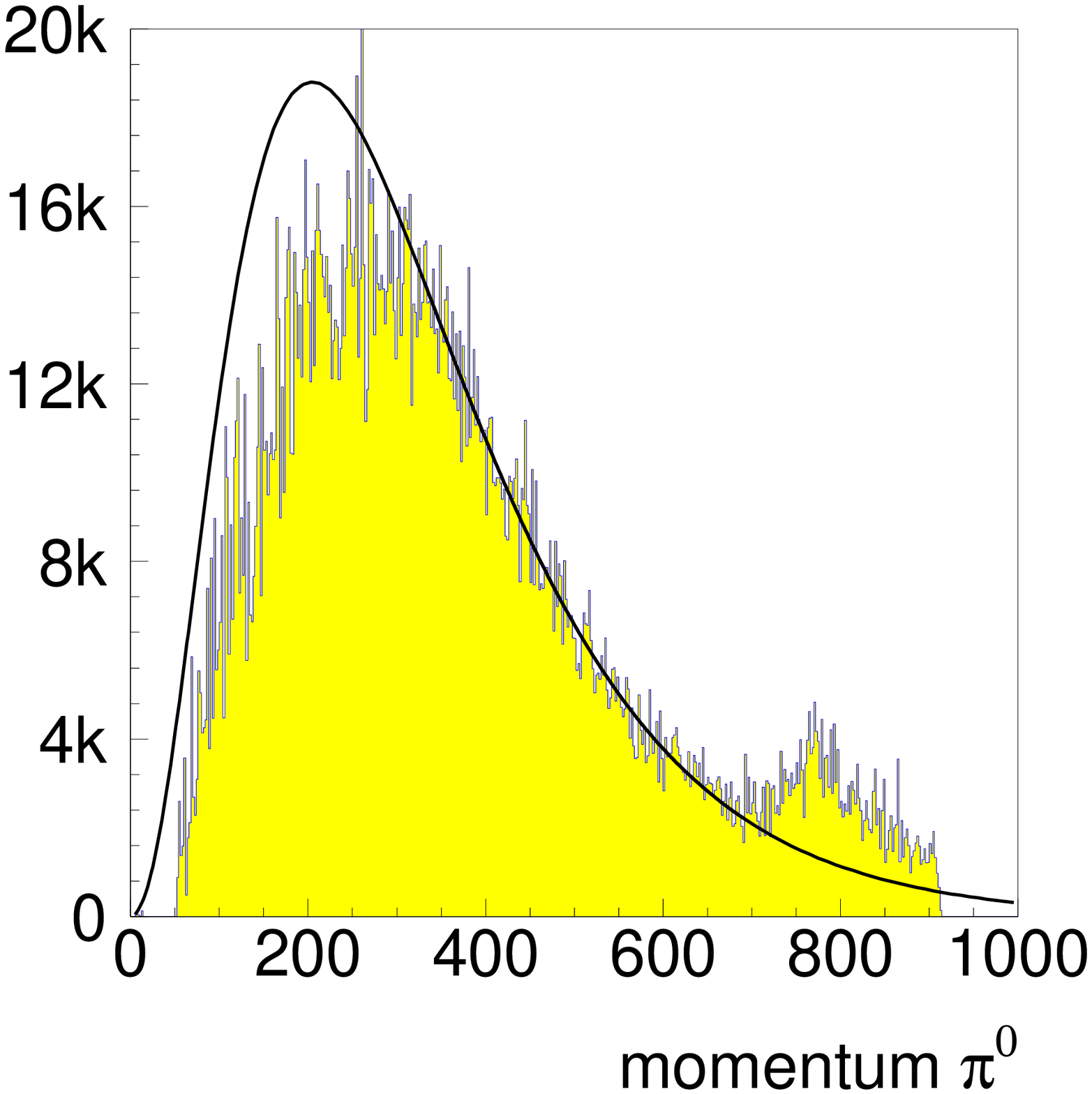}&
\hspace*{-5mm}\includegraphics[width=0.35\textwidth]{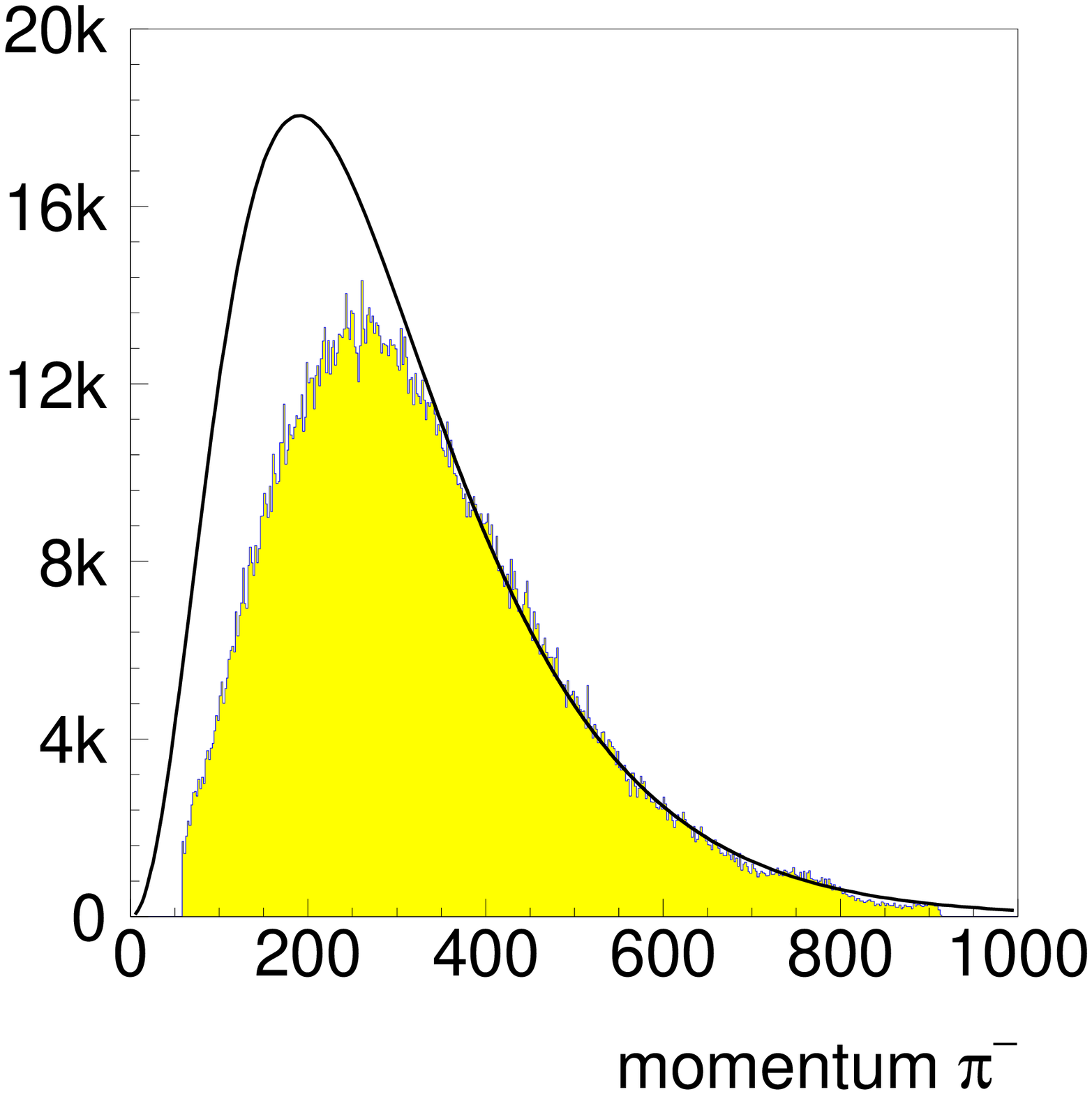}
\end{tabular}
\caption{\label{glob:fig:pion}%
The momentum distribution of charged and neutral pions.}
\end{figure}

The inclusive momentum spectrum of charged particles (mostly pions) is
shown in Fig.~\ref{glob:fig:pion}. The data are from the Crystal
Barrel collaboration. The distribution reveals no significant
structure, except for $\rho\pi$ production which identifies itself as
a peak at 773\,MeV\,/\,c in the momentum distribution. The $\rho^0$
signal seems much more pronounced; this is an artefact of the
experimental resolution which is better for the recoiling $\pi^0$ than
for $\pi^{\pm}$ mesons.

The absence of narrow signals in the momentum spectrum -- which would
indicate production of narrow quasinuclear bound states -- is
evident. Experiments in the early phase of LEAR confirmed the absence
of narrow states against which charged or neutral pions would recoil
\cite{Ahmad:1984bp,Playfer:1985ey,Adiels:1986tr}.

 From the number of events in Fig.~\ref{glob:fig:pion} and the number
of annihilation events, the average multiplicities
\begin{equation}
n_{\pi} = 5.19\pm 0.15~, \qquad n_{\pi^{\pm}} = 3.12\pm 0.12~,
\qquad n_{\pi^{0}} = 2.07\pm 0.08~,
\label{glob:eq:n3neutral}
\end{equation}
are found, consistent with those given in
Eqs.~(\ref{glob:eq:n1neutral}) and~(\ref{glob:eq:n2neutral}).

The fit in Fig.~\ref{glob:fig:pion} corresponds to a
Maxwell--Boltzmann momentum distribution as proposed by Orfanidis and
Rittenberg \cite{Orfanidis:1973ix}. In the high-momentum range the fit
follows the experimental distribution adequately thus defining a
temperature of about 120 MeV.  There is a mismatch at low momenta
which is more pronounced for charged than for neutral pions. This is
not due to a reduced efficiency of the detector for low momentum
particles. Otherwise, the frequencies~(\ref{glob:eq:n3neutral}) would be
smaller than those in~(\ref{glob:eq:n1neutral})
and~(\ref{glob:eq:n2neutral}); rather, pions at low momenta do not
follow a simple thermal distribution law.

The temperature of 120 MeV should not be interpreted as an annihilation
temperature, nor its inverse as an annihilation range of 1.7\,fm.
Pion interferometry gives a similar range; we will discuss the reasons
for this wide range below.
\subsection{Pion interferometry\label{glob:sub:infero}}
In the fireball picture of \pbar p annihilation, pions are emitted
stochastically from an extended source. It can be argued that the pion
fields emitted from different space-time points 1 and 2 superpose and
interfere. The probability to detect a pion is then given by
%
%
\begin{equation}
dP_{12} \propto dr_1dr_2\,\left\vert e^{\displaystyle 
i[p_1(x_1-r_1)+p_2(x_2-r_2)]} +
                          e^{\displaystyle 
i[p_1(x_1-r_2)+p_2(x_2-r_1)]}\right\vert^2~.
\end{equation}
It is worthwhile to recall that such interferences were first used in
astronomy, by Hanbury-Brown and Twiss (HBT)
\cite{HanburyBrown:1954wr}, to determine the
diameter of Sirius from a measurement of the correlated intensity
fluctuations in two optical telescopes. G.\ and S.\ Goldhaber, Lee and
Pais \cite{Goldhaber:1960sf} applied the effect to estimate the
size in space--time of the source emitting pions in \pbar p
annihilation.

Different correlation signals have been defined to extract the correlation
between pions due to the HBT effect. We mention here the
two-pion correlation function
\begin{equation}
C(p_1,p_2) = \frac{\rho_2(p_1,p_2)}{\rho_0(p_1,p_2)}\,,
\end{equation}
where $\rho_2(p_1,p_2)$ represents the distribution of two mesons
correlated due to the HBT effect, and $\rho_2(p_1,p_2)$
an uncorrelated sample. The uncorrelated distribution can be chosen
as the product of two single particle distributions
$\rho_2(p_1,p_2) = \rho_1(p_1)\rho_1(p_2)$, where
\begin{equation}\eqalign{%
\rho_1(p_i)
&=\frac{1}{\sigma}\frac{d\sigma}{d^3P_i/(2E_i)},\qquad i=1,2~,   \cr
\rho_2(p_1,p_2)
&= \frac{1}{\sigma}\frac{d\sigma}{d^3P_1/(2E_1)d^3P_2/(2E_2)}\,.
}
\end{equation}
The correlated sample may be pion pairs of equal charge;
for the uncorrelated pion pair, pions of different charge or from
different events can be used.

In $\pbarp$ and $\pbard$ annihilation, the HBT effect was used by
different groups. The precise source parameters depend on the data
(annihilation into $2\pi^+2\pi^-$ or $2\pi^+2\pi^-n\pi^0$) and the
model (with unlike-sign pions or different events to determine the
uncorrelated dipion spectrum). The results given in
\cite{Deutschmann:1982vz} and \cite{Riedlberger:1989kn} are not compatible
with each other within the quoted errors; the value for
the Bose--Einstein correlation parameter,
\begin{equation}
r_{\rm BE} = c\tau_{\rm BE} = 1.5\pm 0.3\,{\rm fm}\,,
\label{glob:eq:hbt}
\end{equation}
seems to be a fair estimate of the dimension of the pion source in
space and time. It corresponds approximately to the pion Compton wave
length $\lambda_{\pi}$. Again, this is not the size of the
annihilation source. Rather, it is the size of the source from which
pions are emitted. Pions may be emitted as initial- or final-state
radiation (even though there is no experimental support for these
processes); but certainly, pions are produced in secondary decays over
a wide range of distances. Equation~(\ref{glob:eq:hbt}) indicates that
interference between pions is strong for distances corresponding to
$\lambda_{\pi}$.  This is a trivial statement as long as the
annihilation range is small compared to $\lambda_{\pi}$.

The result (\ref{glob:eq:hbt}) is in rather strong disagreement with
recent findings by Locher and Markushin using CPLEAR
\cite{Angelopoulos:1998ix,Apostolakis:1998fy} and Crystal Barrel
\cite{Kortner:2002mp} data. These authors go beyond the conventional HBT
analysis by plotting the double-differential cross-sections as a function
of the invariant masses $M_{++}$ and $M_{--}$, where the sign stands for
the charge of the dipion.  In a first article, devoted to $\pbarp
\to2\pi^+2\pi^-$, a source dimension of the order of
0.4\,fm is quoted, in fair agreement with the value we will derive in
Sec.~\ref{se:phe} from the systematics of two-body annihilation
frequencies. Locher and Markushin warn the reader that the
interpretation of the observed correlation signal by the conventional
HBT effect is questionable.  They observe strong enhancements at low
values of $M_{++}$ and $M_{--}$, and notice that the signal
for pion pairs of large momenta ($\sim 800$\,MeV/c) -- which should be
sensitive to the source dimension of 0.4\,fm -- gives a coherence
which is unreasonably large within the HBT framework.  In the analysis
of the $\pbp\to 2\pi^+2\pi^-\pi^0$ reaction, they show that the
enhancement at low values of $M_{++}$ and $M_{--}$ can be simulated as
the effect of the trivial need to symmetrise isobar amplitudes. Best
suited is an interference between $\rho\sigma\pi$ and $\rho\rho\pi$
amplitudes.  Such an ansatz reproduces also the momentum dependence of
the correlation signal. The author thus refuse to provide simple
numbers on the size and life time of the hypothetical $\pbarp$
fire-ball.  Also in the case of the reaction $\pbp\to 4\pi^0$, an
interpretation within the conventional isobar model is possible, and
there is no evidence for an additional signal due to the HBT effect.
It may be worthwhile to recall that the HBT effect has not been taken into
account explicitly in partial-wave analyses of bubble-chamber or
LEAR data; it is only partly accounted for by Bose--Einstein
symmetrisation of the amplitudes.  There is no parameter to describe
the source dimension.
\subsection{Strangeness production\label{glob:sub:strangeprod}}
The full data sets at BNL and CERN were scanned in searches for $\Ks$ decays into $\pi^+\pi^-$, leading to secondary vertices. These
data samples comprised 40,000 and 20,000 events, respectively, with
strange particles in the final state. In the presence of one $\Ks$, the ionisation density of the tracks was used to identify the charged-kaon track.  Table \ref{glob:tab:kmult} gives annihilation
frequencies with an observed $\Ks$.
\begin{table}[hbt!]
\caption{\label{glob:tab:kmult}%
Frequencies of \ppb\ annihilation at rest in liquid H$_2$ into kaonic
final states (in units of $10^{-3}$) from \cite{Armenteros69}. The
data are corrected for unseen decay modes of the $\Ks$ and for
the $\Ks$ reconstruction efficiency. $\K^0$ stands for the sum of
$\Ks$ and $\Kl$. For the derivation of the $\Ks\Kl\pi^0$ frequency see \cite{Amsler:2003bq}.}
\renewcommand{\arraystretch}{1.2}
\begin{tabular}{lccl}
\hline\hline
\quad Final state 
&        BNL              &   CERN  \quad \\
\hline
\quad $\K^+\K^-$ 
& $1.10\pm 0.10$ & $0.96\pm 0.08$ \\
\quad $\Ks\Ks+\Kl\Kl$&\quad 
$0.010^{+0.012}_{-0.010}$
                      \quad\quad &\quad 
$0.008\pm 0.008$\quad\quad \\
\quad $\Ks\Kl$           & $0.71\pm 
0.10$ & $0.80\pm 0.05$ \\
\quad $(\Ks\Ks+\Kl\Kl)\pi^0$&$1.46\pm 
0.20$&$1.56\pm 0.12$\\
\quad $\Ks\Kl\pi^0$      & $0.67\pm 0.07$ & 
$0.67\pm 0.07$ \\
\quad $(\Ks\Ks+\Kl\Kl)$MM 
 
& $1.28\pm 0.16 $& $1.42\pm 0.26$  \\
\quad $\Ks\K^{\pm}\pi^{\mp}$ 
& $4.25\pm 0.55$ & $4.25\pm 0.20$ \\
\quad 
$(\Ks\Ks+\Kl\Kl)\pi^-\pi^+$ 
                                 & 
$4.02\pm 0.52$ & $3.90\pm 0.46$ \\
\quad $\Ks\Kl\pi^-\pi^+$ & 
$2.41\pm 0.36$ & $2.26\pm 0.45$ \\
\quad 
$\K^0\K^{\pm}\pi^{\mp}\pi^0$\quad\quad 
 
& $8.94\pm 1.06$ & $9.38\pm 1.10$ \\
\quad 
$(\Ks\Ks+\Kl\Kl)\pi^-\pi^+\pi^0$ 
                                 & 
$2.98\pm 0.44$ & $2.20\pm 0.28$ \\
\quad 
$\K^0\K^{\pm}\pi^{\mp}\pi^-\pi^+$\quad\quad 
 
& $0.59\pm 0.08$ & $0.71\pm 0.07$ \\
\quad 
$\K^0\K^{\pm}4\pi$\quad\quad 
                                 & 
$\sim 0$       & $\sim 0$       \\
\hline 
\quad Sum 
& $28.4\pm 1.5$ & $28.1\pm 1.4$ 
\\
\hline\hline
\end{tabular}
\renewcommand{\arraystretch}{1.0}
\end{table}
 From Table \ref{glob:tab:kmult} the fraction of events with at least one
neutral  $\Ks$ is determined to be
\begin{equation}
\BR(\pbp\to\Ks + {\rm anything}) = (1.55\pm 0.06)\%\,.
\label{glob:eq:kfrequ}
\end{equation}

The rate for  \Kl\ production is obviously the same.
The total yield of strange particle production has to include
final states with $\K^+\K^-$ pairs. Based on a scan searching for high
ionisation-density tracks, Armenteros et al. found a contribution
of $(6.82\pm 0.25)$\% to \pbp\ annihilation. In the CERN
list of pionic annihilation modes, these events are included
in the {\it missing mass} class of events. The BNL group assigns
$(4.6\pm 1.8)$\% of all annihilations to strangeness production, which
agrees with the estimate $(4.74\pm 0.22)$\%
of Batusov \cite{Batusov:1990rw} assuming
that annihilation frequencies for $\pbp\to\Kp\Km n\pi$ are similar to
those into $\Kn\Kb^0 n\pi$. A statistical treatment of these
three numbers being not plausible, we quote here their linear mean and spread
\begin{equation}
\BR(\pbp\to {\rm kaons} + {\rm anything}) = (5.4\pm 1.7)\%\,.
\label{glob:eq:kfrequ1}
\end{equation}
In short, one event out of 20 contains strange particles in the final state.
\subsection{Annihilation on neutrons\label{glob:sub:neutron}}
Antiproton--neutron or antineutron--proton interactions at rest offer
additional opportunities to study annihilation dynamics. Obviously,
both systems have isospin $I=1$. Under the hypothesis that annihilation
at rest takes place when the $\NNb$ system is in S-wave, $G$-parity
fixes the total spin: a triplet $S=1$ has $G=(-1)^{L+S+I}=+1$ and
leads to an even number of pions, while the spin singlet
system gives an odd number of pions.

Since there exists no free-neutron target, the cleanest way to study
pure isovector annihilation is to build a antineutron beam line
(produced by charge exchange $\ppb\to\nbar\n$) at very low
energies. The Obelix collaboration has made extensive use of
this possibility \cite{Bressani:2003pv}
to study specific reactions as a function of the
\nbar\ beam momentum. These results will be
presented in Sec.~\ref{se:tm}. The topological annihilation frequencies
agree within errors with those for \pbn\ \cite{Caro:1975qh}
even though in the latter case, one has to worry about the role of the
proton or neutron surviving the annihilation process.
Global features of annihilation were derived by use of a deuterium target.

\begin{figure}[tbh]
\includegraphics[width=.35\textwidth]{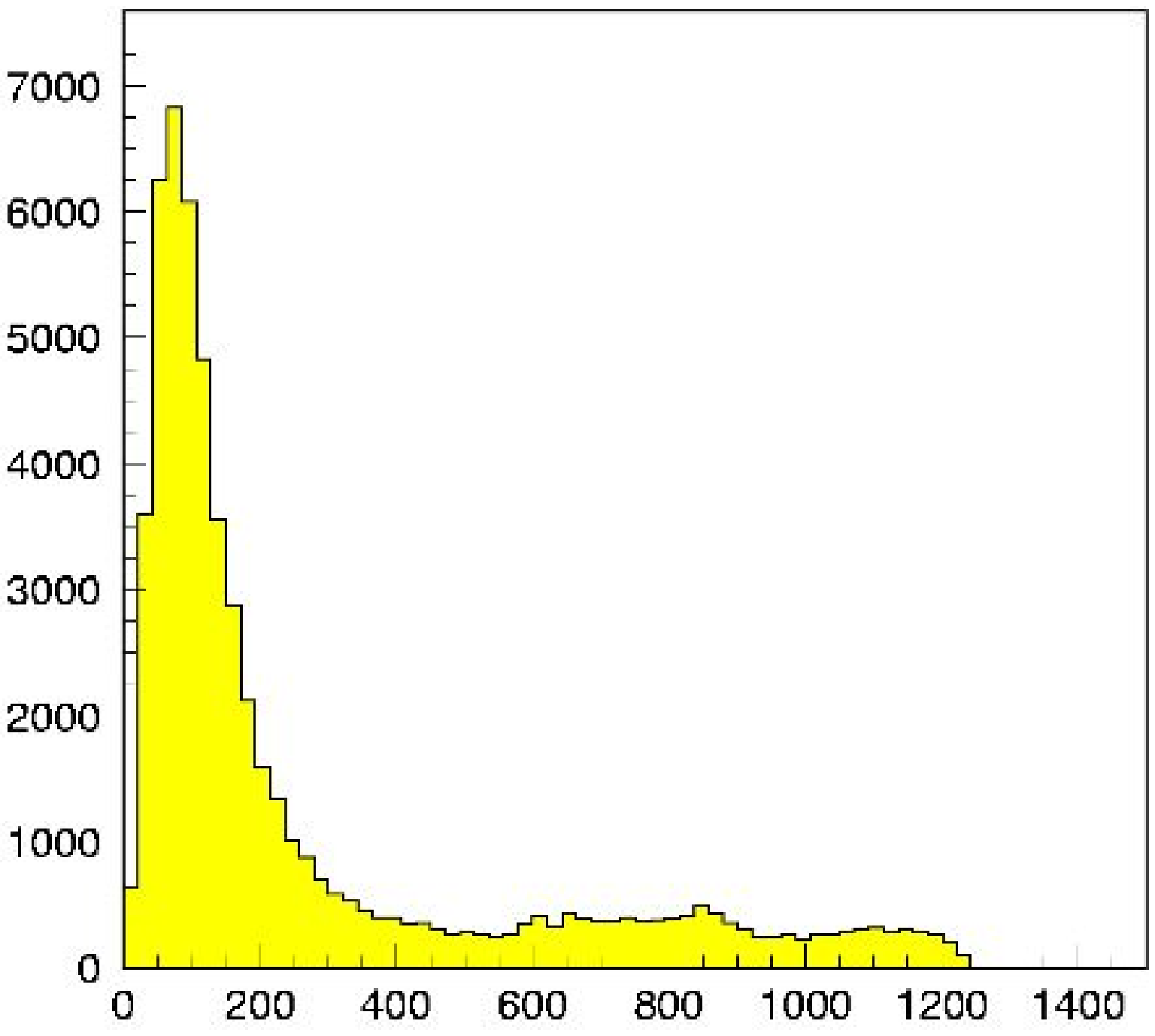}
\hspace*{1cm}
\raise 9pt\hbox{\includegraphics[width=.345\textwidth]{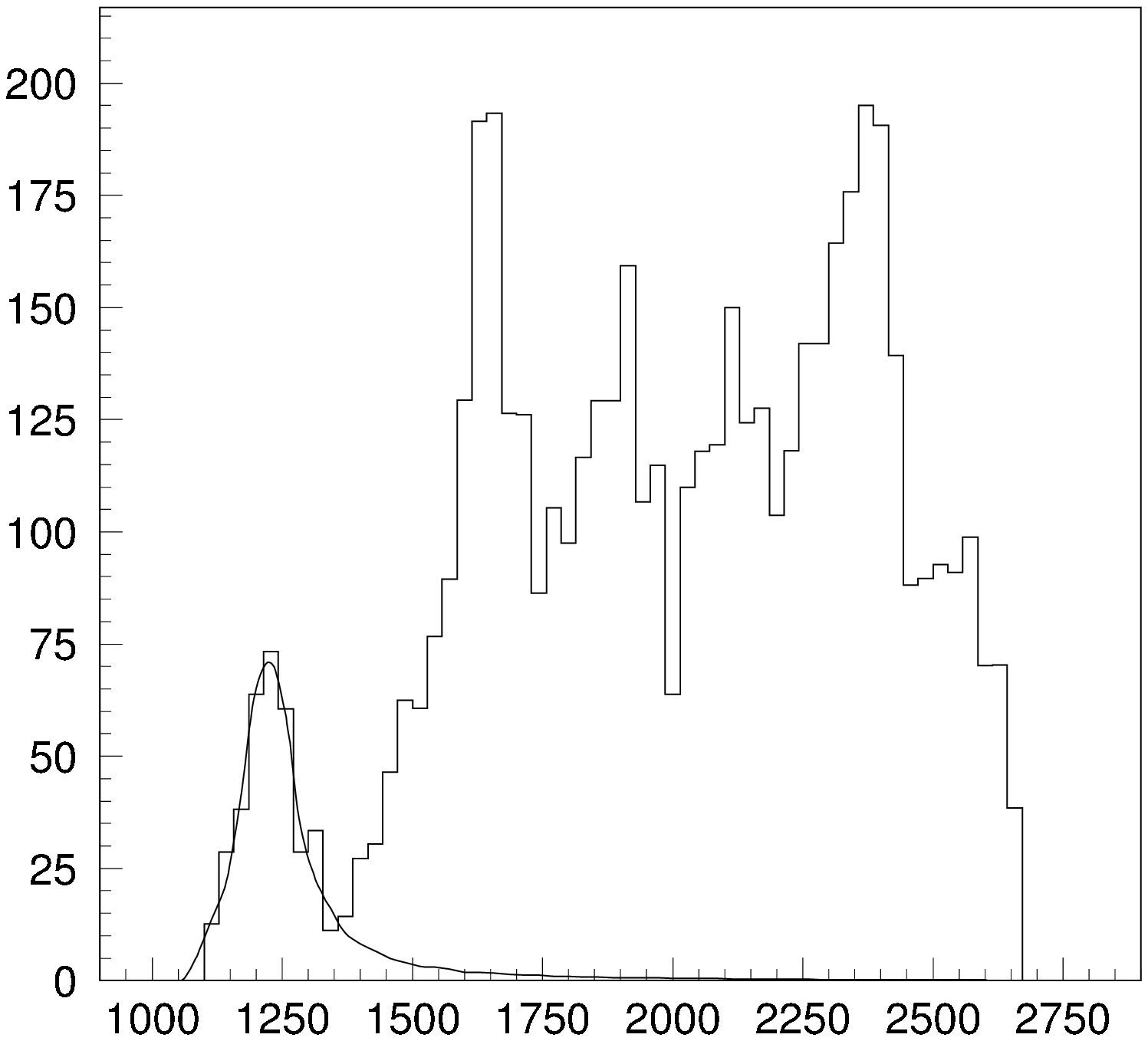}}
\caption{\label{glob:fig:spec}Left: Neutron momentum distribution; right:
p$\pi^0$ invariant mass distribution from
the reaction $\rm\bar pd\to n2\pi^0$.}
\end{figure}

Naively, it may be expected that \pbn\ annihilation on a deuteron
can be viewed at as two-step process:
\be
\pbd \to (\pbn) +\p~,  \qquad {\rm or} \qquad \pbd \to \pbp + \n~,
\ee
with subsequent annihilation of the \pbn\ or \pbp\ system. This
is, however, a crude approximation. Figure~\ref{glob:fig:spec} shows 
the neutron
momentum distribution for the reaction $\pbd\to\n 2\pi^0$ from
the Crystal Barrel experiment \cite{Amsler:1995nw}.
The neutron momentum distribution does not follow the Hulth{\'e}n
function. There is a significant excess of high-momentum neutrons.
The reason can be understood once the p$\pi^0$ invariant mass is
plotted, see Fig.~\ref{glob:fig:spec}, right.  Baryon resonances are
produced, e.g., in the reaction
\begin{equation}
\pbd \to \Delta^0 \pi^0 ; \ \Delta^0 \to \n \pi^0~.
\label{glob:eq:ponte-delta}
\end{equation}
This annihilation mode is called a Pontecorvo reaction; the surviving
nucleon, proton or neutron, can be produced at a rather high
momentum. In this case, a high-momentum $\Delta (1232)$ is produced.
The systematics of Pontecorvo reactions will be discussed in
Sec.~\ref{tm:sub:Ponte}.

In the low-momentum part in Fig.~\ref{glob:fig:spec}, the neutron has acquired
little momentum; it acted as a spectator and was not involved in the
annihilation process. A cut on this momentum, at 200 to 250 MeV/c in
the analysis of bubble chamber data or at about 100 MeV/c in data
taken at LEAR, makes sure that the annihilation took place on a
quasi-free nucleon.
\par
In bubble chambers, antiprotons annihilating on neutrons
lead to odd numbers of visible pion tracks; the tracks of 'spectator'
protons with very low momenta (less than 80\,MeV/c) are not separable
from the primary ionisation spot produced by  the stopped
antiproton. Protons with momenta up to 250 \,MeV/c\ are easily
identified by their large ionisation losses. Thus it was easy to
separate annihilations on quasi-free protons and neutrons.
Bizzarri \cite{Bizzarri68} deduced probabilities
$s_{\rm p} = 0.571 \pm 0.005$ and $s_{\rm n} = 0.429 \pm 0.005$ for
annihilation on protons and neutrons, respectively.
Antiprotons stopped in liquid D$_2$ annihilate more often on
protons than on neutrons, with the ratio
\begin{equation}
\label{glob:eq:iso-deut}
s_{\rm p}/s_{\rm n} = 1.331\pm 0.019~.
\end{equation}
If the isospin content is denoted
\begin{equation}
\label{tm:eq:isod1} s_\p=s_\p^0+s_\p^1~,\quad s_\n=s_\n^1~,
\end{equation}
we have $s_p^1=s_\n^1/2$, thus ensuring model-independent relations due
to overall isospin conservation, such as
\begin{equation}\label{tm:eq:isod2}
\frac{\BR(\pbd\to\pim\omega\p)}{\BR(\pbd\to\piz\omega\n)}=2~.
\end{equation}
Then $s_\p^0>s_\p^1$, more precisely, 
$s_\p^0=(1.662\pm0.038)\,s_\p^1$, is needed
to get the observed $s_{\rm p}/s_{\rm n}$. This means that in \DT,
isoscalar \pbp\ annihilation
is significantly more frequent than isovector annihilation.

Table \ref{glob:tab:deut-k} lists the global annihilation frequencies
for \pbd\ annihilation into pions and kaons
for events with  spectator protons. Frequencies for final states with more than one $\pi^0$ are from \cite{Abele:2000xt,Amsler:2004rd} the frequency for $\pbn\to\pi^-\pi^0$ is derived in \cite{Abele:2000xt,Gaspero:1994tw}. The Crystal Barrel collaboration  \cite{Abele:2000xt,Amsler:2004rd} used no spectator cut to determine frequencies.

\begin{table}[!t]
\caption{Frequencies of $\pbn$ annihilation at rest in liquid D$_2$
into pionic final states \protect\cite{Bettini:1967bb}
and into final states with strangeness \protect\cite{Bettini:1969bc}
after a cut on the spectator momentum of
the proton at 250\,MeV/c .}
\renewcommand{\arraystretch}{1.2}
\begin{tabular}{ccc}
\begin{tabular}{lr}
\hline\hline
Final state                  &  Frequency (in\%)\\
\hline
$\pi^- n\pi^0$                &  $16.4\pm 0.5$ \quad  \\
\quad $\pi^-\pi^0$           & $0.40\pm0.04$ \\
\quad $\pi^-2\pi^0$          &$0.68\pm0.07$\\
\quad $\pi^-4\pi^0$          & $1.32\pm0.20$ \\
\hline
$2\pi^-\pi^{+} n\pi^0$        &  $59.7\pm 1.2$ \quad \\
\quad $2\pi^-\pi^+$          &  $1.57\pm 0.21$ \\
\quad $2\pi^-\pi^+\pi^0$     &  $21.8\pm 2.2$ \\
\quad $2\pi^-\pi^{+}2\pi^0$  & $7.0\pm 1.1$ \\
\hline
$3\pi^-2\pi^{+}n\pi^0$       &  $23.4\pm 0.7$ \quad \\
\quad $3\pi^-2\pi^+$         & $5.15\pm 0.47$ \\
\quad $3\pi^-2\pi^+\pi^0$    & $15.1\pm 1.0$ \\
\hline
$4\pi^-3\pi^{+}n\pi^0$       &  $0.39\pm 0.07$ \quad \\
\hline
Sum                     &     $95.5\pm 1.5$\%                      \\
\hline
\hline
\end{tabular}& \qquad\qquad &
\begin{tabular}{lr}
\hline\hline
Final state                  &  Frequency (in $10^{-4}$)\\
\hline
$\K^0\K^-$                   &  $14.7\pm 2.1$ \quad 
\\
$\K^0\K^-\pi^0$              &  $36.0\pm 4.2$ \quad 
\\
$\Ks\Ks\pi^-$              &  $14.7\pm 2.0$ \quad 
\\
$\Ks\K_l\pi^-$              &  $21.2\pm 3.6$ \quad  \\
\hline 

$\K^0\K^+\pi^-\pi^-$         &  $24.8\pm 2.6$ \quad 
\\
$\K^0\K^-\pi^+\pi^-$         &  $34.2\pm 3.5$ \quad 
\\
$\Ks\Ks\pi^-\pi^0$         &  $25.6\pm 2.8$ \quad  \\
\hline 

$\K^0\K^+\pi^-\pi^-\pi^0$    &  $1.6\pm 0.9$ \quad 
\\
$\Ks\K^-\pi^+\pi^-\pi^0$    &  $33.6\pm 3.8$ \quad 
\\
$\Ks\K^-\omega$             &  $35.0\pm 5.2$ \quad 
\\
$\Ks\Ks\pi^+\pi^-\pi^-$    &  $2.8\pm 1.2$ \quad 
\\
$\Ks\Kl\pi^+\pi^-\pi^-$    &  $1.9\pm 1.2$ \quad  \\
\hline
Sum & $2.5\pm 0.1$\%\\
\hline\hline
\end{tabular}
\end{tabular}
\label{glob:tab:deut-k}
\renewcommand{\arraystretch}{1.0}
\end{table}

\clearpage\markboth{\sl Annihilation dynamics}%
 {\sl Two-meson annihilation}
\clearpage\setcounter{equation}{0}

\section{Annihilation into two mesons}\label{se:tm}
\FloatBarrier\subsection{Introduction}\label{tm:sub:intro}
The study of \NNb\ annihilation at rest
into two mesons is a rich source of information about annihilation
dynamics. The role of symmetries, the topology of the dominant quark
diagrams, the violation of the Zweig rule, etc., can be inferred
from the knowledge of the two-meson frequencies. Detailed
information on the density dependence of annihilation frequencies is
required if these frequencies are to be assigned to specific
states of the  \pbp\ or \pbd\ atom.
This arises 
from the sensitivity of two-meson annihilation to the initial state of the 
nucleon-antinucleon system. For example, from Table
\ref{kin:tab:allowed}, we see that the process 
$\pbp \rightarrow\pip\pim$ takes place from initial S- and P-
states whilst the reaction 
$\pbp \rightarrow \piz\piz$ only originates from P-states. At first
glance, it would seem 
that the fraction of P-state annihilation could be obtained from the relevant 
$\pip\pim$ and \piz\piz\ annihilation frequencies 
using isospin invariance and simple arithmetic.
The derivation of the P-state fraction is actually more subtle 
and unfortunately, more complex. In particular it depends on details of 
the atomic cascade process and on the amount of Stark mixing. 
On the other hand,
a thorough analysis of these cascade effects enables us to obtain
the fraction of \pbp\ annihilation into a variety of channels from different 
fine-structure states of the \pbp\ system. The information which is 
directly given by the $\pip\pim$ and \piz\piz\ frequencies is 
the fraction of annihilations for the $\pip\pim$ and \piz\piz\ 
{\it channels} which take place from S- and P-states. This is {\it not} 
the same as the fraction of S- and P-state annihilations for 
the \pbp\ system.

We first start with a few definitions. The {\it annihilation frequency}
$\AF(ch,\rho$), often also called the branching ratio 
in the literature, is the probability that a
particular channel
$ch$ will be produced in a \pbp\ or \pbd\ annihilation at rest. It 
is usually the quantity measured experimentally and is a function of the 
target density $\rho$. We restrict the term 
{\it branching ratio}
$\BR(ch,\slj$) to the probability that the
channel $ch$ is produced by an annihilation from the initial state 
\slj\ of the \pbp\ or \pbd\ atom. These branching ratios
are independent of the atomic physics effects occurring during the
cascade of the \pbp\ or \pbd\ atom 
and hence do not depend on the target density.  It is this quantity 
which normally should be compared with predictions from theoretical 
models. In Sec.~\ref{se:dsr}, we shall define {\it dynamically corrected 
branching ratios}, DR, including phase space corrections,  
orbital-angular-momentum-barrier effects, and a
phenomenological factor favouring the production of high-mass mesons.

Many of the earlier experiments measured two-body annihilation
frequencies by the detection and reconstruction of a single \piz\ or
$\eta$ with a small solid-angle detector and observed peaks in the
resulting \piz\ or $\eta$ inclusive momentum spectrum to identify the
second meson (see Table \ref{kin:tab:momenta}).  Since the
annihilation frequencies are small, these early data are often
statistically weak or subject to very considerable uncertainties in
the background subtraction from the inclusive spectra. In more recent
experiments, using detectors covering a large solid angle, both of the
mesons in the two-body events are fully reconstructed and any
background is much reduced.

At a given target density $\rho$, the annihilation frequency for the channel
$ch$ is given by
\begin{equation}
\AF(ch) = \frac{N_{evts}(ch)}{N_{tot}\epsilon_{det}VF_{1}F_{2}}~,
\end{equation}
where $\epsilon_{det}$ is the detection efficiency given by
\begin{equation}
\epsilon_{det} = \frac{N^{MC}_{evts}(ch)}{N^{MC}_{tot}}~.
\end{equation}
$N_{tot}$ is the total number of antiprotons stopped in the target,
corrected for
pile-up, and $N_{evts}(ch)$ is the number of events, corrected for background,
attributed to channel $ch$.
$N^{MC}_{tot}$ and $N^{MC}_{evts}(ch)$ are the corresponding numbers for 
the Monte-Carlo simulation and reconstruction. $V$ is the fraction of
antiproton events
which stop in the target volume and $F_1$ and $F_2$ are the correction factors 
for the probability of the observed decay 
for each of the two mesons in channel $ch$; 
for example $F = 0.3943\pm0.0026$ for $\eta\ \rightarrow\ 2\gamma$ in the case where 
the two photons are measured so as to reconstruct the $\eta$ meson.

Here we will particularly discuss two-body annihilation frequencies 
for channels 
involving {\em narrow} ($\Gamma\ \leq$ 10\,MeV) mesons ($\pi, \PK,
\eta,\ \eta^{\prime},\ \omega$ and $\phi$). Measurements 
of annihilation frequencies for broader states from 
inclusive spectra usually have 
considerable uncertainties due to the need for 
background subtraction. In addition, reflections can distort the 
spectrum in an unpredictable way. For measurements of two-body annihilation
frequencies of broad states a full partial wave analysis involving 
all intermediate states is required. For completeness, we include two-body
branching ratios involving the K$^*$ ($\Gamma$ = 50\,MeV) and 
$\rho$ ($\Gamma$ = 151\,MeV) mesons. 
Some two-meson annihilation frequencies obtained from partial wave analyses 
involving pseudovector or tensor mesons are also discussed. 

In measurements of frequencies for two-body annihilation from \pbp\
atoms, the selection of events is straightforward, requiring just two
back-to-back particles each of the same well-defined momentum. 
For \pbard\ atoms 
the selection is less straightforward due to the presence of the recoil 
nucleon which, in some cases, can have momenta up to 
1.2 \gevc\ \cite{Bizzarri:1974hr,Abele:2000xt}.
Measurement of their frequencies requires that all events over the full range
of spectator momenta are included. An additional effect is that the spectator
momentum distribution depends \cite{Bizzarri:1974hr} on whether 
the annihilation 
occurs from an atomic S- or P-state of the \pbard\ system and whether 
the \pbar\ is in an S- or P-state relative to the nucleon with which it 
annihilates. In particular the momentum distribution for P-state 
\pbar N annihilation from an atomic S-state is particularly broad. It 
was pointed out \cite{Reifenroether:1990tq} that tight cuts on the 
momentum of the 
spectator nucleon therefore suppresses P-state annihilation from 
atomic S-state orbitals.

For \pbd\ annihilations at rest, the annihilation frequency is
determined by normalising either to the total number of annihilations on a
deuteron
or to the number of annihilations on a proton or neutron. In the following, 
subscripts d and N (p or n), will be used to distinguish
between frequencies measured for annihilation on 
deuterons or on a nucleon (proton or neutron) in the deuteron.
Sometimes, it has been assumed that there is equal probability for 
annihilation on a proton or neutron in the deuteron. In this case 
the branching ratio for \pbd\ annihilation would be obtained simply 
by dividing by a factor of two. Here, we use the probabilities
$s_{\rm p}=0.571\pm0.005$ and $s_{\rm n}=0.429\pm0.005$ deduced by 
Bizzarri \cite{Bizzarri68}, as discussed already in 
Sec.~\ref{glob:sub:neutron}. The  annihilation frequency on a deuteron,
for reactions involving a spectator neutron, is then given by
\begin{equation}
\AF_{\rm d}(ch,{\rm n},\rho) = s_{\rm p}\AF_{\rm p}(ch,{\rm n},\rho)\,,
\label{tm:equ:BRd}
\end{equation}
and similarly for reactions involving a spectator proton. 

In  Sec.~\ref{tm:sub:stark}, we discuss the \pbp\ and \pbd\ atomic cascade, the
effects of Stark mixing, the role of the fine-structure levels and the need for
{\em enhancement factors}. Atomic cascade calculations, and 
in particular the prediction of enhancement factors is discussed in 
Sec.~\ref{tm:sub:casc} for both \pbp\ and \pbd\ atoms. 
Published annihilation frequencies for \pbp\ atoms are reviewed in
Sec.~\ref{tm:sub:BRexp} which also contains a detailed discussion of  the
\pbp$\to$\piz\piz\ annihilation frequency. 
A new analysis of two-body annihilation frequencies for \pbp\ atoms is
presented in Sec.~\ref{tm:sub:BRanalpbp} with an emphasis on determining the
fraction of P-state annihilation and branching ratios for annihilation from
specific atomic states. Section \ref{tm:sub:AN} reviews antineutron 
annihilation on protons whilst section \ref{tm:sub:BRexpd} reviews 
annihilation frequencies from \pbard . 
The determination of the P- state fraction in \pbard\ annihilation 
is presented in Sec.~\ref{tm:sub:BRanalpbd}. A compilation of data on 
Pontecorvo reactions is given in Sec.~\ref{tm:sub:Ponte}. Some final 
remarks on aspects of 
two-meson annihilation are made in Sec.~\ref{tm:sub:Discuss}. 
\FloatBarrier
\subsection{Stark mixing and density dependence of annihilation frequencies}
\label{tm:sub:stark}
\subsubsectionb{\pbp\ atoms}\label{tm:sub:starkpbp}
 Formation of the \pbp\ atom and its atomic cascade has been discussed
in the earlier review \cite{Klempt:2002ap} and 
elsewhere \cite{Batty:1989rpp,Gotta:2004rq}. Briefly, the capture of
the \pbar\ typically occurs at a principal quantum number $n \approx 30$
\cite{Cohen:2004aa}.
De-excitation then takes place by a number of processes including radiative
transitions with the emission of X-rays, and the external Auger effect
involving
the ionisation of a neighbouring \HT\ molecule. Finally the \pbar\ reaches 
an atomic state with angular momentum $L= 0$ or 1 when annihilation
occurs. Annihilation from states with $L\ge2$ can be ignored due to the
negligible overlap of \pbar\ and p in the atomic wavefunction.

In addition, except at very low target densities, the Stark effect gives mixing
of the angular momentum states $L$ at high $n$ allowing the protonium atoms
to transfer to S- and P-states where they can annihilate before reaching the
low-$n$ states. The Stark mixing increases with the target density;  for
liquid targets the rate is very high.
When considering frequencies for \pbp\ annihilation, 
it is important to consider the effects of the fine-structure of the atomic
states. For \pbp\ atoms, the states with $L < 2$ are \ssz, \tso, \spo,
\tpz, \tpo\ and \tpt\ with the corresponding \jpc = $0^{-+}$, $1^{--}$,
$1^{+-}$, $0^{++}$, $1^{++}$ and $2^{++}$. Table~\ref{tm:tab:siwidthpp} 
shows the 
predictions of Carbonell \etal \cite{Carbonell:1989cs} for the widths
of these states. They were obtained using potentials for the
\pbp\ interaction due to Dover and Richard (DR1 and DR2) and Kohno and
Weise (KW). A particular feature of these predictions is the very
large width of the \tpz\ state which has been confirmed in recent
experiments. Gotta et al \cite{Gotta:1999vj}, assuming the widths of
the \spo, \tpo\ and \tpt\ states to be equal, obtain $\Gamma(\tpz) =
120 \pm 25$\,meV and $\Gamma(\spo,\tpo,\tpt) = 30.5 \pm 2.8$\,meV for the states with principal quantum number $n=2$, in
good agreement with the values listed in Table
\ref{tm:tab:siwidthpp}. The average value of the width of the 1S state
given by four experiments is quoted in the earlier review
\cite{Klempt:2002ap} as $\Gamma(1\mbox{S}) = 1.060 \pm 0.080$\,keV
which again is in good agreement with the values of 
Table~\ref{tm:tab:siwidthpp}.

\begin{table}[!h]
\caption{Widths for \pbp\ atoms as predicted by three potential
models}
\label{tm:tab:siwidthpp}
\begin{center}\renewcommand{\arraystretch}{1.2}
\begin{tabular}{ccccccc}
\hline\hline
State & \ssz   & \tso   & \spo   & \tpz   & \tpo   & \tpt    \\ 
Units &  keV &  keV  &  meV  &  meV  &  meV  & meV   \\  \hline
DR1    & 1.02  & 0.90   & 26      & 114    & 20    & 30      \\
DR2    & 1.04  & 0.92   & 28      & 80     & 18    & 32      \\
KW     & 1.26  & 0.98   & 26      & 96     & 22    & 36      \\   
\hline\hline
\end{tabular}\renewcommand{\arraystretch}{1.0}
\end{center}\end{table}

At high $n$, in the case where Stark mixing is important, the fine-structure
levels are continually and rapidly repopulated according to their statistical
weight. A fine structure level with a large annihilation width will therefore
contribute more to annihilation than would be expected from its statistical
weight only. This effect is particularly important for the \tpz\ level. 
Similar, but smaller, effects will also occur for the other fine 
structure P-states. The effects for S-states are typically less than 5\,\%.

These deviations of the population of the fine-structure states have been
described \cite{Batty:1996uf} in terms of {\em enhancement factors} 
$E(\slj,\rho)$ which are  functions of the initial state \slj\ and 
target density $\rho$. Values of $E(\slj,\rho) < 1 \ ( >  1)$ correspond 
to a fraction of annihilations smaller (larger) than that expected on
the basis of a purely statistical population of the level. 

The annihilation frequency $\AF(ch,\rho)$  can then be written
\cite{Batty:1996uf} 
in terms of the branching ratios \BR($ch,\slj$) in the form
\begin{equation}\eqalign{
\AF(ch,\rho)  = \left (1 - \fp\right)\Bigl[&\frac{1}{4} E(\ssz,\rho)
\BR(ch,\ssz) +
\frac{3}{4} E(\tso,\rho) \BR(ch,\tso)\Bigr] \cr
{}+\fp\;\Bigl[ &\frac{3}{12} E(\spo,\rho) \BR(ch,\spo) + 
\frac{1}{12} E(\tpz,\rho) \BR(ch,\tpz)\cr
 {}+{}&{}\frac{3}{12} E(\tpo,\rho) \BR(ch,\tpo) + 
\frac{5}{12} E(\tpt,\rho) \BR(ch,\tpt)\Bigr] ~,}\label{tm:equ:BRp}
\end{equation}
where \fp\ is the fraction of P-state annihilation and the factors
1/4, 5/12, etc., are the statistical weights of the states. In those cases where 
production of the channel $ch$ is forbidden from an initial state 
\slj\ due to selection rules, then $\BR(ch,\slj) = 0$.

The enhancement factors are normalised \cite{Batty:1996uf} so that
\begin{equation}
\frac{1}{4} E(\ssz,\rho) + \frac{3}{4} E(\tso,\rho) = 1~,
\end{equation}
and
\begin{equation}
\frac{3}{12} E(\spo,\rho)+ \frac{1}{12} E(\tpz,\rho) + 
\frac{3}{12} E(\tpo,\rho) + \frac{5}{12} E(\tpt,\rho) = 1~.
\end{equation}

From Eq.~(\ref{tm:equ:BRp}) it can be seen that the density dependence of
$\AF(ch,\rho)$ arises from two factors. The first and larger one is the 
density dependence of the fraction of P-state annihilation, \fp, which 
is directly due to the Stark effect. The second and more subtle one arises 
from the enhancement factors $E(\slj,\rho)$. In cases where these differ 
significantly at a given $\rho$ for various $\slj$, the 
branching ratios for these states, $\BR(ch,\slj)$ can be
determined.

Attempts have been made to determine the values of the enhancement 
factors from experimental data. A study \cite{Abele:1998kv} of the 
reaction $\pbp\rightarrow\eta\piz\piz\piz$ at rest, both in liquid and 
in gas at 12 \rstp, where STP indicates Standard Temperature and Pressure, finds
\begin{equation}
r = \frac{E(\tpz,{\rm liq.})/E(\tpz,12\ \rstp)}{E(\tpt,{\rm liq.})/E(\tpt,12\ \rstp)}
= 2.46 \pm 0.15~,
\end{equation}
to be compared with the value $r \approx\ 1.7$ obtained using predicted
values for the enhancement factors from
Table \ref{tm:tab:efacp} which will be discussed later.

Salvini et al.\ \cite{Salvini:2001ss} have attempted to determine
enhancement 
factors directly from a best fit to two-body annihilation frequencies. 
The fit 
procedure was not straightforward since they obtained different 
local minima with more or less equivalent $\chi^2$ values. 
This is 
probably due to the fact that the values of $\BR(ch,\slj)$ and 
$E(\slj,\rho)$ are strongly correlated, appearing in the form
$E(\slj,\rho)\BR(ch,\slj)$ in Eq.~(\ref{tm:equ:BRp}). The 
solutions were constrained to those that were regarded as 
physically meaningful, the remaining parameter space was
explored in detail. In most cases values for $E(\slj,\rho)$  were
obtained consistent with unity except for $E(\tpz,{\rm liq.})$
which was in the range $1.7 \pm 0.3$ to $3.6 \pm 0.1$, depending on the 
value for $\AF(\piz\piz,{\rm liq})$ used in the data analysis. The 
discrepancies between the different experimental values for the
$\pbp \rightarrow \piz\piz$ annihilation frequency in liquid \HT\ will be 
discussed in Sec.~\ref{tm:sub:BRpi0pi0}.


More recently Bargiotti et al.\ \cite{Bargiotti:2004aa} have determined
enhancement factors $E(\slj,\rho)$ for all S- and P- fine-structure states in
liquid and gas (\rstp) targets by fitting measurements for two-body and and also 
for resonant two-body channels obtained from partial wave analyses of the
channels $\pip\pim\piz$, $\K^\pm\Kn_{\rm s}\pi^\mp$ and  $\K^+\K^-\piz$
in hydrogen targets at three different densities $(\rho = 0.005\rstp,$
\rstp\ and liquid). They fix the enhancement factors at low density 
$E(\slj,0.005\rstp) = 1$. With this assumption they were able to obtain 
a good fit to the measurements with enhancement factors different from one 
for the \spo, \tpz\ and \tpo\ levels in liquid hydrogen . Comparison with the 
predictions of cascade calculations (see Table \ref{tm:tab:efacp}) shows good 
agreement for the \tpz\ level but disagreement for the \spo\ and \tpo\ levels 
with the \spo\ level, $E(\spo,{\rm liq.}) = 0.09 \pm 0.17$, being particularly 
strongly suppressed. The reason for this discrepancy is not understood.


An alternative approach \cite{Batty:1996uf} is to calculate values of 
$E(\slj,\rho)$ using an atomic cascade calculation \cite{Batty:1989rpp} 
for the \pbp\ atom whose parameters 
are obtained by fits to \pbp\ atomic X-ray data. This method will be 
discussed in detail in Sec.~\ref{tm:sub:casc}.
\subsubsectionb{\pbd\ atoms}\label{tm:sub:starkpbd}
The general features of the atomic cascade for \pbd\ atoms are very similar
to those for \pbp\ atoms discussed above, except that the \pbar\ is now 
captured at a principal quantum number $n \approx\ 45$ and that the
much larger predicted widths \cite{Wycech:1985py} for the 2P levels 
in \pbd\ atoms give rise to increased probabilities for annihilation 
from atomic P-states. D-state annihilation also becomes significant. 
Wycech et al.\ \cite{Wycech:1985py} predict that the widths of the 
fine-structure components for P-states are approximately equal. It is 
therefore to be expected that for \pbd\ atoms the enhancement factors 
will have values $E(\slj,\rho) \approx 1$ and their effect can be neglected.

Due to the Fermi motion of the nucleons in deuterium, the \pbar-nucleon angular
momentum can be different from the angular momentum of the \pbar\ with respect
to the deuteron centre of mass. Even when the \pbar\ is in an atomic S-state,
it may be in a P-state relative to the nucleon with  which it annihilates.
This process, sometimes called ``induced'' P-state annihilation, 
was first considered by Bizzarri et al.\ \cite{Bizzarri:1974hr}
and has recently been discussed by Bugg \cite{Bugg:2000br}. It is an
assumption of these models that the antiproton interacts with only one of the
two nucleons in the deuteron, the other behaving as a ``spectator''. As a
consequence, the annihilation frequencies in deuterium are directly 
related to the frequencies
for capture on neutrons and protons. For the
remainder of this article we use the following notation: 
capital letters denote 
the initial \pbd\ atomic states whilst the subscript ``ann'' (e.g.,
\pann) is used to describe 
the annihilation process. This differs from the notation of 
Bugg \cite{Bugg:2000br}, 
which was also used by Batty \cite{Batty:2002sk}, where small 
letters denote the
initial \pbd\ atomic states and capital letters
were used to describe the annihilation process. For the \pbp\ system 
the fraction of
annihilation from P-states of the \pbp\ atom \fp, and the fraction of P-state 
\pbp\ annihilation \fpa\ are, of course, the same.
\FloatBarrier
\subsection{Cascade calculations}\label{tm:sub:casc}
\subsubsectionb{\pbp\ atoms}\label{tm:sub:cascpbp}
In the calculations of enhancement factors 
\cite{Batty:1996uf}, measurements of atomic X-ray yields 
for \pbp\ atoms, covering the
range of target densities from 0.016 to 10.0 \rstp, 
were analysed using a cascade 
calculation 
based on the method of Borie and Leon \cite{BorieL80}. In this method the 
effects of Stark mixing are calculated using an impact parameter technique
in which, in its interaction with the electric field of the neighbouring
\HT\ molecules, the exotic atom is treated as moving along a straight line
trajectory. If an atomic S-state is involved, the removal of the degeneracy due 
to the energy shift caused by the strong interaction hinders the Stark 
mixing, since the electric field must overcome the energy difference 
between S- and P-states. This is taken into account by using a smaller 
impact parameter. Because of the uncertainties in the absolute rate for 
Stark mixing, an overall normalisation parameter $k_{\rm STK}$ is usually 
used and its value determined by fitting the X-ray yields.

An alternative method, usually referred to as the ``Mainz'' model, has 
been developed in 
\cite{Reifenroether:1989nq,Reifenroether:1990tq}. Here the collisions 
of the \pbp\ atoms with the neighbouring \HT\ molecules are simulated 
using 
\begin{figure}[th!]
\begin{center}
\includegraphics[height=8.0cm]{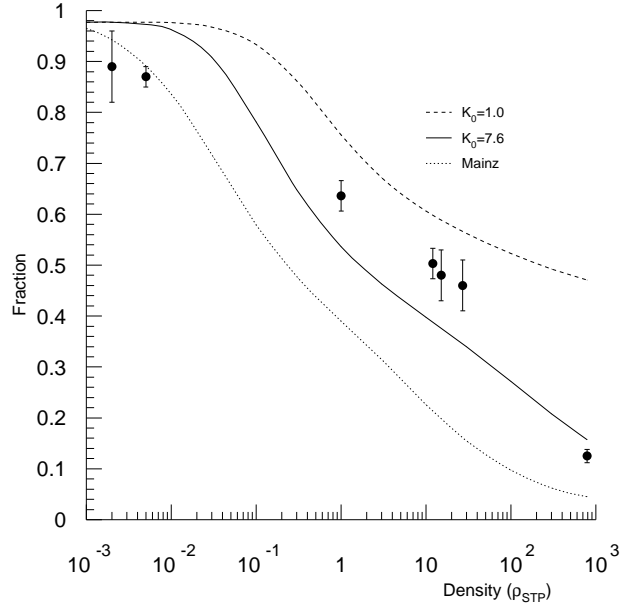}
\caption{Fraction of P-state annihilation predicted by the Borie--Leon
model \protect\cite{BorieL80} giving a best fit to the X-ray data with
$K_0 =1$ (dashed line) or $K_0 = 7.6$ (full line). Also shown are the
predictions of the Mainz
\protect\cite{Reifenroether:1989nq,Reifenroether:1990tq} model (dotted
line) and the values derived in the present work from annihilation
frequencies and annihilation widths from the DR1 potential}
\label{tm:fig:pstatep}
\end{center}
\end{figure}
\noindent
the classical-trajectory Monte-Carlo method which only 
allows Stark mixing when the 
\pbp\ atom experiences strong electric 
fields during collisions with neighbouring hydrogen molecules. This
avoids the uncertainties associated with the straight line
approximation used by Borie and Leon \cite{BorieL80} at the expense of
a considerable increase in computational complexity. Further
information about cascade calculations is given in earlier reviews
\cite{Klempt:2002ap,Batty:1989rpp}.

Once the parameter $k_{\rm STK}$ in the Borie--Leon model has been
adjusted to fit the X-ray data \cite{Batty:1996uf}, the two models
give similar predictions for the overall variation in the X-ray yields
as a function of target density.  However the Borie and Leon model
predicts a significantly larger fraction for P-state annihilation in
\pbp\ atoms than the ``Mainz'' model as shown in
Fig.~\ref{tm:fig:pstatep}.  It has been shown \cite{Batty:1996uf} that
the introduction of an additional parameter $K_0$, as a further
normalisation factor to the rate of Stark transitions between atomic
S- and P-states, gives significantly improved fits to the X-ray yield
data for \pbp\ atoms and much reduced values for the fraction of
P-state annihilation in
\pbp\ atoms. Setting $K_0$ = 1 gives the usual form of the Borie and 
Leon model. 

\begin{figure}[!hp]
\includegraphics[height=8.0cm]{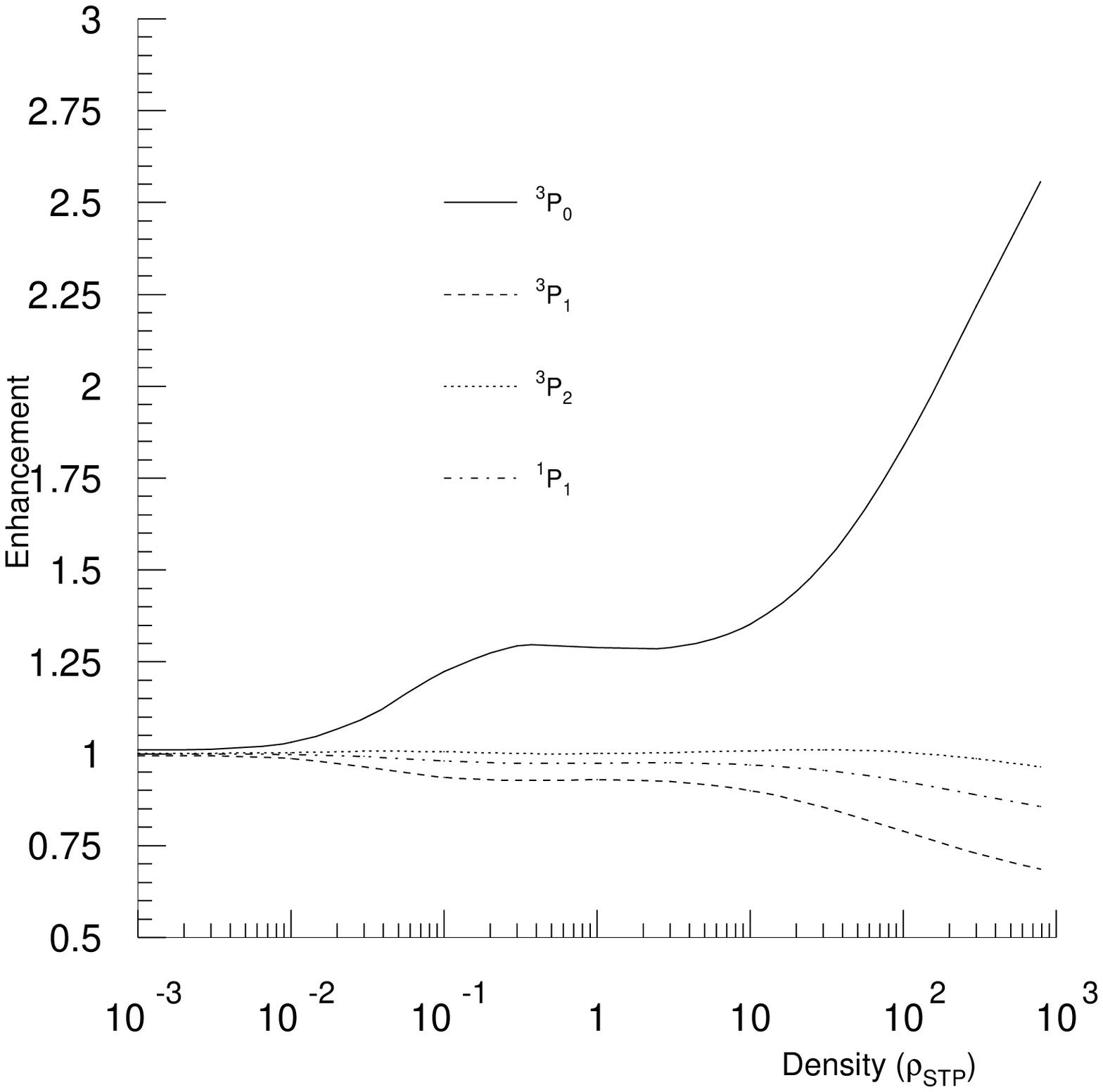}
\caption{P-state enhancement factors from a cascade calculation \protect\cite{Batty:1996uf} 
using $K_0 = 7.6$ and annihilation widths from the {DR1} potential}
\label{tm:fig:efacp}
\end{figure}

Values of $E(\slj,\rho)$ for P-states calculated \cite{Batty:1996uf}
using the Borie--Leon cascade model with the DR1 annihilation widths
of Table \ref{tm:tab:siwidthpp} are shown in Fig.~\ref{tm:fig:efacp}.
The corresponding enhancement factors for the S-states are generally
close to 1. Table \ref{tm:tab:efacp} gives numerical values for the
enhancement factors, for all target pressures at which annihilation
frequencies have been measured, for the three annihilation models of
Table \ref{tm:tab:siwidthpp}.  The errors quoted are those due to the
uncertainties in the parameters $K_0=7.6\pm2.6,\ k_{\rm
STK}=1.19\pm0.06$ used to fit the X-ray yield data for \pbp\ atoms.

\begin{table}[!hp]
\caption{Calculated enhancement factors, as a function of the
density $\rho$ (in units of $\rho_\mathrm{STP}$).}
\label{tm:tab:efacp}
\renewcommand{\arraystretch}{1.0}
\begin{tabular}{crrrrrrrc}  \hline\hline
State 
& $\rho=0.002$ & $\rho=0.005$ & $\rho=1$  &
$\rho=12$
 & $\rho=15$ & $\rho=27$& Liquid &  Model\\
\hline
&  1.044&  1.046&  1.020&  1.011&  1.012&  1.013&  1.032 &
\multirow{2}{.4cm}{DR1}\\[-3pt]
&$ {}\pm  0.001$&$ {}\pm  0.001$& ${}\pm  0.002$&$ {}\pm  0.001$&$ {}\pm  0.001$&$ {}\pm  0.001$&$ {}\pm  0.001$ &  \\[3pt]
  \multirow{2}{.7cm}{\ssz} &  0.973&  0.975&  1.007&  1.005&  1.005&  1.007&  1.028
& 
\multirow{2}{.4cm}{DR2}
\\[-3pt]
     &$ {}\pm  0.001$&$ {}\pm  0.001$&$ {}\pm  0.001$&$ {}\pm  0.001$&$ {}\pm  0.001$
 &$ {}\pm  0.001$&$ {}\pm  0.001$ &  \\[3pt]
     &  1.138&  1.139&  1.030&  1.019&  1.020&  1.022&  1.060 &
\multirow{2}{.4cm}{KW}
\\[-3pt]
     &$ {}\pm  0.001$&$ {}\pm  0.001$& ${}\pm  0.003$&$ {}\pm  0.001$&$ {}\pm  0.001$
 &$ {}\pm  0.001$&$ {}\pm  0.001$ &  \\  \hline
     &  0.985&  0.985&  0.993&  0.996&  0.996&  0.996&  0.989 & \multirow{2}{.4cm}{DR1}
\\[-3pt]
     &$ {}\pm  0.001$&$ {}\pm  0.001$&$ {}\pm  0.001$&$ {}\pm  0.001$&$ {}\pm  0.001$
 &$ {}\pm  0.001$&$ {}\pm  0.001$ &  \\[3pt]
 \multirow{2}{.7cm}{\tso}&  1.009&  1.008&  0.998&  0.998&  0.998& 
0.998&  0.991 & \multirow{2}{.4cm}{DR2}
\\
[-3pt]
     &$ {}\pm  0.001$&$ {}\pm  0.001$&$ {}\pm  0.001$&$ {}\pm  0.001$&$ {}\pm  0.001$
 &$ {}\pm  0.001$&$ {}\pm  0.001$ &  \\[3pt]
     &  0.954&  0.954&  0.990&  0.994&  0.993&  0.993&  0.980 &
\multirow{2}{.4cm}{KW} 
\\[-3pt]
     &$ {}\pm  0.001$&$ {}\pm  0.001$&$ {}\pm  0.001$&$ {}\pm  0.001$&$ {}\pm  0.001$
 &$ {}\pm  0.001$&$ {}\pm  0.001$ &  \\  \hline
     &  0.999&  0.999&  0.974&  0.968&  0.966&  0.958&  0.856 & \multirow{2}{.4cm}{DR1}
\\[-3pt]
     &$ {}\pm  0.001$ & $ {}\pm  0.001$ & ${}\pm  0.002$& ${}\pm  0.002$&$ {}\pm  0.002$
 &$ {}\pm  0.002$&$ {}\pm  0.005 $&  \\ [3pt]
 \multirow{2}{.7cm}{\spo}&  1.000&  1.001&  0.991&  0.992&  0.991& 
0.988&  0.933 &\multirow{2}{.4cm}{DR2}
\\
[-3pt]
     &$ {}\pm  0.001$&$ {}\pm  0.001$&$ {}\pm  0.001$&$ {}\pm  0.001$&$ {}\pm  0.001$
 &$ {}\pm  0.001$&$ {}\pm  0.003 $&  \\ [3pt]
     &  0.998&  0.997&  0.960&  0.945&  0.941&  0.928&  0.809 &
\multirow{2}{.4cm}{KW} 
\\[-3pt]
     &$ {}\pm  0.001$&$ {}\pm  0.001$&$ {}\pm  0.003$&$ {}\pm  0.002$& ${}\pm  0.003$
 & ${}\pm  0.003$&$ {}\pm  0.004 $&  \\  \hline
     &  1.011&  1.016&  1.288&  1.372&  1.399&  1.487&  2.556 & \multirow{2}{.4cm}{DR1}
\\[-3pt]
     &$ {}\pm  0.001$&$ {}\pm  0.002$&$ {}\pm  0.023$&$ {}\pm  0.016$&$ {}\pm  0.017$
 &$ {}\pm  0.021$&$ {}\pm  0.046$ &  \\[3pt]
 \multirow{2}{.7cm}{\tpz}&  1.010&  1.014&  1.206&  1.280&  1.302& 
1.370&  2.076 & \multirow{2}{.4cm}{DR2}
\\
[-3pt]
     &$ {}\pm  0.001$&$ {}\pm  0.001$&$ {}\pm  0.015$& ${}\pm  0.011$& ${}\pm  0.012$
 &$ {}\pm  0.016$&$ {}\pm  0.027$ &  \\[3pt]
     &  1.009&  1.013&  1.227&  1.296&  1.318&  1.388&  2.176 &
\multirow{2}{.4cm}{KW} 
\\[-3pt]
     &$ {}\pm  0.001$&$ {}\pm  0.001$&$ {}\pm  0.017$& ${}\pm  0.012$&$ {}\pm  0.013$
 &$ {}\pm  0.016$&$ {}\pm  0.032$ &  \\  \hline
     &  0.995&  0.993&  0.929&  0.894&  0.886&  0.862&  0.685 & \multirow{2}{.4cm}{DR1}
\\[-3pt]
     &$ {}\pm  0.001$&$ {}\pm  0.001$&$ {}\pm  0.005$& ${}\pm  0.004$&$ {}\pm  0.004$
 &$ {}\pm  0.005$&$ {}\pm  0.005$ &  \\ [3pt]
 \multirow{2}{.7cm}{\tpo}&  0.993&  0.990&  0.914&  0.866&  0.856& 
0.826&  0.641 & \multirow{2}{.4cm}{DR2}
\\
[-3pt]
     &$ {}\pm  0.001$&$ {}\pm  0.001$&$ {}\pm  0.006$&$ {}\pm  0.005$&$ {}\pm  0.005$
 &$ {}\pm  0.006$&${}\pm  0.004 $&  \\[3pt]
     &  0.995&  0.993&  0.932&  0.899&  0.891&  0.868&  0.703 &
\multirow{2}{.4cm}{KW} 
\\[-3pt]
     &$ {}\pm  0.001$&$ {}\pm  0.001$& ${}\pm  0.005$&$ {}\pm  0.004$&$ {}\pm  0.004$
 & ${}\pm  0.005$&$ {}\pm  0.004$ &  \\  \hline
     &  1.001&  1.002&  1.000&  1.008&  1.009&  1.011&  0.964 & \multirow{2}{.4cm}{DR1}
\\[-3pt]
     &$ {}\pm  0.001$&$ {}\pm  0.001$&$ {}\pm  0.001$&$ {}\pm  0.001$&$ {}\pm  0.001$
 &$ {}\pm  0.001$&$ {}\pm  0.004$ &  \\[3pt]
\multirow{2}{.7cm}{ \tpt}&  1.002&  1.003&  1.016&  1.029&  1.032& 
1.038&  1.041 & \multirow{2}{.4cm}{DR2}
\\
[-3pt]
     &$ {}\pm  0.001$&$ {}\pm  0.001$&$ {}\pm  0.001$&$ {}\pm  0.001$&$ {}\pm  0.001$
 &$ {}\pm  0.001$&$ {}\pm  0.001$ &  \\ [3pt]
     &  1.002&  1.003&  1.019&  1.035&  1.037&  1.045&  1.058 &
\multirow{2}{.4cm}{KW} 
\\[-3pt]
     &${} \pm  0.001$&${} \pm   0.001$&${} \pm  0.001$&${}\pm  0.002$&${} \pm  0.001$
 &${} \pm  0.001$&${} \pm  0.001$ &  \\  \hline\hline
\end{tabular}\renewcommand{\arraystretch}{1.0}
\end{table}
\subsubsectionb{\pbd\  atoms}\label{tm:sub:cascpbd}
Calculations of the atomic cascade for \pbd\ atoms  follow those made 
for \pbp\ atoms but with a starting value for $n = 45$ and with strong 
interaction widths appropriate for \pbd\ atoms. Recently proposed 
values for the energy shift and width of the 1S ground state
\cite{Augsburger:1999sx}  
$\Delta E_{\rm 1S} = -1050 \pm 250$\,eV,
$\Gamma_{\rm 1S} = 1100 \pm 750$\,eV and for the width of the 2P-state 
\cite{Gotta:1999vj}
$\Gamma_{\rm 2P} = 489 \pm 30$\,meV were used (see, however, also \cite[page
251]{Klempt:2002ap}).
This latter value is in
 reasonable agreement with the average 2P width $\Gamma_{\rm 2P} = 422$\,meV
calculated by Wycech et al.\ \cite{Wycech:1985py}. In the case of \pbd\ atoms, 
D-state annihilation has to be considered and  
as measurements of the width of the
3D state are not available, the value $\Gamma_{3{\rm D}} = 5\,\mu$eV from the
calculations of Wycech et al.\ \cite{Wycech:1985py} was used.

In almost all cases, the only available yield data for \pbd\ atoms is for 
the L X-ray lines ($n$D -- 2P) at eight target densities
from 0.016 to $10.0\,\rstp$, (where \rstp\ is the density of \DT\ gas
at STP).
As a result, it is only possible to determine the value of $k_{\rm STK}$
from the fit to the X-ray yields. Data for the K-lines ($n$P-1S) is
required 
if the value of  $K_0$ is to be determined. Equally good fits to the 
\pbd\ X-ray data are obtained \cite{Batty:2002sk} both with $K_0 = 1$ 
(Borie--Leon model) and with $K_0 = 7.6$ as determined from fits to 
\pbp\ X-ray yield data.

\begin{figure}[!h]
\includegraphics[height=8.0cm]{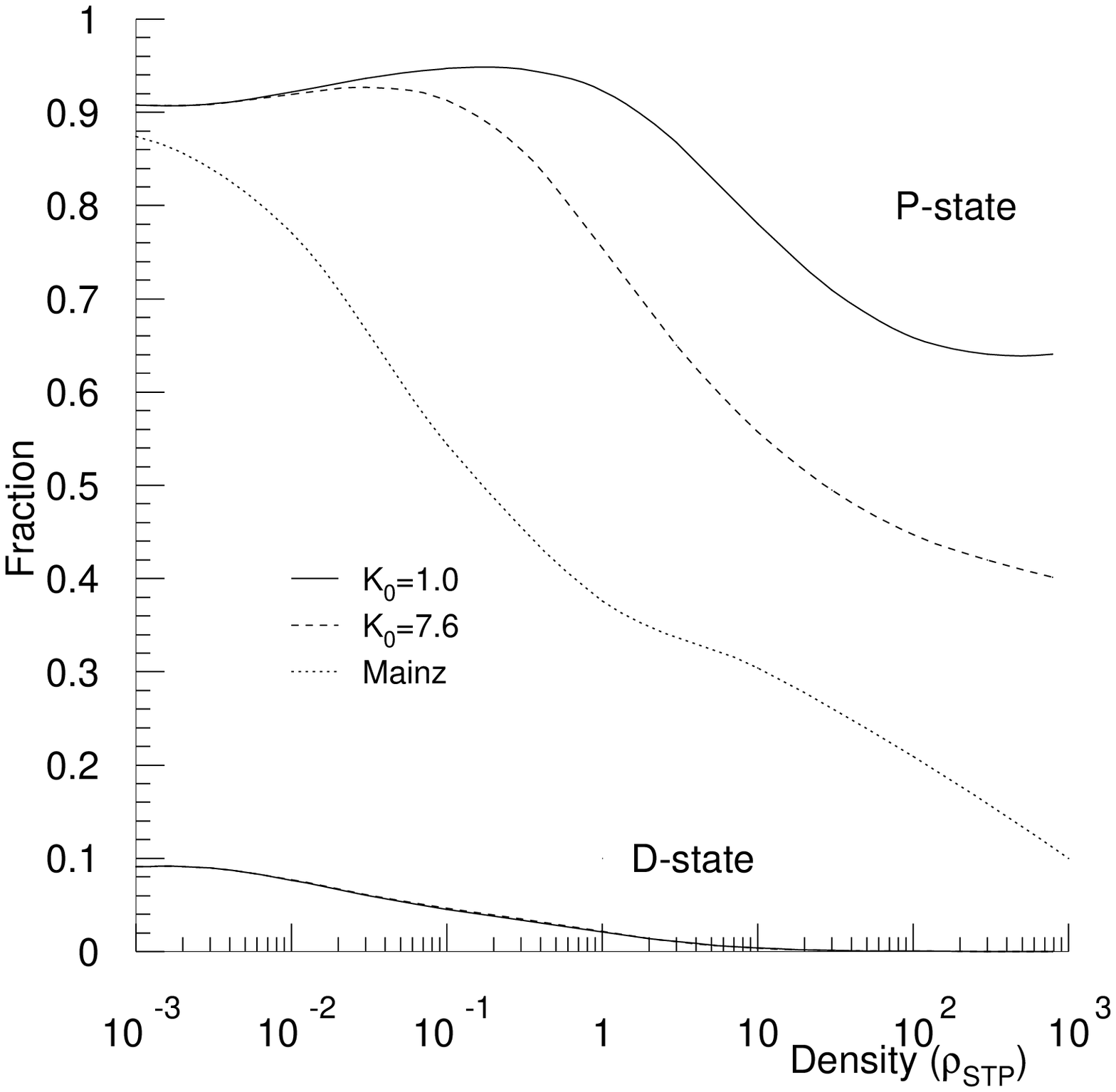}
\caption{Fractions of annihilation from atomic P- and D-states of \pbd\ atoms
predicted by the Borie--Leon model 
\protect\cite{BorieL80} giving a best fit to the X-ray data with $K_0 =1$ ( dashed line) 
or $K_0 = 7.6$ (full line). Also shown are the predictions of the 
Mainz \protect\cite{Reifenroether:1989nq,Reifenroether:1990tq} model (dotted line).}
\label{tm:fig:pdstated}
\vspace*{-4mm}
\end{figure}

The calculated fractions of annihilation, \fp\ and f$_{\rm{D}}(\rho)$, from atomic 
P- and D-states as a function of \DT\ target density, are plotted in 
Fig.~\ref{tm:fig:pdstated} for the best fits with $K_0 = 1$ and $K_0 = 7.6$. 
The values of \fp\ for $K_0 = 7.6$ are seen to be significantly reduced 
at higher target densities compared to those obtained with $K_0 = 1$.
This is in general accord with the results for \pbp\ atoms \cite{Batty:1996uf}, 
 shown in 
Fig.~\ref{tm:fig:pstatep}. 
Also plotted in Fig.~\ref{tm:fig:pdstated} are the calculated 
values \cite{Reifenroether:1989nq,Reifenroether:1990tq} obtained using the ``Mainz'' model.
As in the \pbp\ case shown in Fig.~\ref{tm:fig:pstatep} these values of 
\fp\ are seen to be 
significantly smaller at the higher target densities than those
predicted by the modified Borie--Leon model 
\cite{Batty:1996uf,BorieL80} with $K_0 = 7.6$.

Table \ref{tm:tab:efacd} gives enhancement factors 
for P-states of the \pbd\ atom calculated 
from this cascade calculation with a value $K_0 = 7.6$, 
together with the calculated P-state widths of 
Wycech et al.\ \cite{Wycech:1985py}. Even in the case of the 
$^{4}{\rm P}_{1/2}$
state, which has a width 20\% larger than the spin-averaged value, the
enhancement factor is only an 11\% effect in liquid deuterium. In view of the
relative smallness of this effect and so as to make a significant
simplification in the following discussion, for \pbd\ atoms the enhancement
factors are set to 1. Using Eq.~(\ref{tm:equ:BRp}) the annihilation 
frequency for production of the two-body channel $ch$ in a \DT\ target 
of density $\rho$ is then given by:
\begin{equation}\eqalign{%
\AF_{\rm N}&(ch,{\rm N}_{s},\rho) = {1 -
\fpa\over4}\,\left[\BR(ch,\ssz) + 
3\BR(ch,\tso)\right]\cr
&{}+ {\fpa\over12}\left[3 \BR(ch,\spo) + \BR(ch,\tpz) +  3  \BR(ch,\tpo) + 
5 \BR(ch,\tpt)\right] ~,}\label{tm:equ:BRpd}
\end{equation}
where N$_{s}$ is the spectator nucleon. Note that 
$\AF_{\rm N}(ch,{\rm N}_{s},\rho)$ is the 
frequency for annihilation on a single nucleon.
 
\begin{table}[!hp]
\caption{Calculated enhancement factors for \pbd\ atoms \label{tm:tab:efacd}}
\renewcommand{\arraystretch}{1.1}
\begin{tabular}{ccrr}  \hline\hline
State & Width & \multicolumn{2}{c}{Enhancement factor} \\ 
\slj  & meV   & \rstp    & liq.  \\
\hline
$^{2}{\rm P}_{1/2}$  & 398 & 0.99 & 0.98 \\
$^{2}{\rm P}_{3/2}$  & 386 & 0.98 & 0.96 \\
$^{4}{\rm P}_{1/2}$  & 512 & 1.04 & 1.11 \\
$^{4}{\rm P}_{3/2}$  & 430 & 1.00 & 1.01 \\
$^{4}{\rm p}_{5/2}$  & 420 & 1.00 & 1.00 \\
\hline\hline
\end{tabular}\renewcommand{\arraystretch}{1.0}
\end{table}
\FloatBarrier\subsectionb{Frequencies for \pbp\ annihilation at rest}\label{tm:sub:BRexp}
In this section, we present a compilation of 
annihilation frequencies. Measurements in which peaks in the 
inclusive recoil energy spectrum are used to identify the second 
meson, will be indicated by R (recoil). Measurements in which both of 
the two mesons in the two-body final state are fully reconstructed, 
which are expected to be more reliable, will be denoted by E (event).
In some cases \cite{Amsler:1993kg,Amsler:2003bq}, for all neutral 
channels, the absolute annihilation frequencies
are obtained by normalisation relative to the  
frequency for another channel, usually $\pbp\to\piz\piz$. These 
measurements will be indicated by N (normalisation). For the majority of 
entries in the tables, the detection technique is also identified.
$\mathcal{A}$, $\mathcal{C}$ and $\mathcal{O}$ are used to identify the
Asterix
(see Sec.~\ref{exp:sub:AST}),
Crystal Barrel (see Sec.~\ref{exp:sub:CBAR}) and Obelix
(see Sec.~\ref{exp:sub:OBX})
detectors at the LEAR facility. Other LEAR experiments
(see Sec.~\ref{exp:sub:cool}), in particular CPLEAR
are indicated in the tables by $\mathcal{L}$ , whilst $\mathcal{K}$ are
experiments carried out at 
KEK (see Sec.~\ref{exp:sub:KEK}). $\mathcal{B}$ is used to indicate
early bubble experiments ({see Sec.~\ref{exp:sub:early}), whilst $\mathcal{T}$ indicates
experiments carried out using electronics detection techniques with early
separated antiproton beams (see Sec.~\ref{exp:sub:cool}). 
The notation scheme is summarised in
Table~\ref{tm:tab:not}, for easier reading of the following
tables.

In a few cases, the frequency has been measured for the annihilation 
channel ($\pip\pim$, \Kp\Km, \Ks\Ks, $\phi\pi$, $\phi\eta$, $\phi\omega$ 
and $\rho^0\phi$) in coincidence with L X-rays,
so that annihilation only occurs from 2P states. For these measurements,
the frequency $\AF(ch)_{\rm X}$, no longer depends on the target
density 
and following Eq.~(\ref{tm:equ:BRp}) can be written in the form
\begin{equation}
\AF(ch)_{\rm X} =\frac{3}{12}\BR(ch,\spo) + \frac{1}{12}\BR(ch,\tpz) +
 \frac{3}{12}\BR(ch,\tpo) + \frac{5}{12}\BR(ch,\tpt) \label{tm:equ:BRx}
\end{equation}
It has been confirmed \cite{Batty:1996uf} that enhancement factors for 
2P-state annihilation, as observed in experiments using an L X-ray trigger, 
are within $\pm$1\%\ of the value 1.0 assumed in the derivation of this 
equation.

The labels S or P, in Table \ref{tm:tab:not} and the following tables,
indicate the results of a partial-wave analysis.

\begin{table}[!h]
\caption{Notation scheme for the experimental results: name of the
experiment and method of analysis\label{tm:tab:not}}
\begin{center}\renewcommand{\arraystretch}{1.1}\begin{tabular}{cll}
\hline\hline
  & notation & meaning\\
  \hline
\multirow{14}{.3cm}{\rotatebox{90}{\makebox[0pt][c]{\ data and data analysis}}}&
E
 & event \\
& E$_{\rm mb}$ &event with minimum bias trigger \\
&E$_{\rm an}$ & event  with all neutral trigger\\
&E$_{\phi \rm c}$ & $\phi\to\Kp\Km$\\
&E$_{\phi \rm n}$ & $\phi\to\Ks\Kl$\\
&E$_{\omega \rm n}$ & $\omega\to\piz\gamma$\\
&E$_{\omega \rm c}$ & $\omega\to\pip\pim\piz$\\
&E$_{\rm Kc}$ & $\Ks\to\pip\pim$\\
&E$_{\rm Kn}$ & $\Ks\to\piz\piz$\\
 &N& normalisation with respect to another channel\\
 &P& P-wave annihilation\\
 &R& recoil\\
 &S& S-wave annihilation\\
 &X& X-ray trigger\\
\hline
\multirow{6}{.3cm}{\rotatebox{90}{\makebox[0pt][c]{\ \ experiment}}}%
 & $\mathcal{A}$ & Asterix experiment at LEAR\\
 & $\mathcal{B}$ & Bubble Chamber experiments\\
 & $\mathcal{C}$ & Crystal Barrel experiment at LEAR\\
  & $\mathcal{K}$ & Experiments at KEK \\
& $\mathcal{L}$ & Other LEAR experiments (CPLEAR, $\ldots$)\\
  & $\mathcal{O}$ & Obelix experiment at LEAR\\
  & $\mathcal{T}$ & Electronics experiments using separated \pbar\ beams\\
  \hline\hline
  \end{tabular}\renewcommand{\arraystretch}{1.0}\end{center}\end{table}
\subsubsection{Frequencies for annihilation into two charged
mesons}\label{tm:sub:twoch}
The measurement of frequencies for annihilation into two charged 
mesons is much easier experimentally than for channels involving 
neutral mesons where the problems of measurement and identification 
are more challenging. This is reflected in the results in Table
\ref{tm:tab:BRcharge}, where
 there is good agreement between the values obtained from 
different experiments and experimental techniques. The weighted mean 
value of the annihilation frequency for $\pbp\to\pip\pim$ from 5 experiments 
is $\AF(\pip\pim,{\rm liq.}) = (3.27 \pm 0.05) \times 10^{-3}$ with
$\chi^2/N = 3.01/4$, 
where $N$ is the number of degrees of freedom. For the $\pbp\to\Kp\Km$ 
channel, the weighted mean $\AF(\Kp\Km,{\rm liq.}) = (1.03 \pm 0.03)
\times 10^{-3}$ 
with $\chi^2/N = 2.28/3$. For both channels the measured 
annihilation frequencies vary relatively smoothly as a function of target 
density \cite[Figs. 7(c) and 7(d)]{Abele:2001ek}.
\renewcommand{\bs}{\hspace*{2pt}}
\begin{table}[!h]
\caption{Frequencies for annihilation into two charged mesons}
\label{tm:tab:BRcharge}
\renewcommand{\arraystretch}{1.1}
\begin{tabular}{l@{\bs}c@{\bs}c@{\bs}c@{\bs}l%
@{\bs}|@{\bs}c@{\bs}c@{\bs}c@{\bs}l@{\bs}}  \hline\hline
Chan. \ \ & Dens. & Ann. Freq. & Ref. & Type&Dens. & Ann. Freq. & Ref.
&Type\\ \hline
\multirow{5}{.6cm}{$\pip\pim$}%
& Liq.    & $(3.20 \pm 0.30 ) \times 10^{-3}$ & \cite{Baltay65a} &$\mathcal{B}$ E&
Liq.    & $(3.33 \pm 0.17 ) \times 10^{-3}$ & \cite{Armenteros69} &$\mathcal{B}$ E  \\
& Liq.    & $(3.07 \pm 0.13 ) \times 10^{-3}$ &\cite{Amsler:1993kg} & $\mathcal{C}$ E&
Liq.    & $(3.30 \pm 0.20 ) \times 10^{-3}$ & \cite{Abele:2001ek} &$\mathcal{C}$ E 
\\
& Liq.    & $(3.31 \pm 0.06 ) \times 10^{-3}$ & \cite{Bargiotti:2002mu} &$\mathcal{O}$ E&
12      & $(4.05 \pm 0.23 ) \times 10^{-3}$ & \cite{Abele:2001ek} &$\mathcal{C}$ E \\
& 1       & $(4.30 \pm 0.14 ) \times 10^{-3}$ & \cite{Doser:1988yi} &$\mathcal{A}$ E &
1       & $(4.27 \pm 0.23 ) \times 10^{-3}$ & \cite{Ableev:1994vs} &$\mathcal{O}$ E \\
& 0.005   & $(4.26 \pm 0.11 ) \times 10^{-3}$ & \cite{Ableev:1994bg}  &$\mathcal{O}$ E&
1   & $(4.81 \pm 0.49 ) \times 10^{-3}$ & \cite{Doser:1988yi}  &$\mathcal{A}$ X
E\\ \hline
\multirow{4}{.6cm}{\Kp\Km }
& Liq.    & $(1.10 \pm 0.10 ) \times 10^{-3}$ & \cite{Baltay65a}  &$\mathcal{B}$ E&
Liq.    & $(1.01 \pm 0.05 ) \times 10^{-3}$ & \cite{Armenteros69} &$\mathcal{B}$ E  \\
& Liq.    & $(0.99 \pm 0.05 ) \times 10^{-3}$ &\cite{Amsler:1993kg}  & $\mathcal{C}$ E&
Liq.    & $(1.10 \pm 0.07 ) \times 10^{-3}$ & \cite{Abele:2001ek}  &$\mathcal{C}$ E \\
& 12      & $(9.07 \pm 0.59 ) \times 10^{-4}$ & \cite{Abele:2001ek}  &$\mathcal{C}$ E&
1       & $(6.92 \pm 0.41 ) \times 10^{-4}$ & \cite{Doser:1988yi}  &$\mathcal{A}$ E \\
& 0.005   & $(4.6  \pm 0.3  ) \times 10^{-4}$ & \cite{Ableev:1994bg}  &$\mathcal{O}$ E &
1   & $(2.87 \pm 0.51 ) \times 10^{-4}$ & \cite{Doser:1988yi}  &$\mathcal{A}$ X E \\
\hline\hline
\end{tabular}\renewcommand{\arraystretch}{1.0}\end{table}
\subsubsection{Frequencies for annihilation into two neutral 
  strange mesons}
  The  measurements of frequencies for annihilation into two
neutral kaons are shown in Table \ref{tm:tab:BRstrange0}. The five
measurements for $\pbp\to\Ks\Kl$ in liquid \HT\
are in moderate agreement with each other and give a weighted mean value 
$\AF(\Ks\Kl,{\rm liq.})= (7.86 \pm 0.40) \times 10^{-4}$ with $\chi^2/N
= 8.45/4$, where $N$ is the number of 
degrees of freedom. The two  values obtained by the 
Crystal Barrel experiment \cite{Abele:2001ek} for this same channel in 
a gas target at 12\rstp\ disagree by $2.4\sigma$. The lower value seems 
to be in better agreement with the overall trend of results for the 
\Ks\Kl\ annihilation frequency as a function of target density 
\cite[Fig.~7(b)]{Abele:2001ek}.
\begin{table}[!h]
\caption{Frequencies for annihilation into two neutral strange mesons}
\label{tm:tab:BRstrange0}
\begin{center}
\renewcommand{\arraystretch}{1.1}
\begin{tabular}{l@{\bs}c@{\bs}c@{\bs}c@{\bs}l%
@{\bs}|@{\bs}c@{\bs}c@{\bs}c@{\bs}l@{\bs}}  \hline\hline
Chan.\ \  & Dens. & Ann. Freq. & Ref. & Type&Dens. & Ann. Freq. & Ref.
&Type\\ \hline
\multirow{5}{.5cm}{\Ks \Kl}%
& Liq.    & $(6.10 \pm 0.90 ) \times 10^{-4}$ &\cite{Baltay65a}  & B
E$_{\mathrm Kc}$&
Liq.    & $(7.60 \pm 0.40 ) \times 10^{-4}$ & \cite{Armenteros69} &$\mathcal{B}$ E \\
& Liq.    & $(9.00 \pm 0.60 ) \times 10^{-4}$ & \cite{Amsler:1995up}
 & $\mathcal{C}$ E$_{\mathrm Kn,an}$&
Liq.    & $(7.80 \pm 0.76 ) \times 10^{-4}$ & \cite{Bertin:1996pb}  &
$\mathcal{O}$ $E_{\mathrm Kc}$ \\ 
& Liq.&$(8.64 \pm 1.02 ) \times 10^{-4}$ & \cite{Abele:2001ek}  &
$\mathcal{C}$ E$_{\mathrm Kn,an}$&
12      & $(7.04 \pm 0.74 ) \times 10^{-4}$ & \cite{Abele:2001ek}  &
$\mathcal{C}$ E$_{\mathrm Kn,mb}$\\
& 12      & $(4.89 \pm 0.56 ) \times 10^{-4}$ & \cite{Abele:2001ek}  &
$\mathcal{C}$ E$_{\mathrm Kn,an}$&
1     & $(3.60 \pm 0.60 ) \times 10^{-4}$ & \cite{Doser:1988fw}  & 
$\mathcal{A}$ E$_{\mathrm Kc}$ \\
& 1     & $(3.50 \pm 0.54 ) \times 10^{-4}$ & \cite{Bertin:1996pb}%
 & $\mathcal{O}$ E$_{\mathrm Kc}$&
 0.005   & $(1.00 \pm 0.32 ) \times 10^{-4}$ & \cite{Bertin:1996pb}  & 
$\mathcal{O}$ E$_{\mathrm Kc}$ \\ \hline
\multirow{2}{.5cm}{\Ks \Ks}%
 & Liq.    & $(4.0 \pm 3.0 ) \times 10^{-6}$ &\cite{Baltay65a,Doser:1988fw}  & 
$\mathcal{B}$ E$_{\mathrm Kc}$ &
Liq.    & $(7.0 \pm 3.5 ) \times 10^{-6}$ & \cite{Amsler:2003bq}  & 
$\mathcal{B}$ E$_{\mathrm Kc}$  \\
& 1       & $(3.00 \pm 1.00 ) \times 10^{-5}$ & \cite{Doser:1988fw}  & 
$\mathcal{A}$ E$_{\mathrm Kc}$  &
1   & $(3.70 \pm 1.40 ) \times 10^{-5}$ & \cite{Doser:1988fw}  & 
$\mathcal{A}$ X E$_{\mathrm Kc}$  \\
\hline\hline
\end{tabular}\renewcommand{\arraystretch}{1.0}
\end{center}
\end{table}

The annihilation frequency for $\pbp\to\Ks\Ks$ is small ($\sim 10^{-5}$) and 
hence difficult to measure. Values in a gas target (\rstp), both with 
and without an atomic L X-ray trigger, have been obtained by the 
Asterix experiment. Values for liquid \HT, where the annihilation frequency
is even smaller, have only been determined on the basis of four
\Ks\Ks\ events observed in the bubble chamber data at BNL and CERN.
\subsubsectionb{Annihilation frequencies for $\pbp\to\piz\piz$.}
\label{tm:sub:BRpi0pi0}
As can be seen from Table \ref{tm:tab:BRpi0pi0}, there are widely 
differing measurements for the \piz\piz\ annihilation frequency in liquid 
targets, particularly so for the Crystal Barrel 
\cite{Amsler:1993kg,Abele:2001ek} 
and Obelix \cite{Bargiotti:2002mu} experiments. Whilst the Obelix 
measurement agrees with some earlier data \cite{Adiels:1987tn,Chiba:1988cg} 
where only 2 or 3 photons were observed, it should be noted that Obelix,
Crystal Barrel and the work of Devons et al.\ \cite{Devons:1971rn} 
are the only experiments to measure all 4 photons 
and to reconstruct 2\piz\ events fully. The fact that the 
$\pip\pim$ annihilation frequencies obtained by 
the Crystal Barrel and Obelix experiments are in good agreement, suggests 
that the origin of the problem lies in the determination of the 
reconstruction efficiency for detecting \piz\piz\ events. This has been 
discussed recently by the Crystal Barrel collaboration 
in a comment \cite{Amsler:2002ua} on the Obelix paper 
\cite{Bargiotti:2002mu}. It was pointed out that there is good 
agreement of the annihilation frequencies 
measured by the Crystal Barrel experiment
for several reactions  with different numbers of photons in the final state. 
In \cite{Amsler:1993kg} various final states with 4 (or 5) and with 
8 (or 9) photons in the final state were measured.
\begin{table}[!h]
\caption{Annihilation frequencies for $\pbp\to\piz\piz$  \label{tm:tab:BRpi0pi0}}
\begin{center}\renewcommand{\arraystretch}{1.2}
\begin{tabular}{ccclcccl}  \hline\hline
 Dens. & Ann. Freq. & Ref. & Type && Ann. Freq. & Ref. &
Type\\ 
\hline
 Liq.    & $(4.8  \pm 1.0)  \times 10^{-4}$ & \cite{Devons:1971rn}
&$\mathcal{T}$ E&&
 $(1.4  \pm 0.3) \times 10^{-4}$ & \cite{Bassompierre79} &$\mathcal{T}$ R \\
Liq.    & $(6.0  \pm 4.0)  \times10^{-4}$ & \cite{Backenstoss:1983gu} &$\mathcal{T}$ R &&
$(2.06 \pm 0.14) \times 10^{-4}$ & \cite{Adiels:1987tn} &$\mathcal{L}$ R \\
Liq.    & $(2.5  \pm 0.3)  \times 10^{-4}$ & \cite{Chiba:1988cg} &$\mathcal{K}$  R&& 
$(6.93 \pm 0.43) \times 10^{-4}$ & \cite{Amsler:1992ku,Amsler:1993kg} &$\mathcal{C}$ E\\
Liq.    & $(6.14 \pm 0.40) \times 10^{-4}$ & \cite{Abele:2001ek}
&$\mathcal{C}$ E&&
$(2.8  \pm 0.4)  \times 10^{-4}$ & \cite{Bargiotti:2002mu} &$\mathcal{O}$ E \\
12      & $(1.54 \pm 0.09) \times 10^{-3}$ & \cite{Abele:2001ek} &
$\mathcal{C}$ E$_{\rm mb}$&&
$(1.62 \pm 0.15) \times 10^{-3}$ & \cite{Abele:2001ek} & 
$\mathcal{C}$ E$_{\rm an}$  \\
 1       & $(1.27 \pm 0.21) \times 10^{-3}$ & \cite{Agnello:1994kg}  &
$\mathcal{O}$ E \\ 
\hline\hline
\end{tabular}\renewcommand{\arraystretch}{1.0}
\end{center}\end{table}

The value of $\AF(\piz\piz,{\rm liq})$ has been used by 
the Crystal Barrel collaboration  to normalise other 
annihilation frequencies with only neutral particles
in the final state \cite{Amsler:1993kg,Amsler:2003bq}.
Using the frequency for 
$\AF(\piz\piz,{\rm liq})$ \cite{Amsler:1993kg} to
normalise the annihilation frequency
for the $\omega\omega$ final state gives 
$\AF(\omega\omega,{\rm liq}) = (3.32\pm0.34)\%$ 
\cite{Amsler:1993kg} and more recently 
$\AF(\omega\omega,{\rm liq}) = (2.95\pm0.15)\%$
\cite{Amsler:2003bq}.
The absolute \pbp$\to\omega_1\omega_2$ annihilation frequency 
with $\omega_1\to\pip\pim\piz$,
$\omega_2\to\piz\gamma$ has also
been measured by them \cite{McCrady:1998th}. 
$\AF(\omega\omega,{\rm liq}) = (3.23\pm0.25)\%$ was obtained,
in agreement with the
normalised results, giving confidence in the measured 
$\AF(\piz\piz,{\rm liq})$.
 
Other support for the efficiency determination for the Crystal Barrel 
detector can be made by a comparison of
annihilation frequencies both for all neutral and for final states 
involving charged particles.
These include measurements by the Obelix
collaboration \cite{Bertin:1996pb} for the reaction $\pbp\to \Ks\Kl$;
$\Ks\to\pip\pim$ at three target densities.
The Obelix value $\AF(\Ks\Kl,{\rm liq}) = (7.8\pm0.7\pm0.3)\times 10^{-4}$
is in good
agreement with those measured \cite{Abele:2001ek,Amsler:1995up} by the Crystal 
Barrel experiment, $\AF(\Ks\Kl,{\rm liq}) = (9.0\pm 0.6)\times 10^{-4}$
and 
$\AF(\Ks\Kl,{\rm liq}) = (8.6\pm 1.0)\times 10^{-4}$, 
for the same reaction but with $\Ks\to\piz\piz$. 

Further evidence as 
to the consistency of the Crystal Barrel results is given by recent 
measurements \cite{Amsler:2003bq} of annihilation frequencies 
for final states 
containing up to five $\piz$ and  $\eta$ mesons,
 using a liquid target. Events with between 4 and 11 photons were processed 
and frequencies obtained by normalisation to the Crystal Barrel 
result \cite{Amsler:1993kg} for the \piz\piz\ channel. Including 
annihilation frequencies for the production of neutral kaons from 
other experiments, the identified channels from annihilation in liquid 
hydrogen add up to a total frequency for all neutral annihilations of 
$(3.56 \pm 0.28)$\% per 
annihilation compared to a value of $(3.5 \pm 0.3)$\% for all neutral 
annihilations measured inclusively in the same experiment without using the 
\piz\piz\ normalisation. Both values are in 
good agreement with the values $(3.2 \pm 0.5)$\% and $(4.1^{+0.2}_{-0.6})$\% 
obtained \cite{Armenteros69} with bubble chambers at BNL and CERN 
respectively. Using the Obelix value for the \piz\piz\ frequency to
normalise the Crystal Barrel results 
would give a total all-neutral annihilation frequency of $(1.6 \pm 0.25)$\% 
in marked disagreement with the other measurements.

In the above discussion we have emphasised the internal and external 
consistency of the Crystal Barrel results relating to their  
annihilation frequencies for the \piz\piz\ channel. Unfortunately no similar 
checks are available for the Obelix results, although the authors do 
point out \cite{Bargiotti:2002mu} that measurements of
$\pbp\to\pip\pim\piz$, both with 
and without the \piz\ being detected are in good agreement showing that 
their estimation of the Obelix efficiency for detection of a 
single \piz\ is reliable. We conclude 
however, there is a large body of evidence that the
annihilation frequencies measured with the Crystal Barrel detector for 
a variety of different final states are consistent.

The Obelix collaboration has also measured \cite{Agnello:1994kg} the
\piz\piz\ annihilation frequency with a gaseous \HT\ target at STP. This 
measurement has again been discussed recently by both the 
Obelix \cite{Bargiotti:2002mu} and Crystal Barrel \cite{Amsler:2002ua} 
collaborations. The latter suggest that, on the basis of a 
fit \cite[Fig.~7(a)]{Abele:2001ek} with 
Eq.~(\ref{tm:equ:BRp}) to Crystal Barrel and Asterix data,
the measured $\AF(\piz\piz,\rstp)$ by Obelix \cite{Agnello:1994kg}
could be too low by a factor of about $1.4$. 
\FloatBarrier\subsubsection{Other two neutral pseudoscalar channels}
\label{tm:sub:BR2ps}
Most of the  measurements for two-body annihilation frequencies
into other neutral pseudoscalar mesons, 
$\pbp\to\piz\eta, \piz\eta^{\prime}, \eta\eta$ and $\eta\eta^{\prime}$,
have been made by the Crystal Barrel collaboration 
\cite{Amsler:1993kg,Abele:2001ek,Amsler:2003bq}. As for the case of the 
$\AF(\piz\piz,{\rm liq})$ results, measurements for $\AF(\piz\eta,{\rm
liq})$ made
by the Crystal Barrel and Obelix \cite{Bargiotti:2002mu} experiments again 
differ by a factor about 2. However it has been pointed out that the ratio 
$\AF(\piz\eta,{\rm liq})/\AF(\piz\piz,{\rm liq})$ obtained by the two
experiments, 
$0.32\pm0.07$ (Obelix \cite{Bargiotti:2002mu}) and $0.303\pm0.010$ 
(Crystal Barrel \cite{Amsler:1993kg}) are in good agreement.

\begin{table}[!h]
\caption{Frequencies for annihilation into two neutral pseudoscalar mesons  \label{tm:tab:BR2ps}}
\renewcommand{\arraystretch}{1.1}
\renewcommand{\bs}{\hspace*{1pt}}
\begin{tabular}{l@{\bs}c@{\bs}c@{\bs}c@{\bs}l@{\bs}|@{\bs}%
c@{\bs}c@{\bs}c@{\bs}l@{\bs}}
 \hline\hline
Chan.\ \ & Dens. & Ann. Freq. & Ref. & Type&Dens.& Ann. Freq. & Ref. & Type
\\
\hline
\multirow{5}{.4cm}{$\piz\eta$}& Liq.& $(82.0 \pm 10.0) \times 10^{-4}$
&\cite{Backenstoss:1983gu}& $\mathcal{T}$ R &
Liq.& $(4.6  \pm 1.3)  \times 10^{-4}$ & \cite{Chiba:1988cg}  & $\mathcal{K}$ R \\
& Liq.& $(1.33 \pm 0.27) \times 10^{-4}$ & \cite{Adiels:1989ih}  & $\mathcal{L}$ R
&Liq.&$(2.12 \pm 0.12) \times 10^{-4}$ & \cite{Amsler:1993kg}  & $\mathcal{C}$ E N \\
& Liq. & $(2.50 \pm 0.30) \times 10^{-4}$ & \cite{Abele:2001ek}  & $\mathcal{C}$ E
&Liq.&$(0.90 \pm 0.22) \times 10^{-4}$ & \cite{Bargiotti:2002mu} 
& $\mathcal{O}$ E\\
& Liq.  & $(2.09 \pm 0.10) \times 10^{-4}$ & \cite{Amsler:2003bq}  &
$\mathcal{C}$ E N & 12      & $(5.63 \pm 0.43) \times 10^{-4}$ & \cite{Abele:2001ek}
 &  $\mathcal{C}$ E$_{\rm mb}$\\
&12&$(4.57 \pm 0.30) \times 10^{-4}$ & \cite{Abele:2001ek}  & $\mathcal{C}$ E$_{\rm
an}$& 12      & $(4.78 \pm 0.21) \times 10^{-4}$ & \cite{Amsler:2003bq}  &
$\mathcal{C}$ E N \\ \hline
\multirow{2}{.4cm}{$\piz\eta^{\prime}$}
& Liq.    & $(5.0  \pm 1.9)  \times 10^{-4}$ & \cite{Chiba:1988cg}  &
$\mathcal{K}$ R&Liq.&
 $(1.23 \pm 0.13) \times 10^{-4}$ & \cite{Amsler:1993kg}  &
$\mathcal{C}$ E N \\
& Liq.    & $(0.98 \pm 0.24) \times 10^{-4}$ & \cite{Amsler:2003bq}  &
$\mathcal{C}$ E N & 12      & $(2.03 \pm 0.13) \times 10^{-4}$ & \cite{Amsler:2003bq}  &
$\mathcal{C}$ E N \\ \hline
\multirow{3}{.4cm}{$\eta\eta$}
& Liq.& $(1.60 \pm 0.80) \times 10^{-4}$ & \cite{Chiba:1989yw}  &
$\mathcal{K}$ R&Liq.&
$(0.81 \pm 0.31) \times 10^{-4}$ & \cite{Adiels:1989ih}  &
$\mathcal{L}$ R \\
& Liq.& $(1.64 \pm 0.10) \times 10^{-4}$ & \cite{Amsler:1993kg}  &
$\mathcal{C}$ E N&Liq.&
$(1.53 \pm 0.08) \times 10^{-4}$ & \cite{Amsler:2003bq}  &
$\mathcal{C}$ E N \\
& 12& $(3.17 \pm 0.14) \times 10^{-4}$ & \cite{Amsler:2003bq}  &
$\mathcal{C}$ E N \\ \hline
\multirow{2}{.4cm}{$\eta\eta^{\prime}$}&
 Liq.& $(2.16 \pm 0.25) \times 10^{-4}$ & \cite{Amsler:1993kg}  &
$\mathcal{C}$ E N&Liq.&
$(2.49 \pm 0.33) \times 10^{-4}$ & \cite{Amsler:2003bq}  &
$\mathcal{C}$ E N \\
&12&$(3.81\pm 0.28)\times 10^{- 4}$ &\cite{Amsler:2003bq}  &
 $\mathcal{C}$ E N \\
\hline\hline
\end{tabular}\renewcommand{\arraystretch}{1.0}
\vspace*{-4mm}
\end{table}

Whilst the recent Obelix \cite{Bargiotti:2002mu} value for 
$\AF(\piz\eta,{\rm liq})$ is supported by the measurement of Adiels
\etal\ \cite{Adiels:1989ih}, it is not in agreement 
with other early results \cite{Backenstoss:1983gu,Chiba:1988cg} which
show particularly large fluctuations in value. This may be compared to
the situation for the $\pbp\to\piz\piz$ frequency discussed in
Sec.~\ref{tm:sub:BRpi0pi0} above which has been used by the Obelix
collaboration \cite{Bargiotti:2002mu} as support for their low value
of $\BR(\piz\pi\,{\rm liq.})$. We note that all these early results,
which were obtained from measurements of the inclusive recoil energy
spectrum, are expected to be less reliable than more recent results in
which the events were fully reconstructed.
\subsubsection{Frequencies for annihilation into a 
pseudoscalar and a vector meson}
\label{tm:sub:BRpsv}
For the $\omega\piz, \omega\eta$ and $\omega\eta^{\prime}$ channels,
the majority of the annihilation frequencies (Table
\ref{tm:tab:BRpsv}) were measured by the Crystal Barrel
collaboration. The results for liquid \HT\ targets are only in
moderate agreement with some early measurements. These latter results
were obtained from measurements of the inclusive recoil energy
spectrum.
As the $\omega\piz, \omega\eta$ and $\omega\eta^{\prime}$ channels are
produced solely from \tso\ and \spo\ initial \pbp\ states, the recent
Crystal Barrel measurements \cite{Amsler:2003bq} at a target density
of 12\rstp, allow the separate contributions from these two initial
states to be determined.
\renewcommand{\bs}{\hspace*{1pt}}
\begin{table}[!h]
\caption{Frequencies for annihilation into a pseudoscalar and a vector meson  \label{tm:tab:BRpsv}}
\begin{minipage}[t]{\textwidth}
\renewcommand{\footnoterule}{}
\renewcommand{\arraystretch}{1.06}
\begin{tabular}{l@{\bs}c@{\bs}c@{\bs}c@{\bs}l%
@{\bs}|@{\bs}c@{\bs}c@{\bs}c@{\bs}l@{\bs}}  \hline\hline
Chan.\ \  & Dens. & Ann. Freq. & Ref. & Type&Dens. & Ann. Freq. & Ref.
&Type\\ \hline
\multirow{3}{.4cm}{$\omega\piz$}& Liq.    & $(23.8  \pm 6.5  ) \times
10^{- 3}$ &\cite{Backenstoss:1983gu}  & $\mathcal{T}$ R 
&Liq.&$(5.2  \pm 0.5  ) \times 10^{-3}$ & \cite{Chiba:1988cg}  &$\mathcal{K}$  R \\
& Liq.    & $(5.73 \pm 0.47 ) \times 10^{-3}$ & \cite{Amsler:1993kg}  &
$\mathcal{C}$ E N& 
Liq.    & $(6.16 \pm 0.44 ) \times 10^{-3}$ & \cite{Schmid91phd,Amsler:1998up}  &
$\mathcal{C}$ E \\
& Liq.    & $(6.00 \pm 0.30 ) \times 10^{-3}$ & \cite{Amsler:2003bq}  &
$\mathcal{C}$ E N& 
12     & $(4.60 \pm 0.24 ) \times 10^{-3}$ & \cite{Amsler:2003bq}  &
$\mathcal{C}$ E N \\
\hline
\multirow{3}{.4cm}{$\omega\eta$}& Liq.    & $(0.46 \pm 0.14 ) \times
10^{- 2}$ &\cite{Chiba:1989yw}  &$\mathcal{K}$  R&
Liq.    & $(1.04 \pm 0.10 ) \times 10^{-2}$ & \cite{Adiels:1989ih} 
&$\mathcal{L}$ E\\
& Liq.    & $(1.51 \pm 0.12 ) \times 10^{-2}$ & \cite{Amsler:1993kg}  &$\mathcal{C}$ E N&
 Liq.    & $(1.63 \pm 0.12 ) \times 10^{-2}$ & \cite{Schmid91phd,Amsler:1998up}  &
$\mathcal{C}$ E \\
& Liq.    & $(1.53 \pm 0.06 ) \times 10^{-2}$ & \cite{Amsler:2003bq}  &$\mathcal{C}$  E N&
12      & $(9.31 \pm 0.42 ) \times 10^{-3}$ & \cite{Amsler:2003bq}  &
$\mathcal{C}$ E N \\  \hline
\multirow{2}{.4cm}{$\omega\eta^{\prime}$}& Liq.    & $(7.8 \pm 0.8 )
\times 10^{- 3}$ &\cite{Amsler:1993kg}  &$\mathcal{C}$ E N&
Liq.    & $(8.46 \pm 0.97 ) \times 10^{-3}$ & \cite{Amsler:2003bq}  &$\mathcal{C}$ E N\\
& 12      & $(7.32 \pm 0.56 ) \times 10^{-3}$ & \cite{Amsler:2003bq}  &
$\mathcal{C}$  E N \\   \hline
\multirow{5}{.4cm}{$\phi\piz$}& Liq.    & $(3.3  \pm 1.5  ) \times
10^{- 4}$ &\cite{Chiba:1988cg}  &$\mathcal{K}$  R&
 Liq.    & $(6.50 \pm 0.60 ) \times 10^{-4}$ & \cite{Amsler:1995up}  &
$\mathcal{C}$ E$_{\phi\mathrm{n},\phi\mathrm{c}}$\\
& Liq.    & $(4.88 \pm 0.32 ) \times 10^{-4}$ & \cite{Alberico:1998fr}  &
$\mathcal{O}$ E$_{\phi\mathrm{c}}$&
Liq.    & $(5.20 \pm 0.60 ) \times 10^{-4}$ & \cite{Abele:1999en}  &
$\mathcal{C}$ E$_{\phi\mathrm{c}}$  \\
& 1       & $(1.90 \pm 0.50 ) \times 10^{-4}$ & \cite{Reifenroether:1991ik}  &
$\mathcal{A}$ E$_{\phi\mathrm{c}}$&
1       & $(2.46 \pm 0.24 ) \times 10^{-4}$ & \cite{Ableev:1995aq}  &
$\mathcal{O}$ E$_{\phi\mathrm{c}}$ \\
& 1       & $(2.47 \pm 0.21 ) \times 10^{-4}$ & \cite{Alberico:1998fr}  &
$\mathcal{O}$ E$_{\phi\mathrm{c}}$&
 1   & $(0.30 \pm 0.30 ) \times 10^{-4}$ &
\cite{Reifenroether:1991ik}  & $\mathcal{A}$  X E$_{\phi\mathrm{c}}$  
 \\
&0.005   & $(0.92 \pm 0.10 ) \times 10^{-4}$ & \cite{Alberico:1998fr} 
& $\mathcal{O}$ E$_{\phi\mathrm{c}}$
\\
\hline
\multirow{4}{.4cm}{$\phi\eta$}& Liq.    & $(7.80 \pm 2.10 ) \times
10^{- 5}$ &\cite{Amsler:1993xd,Amsler:1995up} &
$\mathcal{C}$ E$_{\phi\mathrm{n}}$\footnote{Updated by Amsler 
\cite{Amsler:1998up}}&
Liq.    & $(7.10 \pm 0.70 ) \times 10^{-5}$ & \cite{Alberico:1998uj} &
$\mathcal{O}$ E$_{\phi\mathrm{c}}$ \\
& 1       & $(3.70 \pm 0.90 ) \times 10^{-5}$ & \cite{Reifenroether:1991ik}  &
$\mathcal{A}$ E$_{\phi\mathrm{c}}$&
1       & $(8.70 \pm 2.10 ) \times 10^{-5}$ & \cite{Ableev:1995aq}  &
$\mathcal{O}$ E$_{\phi\mathrm{c}}$  \\
& 1    & $(13.3 \pm 1.5 ) \times 10^{-5}$ & \cite{Alberico:1998uj} &
$\mathcal{O}$ E$_{\phi\mathrm{c}}$&
0.005    & $(16.6 \pm 2.0 ) \times 10^{-5}$ & \cite{Alberico:1998uj} &
$\mathcal{O}$ E$_{\phi\mathrm{c}}$ \\
%
& 1   & $(4.10 \pm 1.60 ) \times 10^{-5}$ &
\cite{Reifenroether:1991ik}  & $\mathcal{A}$ X E$_{\phi\mathrm{c}}$  \\
\hline
\multirow{2}{.4cm}{$\rho^0\piz$}& Liq.    & $(1.40 \pm 0.20 ) \times
10^{- 2}$ &\cite{Baltay66} & $\mathcal{B}$ E&
Liq.    & $(1.72 \pm 0.27 ) \times 10^{-2}$ & \cite{Armenteros69} &
$\mathcal{B}$ E  \\
& Liq.    & $(1.6  \pm 0.1  ) \times 10^{-2}$ & \cite{Chiba:1988cg}  &$\mathcal{K}$ R&
Liq.    & $(1.58 \pm 0.12 ) \times 10^{-2}$ & \cite{Abele:1999ac} &$\mathcal{C}$ E
\\ \hline
$\rho^{\pm}\pi^{\mp}$ & Liq.    & $(2.90 \pm 0.40 ) \times 10^{-2}$ & \cite{Baltay66} &
$\mathcal{B}$ E&
Liq.    & $(3.44 \pm 0.54 ) \times 10^{-2}$ & \cite{Armenteros69} &
$\mathcal{B}$ E \\   \hline
\multirow{2}{.4cm}{$\rho^0\eta$}& Liq.    & $(2.20  \pm 1.70  ) \times
10^{- 3}$ &\cite{Baltay66}  & $\mathcal{B}$ E&
Liq.    & $(9.60  \pm 1.60  ) \times 10^{-3}$ & \cite{Chiba:1989yw}  &$\mathcal{K}$  R\\
& Liq.& $(5.30  \pm 2.00  ) \times 10^{-3}$ & \cite{Adiels:1989ih} 
&$\mathcal{L}$ R & \\  \hline
$\rho^0\eta^{\prime}$ & Liq.    & $(1.46  \pm 0.42  ) \times 10^{-3}$ & \cite{Urner95phd}  &
$\mathcal{C}$ E   \\\hline\hline
\end{tabular} \renewcommand{\arraystretch}{1.0}
\end{minipage}
\end{table}

Measurements for the $\phi\piz$ and $\phi\eta$ channels 
(Table \ref{tm:tab:BRpsv}) have mostly 
been made by the Obelix and Asterix collaborations, detecting the 
$\phi$ meson through its decay $\phi\to\Kp\Km$. There are significant 
discrepancies between the results from these two experiments for the 
$\phi\eta$ channel measured with a gas target at a density \rstp, 
although those for the $\phi\piz$ channel are in reasonable agreement.
Results from the Crystal Barrel experiment using a liquid \HT\ target,
detecting the decays $\phi\to\Kp\Km$ or $\phi\to\Kl\Ks\to\Kl\piz\piz$
are generally in good agreement with the other measurements. The Asterix 
collaboration \cite{Reifenroether:1991ik} has also determined the 
annihilation frequencies for these channels with \pbp\ atomic L X-rays in 
coincidence, so selecting annihilation from atomic P-states.

\renewcommand{\bs}{\hspace*{1pt}}
\begin{table}[!h]
\caption{Annihilation frequencies for annihilation into
$\rho\pi$, $\rho\eta$, and $\K^*\Kb$ from partial wave analyses} 
\label{tm:tab:BRrhopi}
\begin{center}
%
\renewcommand{\arraystretch}{1.2}
\begin{tabular}{ccccl|ccl}
\hline\hline
State & Chan. & Ann. Freq. & Ref. & Type& Ann. Freq. &Ref.&Type \\
\hline
\ssz & $\rho^\pm\pimp$  & $<0.14\times10^{-2}$ &\cite{Foster:1968am}&$\mathcal{B}$ E\\
\ssz & $\rho^\pm\pimp$  & $(0.09\pm0.04)\times10^{-2}$
&\cite{May:1990ju}& $\mathcal{A}$ S E
&$(0.19\pm0.03)\times 10^{-2}$ &\cite{Bargiotti:2003ev}&$\mathcal{O}$ S E \\
\tso & $\rho^0\piz$      &$(1.52\pm0.25)\times 10^{-2}$ 
&\cite{Foster:1968am}&$\mathcal{B}$ E\\
\tso & $\rho^0\piz$      & $(1.69\pm0.22)\times10^{-2}$
&\cite{May:1990ju}& $\mathcal{A}$ S E
&$(1.58\pm0.09)\times 10^{-2}$  &\cite{Bargiotti:2003ev}& $\mathcal{O}$ S E\\
\spo& $\rho^0\piz$& $(0.40\pm0.09)\times10^{-2}$ &\cite{May:1990jv}
& $\mathcal{A}$ P E
&  $(0.43\pm0.04)\times 10^{-2}$ &\cite{Bargiotti:2003ev}&$\mathcal{O}$ P E\\ 
${\tpo}_2$ & $\rho^\pm\pimp$  & $(0.69\pm0.11)\times10^{-2}$ &\cite{May:1990jv}& 
$\mathcal{A}$ P E
 &$(0.77\pm0.17)\times10^{-2}$&\cite{Bargiotti:2003ev}&$\mathcal{O}$ P E\\ 
\hline
\tso & $\rho^0\eta$   & $(3.29\pm0.90)\times 10^{-3}$ & \cite{Weidenauer:1990fy}
 & $\mathcal{A}$ S E 
 &$(6.40  \pm 1.40  ) \times 10^{-3}$ & \cite{Foster:1968di}  &$\mathcal{B}$ E\\
\tso & $\rho^0\eta$ & $(5.00  \pm 1.40  ) \times 10^{-3}$ & \cite{Espigat:1972aa}  
 &$\mathcal{B}$ E 
 & $(3.87  \pm 0.29  ) \times 10^{-3}$ & \cite{Abele:1997wg} 
&$\mathcal{C}$ E\\ 
\spo & $\rho^0\eta$   & $(0.94\pm0.53) \times 10^{-3}$ & \cite{Weidenauer:1990fy}  &
$\mathcal{A}$ P E\\  \hline
\tso & $\rho^0\eta^{\prime}$  & $(1.81  \pm 0.44  ) \times 10^{-3}$ & \cite{Foster:1968di,Weidenauer:1990fy}  &
$\mathcal{A}\ \mathcal{B}$ E \\ \hline
\ttso & $\K^*\Kb$   &$(24.6\pm2.4)\times 10^{-4}$ &\cite{Bettini69a}  &$\mathcal{B}$ E
                           &$(16.6\pm2.5)\times 10^{-4}$ &\cite{Abele:1998qd}&$\mathcal{C}$ E\\ 
\ttso & $\K^*\Kb$   &$(18.9\pm2.0)\times 10^{-4}$ &\cite{Bargiotti:2004aa}&$\mathcal{O}$ S E 
                           &$(19.1\pm3.3)\times 10^{-4}$ &\cite{wittmack:2001kw}        &$\mathcal{C}$ E \\
\stso & $\K^*\Kb$   &$(0.8\pm0.4)\times 10^{-4}$  &\cite{Bettini69a}&$\mathcal{B}$ E
                           &$(4.5\pm1.2)\times 10^{-4}$  &\cite{Abele:1998qd}&$\mathcal{C}$ E \\
\stso & $\K^*\Kb$   &$(2.8\pm0.4)\times 10^{-4}$  &\cite{Bargiotti:2004aa}&$\mathcal{O}$ S E 
                           & $(0.4\pm3.1)\times 10^{-4}$ &\cite{wittmack:2001kw}        &$\mathcal{C}$ E  \\
\tssz & $\K^*\Kb$   &$(2.0\pm0.4)\times 10^{-4}$  &\cite{Bettini69a}&$\mathcal{B}$ E 
                           &$(0.9\pm0.3)\times 10^{-4}$  &\cite{Abele:1998qd}&$\mathcal{C}$ E\\
\tssz & $\K^*\Kb$   &$(2.3\pm0.5)\times 10^{-4}$  &\cite{Bargiotti:2004aa}&$\mathcal{O}$ S E  
                           & $(1.1\pm1.5)\times 10^{-4}$ &\cite{wittmack:2001kw}        &$\mathcal{C}$ E  \\
\sssz & $\K^*\Kb$   &$(8.6\pm1.7)\times 10^{-4}$  &\cite{Bettini69a}&$\mathcal{B}$ E
                           &$(6.2\pm0.9)\times 10^{-4}$  &\cite{Abele:1998qd}&$\mathcal{C}$ E\\
\sssz & $\K^*\Kb$   &$(9.0\pm0.9)\times 10^{-4}$  &\cite{Bargiotti:2004aa}&$\mathcal{O}$ S E 
                           & $(11.9\pm1.5)\times 10^{-4}$ &\cite{wittmack:2001kw}        &$\mathcal{C}$ E  \\

\hline\hline
\end{tabular}\renewcommand{\arraystretch}{1.0}
\end{center}\end{table}

For channels involving the $\rho$ meson ($\rho^0$ or $\rho^\pm$),
annihilation frequencies are listed in Tables \ref{tm:tab:BRpsv} and
\ref{tm:tab:BRrhopi}.
The frequencies in Table \ref{tm:tab:BRpsv} were generally obtained from
measurements of the recoil energy spectrum or from an analysis of invariant
mass distributions. Those in Table \ref{tm:tab:BRrhopi} were obtained
by a partial wave analysis of the data; in some cases 
\cite{Espigat:1972aa,Foster:1968am}
for data obtained with a liquid hydrogen target and assuming that
S-state annihilation dominates.
Whilst many of the measurements in Tables \ref{tm:tab:BRpsv} and
\ref{tm:tab:BRrhopi} are rather old and made with the 
bubble-chamber technique, the results are generally in moderate 
agreement with each other and the recent Crystal Barrel values.
Again an exception is the rather high value for the $\rho^0\eta$ channel 
obtained at KEK \cite{Chiba:1989yw} by measuring the inclusive energy 
spectrum. The weighted mean values are $\AF(\rho^0\piz,{\rm liq.}) =
(1.57\pm0.07)\times 10^{-2}$
with $\chi^2/N = 1.1/3$, with N the number of degrees of freedom and 
$\AF(\rho^0\eta,{\rm liq.}) = (3.93\pm0.28)\times 10^{-3}$
with $\chi^2/N = 5.7/5$, where in the latter case the KEK measurement 
\cite{Chiba:1989yw} has been omitted.   

The $\rho^0\piz$ and $\rho^0\eta$ channels are only allowed from
the \tso\ and \spo\ initial states of the \pbp\ system whilst the
$\rho^{\pm}\pimp$ channel is allowed from the \ssz, \tso, \spo,
\tpo\ and \tpt\ states as shown in Table \ref{kin:tab:allowed}.
The Asterix collaboration have measured \cite{May:1990ju,May:1990jv,
Weidenauer:1990fy} 
frequencies for \pbp\ annihilation at rest into $\pip\pim\piz$ 
and into $\pip\pim\eta$ in hydrogen gas (\rstp), both with and 
without a trigger on atomic L X-rays. In this way they obtain two data 
samples with different fractions of S- and P-state annihilation. It is 
then possible from a Dalitz plot analysis to obtain separate frequencies 
for the channels $\pbp\to\rho^0\piz$ \cite{Adler:1995cj},
$\rho^\pm\pimp$ \cite{May:1990ju,May:1990jv} and
for $\pbp\to\rho^0\eta$ \cite{Weidenauer:1990fy} from initial 
S- and P-states. These values are listed in Table 
\ref{tm:tab:BRrhopi}.
The Obelix collaboration \cite{Bargiotti:2003ev}
has performed a coupled-channel partial-wave analysis
of data for  $\pip\pim\piz$, $\Kpm\Kl\pimp$ and  $\Kp\Km\piz$
at three values of the target density. 
Annihilation frequencies for S- and P-states for
$\rho\pi$ channels are also listed in Table \ref{tm:tab:BRrhopi} 
where the S-wave frequencies are derived from
the analysis in liquid $\HT$, the P-wave ones from the analysis at
$\rho=0.005\rstp$. Recently the Obelix collaboration published
a further paper on these experimental results 
\cite{Bargiotti:2004aa}, which gives
annihilation fractions obtained from an average of good fits to the data.
These latest results, which in some cases differ slightly from those
presented in 
Table~\ref{tm:tab:BRrhopi}, give a more realistic estimate of the 
systematic errors.

There are almost no selection rules due to quantum number conservation 
for \pbN\ annihilation into $\K^*\Kb + \mathrm{c.c.}$, only annihilation of the $\tpz$ state
into  3 pseudoscalar mesons is forbidden. Annihilation frequencies 
for $\NNb\to\K^*\Kb + \mathrm{c.c.}$ were determined from 
bubble-chamber data on $\pbp\to\Ks\K{}^{\pm}\pi^{\mp}$ (2000 events) 
and $\Ks\Ks\piz$ (364 events) \cite{Conforto:1967aj}. Later
\cite{Bettini69a}, data on $\pbn\to\Ks\Km\piz$, 
$\Ks\Ks\pim$, $\Ks\Kl\pim$, and $\Kp\Km\piz$ were obtained. 
The latter final states have low statistics only (655 events in four Dalitz plots),
but the data are related by isospin invariance and provide valuable 
constraints for the partial-wave analysis. Since the analysis
\cite{Bettini69a} includes the data from \cite{Conforto:1967aj},
we use only \cite{Bettini69a}. In all cases, the \Ks\ was
identified through its $\pip\pim$ decay mode and its secondary vertex.

The Crystal Barrel collaboration has analysed 11373 events due to 
$\pbp\to \Kl\K^{\pm}\pimp$ with an undetected $\Kl$
\cite{Abele:1998qd}. Results on $\K^*\Kb + \mathrm{c.c.}$ are derived from
a partial wave analysis. The bubble chamber and Crystal Barrel
analyses assume only S-state capture.

The Crystal Barrel Collaboration has also taken data on the reaction
$\pbp\to \Ks\Kpm\pimp$ by triggering on secondary
$\Ks\to\pip\pim$ decays. About 50\,k events were recorded, nearly 
fourfold the statistics obtained in \cite{Abele:1998qd}. Using the larger
data sample, more amplitudes could be included in the fit. In
particular, amplitudes to describe $\K^*_2(1430)\K$ production 
were included~\cite{wittmack:2001kw}. 
Most results are not too different from those reported
in \cite{Abele:1998qd}. We therefore include in Tables~\ref{tm:tab:BRrhopi}
and~\ref{tm:tab:BRrhkstar} the
results on  $\K^*\K$ and $\K^*_2(1430)\K$ production.

As mentioned above,
the Obelix collaboration studied the $\Kpm\Ks\pimp$ 
and  $\Kp\Km\piz$ final states 
at three different target densities \cite{Bargiotti:2003ev}. Revised
annihilation frequencies are again given in their recent paper 
\cite{Bargiotti:2004aa}. The annihilation frequencies for the 
$\K^*\Kb$ channel  
from this latter paper, averaged over the two final states, are again 
given in Table \ref{tm:tab:BRrhopi}.
\subsubsection{Frequencies for annihilation into two vector
mesons}\label{tm:sub:twovec}
As shown in Table \ref{tm:tab:BRvv}, only a limited range of 
frequency measurements are available for 
annihilation into two vector mesons. In Sec.~\ref{tm:sub:BRpi0pi0} 
we commented on the good agreement between the three measurements 
\cite{Amsler:1993kg,McCrady:1998th,Amsler:2003bq} of 
$\AF(\omega\omega,{\rm liq.})$ obtained by the Crystal Barrel
experiment. 

The Asterix collaboration measured \cite{Reifenroether:1991ik} 
the annihilation frequency for $\pbp\to\phi\omega$ with an atomic L X-ray 
trigger as well as the frequency in a gas target (\rstp). Whilst 
the annihilation frequency has also been measured \cite{Bizzarri:1971ax} for
a liquid target, these three results together are insufficient to 
determine partial branching ratios, since the channel occurs from 
four initial \pbp\ states, \tso, \tpz, \tpo\ and \tpt. Note that the possibility of 
$\tpo\to\phi\omega$ was inadvertently omitted in \cite[Table 4c]{Carbonell:1989cs}. 

\begin{table}[!h]
\caption{Frequencies for annihilation into two vector mesons}
\label{tm:tab:BRvv}
\begin{center}
\renewcommand{\arraystretch}{1.2}
\begin{tabular}{@{\bs}c@{\bs}c@{\bs}c@{\bs}r@{\bs}l%
@{\bs}|@{\bs}c@{\bs}c@{\bs}r@{\bs}l@{\bs}}  \hline\hline
Chan. & Dens. & Ann. Freq. & Ref. &\quad Type&Dens. & Ann. Freq. & Ref.
&\quad Type\\ \hline
\multirow{2}{.4cm}{$\omega\omega$}& Liq.    & $(3.32 \pm 0.34 ) \times
10^{- 2}$ &\cite{Amsler:1993kg}  & $\mathcal{C}$ N E$_{\omega \mathrm{n},\omega \mathrm{n}}$&
 Liq.    & $(3.23 \pm 0.25 ) \times 10^{-2}$ & \cite{McCrady:1998th}  &
$\mathcal{C}$ E$_{\omega \mathrm{c},\omega \mathrm{n}}$\\
& Liq.    & $(2.95 \pm 0.15 ) \times 10^{-2}$ & \cite{Amsler:2003bq}  &
$\mathcal{C}$ N E$_{\omega \mathrm{n},\omega \mathrm{n}}$&
12      & $(3.58 \pm 0.19 ) \times 10^{-2}$ & \cite{Amsler:2003bq}  &
$\mathcal{C}$ N E$_{\omega \mathrm{n},\omega \mathrm{n}}$\\
\hline
\multirow{2}{.4cm}{$\phi\omega$}%
&  Liq.   & $(6.30 \pm 2.30 ) \times
10^{- 4}$ &\cite{Bizzarri:1971ax}  & $\mathcal{B}$  E$_{\phi\mathrm{c}}$&
1       & $(3.00 \pm 1.10 ) \times 10^{-4}$ & \cite{Reifenroether:1991ik}  &
$\mathcal{A}$  E$_{\phi\mathrm{c},\omega\mathrm{c}}$  \\
& 1   & $(4.20 \pm 1.40 ) \times 10^{-4}$ & \cite{Reifenroether:1991ik}  &
$\mathcal{A}$ X E$_{\phi\mathrm{c},\omega\mathrm{c}}$ \\ \hline
\multirow{2}{.4cm}{$\rho^0\omega$}%
& Liq.    & $(0.70 \pm 0.30 )\times 10^{-2}$ & \cite{Baltay66} & $\mathcal{B}$ E &
Liq.    & $(2.26 \pm 0.23 ) \times 10^{-2}$ & \cite{Bizzarri:1968aq} &$\mathcal{B}$ E \\
& 1   & $(2.95 \pm 0.72 ) \times 10^{-2}$ & \cite{Weidenauer:1993mv} &$\mathcal{A}$ S E&
1   & $(6.35 \pm 1.14 ) \times 10^{-2}$ & \cite{Weidenauer:1993mv} &A
P E\\ \hline
$\rho^0\phi$& 1     & $(3.40 \pm 0.80 ) \times 10^{-4}$ & \cite{Reifenroether:1991ik} &
$\mathcal{A}$ E$_\phi\mathrm{c}$&
1     & $(4.40 \pm 1.20 ) \times 10^{-4}$ & \cite{Reifenroether:1991ik} &
$\mathcal{A}$ E$_{\phi\mathrm{c}}$\\ \hline
$\rho^0\rho^0 $& Liq.    & $(3.8 \pm 3.0 ) \times 10^{-3}$ & \cite{Baltay66} &
$\mathcal{B}$ E&
Liq.    & $(1.2 \pm 1.2 ) \times 10^{-3}$ & \cite{Diaz:1970as} &
$\mathcal{B}$ E  \\ \hline
{\small$\K^{*+}\K^{*-}$}& Liq. & $(1.5\pm 0.6)\times 10^{-3}$ &\cite{Barash65b}  &
$\mathcal{B}$ E&&&&\\
{\small$\K^{*0}\Kb^{*0}$}& Liq. & $(3.0\pm 0.7)\times 10^{-3}$ &\cite{Barash65b}  &
$\mathcal{B}$ E&&&&\\ \hline\hline
 & State &&&& State & & & \\ \hline

{\small$\K^{*+}\K^{*-}$}&\ssz & $\sim 1.7 \times 10^{-3}$ & \cite{Abele:1997vv} &
$\mathcal{C}$ E& \tso & $\sim 1.3\times 10^{-3}$ &\cite{Abele:1997vv} &
$\mathcal{C}$ E \\

{\small$\K^{*0}\Kb^{*0}$}& \ssz & $\sim 1.5\times 10^{-3}$ &\cite{Abele:1997vv} &
$\mathcal{C}$ E & \tso & $\sim 0.2\times 10^{-3}$ &\cite{Abele:1997vv}  &
$\mathcal{C}$ E \\

\hline\hline
\end{tabular}\renewcommand{\arraystretch}{1.0}
\end{center}\end{table}

The Asterix collaboration also measured \cite{Weidenauer:1993mv} 
annihilation frequencies for \pbp\ annihilation at rest into five pions in 
hydrogen gas (\rstp), both with and 
without a trigger on atomic L X-rays. In this way they obtain two data 
samples with different fractions of S- and P-state annihilation and 
can derive separate frequencies 
for the channel $\pbp\to\rho^0\omega$ from initial 
S- and P-states. These values are also listed in Table \ref{tm:tab:BRvv}.

Values of $\AF(\K^{*+}\Kb^{*-})$ and $\AF(\K^{*0}\Kb^{*0})$ for \ssz\ and \tso\ 
states were obtained by the Crystal Barrel collaboration from a Dalitz 
plot analysis \cite[Table 1]{Abele:1997vv} of \pbp\ annihilation at 
rest into \Kl\Kpm\pimp\piz. The total annihilation frequency for this latter 
final state was taken from Table~\ref{glob:tab:kmult}.
\subsubsection{Two-body annihilation frequencies involving pseudo-vector or 
tensor mesons}\label{tm:sub:BRpvt}
A useful compilation of annihilation frequencies obtained from Dalitz plot 
analyses for two-body \pbp\ annihilation at rest has been given by 
Amsler \cite[Table XIII]{Amsler:1998up}. This includes information for 
final states including either pseudo-vector ($\mathrm{b}_1(1235)$) or
tensor
($\ftmass$, $\atmass$) mesons. Amsler \cite{Amsler:1998up} also 
discusses some of the difficulties associated with determining these 
annihilation frequencies. Note that all these states are broad with widths in 
the range $\Gamma=100$ to $\Gamma=185$ MeV.

The Asterix collaboration measured \cite{Weidenauer:1990fy} 
annihilation frequencies for \pbp\ annihilation at rest into $\pip\pim\eta$ 
and $\pip\pim\eta^{\prime}$ in hydrogen gas (\rstp), both with and 
without a trigger on atomic L X-rays. In this way they obtain two data 
samples with different fractions of S- and P-state annihilation. They are 
then able from a Dalitz plot analysis to obtain separate frequencies 
for the channel $\pbp\to \atmass^{\pm}\pimp$ from initial 
\ssz, \tso\ and P-states, where the latter is averaged over all fine-%
structure states (\spo, \tpo\ and \tpt). The same analysis gives 
separate S- and P-state annihilation frequencies for the channel 
$\pbp\to  \ftmass\eta$.

Similarly, the $\pip\pim\omega$ final state 
was measured \cite{Weidenauer:1993mv}, with $\omega\to\pip\pim\piz$, 
and the S- and P-state annihilation frequencies for the 
annihilation channel $\pbp\to \mathrm{b}_{1}^{\mp}(1235)\pipm$ were
determined. Again the P-state fraction is averaged over all fine-structure 
states and the analysis does not take into 
\renewcommand{\bs}{\hspace*{3pt}}
\begin{table}[!h]
\caption{Annihilation frequencies for annihilation into
a tensor or a pseudoscalar meson 
from partial wave analyses (liquid H$_2$)}
\label{tm:tab:BRrhkstar}
\begin{center}\renewcommand{\arraystretch}{1.2}
\begin{tabular}{cclccl}
\hline\hline
State & Chan. & Ann. Freq. & Final State & Ref. & Type  \\
\hline
\ssz & $\ftmass\piz$  & $(4.3\pm1.2)\times10^{-3}$        &
$\pip\pim\piz$ & \cite{Foster:1968am} & $\mathcal{B}$ E \\
\ssz & $\ftmass\piz$  & $(3.8\pm 1.0)\times10^{-3}$        &
$\pip\pim\piz$ & \cite{May:1990ju} & $\mathcal{A}$ E \\
\ssz & $\ftmass\piz$  & $(3.1\pm 1.1)\times10^{-3}$        &
$3\piz$ & \cite{Amsler:1998up} & $\mathcal{C}$ E \\
\ssz & $\ftmass\piz$  & $(3.7\pm 0.7)\times10^{-3}$        &
$\piz\Kl\Kl$ & \cite{Amsler:1998up} & $\mathcal{C}$ E \\ 
\ssz & $\ftmass\piz$  & $(2.7\pm0.2)\times10^{-3}$        &
$\pip\pim\piz$ & \cite{Bargiotti:2004aa} & $\mathcal{O}$ E \\
\ssz & $\ftmass\piz$  & $(2.6\pm0.8)\times10^{-3}$        &
$\Kp\Km\piz$ & \cite{Bargiotti:2004aa} & $\mathcal{O}$ E \\ \hline
\ssz & $\ftmass\eta$  & $(0.15\pm 0.15)\times10^{-3}$        &
$\pip\pim\eta$ &\cite{Weidenauer:1990fy}  & $\mathcal{A}$ E \\ 
\ssz & $\ftmass\eta$  & $(0.13\pm 0.13)\times10^{-3}$        &
$\pip\pim\eta$ &\cite{Espigat:1972aa}  & $\mathcal{B}$ E \\ \hline
\ssz & $\mathrm{f}_2^{\prime}(1525)\piz$  & $(9.38\pm 1.49)\times10^{-5}$        &
$\piz\Kl\Kl$ & \cite{Abele:1996nn}  & $\mathcal{B}$ E \\ \hline
\ssz & $\atmass\pi$  & $(6.3\pm 0.9)\times10^{-2}$        &
$ 2\pip 2\pim$ & \cite{Diaz:1970as} & $\mathcal{B}$ E \\
\ssz & $\atmass\pi$  & $(1.3\pm 0.4)\times10^{-2}$        &
$ \K^{\pm}\Ks\pi^{\mp} $ & \cite{Conforto:1967aj} & $\mathcal{B}$ E \\
\ssz & $\atmass\pi$  & $(4.8\pm 1.5)\times10^{-2}$        &
$ \pip\pim\eta$ & \cite{Espigat:1972aa} & $\mathcal{B}$ E \\
\ssz & $\atmass\pi$  & $(2.2\pm 0.4)\times10^{-2}$        &
$\K^0\K^{\pm}\pi^{\mp}$ & \cite{Bettini69a,Espigat:1972aa} & $\mathcal{B}$ E \\
\ssz & $\atmass\pi$  & $(2.7\pm 0.8)\times10^{-2}$        &
$ \pip\pim\eta$ & \cite{Weidenauer:1990fy} & $\mathcal{A}$ E \\
\ssz & $\atmass\pi$  & $(3.93\pm 0.70)\times10^{-2}$        &
$ 2\piz\eta$ & \cite{Amsler:1998up} & $\mathcal{C}$ E \\
\ssz & $\atmass\pi$  & $(3.36\pm 0.94)\times10^{-2}$        &
$ 2\piz\eta^{\prime}$ & \cite{Amsler:1998up} & $\mathcal{C}$ E \\
\ssz & $\atmass\pi$  & $(1.55\pm 0.31)\times10^{-2}$        &
$ \piz \Kl\Kl$ & \cite{Amsler:1998up} & $\mathcal{C}$ E \\
\ssz & $\atmass\pi$  & $(2.44\,{}^{+0.44}_{-0.64})\times10^{-2}$        &
$ \K^{\pm}\Kl\pi^{\mp} $ & \cite{Amsler:1998up} & $\mathcal{C}$ E \\
\ssz & $\atmass\pi$  & $(1.2\pm 0.2)\times10^{-2}$        &
\Kp\Km\piz & \cite{Bargiotti:2004aa} & $\mathcal{O}$ E \\
\ssz & $\atmass\pi$  & $(1.2\pm 0.2)\times10^{-2}$        &
$ \K^{\pm}\K^0\pimp$ & \cite{Bargiotti:2004aa} & $\mathcal{O}$ E \\ \hline
\tso & $\atmass\pi$  & $(1.7\pm 0.4)\times10^{-2}$        &
$ 2\pip 2\pim$ & \cite{Diaz:1970as} & $\mathcal{B}$ E \\
\tso & $\atmass\pi$  & $(0.45\pm 0.18)\times10^{-2}$        &
$ \K^{\pm}\Ks\pimp $ & \cite{Conforto:1967aj} & $\mathcal{B}$ E \\
\tso & $\atmass\pi$  & $(0.69\pm 0.34)\times10^{-2}$        &
$ \pip\pim\eta$ & \cite{Espigat:1972aa} & $\mathcal{B}$ E\\
\tso & $\atmass\pi$  & $(0.31\pm 0.16)\times10^{-2}$        &
$ \K^0\K^{\pm}\pimp$ & \cite{Bettini69a,Espigat:1972aa} & $\mathcal{B}$ E \\
\tso & $\atmass\pi$  & $(0.90\pm 0.34)\times10^{-2}$        &
$ \pip\pim\eta$ & \cite{Weidenauer:1990fy} & $\mathcal{A}$ E\\
\tso & $\atmass\pi$  & $(0.58\pm 0.20)\times10^{-2}$        &
$ \K^{\pm}\Kl\pimp $ & \cite{Amsler:1998up} & $\mathcal{C}$ E \\
\tso & $\atmass\pi$  & $(0.33\pm 0.05)\times10^{-2}$        &
$ \K^{\pm}\K^0\pimp$ & \cite{Bargiotti:2004aa} & $\mathcal{O}$ E \\
\hline 
\ttso & $\K^*_2(1430)\K $  & $(5.1\pm 1.8)\times10^{-4}$        &
$ \K^{\pm}\K^0\pimp$ & \cite{wittmack:2001kw} & $\mathcal{C}$ E \\ 
\stso & $\K^*_2(1430)\K $  & $(0.5\pm 1.8)\times10^{-4}$        &
$ \K^{\pm}\K^0\pimp$ & \cite{wittmack:2001kw} & $\mathcal{C}$ E \\ 
\tssz & $\K^*_2(1430)\K$  & $(0.2\pm 0.8)\times10^{-4}$        &
$ \K^{\pm}\K^0\pimp$ & \cite{wittmack:2001kw} & $\mathcal{C}$ E \\ 
\sssz & $\K^*_2(1430)\K$  & $(2.6\pm 0.8)\times10^{-4}$        &
$ \K^{\pm}\K^0\pimp$ & \cite{wittmack:2001kw} & $\mathcal{C}$ E \\ 
\hline\hline
\end{tabular}\renewcommand{\arraystretch}{1.0}
\end{center}\end{table}
\noindent
account any effects due to the 
{\it enhancement factors} discussed earlier in Sec.~\ref{tm:sub:stark}.
The $\pip\pim\piz$ final state has 
been measured \cite{May:1990ju,May:1990jv} and analysed in a similar way 
to give S- and P-state frequencies \cite{Adler:1995cj} for 
the annihilation channel $\pbp\to\ftmass\piz$.

The Obelix collaboration \cite{Bargiotti:2003ev}
has performed a coupled partial-wave analysis
of data for the channels $\pip\pim\piz$, $\Kpm\Kl\pimp$ and  $\Kp\Km\piz$
at three values of the target density. 
An additional paper by the collaboration on this work was 
published later \cite{Bargiotti:2004aa}, which gives
annihilation fractions obtained from an average of good partial-wave fits 
to the experimental results. These measurements are also listed 
in Table \ref{tm:tab:BRrhkstar}.

The annihilation frequencies listed in Table \ref{tm:tab:BRrhkstar}
include all charge states and decay modes; they were calculated 
from the final states which are also listed. This involves in some 
cases using the squares of the isospin Clebsch--Gordan coefficients 
to determine the total annihilation frequencies including all charge 
modes \cite[Table XII]{Amsler:1998up}. The annihilation frequencies 
must also be corrected for all decay modes of the intermediate 
resonances.

The weighted mean value for 
$\AF(\ftmass\piz,\ssz) = (3.70\pm0.47)\times 10^{-3}$ is used in 
the analysis of Sec.~\ref{tm:sub:BRanalpbp}. Measured annihilation frequencies
for $\atmass\pi$ show a very wide spread, with values obtained in 
ref.\cite{Diaz:1970as} appearing to be anomalously big. Excluding these latter
results gives weighted mean values 
$\AF(\atmass\pi,\ssz) = (2.04\pm0.30)\times 10^{-2}$ with $\chi^2/N = 19.9/7$
and $\AF(\atmass\pi,\tso) = (0.48\pm0.09)\times 10^{-2}$ with $\chi^2/N = 3.3/4$. 
Simple (unweighted) averaging gives
$\AF(\atmass\pi,\ssz) = (2.8\pm1.2)\times 10^{-2}$
and $\AF(\atmass\pi,\tso) = (0.59\pm0.23)\times 10^{-2}$, where
the relatively large values for $\sigma$ are again an indication of the
wide spread in experimental values. 

\subsubsection{Other relevant annihilation frequency information}
\label{tm:sub:BRother}
In Table \ref{tm:tab:BRratio} we list a number of cases where the ratio 
of annihilation frequencies for two annihilation channels has been directly 
measured. We have not included values derived from separate frequency 
measurements where in some cases \cite{Amsler:1993kg}, the error 
is reduced since the systematic error is common to the two frequency
measurements and so cancels out. The values of 
$\AF(\Ks\Ks,\rho)/\AF(\Ks\Kl,\rho)$ at densities of 15 and 27 \rstp\  
\cite{Adler:1997it} and 
of $\AF(\Kp\Km,\rho)/\AF(\pip\pim,\rho)$ at 0.002 \rstp\ 
\cite{Ableev:1994bg} are particularly 
useful as they allow the fraction of P-state annihilation at these three 
target densities to be determined. In the absence of other annihilation 
frequency measurements at these densities, this determination relies on the 
availability of branching ratios for these channels which have 
been obtained from measurements of absolute annihilation frequencies 
at other target densities.

\begin{table}[!h]
\caption{Directly measured ratios of annihilation frequencies}
\label{tm:tab:BRratio}
\begin{center}\renewcommand{\arraystretch}{1.2}
\begin{tabular}{ccccl}  \hline\hline
Channels & Dens. & Ratio & Ref. &Type \\ \hline
\Ks\Ks/\Ks\Kl & 15    & $0.041\pm 0.009$ & \cite{Adler:1997it}  &
$\mathcal{L}$  E$_\mathrm{Kc}$ \\
\Ks\Ks/\Ks\Kl & 27    & $0.037\pm 0.002$ & \cite{Adler:1997it}  &
$\mathcal{L}$  E$_\mathrm{Kc}$ \\
\Kp\Km/$\pip\pim$ & 15   & $0.205\pm0.016$ & \cite{Adler:1991wh}  &
$\mathcal{L}$ E \\
\Kp\Km/$\pip\pim$ & 1    & $0.163\pm0.011$ & \cite{Adamo:1992ci}  &
$\mathcal{O}$ E \\
\Kp\Km/$\pip\pim$ & 0.002 & $0.102\pm0.015$ & \cite{Ableev:1994bg}  &
$\mathcal{O}$ E \\
$\rho\omega/\rho\piz$ & 16 & $2.67\pm0.45$  & \cite{Adler:1995cj} &
$\mathcal{L}$ E \\
$\ftmass\piz/\rho^0\piz$ & 16 & $0.83\pm0.06$  & \cite{Adler:1995cj} &
$\mathcal{L}$ E \\
$\ftmass\eta/\ftmass\piz$ & 16 & $0.37\pm0.16$  & \cite{Adler:1995cj} &
$\mathcal{L}$ E \\
\hline\hline
\end{tabular}\renewcommand{\arraystretch}{1.0}
\end{center}\end{table}

The ratio $\AF(\ftmass\piz,\rho)/\AF(\rho^0\piz,\rho)$  has
been estimated \cite{Adler:1995cj} by the 
CPLEAR collaboration at a target density of 16\,\rstp\ and used by them 
together with the separate frequencies 
for annihilation from S- and P-states for the two-body channels
$\ftmass\piz$ and $\rho^0\piz$ to obtain the fraction of P-state annihilation
f$_{\rm{P}}(16\,\rstp) = 0.38\pm0.07$. This estimate is not based
on a partial-wave analysis and
does not take into account the effect of the enhancement factors. For
these reactions the initial states are \ssz, \tso, \tpo, \tpt\ and
\tso, \spo, respectively. With the \tpz\ state not contributing to
either reaction, the enhancement factors for the relevant states are close to 1.0 and 
should have little effect on the derived f$_{\rm{P}}(16\,\rstp)$. The latter 
value is in good agreement with that obtained from the more detailed
analysis of a wide range of data to be discussed in
Sec.~\ref{tm:sub:BRanalpbp}. See also Table~\ref{tm:tab:Pstate} 
and Fig.~\ref{tm:fig:pstatep}.
\renewcommand{\bs}{\hspace*{4.5pt}}
\begin{table}[!h]
\caption{Fraction of P-state annihilation as a function of \HT\ target density $\rho/\rstp$}
\label{tm:tab:Pstate}
\begin{center}\renewcommand{\arraystretch}{1.2}
\begin{tabular}{@{\bs}c@{\bs}c@{\bs}c@{\bs}c@{\bs}c@{\bs}c@{\bs}c@{\bs}c@{\bs}}  \hline\hline
Dens.& 0.002    &0.005    & 1.0      & 12.0     &15.0     & 27.0    &
Liq.\\
P frac.  &            $0.89\er0.07$  & $0.87\er0.02$  & $0.64\er0.03$  &
 $0.50\er0.03$  & $0.48\er0.05$  & $0.46\er0.05$  & $0.125\er0.02$  \\
\hline\hline\end{tabular} \renewcommand{\arraystretch}{1.0}  \end{center}
\renewcommand{\arraystretch}{1.0}\end{table}

In Table \ref{tm:tab:BRother} we list production rates for the channel 
$\pbp\to\eta(1440)\pip\pim$ with $\eta(1440)\to \Kpm\Kl\pimp$.
The values obtained by the Obelix collaboration at three target densities
\cite{Bertin:1996nq}  
are in good agreement with those obtained previously in liquid hydrogen 
using the Bubble Chamber technique \cite{Baillon:1967al} and by the 
Asterix collaboration \cite{Duch:1989sx} in \HT\ gas at STP.
\begin{table}[!h]
\caption{Production rate of the final state $\eta(1440)\pip\pim$,
with $\eta(1440)\to\Kpm\Kl\pimp$}
\label{tm:tab:BRother}
\begin{center}\renewcommand{\arraystretch}{1.2}
\begin{tabular}{cccl|cccl}  \hline\hline
Dens. & Ann. Freq. & Ref. & Type&Dens. & Ann. Freq. & Ref. & Type\\ 
\hline
 Liq.  & $(7.1 \pm 0.7 ) \times 10^{-4}$ & \cite{Baillon:1967al,Armenteros69} &$\mathcal{B}$ E&
Liq.  & $(6.0 \pm 0.5 ) \times 10^{-4}$ & \cite{Bertin:1996nq}
&$\mathcal{O}$ E\\
1     & $(3.0 \pm 0.9 ) \times 10^{-4}$ & \cite{Duch:1989sx} &$\mathcal{A}$  E&
1     & $(2.9 \pm 0.4 ) \times 10^{-4}$ & \cite{Bertin:1996nq} &$\mathcal{O}$ E
\\
 0.005 & $(1.0 \pm 0.2 ) \times 10^{-4}$ & \cite{Bertin:1996nq} &
$\mathcal{O}$ E \\
\hline\hline
\end{tabular}\renewcommand{\arraystretch}{1.0}
\end{center}\end{table}

In the $\eta(1440)\pip\pim$ production reaction, the dipion 
$(\pip\pim)$ and the $\eta(1440)$ resonance are produced with relative 
angular momentum $L=0$ \cite{Bertin:1995fx,Duch:1989sx}. It is expected
that the dipion system will occur mainly with relative orbital angular 
momentum $\ell = 0$, since its invariant mass is less than 500 $\mevc^2$. 
The assumption $L = \ell = 0$ then implies that $\eta(1440)$
production 
is only possible from the \ssz\ initial state. As a consequence the 
$\eta(1440)$ production frequency is directly 
related to the population of the \ssz\ state.
\FloatBarrier\subsubsection{Compilation of two-meson annihilation frequencies}
In view of the discrepancies mentioned above, 
a compilation of preferred measurements of annihilation frequencies 
could be misleading since it requires rejection of some data points
which we considered to be less reliable. Obviously,
personal   preferences could enter here. On the other hand, the reader may
expect a `final' list of annihilation frequencies
which can be used to compare with
model calculations, and also we shall need a table of 
frequencies for further analyses. With this warning we
present in Tables \ref{tm:tab:BRanal1}, \ref{tm:tab:BRanal2} and
\ref{tm:tab:BRrhkstar1} our choice of annihilation frequencies.
Annihilation frequencies listed in Table \ref{tm:tab:BRrhkstar1} are 
weighted means of values taken from Tables \ref{tm:tab:BRpsv}, 
\ref{tm:tab:BRrhopi} and \ref{tm:tab:BRrhkstar}.

\renewcommand{\bs}{\hspace*{4pt}}
\begin{table}[!h]
\caption{Two-meson annihilation frequencies at 
density $\rho$ used in the analysis.}
\label{tm:tab:BRanal1}
\begin{minipage}[t]{\textwidth}
\renewcommand{\arraystretch}{1.2}
\renewcommand{\thefootnote}{\fnsymbol{footnote}}
\renewcommand{\footnoterule}{}
\begin{tabular}{@{\bs}c@{\bs}c@{\bs}c@{\bs}l@{\bs\bs}c@{\bs}l@{\bs\bs}c@{\bs}l@{\bs}}
\hline\hline Channel & Units&$\rho=\rstp$ &Ref.& $\rho=12\rstp$  &
Ref.&Liquid &Ref.  \\
\hline
$\pip\pim$  &$10^{-3}$& $4.30\pm0.15$  & \cite{Doser:1988yi}&$4.05\pm0.23$&\cite{Abele:2001ek}& 
 $3.14\pm0.12\footnotemark[2]$ &\cite{Amsler:1993kg,Abele:2001ek}\\
\piz\piz  &$10^{-3}$& $1.27\pm0.21$  &\cite{Agnello:1994kg}& $1.56\pm 0.08\footnotemark[2]$ & \cite{Abele:2001ek}& 
$0.651\pm 0.029\footnotemark[2]$&\cite{Amsler:1992ku,Amsler:1993kg,Abele:2001ek}\\
\Kp\Km  & $10^{-4}$& $6.92\pm0.41$ &\cite{Doser:1988yi} & $9.07\pm0.59$& \cite{Abele:2001ek}  &
$10.3\pm 0.4\footnotemark[2]$ &     \cite{Amsler:1993kg,Abele:2001ek} \\ 
\Ks\Ks  & $10^{-5}$& $3.0\pm1.0$  & \cite{Doser:1988fw}   &  &  &  $0.70 \pm0.35$ 
& \cite{Amsler:2003bq}   \\
\Ks\Kl & $10^{-4}$& $3.54\pm 0.40\footnotemark[2]$ &
\cite{Bertin:1996pb,Doser:1988fw} & $4.89\pm0.56$& \cite{Abele:2001ek} &
$8.56 \pm 0.43\footnotemark[2]$  &
\cite{Bertin:1996pb,Abele:2001ek,Amsler:1995up} \\
$(\pip\pim)_{\rm X}$  &$10^{-3}$  & $4.81\pm0.49$    & \cite{Doser:1988yi} \\
$(\Kp\Km)_{\rm X}$  & $10^{-4}$ & $2.87\pm0.51$  & \cite{Doser:1988yi}      \\
$(\Ks\Ks)_{\rm X}$  & $10^{-5}$ & $3.70\pm1.40$  &\cite{Doser:1988fw}    \\
$\eta(1440)\pip\pim$   & $10^{-4}$ & $2.9\pm0.4$      &  \cite{Bertin:1996nq} &     
&&$6.0\pm0.5$& \cite{Bertin:1996nq}    \\ 
$\phi\pi$   & $10^{-4}$ & $2.41 \pm 0.15\footnotemark[2]$      
&\multicolumn{2}{l}{\hspace*{-9pt}%
\cite{Reifenroether:1991ik,Ableev:1995aq,Alberico:1998fr}}
& &$6.50\er0.60$  & \cite{Amsler:1995up} \\  
$(\phi\pi)_{\rm X}$ & $10^{-5}$ & $3.0\pm3.0$  &  \cite{Reifenroether:1991ik}    \\
$\phi\eta$  & $10^{-5}$ & $13.3\pm1.5 $      & \cite{Alberico:1998uj}      &&&
$7.10\pm0.70$ &\cite{Alberico:1998uj}    \\ 
$\pi\eta$  &    $10^{-4}$&&    & $4.78\pm0.2$& \cite{Amsler:2003bq}&
$2.09\pm0.10$ & \cite{Amsler:2003bq}    \\
$\eta\eta$  &    $10^{-4}$&&    &  $3.17\pm0.14$     & \cite{Amsler:2003bq} &
$1.53\er0.08$  & \cite{Amsler:2003bq}    \\  
$\pi\eta^{\prime}$ &    $10^{-4}$&&    &   $2.03\pm0.13$  & \cite{Amsler:2003bq}    &
$0.98\pm0.24$  & \cite{Amsler:2003bq} \\
$\eta\eta^{\prime}$  &    $10^{-4}$&&    &    $3.81\pm0.28$   & \cite{Amsler:2003bq}  &
$2.49\pm0.33$   & \cite{Amsler:2003bq}    \\ 
$\omega\pi$  &    $10^{-3}$&&    &   $4.60\pm0.24$    & \cite{Amsler:2003bq} &
 $6.00\pm0.30$  & \cite{Amsler:2003bq}    \\
$\omega\eta$   &    $10^{-3}$&&    &    $9.31\pm0.42$     & \cite{Amsler:2003bq}
 &$15.3\pm0.6$  & \cite{Amsler:2003bq}    \\ 
$\omega\eta^{\prime}$  &    $10^{-3}$&&    & $7.32\pm0.56$   & \cite{Amsler:2003bq}   
& $8.46\pm0.97$  & \cite{Amsler:2003bq}    \\     
\hline\hline
\end{tabular}\footnotetext{$^\dagger\,$Weighted mean value}\end{minipage}
\renewcommand{\arraystretch}{1.0}
\end{table}

\begin{table}[!h]
\caption{Two-meson annihilation frequencies 
used in the analysis (cont). The density is in units of \rstp.}
\label{tm:tab:BRanal2}
\renewcommand{\bs}{\hspace*{5pt}}
\begin{center}\renewcommand{\arraystretch}{1.2}
\begin{tabular}{@{\bs}c@{\bs}c@{\bs}c@{\bs}c@{\bs}c@{\bs}c@{\bs}l@{\bs}}
 \hline\hline
Channel  &Units           & $\rho=0.002$ & $\rho=0.005$ &$\rho=15$ &$\rho=27$&Refs.\\
\hline
$\pip\pim $&$  10^{-3}$  &              & $4.26\pm0.11$&          &         &\cite{Ableev:1994bg}\\
$\Kp\Km     $&$  10^{-4}$  &              & $4.6\pm0.3 $ &          &         & \cite{Ableev:1994bg}     \\
$\Kp\Km/\pip\pim$ &      &$0.102\pm0.015$&             &$0.205\pm0.016$&    & \cite{Ableev:1994bg,Adler:1991wh} \\
$\Ks\Kl     $&$  10^{-4}$  &              & $1.00\pm0.32$ &         &         &  \cite{Bertin:1996pb}   \\
\Ks\Ks/\Ks\Kl &            &              &                &$0.041\pm0.009$&$0.037\pm0.002$ & \cite{Adler:1997it}\\
$\eta(1440)\pip\pim$& $10^{-4}$& &$1.0\pm0.2$ &&         &\cite{Bertin:1996nq}\\
$\phi\eta   $&$ 10^{-4}$   &              & $ 1.66\pm0.20$          &  &       &  \cite{Alberico:1998uj} \\
\hline\hline
\end{tabular}\renewcommand{\arraystretch}{1.0}
\end{center}\end{table}

\renewcommand{\bs}{\hspace*{3pt}}
\begin{table}[!h]
\caption{Annihilation frequencies for
$\rho\pi$, $\rho^0\eta$, $\rho^0\eta^{\prime}$, $\K^*\Kb$, 
\piz\ftmass, $\eta\ftmass$ and $\pi\atmass$, used in the analysis.}
\label{tm:tab:BRrhkstar1}
\begin{center}\renewcommand{\arraystretch}{1.2}
\begin{tabular}{ccl|ccl}
\hline\hline
State & Chan. & Ann. Freq. & State & Chan. & Ann. Freq.  \\
\hline
\ssz & $\rho^\pm\pimp$  & $(0.15\pm0.05)\times10^{-2}$        &
\ttso & $ \K^*\Kb$   &$(20.8\pm4.0)\times 10^{-4}$ \\
\tso & $\rho^0\piz$       &$(1.59\pm0.08)\times 10^{-2}$ &
\stso & $ \K^*\Kb$   &$(1.2\pm1.1)\times 10^{-4}$ \\
\tso & $\rho^0\eta$       & $(3.95\pm0.32)\times 10^{-3}$  &
\tssz & $ \K^*\Kb$   &$(1.3\pm0.5)\times 10^{-4}$ \\
\tso & $\rho^0\eta^{\prime}$&$(1.63\pm0.30)\times 10^{-3}$ &
\sssz & $ \K^*\Kb$   &$(6.7\pm1.0)\times 10^{-4}$ \\
\spo & $\rho^0\piz$       &$(0.43\pm0.04)\times 10^{-2}$ &
\spo & $\rho^0\eta$   &$(1.84\pm0.44)\times 10^{-3}$ \\
\ssz & $\piz\ftmass$ & $(3.70\pm0.47)\times 10^{-3}$ & 
\ssz & $\eta \ftmass$ & $(0.14\pm0.10)\times 10^{-3}$ \\
\ssz & $\pi \atmass$ &$(2.04\pm0.30)\times 10^{-2}$ &
\tso & $\pi \atmass$ & $(0.48\pm0.09)\times 10^{-2}$ \\
\hline\hline
\end{tabular}\renewcommand{\arraystretch}{1.0}
\end{center}\end{table}

\FloatBarrier\subsectionb{Analysis of two-body annihilation frequencies for \pbp\ atoms}
\label{tm:sub:BRanalpbp}
The annihilation frequencies 
summarised in Tables \ref{tm:tab:BRanal1} and 
\ref{tm:tab:BRanal2} were analysed using Eq.~(\ref{tm:equ:BRp}).
The enhancement factors $E(\slj,\rho)$ were fixed at the values  of
Table \ref{tm:tab:efacp},
calculated from cascade calculations.
Calculated annihilation frequencies were then fitted using the least squares 
method to give a best fit to the experimental measurements by varying 
the branching ratios $\BR(ch,\slj)$ and fraction of 
P-state annihilation \fp. The search was constrained so that $\BR(ch,\slj) \geq 0$.
For those frequency measurements, 
$\AF(ch)_{\rm X}$, made in coincidence with atomic L X-rays, Eq.~(\ref{tm:equ:BRx}) 
was used.

Two-body final states are only allowed from certain initial states of the
\pbar N system as shown in Table \ref{kin:tab:allowed}. For 
most channels this 
gives a considerable reduction in the number of free parameters
$\BR(ch,\slj)$ in Eqs. (\ref{tm:equ:BRp}) and (\ref{tm:equ:BRx}) when
fitting the annihilation frequency
data. A further simplification is obtained for
the $\pi\pi$ system by using charge symmetry when
\begin{equation}\label{tm:equ:pizpiz}
\BR(\piz\piz,\tpz) =\frac{1}{2}\BR(\pip\pim,\tpz)
\end{equation}
and
\begin{equation}\label{tm:equ:pizpiz1}
\BR(\piz\piz,\tpt) = \frac{1}{2}\BR(\pip\pim,\tpt)
\end{equation}
Explicit equations for the frequencies for $\pip\pim, \piz\piz,
\Kp\Km, \Ks\Ks$ and \Ks\Kl\ two-body annihilation channels are given 
in the paper by Batty \cite[Eqs. (3.13) to (3.21)]{Batty:1996uf}.

For this review a new analysis has been made, following that made by 
Batty\cite{Batty:1996uf} but using an extended set of annihilation 
frequencies. As well as the measurements used in 
\cite[Table 3]{Batty:1996uf}, the recent two-body annihilation 
frequencies reported 
by the Crystal Barrel collaboration \cite{Abele:2001ek}, 
were used together with a number of other recent measurements.
The Obelix collaboration \cite{Bertin:1996pb} has measured frequencies
for the reaction  $\pbp\to\Ks\Kl$;
$\Ks\to\pip\pim$ at three densities, 0.005\,\rstp, \rstp\ and liquid,
whilst the CPLEAR collaboration have measured \cite{Adler:1997it} the ratio 
$\AF(\Ks\Ks,\rho)/\AF(\Ks\Kl,\rho)$ at densities of 15\,\rstp\ and
27\,\rstp. 
This latter measurement enables the P-state fraction 
\fp\ at 27\,\rstp\ to be determined for the first 
time.
The production of $\eta(1440)$ at three target densities in the reaction
$\pbp\to\eta(1440)\pi^{+}\pi^{-}$ has been determined 
\cite{Bertin:1996nq} by the Obelix collaboration. As discussed in
Sec.~\ref{tm:sub:BRother} this production frequency is directly related to the 
population of the \ssz\ initial state.

\begin{table}[!h]
\caption{Two-body Branching Ratios from S and P wave atomic
states, in units of $10^{-3}$. 
The last column (\tpz $^*$) corresponds to a fit where it is 
assumed that the $\tpt$ state does not
contribute.}
\label{tm:tab:Bpartial}
\begin{minipage}[t]{\textwidth}
\renewcommand{\footnoterule}{}
\renewcommand{\arraystretch}{1.1}
\begin{tabular}{lccccccc}  \hline\hline
Channel & \ssz  & \tso & \spo & \tpz & \tpo & \tpt  & \tpz $^*$ \\  
\hline
$\piz\piz$ &  x  & x & x  & $46.0\er8.9$  & x &$1.7\er1.8$ & $54.3\er1.5$   \\
$\pip\pim$ &  x  & $2.79\er0.18$ & x  & $92\er18$  & x &$3.4\er3.6$ & $108.6\er3.0$   \\
$\pi\eta$ & x & x  & x & $14.3\er2.9$  & x & $0.7\er0.8$   & $17.5\er0.9$ \\
$\pi\eta^{\prime}$  & x & x  & x & $7.2\er0.4$  & x & $0.0\er0.05$   & $7.5\er 0.5$ \\
$\eta\eta$ & x & x   & x  & $11.4\er0.9$   & x & $0.0\er0.01$   & $12.1\er0.6$ \\
$\eta\eta^{\prime}$  & x & x  & x & $14.2\er0.9$  & x & $0.0\er0.01$   & $14.7\er1.1$ \\
\Kp\Km &  x  & $1.47\er0.06$  & x  & $4.25\er0.33$  & x & $0.0\er0.05$ & $4.20\er0.30$ \\  
$\Kn\Kbn$ &  x &$1.31\er0.07$  & x  & $0.41\er0.50$  & x & $0.11\er0.14$ & $0.79\er0.15$ \\
$\rho^0\piz$ & x & $21.2\er1.1$ & $17.2\er1.6$ & x   & x &  x   & x \\
$\rho^{\pm}\pimp$ & $6.0\er2.0$ & $42.4\er2.2$ & - & x   & - &  -   & x \\
$\omega\pi$ & x & $8.75\er0.53$& $12.1\er2.8$ & x   & x &  x   & x \\
$\phi\pi$ & x & $0.87\er0.09$ & $0.06\er0.10$ & x  & x & x    & x \\
$\rho^0\eta$ & x &  $5.3\er0.4$ & $7.4\er1.8$ & x   & x &  x   & x \\
$\rho^0\eta^{\prime}$ & x &  $2.2\er0.4$ & - & x   & x &  x   & x \\
$\omega\eta$ & x & $23.1\er1.1$& $7.0\er5.0$ & x   & x &  x   & x \\
$\omega\eta^{\prime}$ & x & $12.1\er1.6$ & $23.3\er6.9$ & x   & x &  x   & x \\
$\phi\eta$ & x & $0.08\er0.02$ & $0.73\er0.07$ & x & x & x   & x \\
$\K^*\Kb(I=0)$ & $2.7\er0.4$ &  $0.16\er0.15$ & - & x   & - &  - & x \\
$\K^*\Kb(I=1)$ & $0.5\er0.2$ &   $ 2.8\er0.5$ & - & x   & - &  - & x \\
$\piz\atmass^0$ & $27.3\er4.0$ & x & - & x & - & - & x \\
$\pipm\atmass^{\mp}$ & $54.7\er8.0$ & $6.4\er1.2$ & - & x & - & - & x \\
$\eta\ftmass$ & $0.56\er0.40$ & x & x & x & - & - & x \\
\piz\ftmass & $14.8\er1.9$ & x & x & x & - & - & x \\
$\piz\mathrm{f}_2^{\prime}$(1525) & $0.38\er0.06$ & x & x & x & - & - & x \\
$\K^*_2\K(I=0)$ & $1.0\er0.3$ &  $0.07\er0.24$ & - & x   & - &  - & x \\
$\K^*_2\K(I=1)$ & $0.08\er0.32$ &    $0.7\er0.2$ & - & x   & - &  - & x \\
$\eta(1440)\pip\pim$\hspace*{-0mm} & $2.78\er0.19$\hspace*{-0mm} & S & S
& x & S & x   & x \\ 
\hline\hline
\end{tabular}
\renewcommand{\arraystretch}{1.0}
\footnotetext{x:  Channel forbidden from this initial state.}
\footnotetext{S: Channel suppressed by dynamical effects.}
\footnotetext{-: Value not available.}
\end{minipage}
\end{table}

Together this gives a total of 36 frequencies covering 
the $\pip\pim$, \piz\piz, \Kp\Km, \Ks\Ks\ and \Ks\Kl\ 
annihilation channels and $\eta(1440)$ production,  
at a total of 7 densities. In those cases 
where several measurements are available for a particular annihilation 
channel and target density, the weighted mean value was used. This 
reduced to 29 the number of annihilation frequencies to be fitted. 
Seventeen parameters were varied to obtain a least squares best fit; 
10 branching ratios $\BR(ch,\slj)$ and 7 values of the fraction 
of P-state annihilation \fp\ over the range of target densities from 
0.002\rstp\ to liquid \HT.

A least squares fit to this data, which omits the Obelix value
\cite{Bargiotti:2002mu} of $\AF(\piz\piz,{\rm liq.})$ discussed in 
Sec.~\ref{tm:sub:BRpi0pi0}, gave a best fit with a 
$\chi^2$ per degree of freedom $\chi^2/N = 21.7/11$. Rather poorly fitted were the
value of $\AF(\Ks\Kl,12\,\rstp) = (7.04\pm0.74)\times 10^{-4}$
measured by the Crystal Barrel experiment \cite{Abele:2001ek} and
the value of $\AF(\piz\piz,\rstp)$ measured by the Obelix experiment 
\cite{Agnello:1994kg}
which also gave difficulties in the previous analysis \cite{Batty:1996uf}. 
Values of
$\chi^2$ for these two measurements were 6.8 and 4.8 respectively. Omitting
these two annihilation frequencies gave a very good fit to the data with 
$\chi^2/N = 8.72/9$. The values obtained for the branching ratios 
were very similar to those obtained in the earlier analysis 
\cite[Table 6]{Batty:1996uf}. In particular 
$\BR(\Kn\Kbn,\tpz) = (0.0 \pm 0.06)\times 10^{-3}$ and 
$\BR(\Kn\Kbn,\tpt) = (0.20 \pm 0.04)\times 10^{-3}$. The relatively large
value for the \tpt\ branching ratio is in contradiction to the 
results for other channels where the \tpt\ branching ratio is 
consistent with zero. The analysis was 
therefore repeated with the recent value \cite{Amsler:2003bq} for 
$\AF(\Ks\Ks,{\rm liq.}) = (7.0 \pm 3.5)\times 10^{-6}$ replacing the
one \cite{Baltay65a,Doser:1988fw} used in 
the earlier analysis. An equally good fit was obtained 
with $\chi^2/N = 8.38/9$ 
and a value of $\BR(\Kn\Kbn,\tpt)$ consistent with zero. Values obtained
for the fraction of P-state 
annihilation are given in Table \ref{tm:tab:Pstate} and plotted in 
Fig.~\ref{tm:fig:pstatep}. Values for the branching ratios 
obtained from this best fit to the data are given in the first four 
lines of Table \ref{tm:tab:Bpartial}. Note that
\begin{equation}
\BR(\Kn\Kbn,\tpz) = 2\,\BR(\Ks\Ks,\tpz) = D_{10},
\end{equation}
where $D_{10}$ is the parameter used by Batty \cite[Eq. (3.19)]{Batty:1996uf}.
Similar expressions apply for the \tpt\ state.

For these least square fits, enhancement factors calculated with 
strong interaction widths \cite{Carbonell:1989cs} from the DR1 potential 
were used. Very similar results, both for the fraction of P-state 
annihilation and for the branching ratios, were obtained with the 
DR2 and KW enhancement factors. A fit to the same data, but with the 
enhancement factors set equal to 1 (i.e. no enhancement) gave a 
significantly worse fit with $\chi^2/N = 26.3/9$.

As a second stage in the analysis, the range of annihilation frequencies 
was extended to include the $\phi\pi$ and $\phi\eta$ 
channels. For the $\phi\pi$ channel the measurement for a liquid 
\HT\ target by the Crystal Barrel collaboration \cite{Amsler:1995up} 
was used, for \rstp\ gas the average of the Obelix 
\cite{Alberico:1998fr,Ableev:1995aq} and 
Asterix \cite{Reifenroether:1991ik} measurements; the Asterix \cite{Reifenroether:1991ik} 
measurement with \pbp\ 
atomic L X-rays in coincidence was also included. For  the $\phi\eta$ channel, the Obelix 
measurements \cite{Alberico:1998uj} in liquid, \rstp\ and 0.005\rstp\ 
gas were used. A best fit, in which the fraction of P-state annihilation 
and frequencies for the $\pip\pim, \piz\piz, \Kp\Km, 
\Ks\Ks$ and \Ks\Kl\ channels were unchanged, was obtained 
with a very good fit to the data $\chi^2/N = 10.8/11$. The 
branching ratios for the $\phi\pi$ and $\phi\eta$ channels are also 
listed on Table \ref{tm:tab:Bpartial}.

As a final stage of the analysis the measurements by the Crystal Barrel 
experiment \cite{Amsler:2003bq} for the 
$\pi\eta, \eta\eta, \pi\eta^{\prime}, \eta\eta^{\prime}, \omega\pi, 
\omega\eta$ and $\omega\eta^{\prime}$ channels in liquid and 
12\,\rstp\ gaseous \HT\ targets were included. As these channels involve just
two initial \pbp\ states, the two annihilation frequencies at different target 
densities allow the branching ratios to be determined and the 
value of $\chi^2/N$ remains unchanged. Again values of the resulting 
branching ratios are given in Table \ref{tm:tab:Bpartial}. As we have 
commented earlier, and as can be seen from Table \ref{tm:tab:Bpartial}, 
the values of $\BR(ch,\tpt)$ for the annihilation channels included in
this 
analysis are all consistent with zero. Fixing the values for $\BR(ch,\tpt) = 0$, 
and repeating the fit to the data gave an equally good fit to the data 
with $\chi^2/N = 12.5/18$. The corresponding values of $\BR(ch,\tpz)$ are
given in the last column of Table \ref{tm:tab:Bpartial}. 

The $\rho^0\piz$ and $\rho^0\eta$ two-body final states are produced solely from 
the \tso\ and \spo\ states of the \pbp\ system. The Asterix measurements
\cite{May:1990ju,Adler:1995cj,Weidenauer:1990fy} of S- and P-state 
annihilation frequencies
for these channels (Sec.~\ref{tm:sub:BRpsv} and Table \ref{tm:tab:BRrhopi})
can be used to derive the corresponding branching ratios. The S-state
annihilation frequency is derived from data measured in a 
gaseous hydrogen target
of density \rstp\ and using Eq.(\ref{tm:equ:BRp}) is related to the 
branching ratio by $\AF(\rho^0\piz)_{\rm S} = \frac{3}{4}
E(\tso,\rstp)\BR(\rho^0\piz,\tso)$
and correspondingly for the $\rho^0\eta$ channel. 
The P-state annihilation frequency is 
derived from data obtained with an atomic L X-ray trigger 
(Sec.~\ref{tm:sub:BRexp}).
Using Eq.(\ref{tm:equ:BRx}) then gives $\AF(\rho^0\piz)_{\rm P} =
\frac{3}{12} \BR(\rho^0\piz,\spo)$.
The branching ratios obtained using these equations for the S- and P-state 
annihilation frequencies listed in Table \ref{tm:tab:BRrhopi} are given in 
Table \ref{tm:tab:Bpartial}. It should be noted that 
the S- and P-state annihilation frequencies
are calculated neglecting interference effects \cite{May:1990ju} 
from a partial wave fit to the Dalitz plot.
In a similar way, branching ratios for \ssz\ and \tso\ states with
I = 0 and I = 1 can be obtained from Table \ref{tm:tab:BRrhkstar1} for the
${\K^*\Kb}$ channel and also for the \piz\ftmass, $\eta\ftmass$ and 
$\pi\atmass$ channels. These values are again listed in Table \ref{tm:tab:Bpartial}.

After the results of Table \ref{tm:tab:Bpartial} were obtained, a somewhat
similar analysis was made by the Obelix collaboration \cite{Bargiotti:2004aa}.
An essential difference from the present work was that they varied the enhancement
factors at STP and liquid \HT\ densities to give a best fit to the data whilst 
those for target density 0.005 \rstp\ were fixed $E(\slj,0.005\rstp) = 1$. In the 
present work the enhancement factors were fixed at values obtained from a 
cascade calculation (See Sec.~\ref{tm:sub:casc} and Table~\ref{tm:tab:efacp}.)
Two body branching ratios obtained by them for the $\phi\pi, \K^*\Kb,
\rho\pi, \pi\ftmass$ and $\pi\atmass$ channels are in reasonably good agreement
with those obtained in the present work (Table~\ref{tm:tab:Bpartial}).
For the $\rho\pi, {\K^*\Kb}, \pi\ftmass$ and $\pi\atmass$ channels
they were, in addition, able to obtain two-body branching ratios for P-wave 
atomic states.

The Obelix collaboration reported some interesting ratios
of annihilation frequencies~\cite{Bargiotti:2003ev} which are
reproduced in Table~\ref{tm:tab:roaf} and compared to the
results from Table~\ref{tm:tab:Bpartial}. 
 
\begin{table}[!h]
\renewcommand{\arraystretch}{1.3}
\caption{\label{tm:tab:roaf}
Ratios of annihilation frequencies measured by the
Obelix collaboration\ \protect\cite{Bargiotti:2003ev}.}
\renewcommand{\arraystretch}{1.1}
\begin{tabular}{llcl}
\hline\hline
\ Frequency ratio& from \cite{Bargiotti:2003ev}& 
from Table~\protect\ref{tm:tab:Bpartial}& \pbp\\
\hline
 & $0.40\pm0.07$   & $ 0.19\pm0.08$   &\ssz\\
\multirow{3}{44mm}[6mm]{
$\fracd{\AF(\pbp\to \K^*\K, I=1)}{\AF (\pbp\to \K^*\K, I=0)}=\Bigg\{$} 
& $ 0.25\pm0.02$  &                  &\spo\\
& $ 0.10\pm0.01$  &                  & \tpo\\
\hline
\multirow{2}{44mm}[0mm]{
$\fracd{\AF(\pbp\to \K^*\K, I=0)}{\AF (\pbp\to \K^*\K, I=1)}=\bigg\{$}
& $0.17\pm0.03$  & $ 0.06\pm 0.06$ &\tso\\
& $0.20\pm0.01$& & \tpt\\
\hline
 & $0.028\pm 0.006$   & $0.026\pm 0.005$  &\ssz\\
\multirow{3}{44mm}[6mm]{
$\fracd{\AF(\pbp\to \mathrm{f}_2^{\prime}(1525)\piz)}
{\AF (\pbp\to \mathrm{f}_2(1270)\piz)}=\Bigg\{$} 
& $ 0.026\pm0.003$  &                  &\tpo\\
& $ 0.051\pm0.020$  &                  & \tpt\\
\hline\hline
\end{tabular}
\end{table}
\FloatBarrier\subsectionb{Antineutron annihilation on protons}
\label{tm:sub:AN}
The study of \nbar p annihilation 
at very low \nbar\ momenta into two-body final
states gives access to the isospin $I=1$ part of the
annihilation operator. Such experiments were carried out by the
Obelix collaboration; methods and results have been reviewed
in detail in \cite{Bressani:2003pv}. Of interest here are
cross section measurements for two-body final states. 
Data are reported for $\nbp\to\piz\pip$ , $\Ks\K^+$, 
$\phi\pip$, $\omega\pip$, $\eta\pip$ and $\Kb^{*0}\Kp$,
in the momentum range from 50 to 405\,MeV/c. Several 
partial waves can contribute to the final states; their
momentum dependence can be fixed from the Dover--Richard 
potential model \cite{Dover:1980pd}. A system of
linear equations can be formulated (and solved) which
should determine hadronic branching ratios for the
reactions considered, for S- and P-state annihilation.
Only two results are given quantitatively:
\begin{equation}
\label{tm:eq:AN1}\eqalign{
\AF(\tso\to\pip\piz) &= (3.1\pm 0.5)\times 10^{-3}~,\cr
\AF(\rm\tso\to \Kp\Kn_s) &= (1.3\pm 0.2)\times 10^{-3}~.}
\end{equation}
 
To include these results in further analyses, equivalent
branching ratios are calculated. There are, of course, no
enhancement factors due to an atomic cascade, only the
normalisation 4/3 has to be applied (since only 3/4 of
all \nbar p systems have spin 1) and we allow
 for
the undetected $\Kl$ to arrive at 
branching ratios which can be compared to those
of Table~\ref{tm:tab:Bpartial}.
\begin{equation}
\label{tm:eq:AN2pi}\eqalign{
\BR(\tso\to\pip\piz) &= (4.1\pm 0.7)\times 10^{-3}~,\cr
\label{tm:eq:AN2K}
\BR(\tso\to \Kp\Kn) &= (3.5\pm 0.5)\times 10^{-3}~.}
\end{equation}

\par
The collaboration also reported some ratios of
annihilation frequencies~\cite{Filippi:1999xk}. In particular they
find,  
integrating over the full momentum range,
\begin{equation}
\label{tm:eq:ET}
\frac{\AF(\pip\eta^{\prime})}{\AF(\pip\eta)} = 0.63\pm 0.16~.
\end{equation}
The ratio $\phi\pip / \omega\pip$ was determined 
for the full momentum range and for
three momentum bytes \cite{Filippi:1999ym}. The 
results are reproduced in Table~\ref{tm:tab:RAphiomg}.

\begin{table}[!h]
\caption{Cross section Ratio for $\nbp\to\phi\pip /\omega\pip$.}
\label{tm:tab:RAphiomg}
\begin{center}\renewcommand{\arraystretch}{1.2}
\begin{tabular}{ccrr}  \hline\hline
Momentum (\mevc)   & $\sigma (\nbp\to\pip\phi)/\sigma(\nbp\to\pip\omega)$ \\
\hline
50 to 405  & $0.075 \pm 0.008 $ \\
50 to 200  & $0.100 \pm 0.017 $ \\
200 to 300 & $0.074 \pm 0.009 $ \\
300 to 405 & $0.062 \pm 0.009 $ \\
\hline\hline
\end{tabular}\renewcommand{\arraystretch}{1.0}
\end{center}\end{table}
\FloatBarrier\subsectionb{Antiproton annihilation in $\DT$}
\label{tm:sub:DT}
The determination of annihilation frequencies for \pbar\ stopping in
D$_2$ requires some remarks so as to appreciate the meaning of the results. 
We discuss these measurements considering as an example the two
annihilation modes 
\begin{equation}\eqalign{
&(\mathrm{A})\qquad \pbd \to \pim\omega\, \p~, \cr
&(\mathrm{B})\qquad \pbd \to \piz\omega\, \n~. }
\end{equation}
In these reactions, the surviving nucleon can have large momenta;
mostly one is interested in determining the frequency for
annihilation on a {\it quasi-free nucleon}. A nucleon
not participating in the reaction will survive the process with
a Hulth{\'e}n momentum distribution. At sufficiently low momenta, the
nucleon can thus be seen as a {\it spectator} particle. The experimental
difficulty is then to determine the fraction of all annihilations
in which one of the nucleons acts as spectator.

In bubble chamber data, this task is easily met, at least
for annihilations on neutrons. The incoming antiproton
leaves a visible track. If a short secondary track is 
observed, it is likely to be a proton; its momentum is determined 
by its curvature or its range. A cut at 250\,MeV/c was mostly
applied to select events with a spectator proton.
Protons with momenta below 80\,MeV/c leave no visible track
and produce events with an odd number of tracks. These
events are retained. 
In events without short track, the proton had a long track which
requires a large momentum and was not a spectator proton.

In electronic counter experiments, the acceptance for low-momentum
protons has to be well controlled, which is experimentally
challenging. The Obelix collaboration determined annihilation
frequencies in D$_2$ only when using a gas target. Stopping
of protons is then not a problem. The Crystal Barrel collaboration 
\cite{Abele:2000xt} measured first reaction (B), using the all-neutral
final state, with all five photons ($\omega\to\piz\gamma$ and the two
$\piz$ then decaying to $\gamma\gamma$) and the neutron detected
or reconstructed as missing particle. 
Thus reaction (B) was measured over the full kinematic range. Then
it was assumed that the proton momentum distribution in (A) is the
same as the neutron momentum distribution in (B). Thus the Monte-Carlo
proton momentum distribution is very close to the true one, and
the cut, at 100\,MeV/c, leads 
to the same fractional loss in Monte Carlo and real data.
The KEK collaboration observes only one of the particles, a
$\piz$ or $\eta$. A finite momentum transfer to the spectator
nucleon leads to a broadening of the meson recoil momentum 
distribution and, eventually, to a loss of the signal. A 
precise definition of a spectator nucleon is not possible, and
the data could be difficult to interpret.
\par
The Asterix collaboration determined the annihilation frequency only
for $\pip\pim$ and $\Kp\Km$. A cut in collinearity is
equivalent to a cut in the neutron momentum distribution. It
was checked that the final result did not depend on
this cut.

Finally we comment on measurements over the
full kinematical range. 
Bizzarri {\it et al.} \cite{Bizzarri:1974xf} have
measured the frequency of annihilation into \Ks \Kl\ and
\Ks \Ks . The purpose of this study was to demonstrate
the importance of induced P-state annihilation. Indeed,
their \Ks \Ks /\Ks \Kl\ ratio is about 20 times larger 
than in H$_2$. It is hence not suited for a comparison
with the other results.
\subsubsectionb{Two-body annihilation frequencies}\label{tm:sub:BRexpd}
Measurements of two-body annihilation frequencies for the 
annihilation at rest of antiprotons 
on deuterons are listed in Table \ref{tm:tab:BRexpd}. The measurements by 
Chiba et al. \cite{Chiba:2000ev} for the channels 
$\piz\rho^0$, $\piz\rho^-$, $\eta\rho^0$ and $\eta\rho^-$
have not been included. As well as indicating the detection 
technique and method for 
event identification, as in Sec.~\ref{tm:sub:BRexp}, it is also stated
whether the measured annihilation frequencies are normalised to the number of
 \pbd\ (d), \pbp\ (p) or \pbn\ (n) annihilations 
(see Sec.~\ref{tm:sub:intro}). 
Ratios of annihilation frequencies
for two-body channels from \pbd\ annihilation at rest are presented
in Table~\ref{tm:tab:BRratiod}.

A number of checks can be made on the consistency of the 
annihilation frequency data as has been discussed by Batty 
\cite{Batty:2002sk}.  By charge independence \cite{Lipkin:1972ju,Gray:1973dh} 
(see Sec.~\ref{theo:sub:eq3}),
for \pbd\ atoms,
\begin{equation}
\AF_{\rm p}(\pip\pim,{\rm n},\rho) = 
\frac{1}{2} \AF_{\rm n}(\pim\piz,{\rm p},\rho) + 
2 \AF_{\rm p}(\piz\piz,{\rm n},\rho)~.
\label{tm:equ:chind}
\end{equation}
%

For liquid \DT\, using Eq.~(\ref{tm:equ:BRd}) with the measurement of
$\AF_{\rm d}(\piz\piz,{\rm n},{\rm liq.})$ by Amsler et al.\
\cite{Amsler:1995nw} and the
other measurements \cite{Gray:1973dh,Gaspero:1994tw} also 
from Table
\ref{tm:tab:BRexpd}, gives a value of $(4.2 \pm 1.2)\times 10^{-3}$ for the left-hand
side and $(6.47 \pm 0.51)\times 10^{-3}$ for the sum of the two terms on the 
right-hand side of the equation. This latter value becomes $(6.27 \pm
0.58)\times 10^{-3}$ when the measurement of $\AF_{\rm d}(\pim
\piz,{\rm p},{\rm liq.})$ 
of Abele et al.\cite{Abele:2000xt}, together with Eq.~(\ref{tm:equ:BRd}), is used. 
Whilst these two latter values for the right-hand
side are in good agreement, there is a 2 sigma
discrepancy between the left- and right-hand sides of Eq.~(\ref{tm:equ:chind}).

A further check can be made by calculating values of 
\begin{equation}
\label{tm:equ:rvalue}
r = \AF_{\rm d}(\pim\piz,{\rm p},{\rm liq.})/\AF_{\rm d}(\pip\pim
,{\rm n},{\rm liq.}) ~,
\end{equation}
from the annihilation frequencies
of Table \ref{tm:tab:BRexpd} for liquid \DT, using Eq.
(\ref{tm:equ:BRd}) where necessary, and comparing them with
the direct measurements of Table \ref{tm:tab:BRratiod}. Using the 
measurements of
Refs.~\cite{Gray:1973dh,Gaspero:1994tw}, together with Eq.~(\ref{tm:equ:BRd}), 
gives $r = (1.58 \pm 0.49)$. Replacing the $\pim\piz$ frequency
measurement of ref.~\cite{Gaspero:1994tw} by that of
ref.\cite{Abele:2000xt} then gives $r = (1.50 \pm 0.48)$. 
Neither of these deduced values is in 
agreement with the direct
measurements of Table \ref{tm:tab:BRratio}, with the possible 
exception of the
value of $r$
\renewcommand{\bs}{\hspace*{2pt}}
\begin{table}[!h]
\caption{Annihilation frequencies for \pbd\ interactions at rest 
normalised to the number of \pbd\ (d), \pbp\ (p) or \pbn\ (n) 
annihilations. 
}
\label{tm:tab:BRexpd}
\begin{center}\renewcommand{\arraystretch}{1.2}
\begin{tabular}{@{\bs}c@{\bs}c@{\bs}c@{\bs}c@{\bs\bs}l@{\bs}|@{\bs}%
c@{\bs}c@{\bs}c@{\bs\bs}l@{\bs}}
 \hline\hline
Chan.& Dens. & Ann. Freq. & Ref. & Type&Dens.& Ann. Freq. & Ref. & Type
\\
\hline\hline
\multirow{2}{.8cm}{$\piz\piz\n$}%
& Liq.    & $(5.90 \pm 0.32 ) \times 10^{-4}$ &
\cite{Amsler:1995nw} & $\mathcal{C}$ E (d)&
Liq.    & $(6.70 \pm 0.80 ) \times 10^{-4}$ & \cite{Abele:2000xt} &
$\mathcal{C}$ E  (d)\\
& Liq.    & $(1.23 \pm 0.55 ) \times 10^{-4}$ & \cite{Chiba:2000ev} &
$\mathcal{K}$ R (p)  
\\ \hline
$\pip\pim\n$%
& Liq.    & $(4.20 \pm 1.20 ) \times 10^{-3}$ &
\cite{Gray:1973dh} & $\mathcal{B}$ E (p)  &
1       & $(2.01 \pm 0.27 ) \times 10^{-3}$ & \cite{Riedlberger:1989kn} &
$\mathcal{A}$ E  (d) \\  \hline
\multirow{2}{.8cm}{$\pim\piz\p$}%
& Liq.    & $(3.61 \pm 0.49 ) \times10^{- 3}$ &\cite{Abele:2000xt} &
$\mathcal{C}$ E (d)&
Liq.    & $(8.80 \pm 1.00 ) \times 10^{-3}$ & \cite{Gaspero:1994tw} & 
$\mathcal{B}$ E (n)   \\
&Liq.    & $(2.02 \pm 0.31 ) \times 10^{-3}$ & \cite{Chiba:2000ev} &
$\mathcal{K}$ R (n)  
\\ \hline
$\piz\eta\n$%
& Liq.    & $(2.46 \pm 0.12 ) \times 10^{-4}$ &\cite{Amsler:1995nw} & $\mathcal{C}$ E (d)&
 Liq.    & $(5.7 \pm 1.8 ) \times 10^{-4}$ & \cite{Chiba:2000ev} &
$\mathcal{K}$ R (p)  \\ \hline
$\pim\eta \p$%
& Liq.    & $(4.06 \pm 1.00 ) \times 10^{-4}$ &\cite{Abele:2000xt} &
$\mathcal{C}$ E (d) \\ \hline
$\pim\eta^{\prime}\p$%
& Liq.    & $(2.98 \pm 1.52 ) \times 10^{-4}$ &\cite{Abele:2000xt} &
$\mathcal{C}$ E (d)
\\ \hline
\multirow{3}{.8cm}{$\pim\omega\p$}%
& Liq.    & $(6.04 \pm 0.69 )\times 10^{- 3}$ &\cite{Abele:2000xt} & $\mathcal{C}$ E (d)&
Liq.    & $(4.10 \pm 0.80 ) \times 10^{-3}$ & \cite{Bettini:1967bb} &  $\mathcal{B}$ E (n)  \\
&Liq.& $(6.00 \pm 1.00 ) \times 10^{-3}$ & \cite{Bizzarri:1970ta} & 
$\mathcal{B}$ E (n) &
 Liq.    & $(13.2 \pm 4.3 ) \times 10^{-3}$ & \cite{Gaspero:1994tw} &  $\mathcal{B}$ E (n)   \\
&    1    & $(4.97 \pm 0.89 ) \times 10^{-3}$ & \cite{Ableev:1995uu}& $\mathcal{O}$ E (d)
\\
\hline
$\piz\omega\n$%
&Liq.    & $(4.18 \pm 0.39 ) \times 10^{-3}$ &\cite{Abele:2000xt}
&$\mathcal{C}$ E (d)&
Liq.    & $(1.9 \pm 0.9 ) \times 10^{-3}$ &\cite{Chiba:2000ev} & $\mathcal{K}$
R (p)\\  \hline
$\eta \omega\n$%
& Liq.    & $(6.71 \pm 0.81 ) \times 10^{-3}$ &\cite{Abele:2000xt} &
$\mathcal{C}$ E (d)&
Liq.    & $(4.9 \pm 1.2 ) \times 10^{-3}$ &\cite{Chiba:2000ev} &$\mathcal{K}$ 
R (p)  \\ \hline
$\Km \Ks \p$%
& Liq.    & $(1.42 \pm 0.36 ) \times 10^{-3}$ &\cite{Abele:2000xt} & $\mathcal{C}$ E  (d)\\
\hline
$\Ks \Ks\n$%
& Liq.    & $(3.60 \pm 1.00 ) \times 10^{-5}$ &
\cite{Bizzarri:1974xf} & $\mathcal{B}$ E (d)  \\ \hline
$\Ks \Kl \n$%
& Liq.    & $(3.60 \pm 0.40 ) \times 10^{-4}$ &
\cite{Bizzarri:1974xf} & $\mathcal{B}$ E  (d) \\ \hline
$\Kp \Km \n $%
& 1       & $(7.30 \pm 1.60 ) \times 10^{-4}$ &
\cite{Riedlberger:1989kn} & $\mathcal{A}$ E  (d)  \\  \hline
\multirow{2}{.8cm}{$\pim\phi\p$}
& Liq.    & $(8.80 \pm 2.20 ) \times10^{- 4}$ &
\cite{Bettini69a} &  $\mathcal{B}$ E (n)&
Liq.    & $(9.20 \pm 1.10 ) \times 10^{-4}$ & \cite{Bizzarri:1974} &  
$\mathcal{B}$ E (n)   \\
&    1    & $(6.62 \pm 0.49 ) \times 10^{-4}$ & \cite{Ableev:1995uu} &
$\mathcal{O}$ E  (d)   \\
\hline\hline
\end{tabular}\end{center}
\renewcommand{\arraystretch}{1.0}
\end{table}

\begin{table}[!h]
\caption{Ratios of annihilation frequencies for \pbd\ annihilation at rest}
\label{tm:tab:BRratiod}
\renewcommand{\arraystretch}{1.1}
\begin{center}
\begin{tabular}{ccccl}  \hline\hline
Channels & Density & Ratio & Ref. & Type \\ \hline
$\pim\piz\ p/\pim\pip\ $n & Liq. & $0.68\pm0.07$ & \cite{Gray:1973dh}  & 
$\mathcal{B}$ E\\
$\pim(\piz) p/\pim(\pip)\ $n & Liq. & $0.70\pm0.05$ & \cite{Bridges:1986vc}  &
$\mathcal{T}$ R \\
$\pim(\piz) p/\pip(\pim)\ $n & Liq. & $0.55\pm0.05$ & \cite{Bridges:1986vc}  &
$\mathcal{T}$ R \\
$\pim\piz\ p/\pim\pip\ $n & Liq. & $2.07\pm0.05$ & \cite{Angelopoulos:1988xe}  &
$\mathcal{L}$ R \\
$\Kp \Km \ n/\pip\pim\ $n & 1    & $0.36\pm0.08$ & \cite{Riedlberger:1989kn}  &
$\mathcal{A}$ E \\
$\Kp \Km \ n/\pip\pim\ $n & 1    & $0.27\pm0.02$ & \cite{Adamo:1992ci}  &
$\mathcal{O}$ E \\
$\omega\pim \ p/\phi\pim \ $p & Liq. & $7.3\pm1.5$ & \cite{Bizzarri:1974}  &
$\mathcal{B}$ E \\
\hline\hline
\end{tabular}\renewcommand{\arraystretch}{1.0}
\end{center}\end{table}
\noindent
measured by Angelopoulos et al.\ \cite{Angelopoulos:1988xe} 
which is much
larger than the other directly measured values 
\cite{Gray:1973dh,Bridges:1986vc}.
These discrepancies between the annihilation frequencies and 
also between their ratios are discussed further in the
following section. We also note, from Table 
\ref{tm:tab:BRexpd}, that the 
KEK measurements \cite{Chiba:2000ev} are generally in 
disagreement with other work. These 
measurements were therefore excluded from the 
forthcoming analyses. 
\subsubsectionb{Analysis of two-body annihilation frequencies for \pbard\
annihilation at rest}\label{tm:sub:BRanalpbd}
Before discussing the analysis of two-body 
annihilation frequencies in D$_2$, a
few sentences of warning seem appropriate. A final separation of
truly spectator-like events would require a full partial-wave
analysis of the three-body final state. Such fits have never been 
performed. The largest contribution from 3-body Pontecorvo reactions 
(Sec.~\ref{tm:sub:Ponte}) is 
expected from  $\Delta (1232)$ recoiling against a pion in
the reaction $\pbd\to \pi\pi {\rm N}$. Hence we should expect
deviations from simple models. A further problem lies in discrepancies
between measurements of the same annihilation frequency obtained
from different experiments. As we have seen, such discrepancies 
also occurred in data on \pbp\ annihilation frequencies which are
conceptually much easier.

An analysis of the annihilation frequency 
information for \pbd\ annihilation at rest has 
been made by Batty \cite{Batty:2002sk}. The annihilation frequencies 
were analysed using Eq.~(\ref{tm:equ:BRpd}). Calculated annihilation 
frequencies were fitted using 
the least squares method to give a best fit to the 
experimental measurements by 
varying the fraction of P-state annihilation \fpa. 
The branching ratios 
$\BR(ch,\slj)$ were fixed at values obtained from the analysis of \pbp\
annihilation frequencies
by Batty \cite[Table 6]{Batty:1996uf} using either the DR1 or DR2 models. 
For the $\pi\pi$ system, in addition to
Eqs. (\ref{tm:equ:pizpiz}) and (\ref{tm:equ:pizpiz1}), 
charge symmetry can be used 
to give
\begin{equation}\label{tm:equ:pimpiz}
\BR(\pim\piz,\tso) = 2\BR(\pip\pim,\tso)~.
\end{equation}

Explicit equations for the annihilation frequencies 
for the $\pip\pim, \piz\piz, 
\pim\piz, \Kp\Km, \Ks\Ks$ and \Ks\Kl\ channels in terms of branching ratios 
and the fraction of P-state annihilation \fpa\ for \pbd\ 
annihilation have been 
given by Batty \cite[Eq.~(9)]{Batty:2002sk}.
As a first step, single annihilation frequencies 
for these channels were fitted to determine
whether the best fit values of \fpa\ lay, within errors, in the physically
acceptable range $0 \leq \fpa \leq 1$. This preliminary analysis indicated
problems with the two measured values \cite{Abele:2000xt,Gaspero:1994tw} of 
$\AF(\pim\piz,{\rm p},{\rm liq.})$ and the measured value of $r$ (see
Eq.~(\ref{tm:equ:rvalue})) determined by Angelopoulos et al.
\cite{Angelopoulos:1988xe} which were
discussed previously in Sec.~\ref{tm:sub:BRexpd}. 
The measured \Ks\Ks\ annihilation frequency~\cite{Bizzarri:1974xf}  
also gave values of $\fpal > 1$. This
result included the full proton momentum range and cannot be compared
to data selecting events with a spectator proton.

These 4 values were therefore not included in the fit to the measurements 
of Tables \ref{tm:tab:BRexpd} and \ref{tm:tab:BRratiod} for these channels. 
A least squares fit to the 
remaining 6 measurements for a liquid target with $s_{p} = 0.571$, as
discussed in Sec.~\ref{tm:sub:intro}, and branching ratios obtained
with the DR1 model, gave $\fpal = 0.40 \pm 0.02$ with a chi-squared per
degree of freedom $\chi^{2}/N = 10.9/5$. Much of the contribution to $\chi^{2}$
comes from the measurement \cite{Bridges:1986vc} of $r = 0.70 \pm 0.05$. Omitting this
measurement gives little change in the value of $\fpal = 0.42 \pm 0.02$, but
chi-squared decreases significantly and $\chi^{2}/N = 4.8/4$. Using a value of
$s_{p} = s_{n} = 0.5$, i.e. assuming equal annihilation on neutrons and
protons, gives $\fpal = 0.48 \pm 0.02$ and $\chi^{2} = 12.0/5$. Repeating the
analysis with branching ratios obtained using the DR2 model, gave
the same values for \fpal, within errors, whilst $\chi^{2}$ decreased at most by
0.5 in absolute value. This result is not surprising, since the values of 
$\BR(\Kp\Km,\slj)$ and $\BR(\K^0\Kb^0,\slj)$ obtained with these two models
are almost identical. The same is also true for the values of 
$\BR(\pip\pim,\tso)$ and $(\frac{1}{12}\BR(\pip\pim,\tpz) + 
\frac{5}{12}\BR(\pip\pim,\tpt))$ which appear
on the right hand side in Eq.~(\ref{tm:equ:BRpd}).

Repeating the analysis using the four measurements obtained with a gas target gave 
$\fpas = 0.34 \pm 0.04$ with $\chi^{2}/N = 5.7/3$. With $s_{p} = s_{n} = 0.5$
values of $\fpas = 0.34 \pm 0.04$ and $\chi^{2}/N = 9.5/3$ were obtained. For
branching ratios obtained with the DR2 potential and $s_{p} = 0.571$
a best fit was obtained with $\fpas = 0.34 \pm 0.04$ and $\chi^{2} = 5.3/3$.

As we have already mentioned, P-state annihilation on a
nucleon can occur from S-states of a \pbd\ atom, due to the size and
momentum distribution of 
the nucleons in the deuteron.
This is sometimes referred to as ``induced'' P-state annihilation. Bizzarri
\cite{Bizzarri:1974hr} first considered this process in terms of the spectator model, 
using a semi-classical method based
on the Fermi motion of the nucleons in the deuteron and an impact parameter
relative to the deuteron centre of mass. A calculation then gave the amount of
P-state annihilation from atomic S-states to be 0.4 times the rate of S-state
annihilation and predicted that there
should be a relatively large amount of S-state annihilation from atomic P-states.

If $a_{{\rm SS}_{\rm ann}}$ and $a_{{\rm PP}_{\rm ann}}$ are used to denote the fraction of S- and P-state
annihilation from atomic S- and P-states respectively, then the fraction of
P-state annihilation is given by $\fpa = (1 - \fp)(1-a_{{\rm SS}_{\rm ann}}) + \fp a_{{\rm PP}_{\rm ann}}$.
Using the values of \fps\ = 0.75, \fpl\ = 0.40 obtained in Sec.~\ref{tm:sub:cascpbd}, 
and $\fpas\ = 0.34 \pm 0.04$ and
$\fpal\ = 0.40 \pm 0.02$, given above, then 
values of $a_{{\rm SS}_{\rm ann}} = 0.53 \pm 0.06$ and $a_{{\rm PP}_{\rm ann}} = 0.30 \pm 0.07$
are obtained, where the errors shown are purely statistical. The fitted values
of $a_{{\rm SS}_{\rm ann}}$ and $a_{{\rm PP}_{\rm ann}}$ are sensitive to the values of \fps\ and \fpl\ which
were calculated using a cascade calculation as described earlier in
Sec.~\ref{tm:sub:cascpbd}. These were obtained assuming the Stark
mixing parameter 
$K_{0} = 7.6$. If
instead $K_{0} = 1.0$ is used, then the cascade calculation gives \fps\ = 0.92
and \fpl\ = 0.64 (see Fig.~\ref{tm:fig:pdstated}) and a best fit then gives 
$a_{{\rm SS}_{\rm ann}} = 0.46 \pm 0.11$ and $a_{{\rm PP}_{\rm ann}} = 0.32 \pm 0.05$. Whilst the
numerical values for $a_{{\rm SS}_{\rm ann}}$ and $a_{{\rm PP}_{\rm ann}}$ given by these estimates should 
be treated with caution, they nevertheless do indicate the importance 
of ``induced'' S- and P-state annihilation.
\subsubsectionb{Pontecorvo reactions}\label{tm:sub:Ponte}
So far in the discussion of \pbard\ annihilation we have concentrated on the
production of two-meson final states in which the antiproton annihilates
with a single ``quasi-free nucleon'' in the deuteron whilst the other non-participating 
nucleon is a ``spectator'' particle. It was pointed out by 
Pontecorvo \cite{Pontecorvo:1956vx} in 1956 that there is a further class of 
annihilation reactions in which, through a three-body interaction involving
both nucleons, the final state consists of a single meson and a baryon. These
Pontecorvo reactions can conveniently be considered in three classes.

The first of these classes has a final state consisting of a single meson and 
a nucleon, e.g., $\pbard\to\pim\p$. These reactions, in the simplest model, 
can be considered in terms of a two-step (rescattering) process, for example
$\pbard\to\pim\ (\pip\n) \to\pim\p$ in the above case. Other possible reactions
involve the production of $\eta, \eta^{\prime}, \omega$ and $\phi$ mesons.
The second class consists of cases where one of the mesons interacts with
the second nucleon to form a nucleon resonance, e.g., $\pbard\to\pim (\pip\n)
\to\pim\Delta^{+}$.
An alternative to the two-step involves mentioned above is the fireball model
which pictures a compound system (``fireball'') formed by the participating
three antiquarks and six quarks, decaying statistically. This latter model predicts
significant production of the third class of channels with open strangeness, 
e.g., $\pbard\to \Lambda\Kn, \Sigma^{0}\Kn$ or $\Sigma^{-}\Kp$. These latter 
reactions have recently been observed by the Crystal Barrel Collaboration
\cite{Abele:1999uv}.

A compilation of measured frequencies for Pontecorvo reactions in 
\pbard\ annihilation is given in Table \ref{tm:tab:Ponte}. In general
the results are in moderate agreement with each other in 
those cases where several 
measurements have been made for a single channel. In particular 
$\AF(\pim\p) = (1.38 \pm 0.07) \times 10^{-5}$ with $\chi^{2}/N = 4.1/3$, 
where $N$ is the number of degrees of freedom. Similarly
$\AF(\piz\n) = (7.13 \pm 0.71) \times 10^{-6}$ with $\chi^{2} = 0.6$.

A recent comparison of these results with the predictions of dynamical 
(two-step) and statistical (fireball) models has been made by the Crystal 
Barrel Collaboration \cite{Abele:1999uv}. They show that the rates
$\AF(\Lambda\Kn)$, $\AF(\Sigma^{0}\Kn)$ and in particular the ratio
$\AF(\Sigma^{0}\Kn)/\AF(\Lambda\Kn)$, exceed by a large factor the values 
predicted by two-step models, but are in general agreement with the 
predictions of the fireball model.

\begin{table}[!h]
\caption{Frequencies for Pontecorvo reactions in \pbd\ annihilation.
Some final states are produced via the same isobar. In these
cases we give the isobar and the inverse squared Clebsch--Gordan
coefficients to facilitate the comparison of different results.
K$^0$ stands for the sum of \Ks\ and \Kl\ modes.  }
\label{tm:tab:Ponte}
\renewcommand{\arraystretch}{1.1}
\begin{tabular}{lcccrcll}  \hline\hline
Chan. & Dens. & Isobar & C.G.$^{-2}$ 
& {Ann. Freq.}\qquad\ & Ref. & Type \\ 
\hline\hline
\pim     p & 1       & $\pi$N & 2  
& $(1.40 \pm 0.70 )\times 10^{-5}$ & \cite{Riedlberger:1989kn} & $\mathcal{A}$ E \\
\pim     p & 1       & $\pi$N & 2  
& $(1.20 \pm 0.14 )\times 10^{-5}$ & \cite{Ableev:1993uz} & $\mathcal{O}$ E \\
\pim     p & 1       & $\pi$N & 2  
& $(1.46 \pm 0.08 )\times 10^{-5}$ & \cite{Denisov:1999tr} & $\mathcal{O}$ E \\
\pim    p & Liq.     & $\pi$N & 2  
& $(0.90 \pm 0.40 )\times 10^{-5}$ & \cite{Bizzarri:1969pon} & $\mathcal{B}$ E \\
\piz     n & Liq.       & $\pi$N & 2  
& $(7.03 \pm 0.72 )\times 10^{-6}$ & \cite{Amsler:1995ct} & $\mathcal{C}$ E \\
\piz     n & Liq.       & $\pi$N & 2  
& $(1.03 \pm 0.41 )\times 10^{-5}$ & \cite{Chiba:1997tf} & $\mathcal{K}$ R \\
$\eta$    n & Liq.      &&
& $(3.19 \pm 0.48 )\times 10^{-6}$ & \cite{Amsler:1995ct} & $\mathcal{C}$ E \\
$\eta$    n & Liq.      &&
& $ < 8.94 \times 10^{-6}$ & \cite{Chiba:1997tf} & $\mathcal{K}$ R \\
$\omega$     n & Liq.   && 
& $(2.28 \pm 0.41 )\times 10^{-5}$ & \cite{Amsler:1995ct} & $\mathcal{C}$ E \\
$\eta^{\prime}$   n & Liq.&&    
& $\leq 14\times 10^{-6}$ & \cite{Amsler:1995ct} & $\mathcal{C}$ E \\
$\phi$   n & 1          && 
& $(3.56 \pm 0.35 )\times 10^{-6}$ & \cite{Gorchakov:2002dm} & $\mathcal{O}$ E \\
\piz     $\Delta^0$ & Liq. & $\pi\Delta$ & 9/2     
& $(2.21 \pm 0.24 )\times 10^{-5}$ & \cite{Amsler:1995nw} & $\mathcal{C}$ E 
$\Delta^0\to\piz$ n   \\
\piz     $\Delta^0$ & Liq. & $\pi\Delta$ & 3      
& $(4.67 \pm 1.66 )\times 10^{-5}$ & \cite{Chiba:1997tf} &  $\mathcal{K}$ E \\
\piz     $\Delta^0$ & 1 & $\pi\Delta$ & 9        
& $(1.22 \pm 0.20 )\times 10^{-5}$ & \cite{Denisov:1999tr} & $\mathcal{O}$ E 
$\Delta^0\to\pim$ p   \\
$\pim$  $\Delta^+$ & 1 & $\pi\Delta$ & 9       
& $(1.01 \pm 0.08 )\times 10^{-5}$ & \cite{Denisov:1999tr} & $\mathcal{O}$ E
$\Delta^+\to\piz$ p   \\
$\eta$   $\Delta^0$ & Liq.  &&        
& $ < 6.49 \times 10^{-5}$ & \cite{Chiba:1997tf} & $\mathcal{K}$ E \\
$\Lambda$  $\K^0$ & Liq.    &&
& $(2.35 \pm 0.45 )\times 10^{-6}$ & \cite{Abele:1999uv} & $\mathcal{C}$ E \\
$\Sigma^0$ $\K^0$ & Liq.   && 
& $(2.15 \pm 0.45 )\times 10^{-6}$ & \cite{Abele:1999uv} & $\mathcal{C}$ E \\
\hline\hline
\end{tabular}\renewcommand{\arraystretch}{1.0}
\end{table}

\FloatBarrier\subsection{Discussion}
\label{tm:sub:Discuss}
A compilation of two-meson annihilation frequencies 
for \pbp\ annihilation at rest 
has been presented, together with an analysis of some of the data in terms 
of branching ratios $\BR(ch,\slj)$ for the channel $ch$ 
from specific \slj\ states of 
the \pbp\ system. The fraction of P-state annihilation as a function of 
\HT\ target density has also been determined.

A very wide range of measurements are available with some 187 
annihilation frequencies 
covering about 30 different annihilation channels at 7 target densities. 
Unfortunately in some cases there are very significant discrepancies 
between measurements for the same channel and target density; in one 
extreme case by a factor of about 40. In most of these cases, the 
discrepant measurements were obtained using peaks in the inclusive recoil 
energy spectrum to identify the second meson. These experiments should 
generally be regarded as less reliable than those in which the complete 
two-body event is recorded and identified.

A particularly unfortunate discrepancy is that between the recent 
measurements for the $\pbp\to\piz\piz$ channel in liquid \HT, made by 
the Crystal Barrel and Obelix collaborations. Both experiments fully 
reconstruct the \piz\piz\ events but the measured 
annihilation frequencies differ 
by a factor of about 2. This discrepancy is discussed in some detail in 
Sec.~\ref{tm:sub:BRpi0pi0} where the internal and external consistency 
of the Crystal Barrel results relating to their  
annihilation frequency
for the \piz\piz\ channel has been emphasised. In the absence of similar 
checks for the Obelix measurements, the Crystal Barrel results are 
regarded as being more reliable.

For some channels ($\rho^0\piz, \rho^0\eta, \Ks\Kl, \pip\pim$ and 
\Kp\Km) produced from a liquid \HT\ target, there is good consistency between 
several measurements of the annihilation frequencies
 and the weighted mean value is given 
in the text. In Tables \ref{tm:tab:BRanal1}, 
\ref{tm:tab:BRanal2} and \ref{tm:tab:BRrhkstar1} we presented a selection of annihilation
frequencies which is used to derive branching ratios from specific
atom states.

The target density dependence of annihilation frequencies 
has been discussed in 
Sec.~\ref{tm:sub:stark}, where it was shown that the effects of Stark 
mixing are important. In particular, where there is a large contribution 
from Stark mixing, an atomic fine-structure level with 
a large annihilation width will contribute more to annihilation than 
would be expected from its statistical weight only. These deviations, 
which are particularly important for the \tpz\ state, can be described 
\cite{Batty:1996uf} in terms of {\it enhancement factors}, which are a 
function of the initial state and target density $\rho$. In 
Eq.~(\ref{tm:equ:BRp}) the 
annihilation frequency $\AF(ch,\rho)$ is written in
terms of the branching ratios $\BR(ch,\slj)$, enhancement
factors 
$E(\slj,\rho)$ and the fraction of P-state annihilation \fp. Values of 
the enhancement factors (Table \ref{tm:tab:efacp}), have been calculated 
from a cascade calculation using a modified version of the Borie and Leon 
model \cite{BorieL80}, whose parameters $K_0$ and $k_{STK}$ are obtained 
from a least-squares fit to \pbp\ atomic X-ray yields as a function of 
\HT\ target density.

Using these enhancement factors, the annihilation frequencies listed 
in Tables \ref{tm:tab:BRanal1}, \ref{tm:tab:BRanal2} and \ref{tm:tab:BRrhkstar1} 
were fitted using the least-squares method by 
varying the fraction of P-state annihilation and branching 
ratios. The best fit values are given in Table \ref{tm:tab:Pstate} 
(See also Fig.~\ref{tm:fig:pstatep}) and Table \ref{tm:tab:Bpartial}. 
Branching ratios for the $\rho^0\piz$ and $\rho^0\eta$ channels, 
also listed in Table \ref{tm:tab:Bpartial}, 
were obtained directly from measurements of S- and P-state 
annihilation frequencies. 
Branching ratios for \ssz\ and \tso\ states for ${\K^*\Kb}$
and $\K^*_2\Kb$, 
and for \piz\ftmass, $\eta\ftmass$, \piz f$_2(1525)$ and $\pi\atmass$
channels were obtained in a similar way from the results of partial wave 
analyses and are also listed in Table \ref{tm:tab:Bpartial}.
Branching ratios are independent of atomic physics effects and are 
the quantities to be compared with the predictions from models of the 
annihilation process and two-meson production.

A problem with determining branching ratios by using 
Eq.~(\ref{tm:equ:BRp}) to fit experimental annihilation frequencies,
should be mentioned. For a particular annihilation channel which occurs 
from several (say, $n$) annihilation states, measured 
annihilation frequencies for 
at least $n$ different target densities (or in coincidence with atomic 
L X-rays) are required. However the branching ratios can only 
be determined for two sets of cases. Firstly where there is only one 
S- and one P-state (e.g., \tso\ and \spo\ for $\omega\pi$) and the fraction 
of P-state annihilation is known or can be determined from the data. 
Secondly for the case where there is more than one P-state 
(e.g., \tpz\ and \tpt\ for \piz\piz) and the enhancement factors for 
these states differ significantly at a given density for the 
relevant \slj. The latter condition generally implies that data for a 
liquid \HT\ target is available and that one of the initial states is 
\tpz\ which has a large enhancement factor at liquid \HT\ density.

To illustrate this point we consider the reaction $\pbp\to\phi\omega$ 
which occurs from the initial states \tso, \tpz, \tpo\ and \tpt. In 
this case, as can be seen from Fig.~{\ref{tm:fig:efacp}, the enhancement 
factors for the \tpo\ and \tpt\ states are 
similar and the two branching ratios cannot be separated by 
a fit to the annihilation frequencies. In practice, 
as can be seen from Eq.~(\ref{tm:equ:BRp}), the quantity 
determined will be 
\begin{equation}
\frac{3}{12}\overline{E(\tpo,\rho)}\BR(\phi\omega,\tpo) 
+ \frac{5}{12}\overline{E(\tpt,\rho)}\BR(\phi\omega,\tpt)~,
\end{equation}
 where $\overline{E(\slj,\rho)}$ is the enhancement factor averaged in some way 
over the range of target densities. 

A compilation of two-meson annihilation frequencies from 
\pbd\ annihilation at rest  shows that only a limited range of data 
is available and that much of it is inconsistent. A fit to the 
rather limited range of \pbd\ atom X-ray data, assuming that the 
value of the cascade parameter $K_0$ is the same as in the \pbp\ case, 
gives values for the fraction 
of annihilation (\fp) from atomic P-states in gaseous (\rstp) and liquid 
\HT. For the case of 
\pbd\ atoms the values of the enhancement factors are expected to be 
close to one. This is confirmed by a cascade calculation. 

Using the method described earlier for \pbp\ atoms, a fit is made to a 
selected set of annihilation frequencies for \pbd\ annihilation at rest with 
branching ratios given by the analysis of \pbp\ data discussed earlier. 
Values for the fraction of P-state annihilation (\fpa) 
are obtained for gaseous (\rstp) and liquid \HT. The difference in value 
between \fp\ and \fpa\ is explained using a simple model in terms of a 
significant amount of ``induced''  S- and P-state annihilation. This latter 
conclusion is in agreement with the predictions of Bizzarri 
\cite{Bizzarri:1974hr}.

\clearpage\markboth{\sl Annihilation dynamics}{\sl Dynamical selection rules}
\setcounter{equation}{0}
\section{Dynamical selection rules}\label{se:dsr}
Nucleon--antinucleon annihilation reveals a great surprise: some
processes which are perfectly compatible with the rules of
quantum-number conservation
summarised in Sec.~\ref{se:kin} (parity $P$, charge conjugation $C$,
$G$-parity, among others) are observed only at a much reduced
rate. These are {\em dynamical selection rules}.  This property has
often been emphasised; in particular Dover stressed the link between
\dsrs\ and the dynamics of quarks and gluons in \NNb\ annihilation
\cite{Dover:1992vj}.
\subsectionb{The $\rho\pi$ puzzle}\label{dsr:sub:rhopi}
The best known example of a \dsr\ is the so-called
{\em $\rho\pi$ puzzle}. Figure \ref{dsr:fig:3pidp} shows the Dalitz plot of the
reaction $\pbarp\to\pi^+\pi^-\pi^0$ in annihilation at rest
\cite{Abele:1997wg}. The three bands in Fig.~\ref{dsr:fig:3pidp}
show production of $\rho^+\pi^-$, $\rho^-\pi^+$, and $\rho^0\pi^0$
intermediate states.  The intensity distribution along, e.g., the
$\rho^+$ band gives directly the $\rho^+\to\pi^+\pi^0$ angular
distribution in the $\rho^+$\ rest frame. On the right-hand side of
Fig.~\ref{dsr:fig:3pidp}, in (a), the angular distribution is plotted
as a function of $\cos\Theta$ where $\Theta$ is defined as angle between
the $\rho^+$ flight direction and the direction of the $\pi^+$, in the
$\rho^+$ rest frame.  The decay angular distributions give access to
the quantum numbers of the initial state feeding the final
state. Table \ref{dsr:tab:threepiselect} lists the initial states from
which annihilation into $\rho\pi$ is allowed, the $\rho -\pi$ orbital
angular momentum, and the $\rho$ decay angular distribution when all
interferences are neglected.  The sign of the amplitudes is also given
($\rho^+\pi^-\,-\,\rho^-\pi^+$ stands for a negative sign between the
peak amplitudes for $\rho^+\pi^-$ and $\rho^-\pi^+$ production). At lines
of identical $\pi\pi$ masses, i.e., for identical Breit--Wigner
phases, the interference is constructive for annihilation from
isoscalar and destructive for annihilation from isovector initial
states.

\begin{figure}[h!]
\setlength{\unitlength}{1mm}
\begin{picture}(180,100)
\put(-7,0){\includegraphics[width=.57\textwidth]{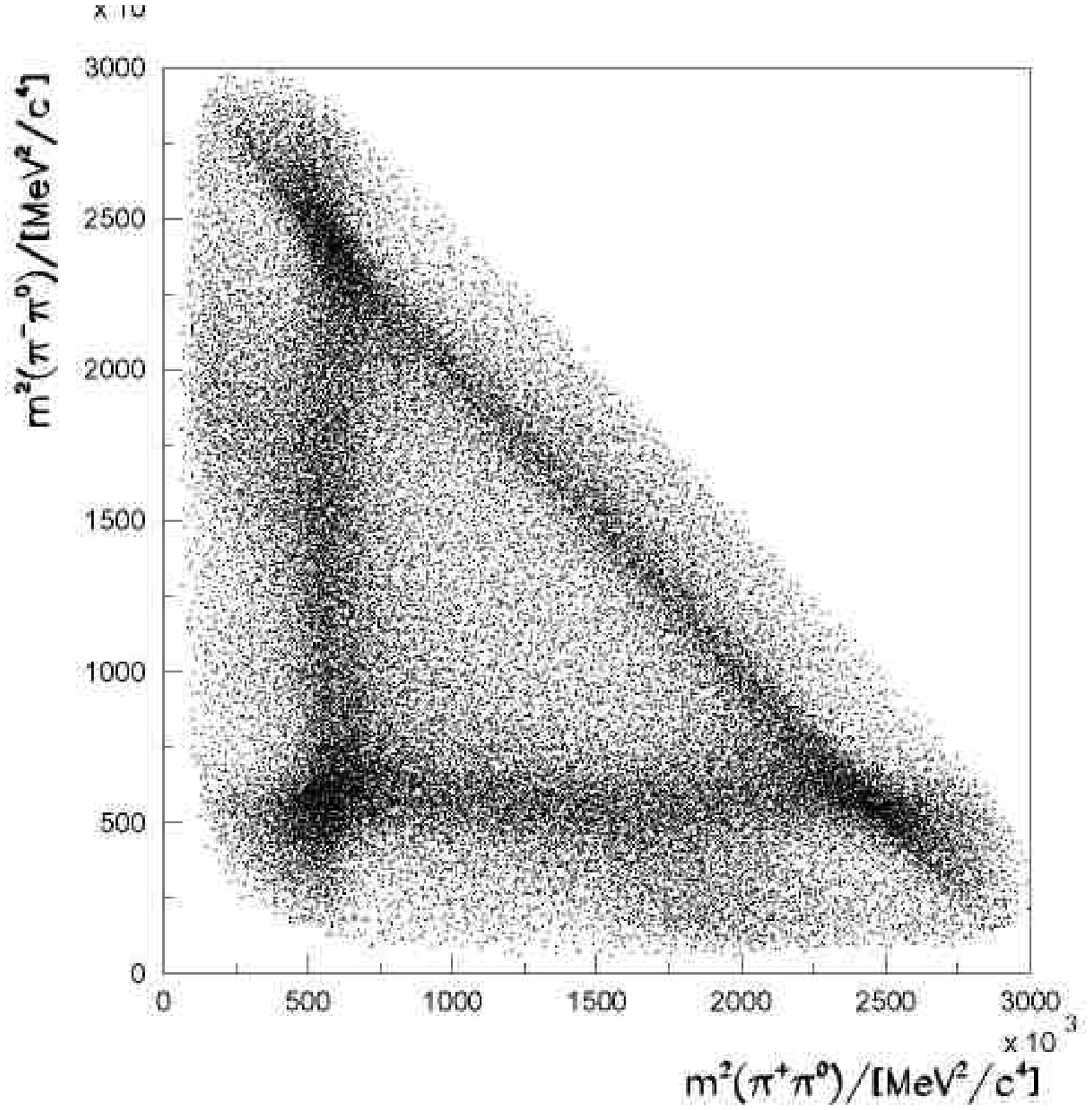}}
\put(0,79){\small $\times 10^3$\normalsize}
\put(65,0){\includegraphics[width=.57\textwidth]{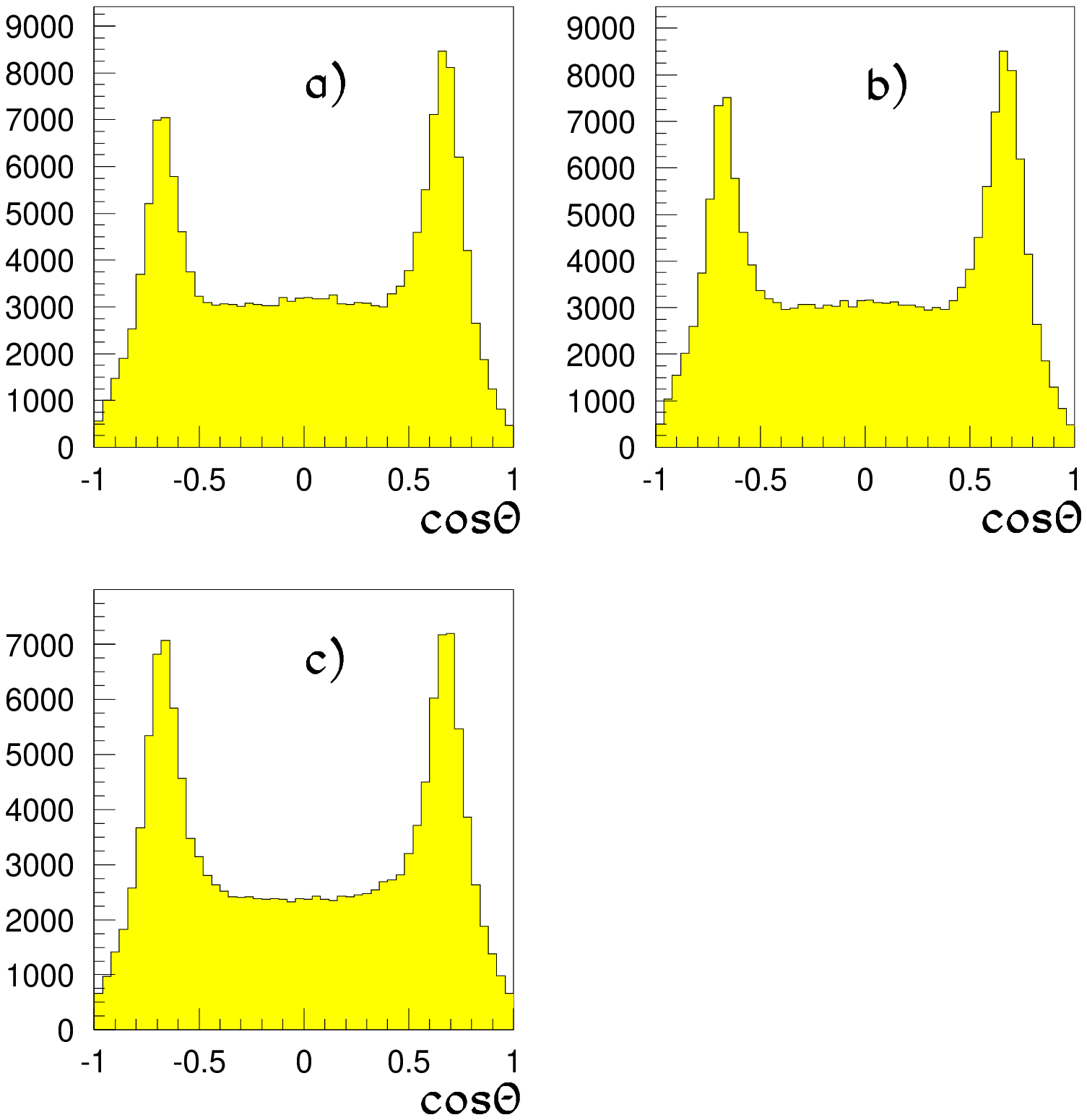}}
\end{picture}
\caption{The $\pi^+\pi^-\pi^0$ Dalitz plot in $\pbp$ annihilation
at rest, and $\rho^+$ (a), $\rho^-$ (b)
and $\rho^0$ (c) decay angular distributions.}
\label{dsr:fig:3pidp}
\end{figure}

A first look at the decay angular distributions reveals two strong
peaks above a $\sin^2\Theta$ distribution. The peaks can be understood
by inspecting the full Dalitz plot: the $\rho^+$ and $\rho^-$ bands
cross in the lower left corner of the Dalitz plot. The amplitudes for
the two processes $\pbarp\to\rho^+\pi^-$ and $\pbarp\to\rho^-\pi^+$
interfere constructively and lead to marked deviations of the observed
angular distribution from the expected one.  The crossing of the
$\rho^+$ band with the $\rho^-$ or $\rho^0$ band leads to an increase
of the intensity by a factor 4 because of quantum mechanical
interference (the amplitudes are added). The dominant $\sin^2\Theta$
distribution beneath the two interference peaks is due to strong
contributions from the $\islj=\stso$ initial state. The
$\tssz\;\pbarp$ initial state generates a $\cos^2\Theta$ which would
lead, with a fraction of the $\sin^2\Theta$ part, to a constant
plateau beneath the $\sin^2\Theta$ distribution.
\begin{table}[h!]
\caption{\label{dsr:tab:threepiselect}%
Angular distributions for $\pbN\to\rho\pi$ annihilation.  The
atomic states are represented as $\islj=\stso$
with $I, S, L, J$ being isospin, spin, orbital and total angular
momenta; $\ell$ is the orbital angular momenta between $\rho$ and
$\pi$, $\Theta$ the angle between the direction of the more positively
charged pion from $\rho$ decay with respect to the $\rho$ direction of
flight. Forbidden transitions are marked by an x. The signs indicate
constructive and destructive interference.  }
\renewcommand{\arraystretch}{1.1}
\begin{tabular}{ccccc}
\hline\hline
State   & $\rho\pi$ content                    &  $\ell=0$  
         &  $\ell=1$               & $\ell=2$ \\
\hline
 \stso &  $\rho^+\pi^-+\rho^0\pi^0+\rho^-\pi^+$& x       
         &$\sin^2\Theta$           &                     \\
\tssz  &  $\rho^+\pi^--\rho^-\pi^+$            & x       
         &$\cos^2\Theta$           &                      \\
\ttpt  &  $\rho^+\pi^--\rho^-\pi^+$            &x       
         &   x                     &  $\cos^2\Theta$      \\
\ttpo  &  $\rho^+\pi^--\rho^-\pi^+$            & flat
         &  x                      &$\sin^2\Theta$        \\
\sspo  &  $\rho^+\pi^-+\rho^0\pi^0+\rho^-\pi^+$& flat
         &  x                      & $\cos^2\Theta+1/3$     \\
\hline\hline
\end{tabular}
\renewcommand{\arraystretch}{1.0}
\end{table}

A second look reveals an even more important aspect: the three $\rho$
peaks have nearly the same strength. Now, $\rho^0\pi^0$ production is
forbidden from initial states with charge conjugation $C=+1$.  From
initial states with $C=-1$, the rates for the three charge modes are
fixed by Clebsch--Gordan coefficients, and are identical. In the limit of
exactly equal strengths, only initial states with $C=-1$
contribute. Because $G=(-1)^I\,C$, $C=-1$ and an odd  number of
pions entails $I=0$.  The three angular distributions give evidence for a small
reduction of $\rho^0\pi^0$ relative to $\rho^+\pi^-$ or
$\rho^-\pi^+$. The smallness of the reduction shows that states with
$C=+1$ do contribute, but only by a small fraction.

The small flat contribution to the decay angular distributions
can be due to a $\cos^2\Theta$ part or due to annihilation from atomic
P-states; both contributions are evidently small.

A partial-wave analysis quantifies these observations, determines
masses and widths of contributing resonances and gives fractional
contributions from the different $\pbp$ initial states to
the $\pi^+\pi^-\pi^0$ final state. Here, only the 
annihilation frequencies for $\rho\pi$ production 
are given (adapted from Table~\ref{tm:tab:Bpartial}).

The $\rho\pi$ channel is produced from  the $\islj=\stso$ $\pbarp$
initial state with a branching ratio $(63.6\pm 3.3)\times 10^{-3}$, and 
with branching ratio $(6.0\pm 2.0)\times 10^{-3}$
from the $\tssz$ state. The ratio
\begin{equation}
\frac{\AF(\tssz\to \rho^+\pi^- )} 
{\AF(\stso \to \rho^+\pi^- )} = 0.094 \pm 0.032~,
\label{dsr:eq:ratio1}
\end{equation}
is rather small. There is no obvious  explanation of the suppression of
$\rho\pi$ from the $\tssz$ state. The preference for
annihilation via the $I=0$ initial state is a dynamical selection
rule. In the literature, it is known as the $\rho\pi$ puzzle.

\begin{figure}[h!]
\includegraphics[width=0.90\textwidth]{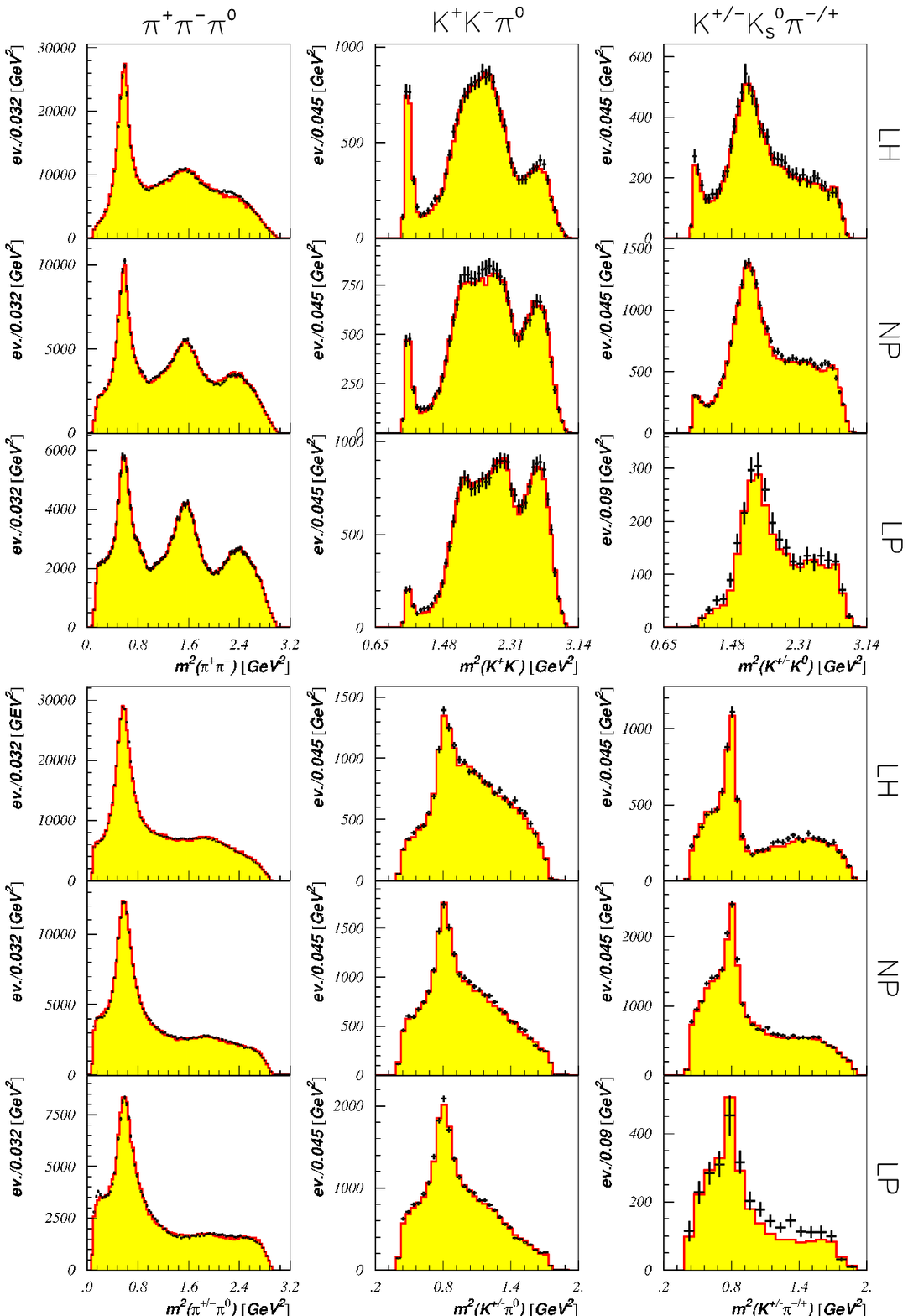}
\caption{\label{dsr:fig:obelix}%
Invariant mass square distributions for $\pbarp$ annihilation at
rest using H$_2$ targets of three different densities
\cite{Bargiotti:2003ev}. Left: $\pbp\to\pip\pim\piz$ , centre:
$\pbp\to\Kp\Km\Kn$ , right: $\pbp\to\Kpm\Ks\pimp$ .  LH: liquid H$_2$;
NP: atmospheric pressure and room temperature; LP: low pressure.  }
\end{figure}

This selection rule is partly confirmed in annihilation from atomic
P-states. Recently, the Obelix collaboration reported a
pressure-dependent study of $\pbp$ annihilation into
$\pi^+\pi^-\pi^0$, $\K^+\K^-\pi^0$, and $\K^{\pm}\Ks\pi^{\mp}$. 
Figure~\ref{dsr:fig:obelix} shows the
$\pi^+\pi^-$ and $\pi^{\pm}\pi^0$ invariant mass
distributions for $\pbp$ annihilation at rest in liquid H$_2$,
with $\sim (87.5\pm 2.0)\%$ S-state, $\sim (12.5\pm 2.0)\%$ P-state capture 
and in gaseous H$_2$ at atmospheric pressure (with $\sim (36\pm 3)\%$ 
S-state, $\sim (64\pm 3)\%$ P-state capture, and at very low pressure, 
with $\sim (11\pm 7)\%$ S-state, $\sim (89\pm 7)\%$ P-state capture.
 
The comparison of the upper two plots showing the $\pi^+\pi^-$ and
$\pi^{\pm}\pi^0$ (squared) invariant mass distributions in liquid
H$_2$ confirms the observations of the Crystal Barrel data: the peak
height for the $\rho^0$ is nearly as large as the peak height of the
charged $\rho^{\pm}$: isoscalar initial states must dominate the
reaction. Since there is S-state capture dominance, the most important
contribution has to come from the $^3\rm S_1$ protonium state. The two
lower plots, from annihilation at low pressure, show a more
significant $\rho^{\pm}$ signal. The $\rho^{0}$ is reduced by
approximately 30\%. This would indicate that 2/3 of all annihilations
at low pressure are from the isoscalar, and 1/3 from the isovector
component of the protonium atom. Inspection of
Table~\ref{dsr:tab:threepiselect} shows that the isovector P-states
have a statistical weight 8, the isoscalar ones a weight 3. Taking
this into account, the $\rho\pi$ puzzle is present also in
annihilation with a large P-state capture probability, even though
less pronounced. More remarkable is, however, the result of the
partial wave analysis that the \sspo\ level decays only weakly into
$\rho\pi$ with $\ell=0$:
\begin{equation}
\label{dsr:eq:rhopiob}
\frac{\rm \AF (\pbp\to \rho^{\pm}\pi^{\mp},\spo, {\ell=2})}
{\rm \AF (\pbp\to \rho^{\pm}\pi^{\mp},\spo, {\ell=0})} = 8.1\pm 2.0~.
\end{equation}
This is a surprise: as seen in Table \ref{kin:tab:allowed}, $\sspo$ is
forbidden to decay into most simple two-meson modes ($\pi\pi$,
$\K\Kb$, $\pi\omega$, \dots), and one would naively expect this
channel to make full use of the $\rho\pi$ mode, in particular in
S-wave ($\ell=0$). The decay 
$\sspo\to\rho\pi$ with $\ell=0$ seems to be dynamically  suppressed.
\subsectionb{Annihilation into \atmass$\pi$}\label{dsr:sub:a2pi}
A further dynamical selection rule similar to the $\rho\pi$ puzzle is
observed in $\pbp$ annihilation into
$\atmass\pi$. Figure~\ref{dsr:fig:a2pi} 
\begin{figure}[b!]
\begin{tabular}{ccc}
\hspace*{-2mm}\includegraphics[width=0.33\textwidth]{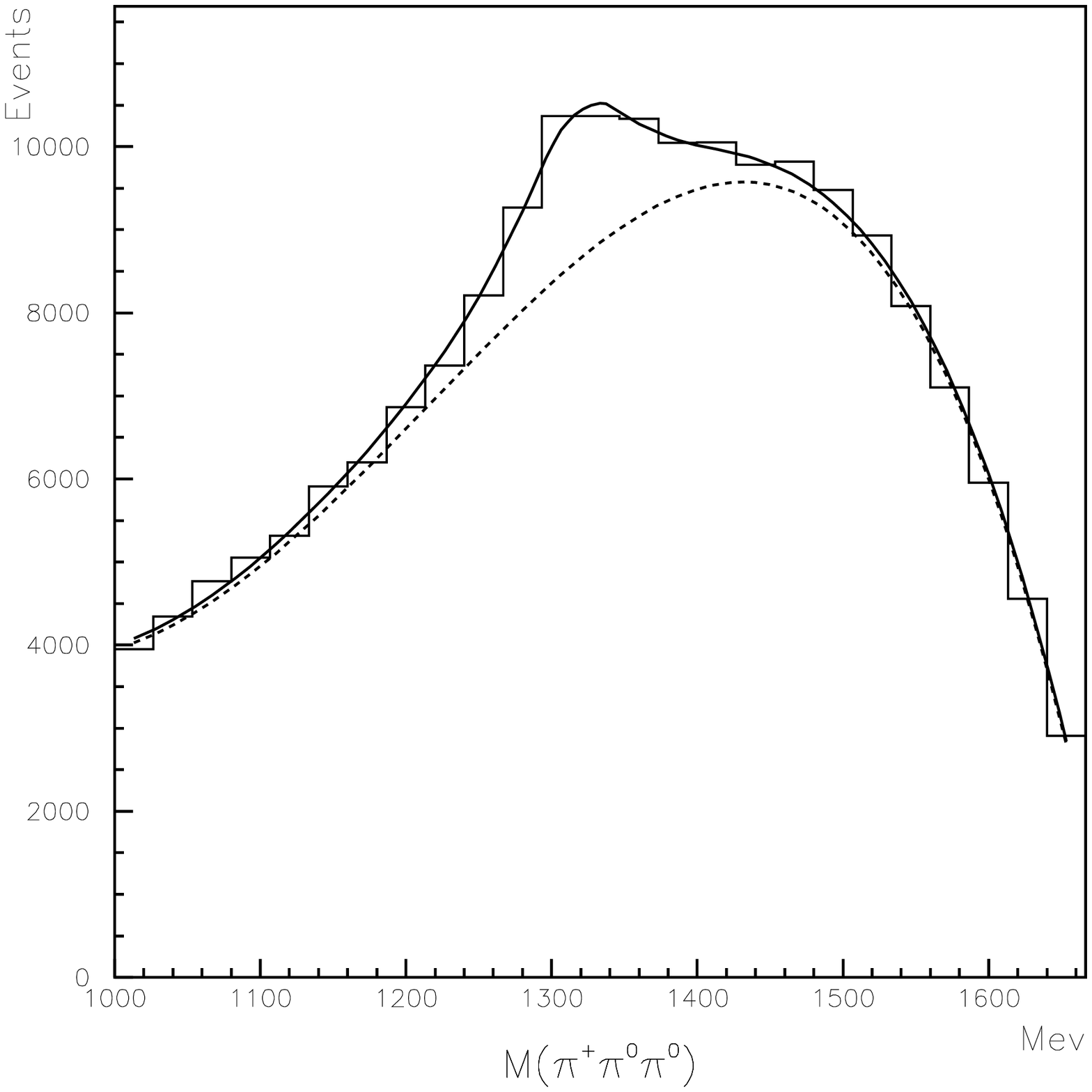}&
\hspace*{-4mm}\includegraphics[width=0.33\textwidth]{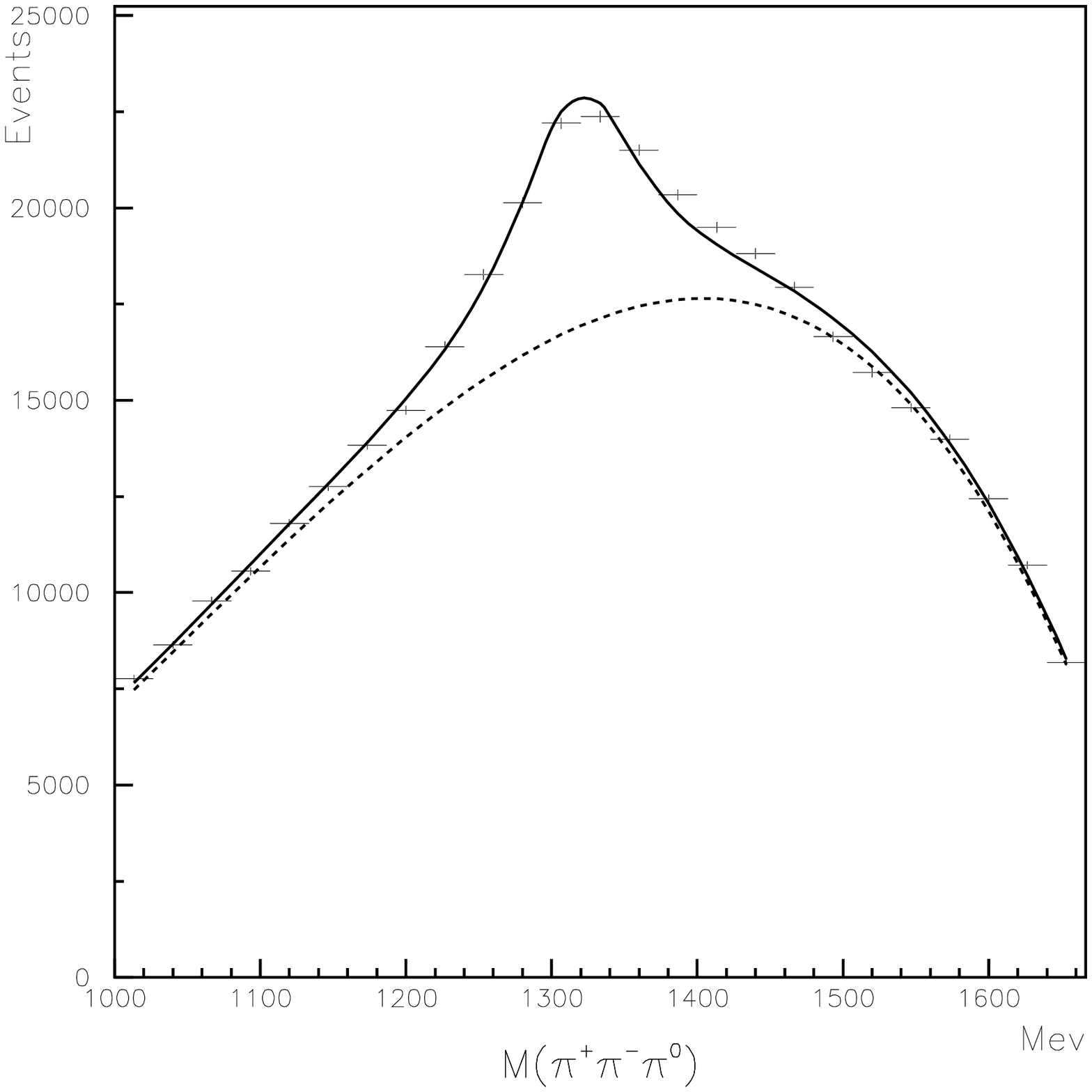}&
\hspace*{-4mm}\includegraphics[width=0.33\textwidth]{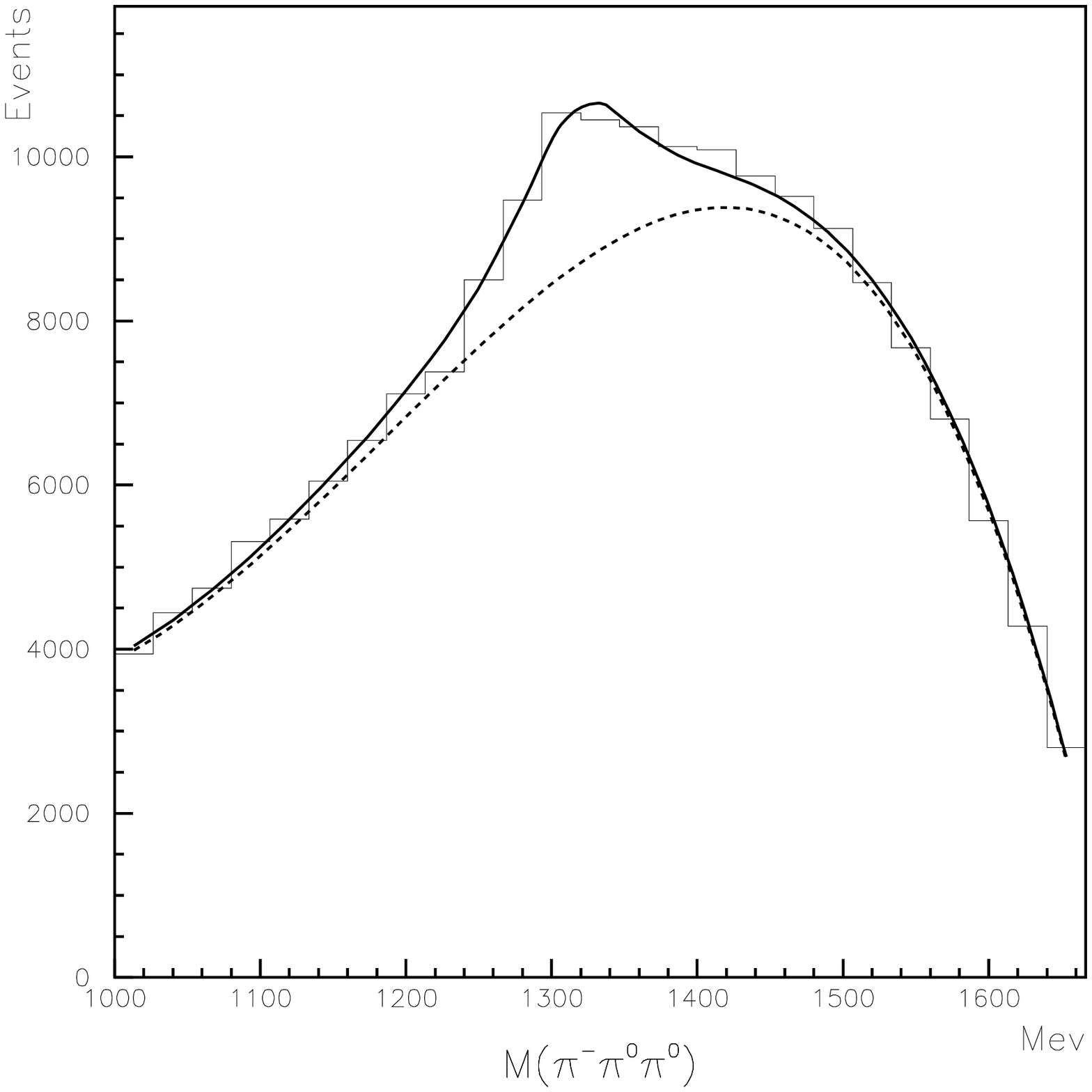}
\end{tabular}
\caption{\label{dsr:fig:a2pi}%
The $\pi^+2\pi^0$, $\pi^+\pi^-\pi^0$, and $\pi^-2\pi^0$ invariant mass
distributions from $\pbp$ annihilation into the
$\pi^+\pi^-2\pi^0$ final state. There are two entries per event for
the $\pi^+\pi^-\pi^0$ mass plot.  All three distributions show a broad
peak due to the $\atmass\pi$ production. The number of neutral
$\atmass$ is not visibly smaller than that of charged
$\atmass$.}
\end{figure}
shows the most direct
evidence for this rule, from Crystal Barrel data on the
$\pi^+\pi^-2\pi^0$ final state. In the 3-pion invariant mass
distribution, clear evidence for the intermediate $\atmass$ decaying
into $\rho\pi$ is seen. The strength of the $\atmass$ in the three
plots is about the same. Now, annihilation into
$\atpmmass\pi^{\mp}$ is allowed from both the \sssz\ and the
\ttso\ initial state while $\atzmass\pi^{0}$ is forbidden from
\ttso . Both processes require $\ell=2$ between
\atmass\ and the pion. The fact that the rates for production of
charged \atmass\ and neutral \atmass\ mesons are not too different
entails that $\atmass\pi$ production is preferred from the isoscalar
\sssz\ state and suppressed from the isovector \ttso\ initial
state. Again, there is no known reason for the suppression of one
initial state compared to another initial state. The suppression is
due to a dynamical selection rule.

The data shown above were not analysed for the different partial-wave
contributions. They are shown here only since they visualise best the
new dynamical selection rule. Partial wave analyses were performed on
bubble-chamber data, with the $\atmass$ decaying into $\rho\pi$,
$\eta\pi$, and $\K\Kb$. The partial wave analyses confirm the qualitative
observation discussed above: annihilation into \atmass$\pi$ is strong
from the $I=0$ initial state (\ssz ) and weak from the $I=1$ component
of the \tso\ state, with a ratio, see Table~\ref{tm:tab:Bpartial}, 
\begin{equation}
\frac{\BR(\ttso\to \at\pi)}
{\, \BR(\sssz\to \at\pi)}  =  \frac{6.4\pm 1.2}{82.0\pm 12.0}~.
\label{dsr:eq:ratio3}
\end{equation}
The \atmass$\pi$ annihilation mode from the isovector 
component of the protonium
atom is suppressed by one order of magnitude compared to 
annihilation from the isoscalar component.
This is the same dynamical selection rule as the
$\rho\pi$ puzzle. 
\subsectionb{Annihilation into $\pi\pi$}\label{dsr:sub:pipi}
There is a related observation in $\pbp$ annihilation into
$\pi\pi$ and into $\K\Kb$. The branching ratios 
into $\K\Kb$ and into $\pi\pi$ depend strongly
on the initial state. In Table \ref{dsr:tab:pipikk} the relevant
branching ratios are listed. Branching ratios determined from
\nbar p scattering are included in the table.

The ratio of $\K\Kb$ versus $\pi\pi$ production is about 1 in
annihilation from the \tso\ state and less than 0.1 for annihilation
from \tpz . This result was interpreted as $\K\Kb$ suppression
in P-wave annihilation \cite{Mundigl:1990cj}.  The (corrected)
branching ratios in Table~\ref{dsr:tab:pipikk} show that, instead,
$\pi\pi$ production is suppressed in annihilation from the \tso\ state
and not $\K\Kb$ in P-wave annihilation.  This interpretation is also
favoured by a comparison of cross sections for annihilation in
flight. The cross sections for annihilation into $\pi\pi$ and into
$\rho\pi$ are very similar in size, while at rest the annihilation
frequencies differ by one order of magnitude.  Hence we interpret the
change in the $\K\Kb/\pi\pi$ ratio as suppression of the
$\pi\pi$ frequency in the \tso\ state.

\begin{table}[h!]
\caption{\label{dsr:tab:pipikk}Branching ratios in units of 10$^{-3}$
for $\pbp$ and $\nbp$ annihilation at
rest into $\pi\pi$ and into $\K\Kb$, from 
Table~\ref{tm:tab:Bpartial} and Eqs.~(\ref{tm:eq:AN2pi}), (\ref{tm:eq:AN2K}).
 An x indicates that the transition is forbidden.}
\renewcommand{\arraystretch}{1.1}
\begin{tabular}{ccccccc}
\hline\hline
\NNb  & $\pi\pi$ &$\Kp\Km$& $\Ks\Kl$& $\Ks\Ks+\Kl\Kl$  & $\pip\piz$ & $\K^+\Kb^0$\\
\hline
\tso&$2.79\pm 0.18$&$1.47\pm
0.06$&$1.31\pm 0.07$  &  x   & $4.1\pm0.7$& $3.5\pm0.5$ \\
\tpz&$54.3\pm 1.5$&$4.2\pm 0.3$
& x &$0.79\pm 0.15^*$  && \\
\hline\hline
\end{tabular}
\renewcommand{\arraystretch}{1.0}
\centerline{$^*$\footnotesize{The branching ratio for $\pbp\to \Ks\Ks+ \Kl\Kl$ is
reduced by a factor of 2 due to Bose symmetry.}\hfill}
\end{table}

Annihilation into $\pi\pi$ from the $\tso$
initial state proceeds via the isovector
component of the $\pbarp$ system: again annihilation from the
isovector part of the protonium wave function into two isovector
particles is suppressed, by about one order of magnitude.

Obviously, there is a common property in these three cases.
Annihilation into two isovector mesons such as $\pi\pi$, $\rho\pi$ and
\atmass$\pi$ is strong from isoscalar initial states.  In the language
of SU(3)$_{\rm F}$ flavour symmetry, to be described later in this
review, the transition from an isoscalar protonium state to two
isovector mesons corresponds to a symmetric $D$ coupling that is
large. The transition from an isovector protonium state to two
isovector mesons, described by an (antisymmetric) $F$ coupling, is
weak.
\subsectionb{Annihilation into $\K\Kb$}\label{dsr:sub:kk}
The annihilation frequencies listed in Table \ref{dsr:tab:pipikk}
reveal another surprising fact: the annihilation frequencies into $\Kp\Km$
and $\Ks\Kl$ differ only by 10\%. Assuming an
isospin-blind annihilation potential, annihilation into $\K^+\K^-$ or into $\Ks\Kl$ (depending on the relative phase of the $I=0$ and $I=1$ amplitudes) could 
be suppressed in the same way as the charge exchange
 reaction $\pbp\to \bar{\n}\n$ is in a scattering situation.  
On the other hand, if only one isospin component of the protonium wave
function contributes to the process, the two annihilation
frequencies must be identical.

Assuming that one isospin component dominates annihilation into a $\rm
K\Kb$ pair, it can be identified by comparison with the reaction
$\nbp\to \K^+\Kb^0$. The branching ratio for the latter
reaction should be either close to zero (for an isoscalar transition
operator), or close to the sum of annihilation frequencies into $\rm
K^+K^-$ and $\Ks\Kl$ $(2.78\pm 0.09)\times 10^{-3}$ if the
transition operator is iso-vectorial. The experimental value
$(3.47\pm0.53)\times 10^{-3}$ requires $I=1$. Why is the annihilation
reaction
\begin{equation}
\pbarp\,(\ttso)\to \K^+\K^- - \K^0\Kb^0~,\quad \hbox{strong}~,
\end{equation}
and
\begin{equation}
\pbarp\,(\stso)\to \K^+\K^- + \K^0\Kb^0~, \quad \hbox{weak?}
\end{equation}
There is no known reason for the suppression of one isospin channel
with respect to the other channel. It is a dynamical selection rule. 
\subsectionb{Annihilation into $\K^{\pm}\Ks\pi^{\mp}$
and $\K^{\pm}\Kl\pi^{\mp}$}\label{dsr:sub:kks}
A similar selection rule is seen in $\pbarp\to
\K^{\pm}\Ks\pi^{\mp}$. Figure~\ref{dsr:fig:kkbpi} 
shows (on the right) the Crystal Barrel data on this reaction. The
most striking feature are the two $\K^*$ bands ($\K^{\pm}\pi^{\mp}$
and $\Ks\pi^{\mp}$). The two bands cross but show little intensity at
the crossing on the diagonal, where the two $\K^*$ masses are
identical. Obviously, the two amplitudes (for $\K^{\star 0}$ and
$\K^{\star\pm}$ production) interfere destructively.

\begin{figure}[h]
\setlength{\unitlength}{.01\textwidth}
\begin{picture}(100,45)(38,5)
\put(0,22.5){\vbox{\includegraphics[width=0.27\textwidth]{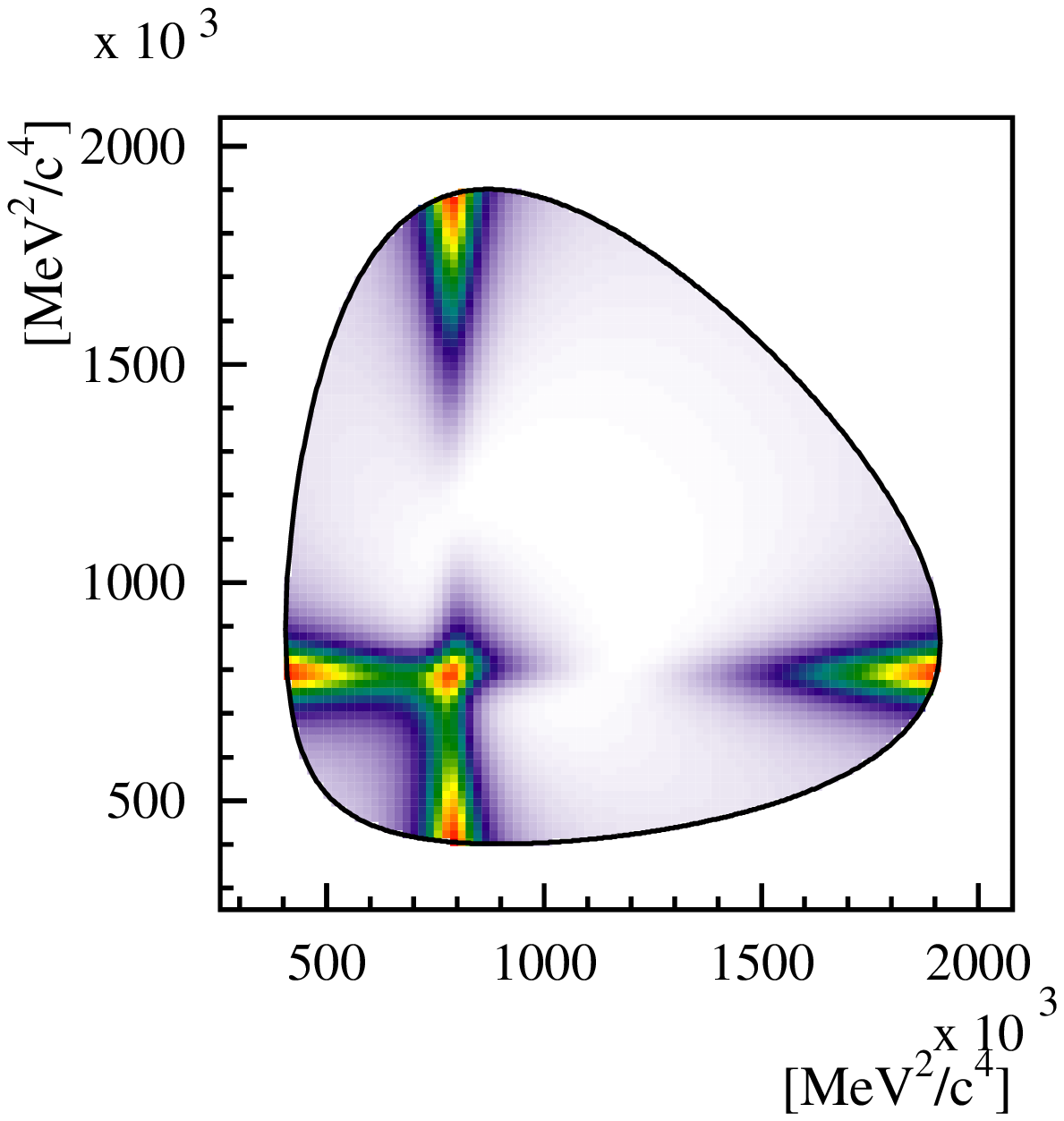}}}
\put(0,0){\vbox{\includegraphics[width=0.27\textwidth]{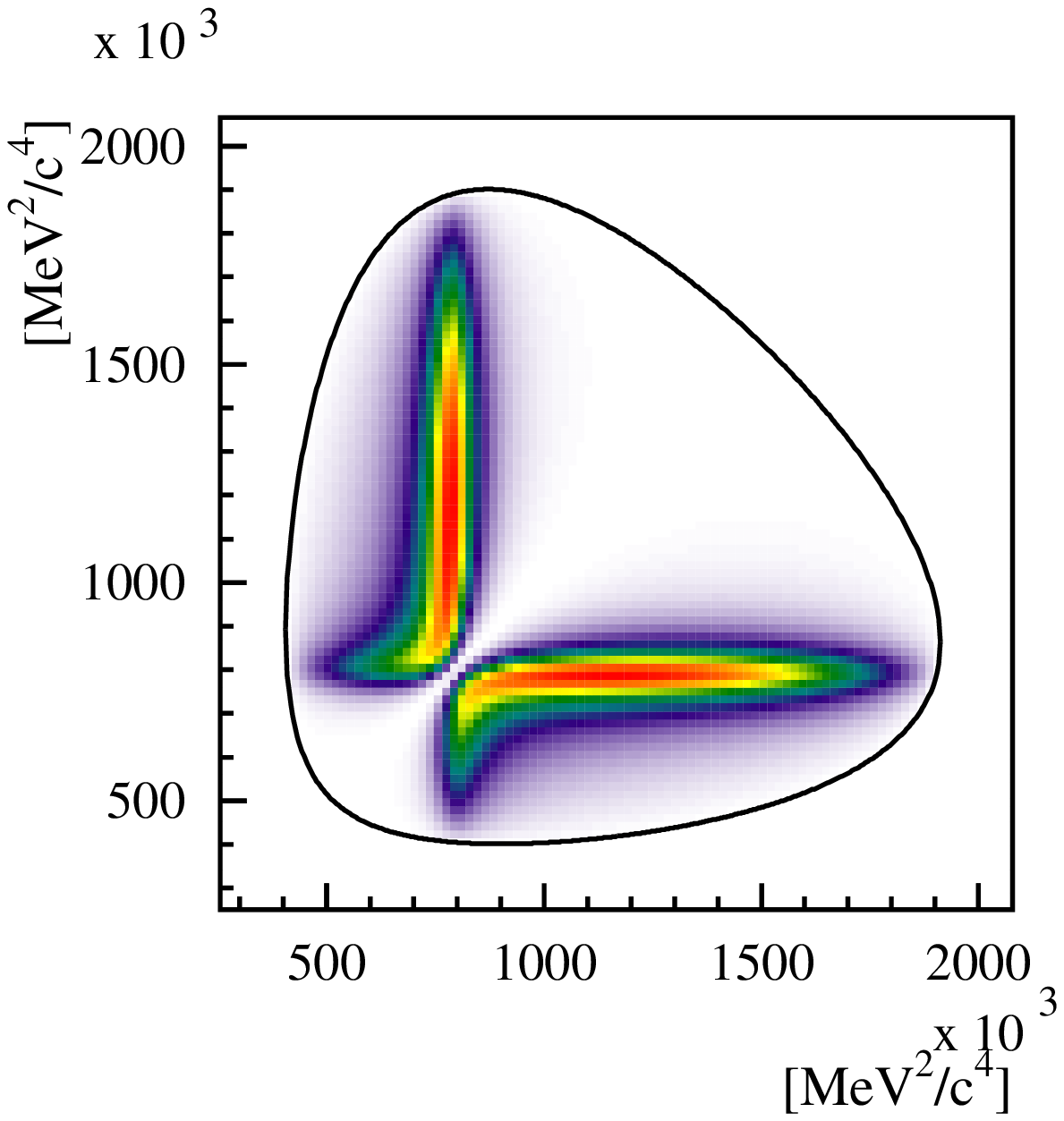}}}
\put(25,22.5){\vbox{\includegraphics[width=0.27\textwidth]{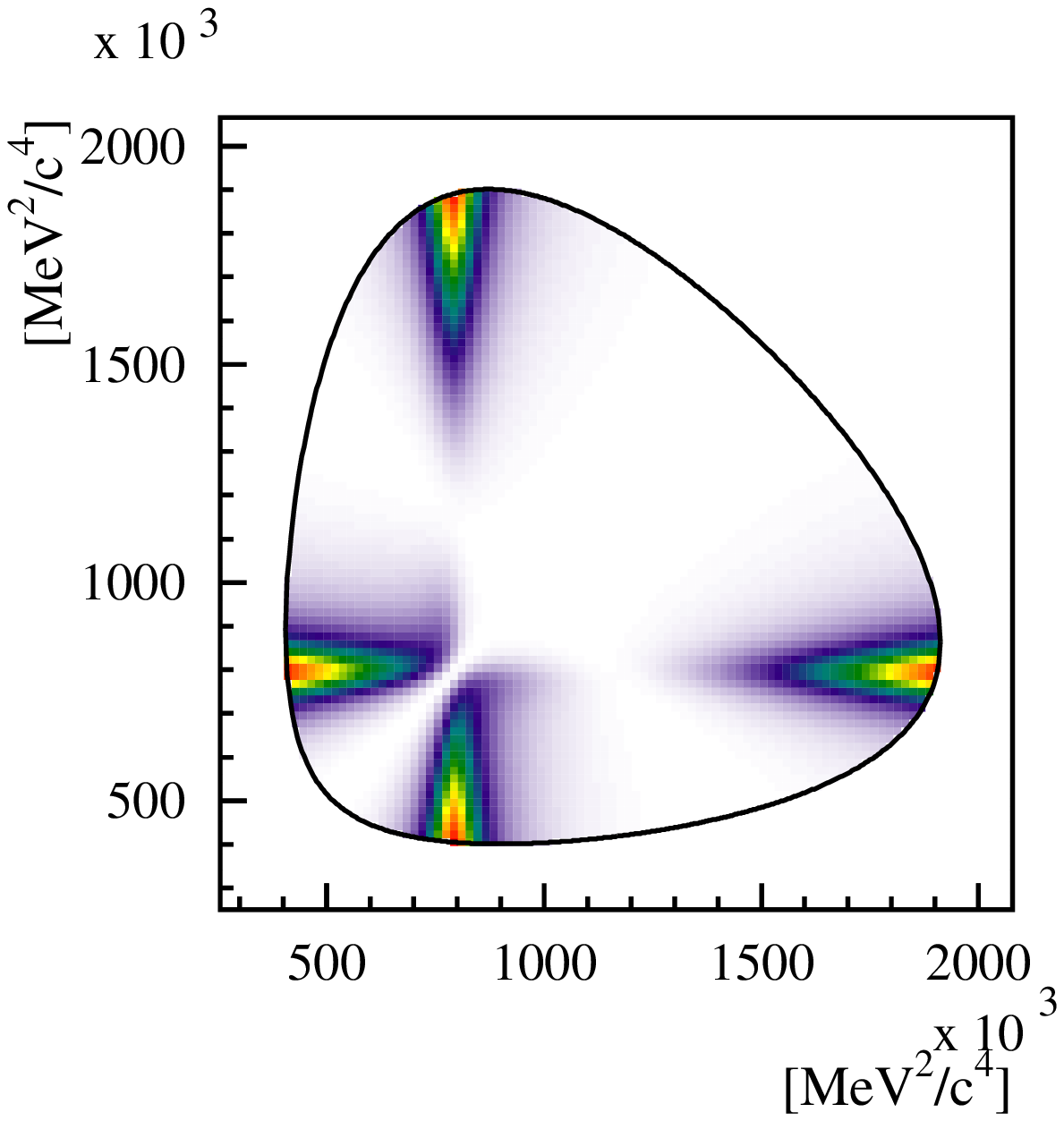}}}
\put(25,0){\vbox{\includegraphics[width=0.27\textwidth]{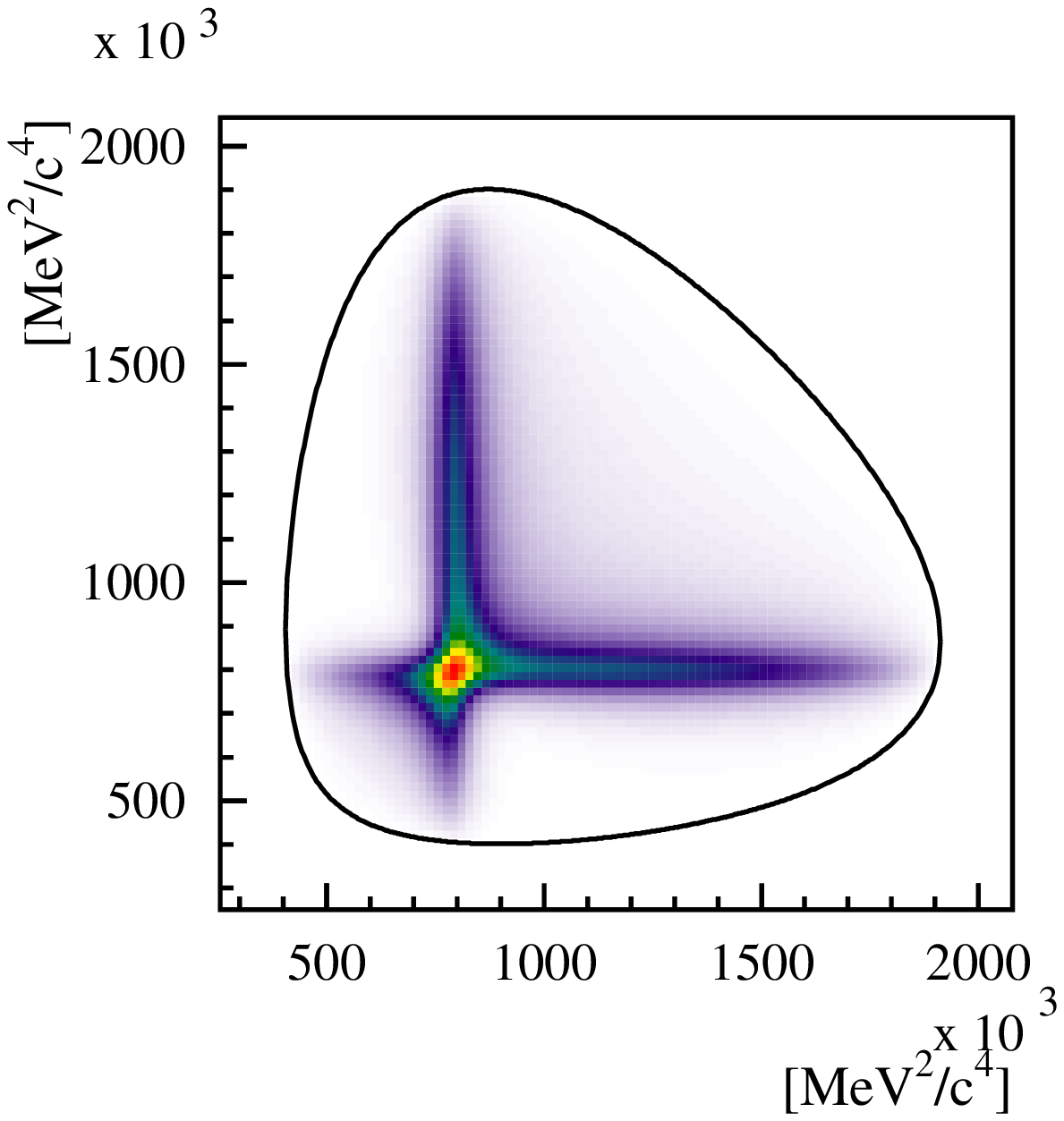}}}
\put(62,0){\vbox{\includegraphics[width=.52\textwidth]{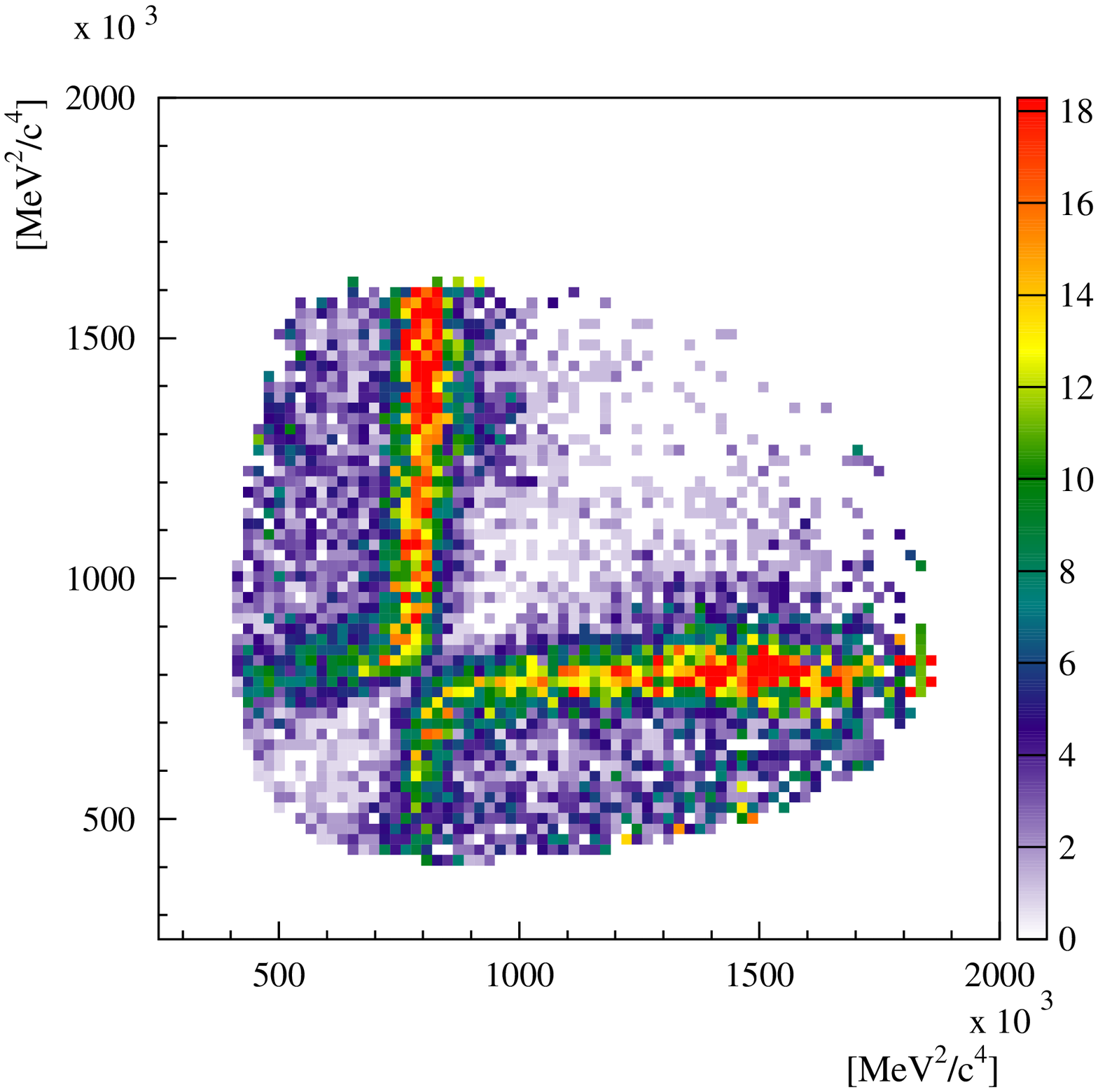}}}
\put(52.5,43){$\tssz$}
\put(77.5,43){$\sssz$}
 \put(52.5,20.5){$\ttso$}
\put(77.5,20.5){$\stso$}
\end{picture}
\caption{\label{dsr:fig:kkbpi}%
The experimental Dalitz plot for the reaction
$\pbp\to\K^{\pm}\Ks\pi^{\mp}$ (liquid H$_2$). The
experimental data (from Crystal Barrel) are shown on the right; 
on the left are Dalitz plots expected for annihilation from 
the various possible S-wave components of protonium.}
\end{figure}

The following discussion is restricted to S-state capture.  The final
state $\K^{\pm}\Ks\pi^{\mp}$ does not have defined $G$-parity, and
both isospin components may contribute to the annihilation
process. There are four initial states which may contribute to
the Dalitz plot in Fig.~\ref{dsr:fig:kkbpi} (right).  On the left,
Fig.~\ref{dsr:fig:kkbpi} shows the Dalitz plots expected for the four
allowed initial states. The \ssz\ initial state is
characterised by a $\cos^2\Theta$ angular distribution, with high
intensity at the ends with $\cos\Theta =\pm 1$. The
\tso\ state produces a $\sin^2\Theta$ distribution and
the intensity vanishes at the ends. 
The Dalitz plot expected for the isovector and the
isoscalar component of the protonium wave function differ
by their interference pattern.

It is obvious from the comparison of data and theoretical Dalitz plots
that the two theoretical Dalitz plots with destructive interference
between the two $\K^*\K$ amplitudes are the ones contributing most to the
final state.  This observation is confirmed in partial wave analyses,
see Table~\ref{tm:tab:BRrhkstar}, where the contributions to $\K^*\K$
production are listed.  The sum of the two partial waves with
destructive interference exceeds the sum of constructive interference
by one order of magnitude.

This dynamical selection rule can be interpreted at the level of the
produced $\K^*$. For this purpose the decomposition of the
$\K^*$ contribution into the final-state particles needs to be
considered:
\begin{equation}\label{dsr:eq:kdecay1}
\eqalign{%
&\left|\ttso\right>\to\left|I^C=1^-\right>=
                      \tfrac{1}{2}\left(
                      \left|\K^{*+}\Km\right> -
                      \left|\K^{*0}\Kb{}^0\right>  + 
                      \left|\K^{*-}\Kp\right>  - 
                      \left|\Kb{}^{*0}\K^0\right>
\right) \quad\rm strong~,\cr
&\left|\tssz\right>\to\left|I^C=1^+\right>= 
                      \tfrac{1}{2}\left(
                      \left|\K^{*+}\K^-\right>  - 
                      \left|\K^{*0}\Kb{}^0\right>  - 
                      \left|\K^{*-}\K^+\right>  + 
                      \left|\Kb{}^{*0}\K^0\right>
\right) \quad\rm weak~, \cr
&\left|\stso\right>\to\left|I^C=0^-\right>= 
                      \tfrac{1}{2}\left(
                      \left|\K^{*+}\K^-\right>  + 
                      \left|\K^{*0}\Kb{}^0\right>  + 
                      \left|\K^{*-}\K^+\right>  + 
                      \left|\Kb{}^{*0}\K^0\right>
\right) \quad\rm weak~,\cr
&\left|\sssz\right>\to\left|I^C=0^+\right>= 
                      \tfrac{1}{2}\left(
                      \left|\K^{*+}\K^-\right>  + 
                      \left|\K^{*0}\Kb{}^0\right>  - 
                      \left|\K^{*-}\K^+\right>  - 
                      \left|\Kb{}^{*0}\K^0\right>
\right) \quad\rm strong~,}
\end{equation} 

In the Dalitz plots, Fig.~\ref{dsr:fig:kkbpi}, interference between the
$\K^{*\pm}$ and $\Kb{}^{*0}$ is observed.  Interference occurs in a
given final state; the $\K^*$ has to be expanded into its decay
products to appreciate the meaning of the interference pattern. The
next two lines give the final states produced in the above reactions.
\begin{equation}\label{dsr:eq:kdecay2}
\eqalignt{%
& \phantom{-}\left|(\pi^0\K^{+})\K^-\right> \qquad    \left|(\pi^-\K^{+})\Ks\right>  \qquad  
&&\left|(\pi^0\K^{-})\K^+\right>   \qquad      \underline{\left|(\pi^+\K^{-})\Ks\right>} \cr 
& \underline{-\left|(\pi^+\Ks)\K^-\right>}\qquad\left|(\pi^0\,\Ks)\,\Ks\right> \qquad
 -&&\left|(\pi^-\Ks)\K^+\right>  \qquad \left|(\pi^0\,\Ks)\,\Ks\right>
}~.
\end{equation}

A pair of final states in which the $\K^*$ can interfere is underlined
in (\ref{dsr:eq:kdecay2}).  These two states have $\K^{*+}\K^-$ and
$\Kb{}^{*0}\K^0$ intermediate states, the first and the last entries in
Eq.\ (\ref{dsr:eq:kdecay1}).  The two annihilation modes marked strong
in (\ref{dsr:eq:kdecay1}) produce the $\K^*\K$ in the form
$\K^{*+}\K^- - \Kb{}^{*0}\K^0$ and the two weak modes in the form
$\K^{*+}\K^- + \Kb{}^{*0}\K^0$.  It is not the isospin which drives this
dynamical selection rule: the two initial states contributing strongly
are
\begin{equation}\eqalign{%
&\ttso\ \to\ \K^{*+}\K^- - \K^{*0}\Kb{}^0 + 
\hbox{c.c.}\quad \hbox{strong}~,\cr
&\sssz\ \to\ \K^{*+}\K^- - \K^{*0}\Kb{}^0 + 
\hbox{c.c.}\quad \hbox{strong} ~,   }
\end{equation}
while the two  states
\begin{equation}\eqalign{%
&\stso\ \to\ \K^{*+}\K^- + \K^{*0}\Kb{}^0 + 
\hbox{c.c.}\quad \hbox{weak}~,
\cr
&\tssz\ \to\ \K^{*+}\K^- + \K^{*0}\Kb{}^0 + 
\hbox{c.c.}\quad \hbox{weak}   ~,     }
\end{equation}
contribute at most weakly. The charged and neutral $\K^*\,\K$
combinations are produced strongly with a relative minus sign.

Inspecting Table~\ref{tm:tab:Bpartial} shows that $\K^*_2(1430)$K 
production exhibits a similar selection rule: 

\begin{equation}
\eqalign{%
\ttso\ \to\ \K_2^{*+}\Km - \K_2^{*0}\Kb^0 + {\rm c.c.}&\quad \hbox{strong}~,\cr
\sssz\ \to\ \K_2^{*+}\Km - \K_2^{*0}\Kb^0 + {\rm c.c.}&\quad \hbox{strong}~,\cr
\stso\ \to\ \K_2^{*+}\Km + \K_2^{*0}\Kb^0 + {\rm c.c.}&\quad \hbox{weak}~,\cr
\tssz\ \to\ \K_2^{*+}\Km + \K_2^{*0}\Kb^0 + {\rm c.c.}&\quad \hbox{weak}~.
}
\end{equation}  
The dominance of one isospin channel for these decay modes may be
surprising since the nominal mass of the $\K_2^{*}(1430)$ plus
the mass of the $\K$ exceed the available energy. Hence
rescattering of the final-state particles is expected to play
a large role.

Equations~(\ref{dsr:eq:kdecay1}) suggest a pattern of strongly and
weakly produced $\K$ and $\K{}^*$ which 
depends on charge conjugation and isospin of the initial state.
This generalisation can be tested in annihilation from atomic
P-states. 
The Obelix collaboration has measured the density dependence of
$\K\Kb\pi$ production, see Fig.~\ref{dsr:fig:obelix}. Also here,
a preference for one isospin for  $\K \K{}^*$ production
is observed; the prevailing isospin channel is shown in
Table~\ref{tm:tab:roaf}. For the $\tpo$ initial state, the
expectation is fulfilled, for the $\spo$ and $\tpt$ not. 
Annihilation from P-states seems not to follow the same simple 
pattern as observed in S-state annihilation.

In the
$\K^+\K^-\pi^0$ final state, a clear signal due to $\phi$ production
is seen. The final state $\phi\pi$ can be produced only from states
with negative charge conjugation and with isospin $I=1$. Thus only the
\ttso\ and \tspo\ states can contribute. At high density, the $\phi$
is strong, but weak at low densities, indicating a preference for the
$\ttso$ initial state.  If $\phi$'s are produced as two-step process,
through $\K^*\Kb$ and rescattering of the two kaons into a $\phi$,
then $\K^*\Kb$ should be strong in \ttso\ (which is the case) and
weak in \tspo . This finding is supported by the partial wave analysis
of the data of Fig.~\ref{dsr:fig:obelix} where the \tspo\ couples
weaker to $\K^*\Kb$ than \sspo .  This question will be addressed
again in Sec.~\ref{phe:sub:OZI}.
\subsectionb{Annihilation into \protect$\K^{*}\Kb{}^{*}$ }\label{dsr:sub:ksks}
The frequency for \pbp\ annihilation into
$\K^{*}\Kb^{*}$ was determined in the BNL bubble chamber  
experiment \cite{Barash65b}: 
\begin{equation}\label{dsr:eq:kprodg}
\eqalign{%
\AF(\pbarp\to \K^{*+}\K^{*-})  &=  (3.3\pm 1.1)\times 10^{-4} ~,\cr     
\AF(\pbarp\to \K^{*\,0\,}\Kb^{*\,0\,}) &=  (7.3\pm 1.5)\times 10^{-4}~. } 
\end{equation}
The more recent annihilation frequencies from Crystal Barrel are
at variance with these findings.

The Crystal Barrel collaboration has studied the dynamics of
annihilation into two $\K^{*}$, in the final states $\Ks\Kl\pi^0\pi^0$
\cite{Abele:1997vu} and $\K^{\pm}\Ks\pi^{\mp}\pi^0$
\cite{Abele:1997vv}.  The former is forced to proceed via the $\tso$
initial state, and is dominated by $\K^*\Kb{}^*$ production.  The two
$\K^*$ are produced with $\ell=1$ between them; the two $\K^*$ spins
add up to $S=0$ or $2$ with about equal amplitudes and opposite
phases; the contribution from $S=1$ is small. 
The absolute annihilation frequencies were not determined.

In the reaction $\K^{\pm}\Ks\pi^{\mp}\pi^0$, $\K^*\Kb^*$ production
plays a less significant role. The fractional contributions in the
partial wave analysis can be normalised to the annihilation frequency
of the reaction channel (given in Table \ref{glob:tab:kmult}) to
arrive at the following branching ratios:
\begin{equation}\label{dsr:eq:kprodcb}
\eqalign{
\BR(\pbarp(\tso )\to \K^{*+}\K^{*-})   &\sim \  1.3\times  10^{-3}~, \cr
\BR(\pbarp(\tso )\to \K^{*\,0\,}\Kb^{*\,0\,})  &\sim \   0.2\times  10^{-3}~,\cr 
\BR(\pbarp(\ssz )\to \K^{*+}\K^{*-})   &\sim \   1.7\times  10^{-3}~,\cr
\BR(\pbarp(\ssz )\to \K^{*\,0\,}\Kb^{*\,0\,})  &\sim \   1.5\times 10^{-3}~.}
\end{equation} 

Annihilation from the \tso\ state requires $\ell=1$; spin states $S=0$
and $S=2$ both contribute. As above, the amplitudes are similar in
magnitude but opposite in phase. However, the dominance of
$\K^{*+}\K^{*-}$ over $ \K^{*\,0\,}\Kb^{*\,0\,}$ shows that both
isospin channels contribute.

Annihilation from the \ssz\ initial state may proceed with $\ell=0$
and with one total spin only, $S=1$. The frequency for $\K^*\Kb^*$
production is larger, and the two charge modes are about equal in
size. The measured phase between the two amplitudes is
$-50^{\circ}$. Hence we do not find the dynamical selection rules as
in the other annihilation modes into strange final-state mesons.

We note that in $\pbarp\to \K^{*}\K^{*}$, the $\K^*$ momenta are
$285\mevc$ and the two $\K^*$ decay before having left the strong
interaction volume.  $\K^*\Kb^*$ production constitutes only a small
fraction of the final state and rescattering of the final-state particles presumably has
a decisive influence. Hence it is not surprising that this reaction
does not exhibit any striking dynamical selection rule.
\FloatBarrier\subsection{Discussion of the dynamical selection rules}
\label{dsr:sub:disc}
Table \ref{dsr:tab:dsr} summarises the most important selection
rules. Weak and strong are meant as relative weights; annihilation
into \atmass$\pi$ from $I=1$ is listed as weak even though it is a
larger branching ratio than the one for $\K\Kb$ annihilation.
But for a given final state, a strong decay mode has a branching ratio
exceeding its associate weak decay mode by about one order of
magnitude.

\begin{table}[h!]
\caption{\label{dsr:tab:dsr}%
Summary of the most important {\em dynamical selection rules}.
A cross x indicates $G$-parity forbidden reactions.}
\renewcommand{\arraystretch}{1.1}
\begin{tabular}{ccc|ccc}
\hline\hline
$\slj\, (\pbarp)$ &\quad $I=0$\quad\ &\quad $I=1$\quad\ &\quad
$\slj\, (\pbarp)$ &\quad $I=0$\quad\ &\quad $I=1$\quad\ \\
\hline
\tso $\to\rho\pi$              & strong              &   x&\quad
\tso$\to\K\Kb$            &  weak               & strong \\
\ssz$\to\rho\pi$               &   x                 & weak   &
                           &                     &        \\
\hline
\ssz$\to\atmass\pi$            & strong              &   x    &\quad
\tso$\to\K^*\Kb +\mathrm{c.c.}$    &  weak   & strong \\
\tso$\to\atmass\pi$            &   x                 & weak   &\quad
\ssz$\to\K^*\Kb +\mathrm{c.c.}$    &  strong  & weak   \\
\hline
\tpz$\to\pi\pi$                & strong              &   x    &
\tso$\to\K^*_2(1430)\Kb +\mathrm{c.c.}$    &  weak   & strong \\
\tso$\to\pi\pi$                &   x                 & weak   &
\ssz$\to\K^*_2(1430)\Kb +\mathrm{c.c.}$    &  strong  & weak   \\
\hline\hline
\end{tabular}
\renewcommand{\arraystretch}{1.0}
\end{table}
In any reasonable model, the transition rate for annihilation of the
$\pbarp$ system to a two-meson final state is proportional to
\begin{enumerate} \itemsep -2pt
\item the probability to find the needed isospin component in the
protonium wave function (initial state interaction), 
\item the strength of the hadronic transition operator,
\item the probability that the final state is formed.
\end{enumerate}
Are the \dsrs\ an effect of initial state interaction?  Annihilation
of the \stso\ level to $\rho\pi$ is strong, hence the isoscalar
component of the protonium \tso\ wave function must be enhanced. 
Annihilation from $\ttso\to\pi\pi$ is weak, hence the isotriplet
component of the wave function must be small at this momentum.
From
\stso\ and \ttso\ annihilation into open strangeness we conclude that
the \tso\ isoscalar component must be small. A component can hardly be
enhanced and suppressed at the same time. If the initial state interaction
is responsible, there must be a strong momentum dependence of this
suppression and enhancement. Note, however, that the protonium wave
function, as calculated in potential models, sometimes exhibit
oscillations, see, e.g., \cite{Dover:1991mu}.

Table \ref{phe:tab:dsr-p} lists the dominant isospin
component of protonium wave functions, the momentum at which 
it should prevail, and the final state from which the conclusion 
is drawn. In particular for the \tso\ state, an oscillating
wave function would result. Translated into a spatial wave function,
the isoscalar part should dominate at 0.21; 0.26; and 0.43\,fm, 
the isovector one at 0.25 and 0.32\,fm. Extremely sharp oscillations
between isovector and isoscalar components would be needed.
\begin{table}[h!]
\caption{\label{phe:tab:dsr-p}%
Interpretation of the \dsrs\ as an effect of the initial state
interaction. The Table gives the prevailing isospin component of
protonium at a given momentum if the origin of the \dsrs\ is assigned
to the initial state interaction. The reaction from which the dominant
isospin component is determined is given in the last column.}
\renewcommand{\arraystretch}{1.4}
\begin{tabular}{cccccc}
\hline\hline
$\slj\,(\pbarp)$&dominant isospin& 
$\slj\,(\pbarp)$&dominant isospin&at momentum& From  \\  \hline
\tso & $I=0$ & \tpz & $I=0$ & 928 \mevc& $\pi\pi$           \\ 
\tso & $I=1$ & \ssz & $I=1$ & 797 \mevc& $\K\Kb + \mathrm{c.c.}$   \\
\tso & $I=0$ & \ssz & $I=0$ & 773 \mevc& $\rho\pi$          \\
\tso & $I=1$ & \ssz & $I=0$ & 616 \mevc& $\K^*\Kb +  \mathrm{c.c.}$ \\
\tso & $I=0$ & \ssz & $I=0$ & 460 \mevc& $\atmass\pi$       \\
\tso & $I=1$ & \ssz & $I=0$ & $\sim 120$ \mevc& $\K^*_2\Kb +  \mathrm{c.c.}$ \\
\hline\hline
\end{tabular}
\renewcommand{\arraystretch}{1.0}
\end{table}

However, tensor forces induce $\ISLJ13D1$ and $\ISLJ33D1$ components in the 
protonium wave function, which contribute to annihilation. 
Maruyama et al. \cite{Maruyama:1988cb}, for instance found that in the framework 
of a specific model with planar diagrams, ``inclusion of D states  solve the $\rho\pi$ puzzle.''

The role of final-state interaction in the \rpp\ was underlined by
Mull et al.\ \cite{Mull:1994vn}. For the $\tssz:\stso$ ratio they found
about 1:5 without final-state interaction and 1:25 when it is
added. Surprisingly, the effect is not due to the $\rho\pi$ interaction
of their model being different in these two states, but simply to the
existence of this interaction. The effect is almost negligible 
for P-wave annihilating into $\rho\pi$. 

Anyhow, the interaction between final state mesons is not well known, 
and thus one must rely on models. In the case of a well pronounced 
resonance, one could expect all channels, e.g., $\pi\pi$, $\K\Kb$, 
etc, to feel attractive forces. In other circumstances, channels 
that are weakly coupled might behave differently.
We conclude that final state interaction cannot be neglected, but is 
not demonstrated to explain all observed \dsrs.

Alternatively, we may try to understand the \dsrs\ as the effect of an
hadronic transition operator. Then we find that the transition to two
isovector mesons is small from the isovector component of protonium
wave functions (at least for \tso\ and \ssz ) and large for the
isosinglet component. The remaining \dsrs\ are condensed into the
observation that in $\pbp$ annihilation, kaons (and $\K^*$,
..) are produced in the form $\Kp\Km$ and in $\K^0\Kb{}^0$.
Whenever we probe the annihilation potential with two isovector mesons
in the final state, we find a large isospin $I=0$ contribution. When
we probe the annihilation potential with final states with open
strangeness, we find the combination $\pbarp\to \Kp\Km - \K^0
\Kb{}^0$ to be large compared to $\pbarp\to \K^+\K^- + \K^0\Kb{}^0$.

We conclude that the dynamical selection rules can economically be
interpreted as suppression of specific hadronic transition
operators. The hadronic operator acts in a similar way for two
pseudoscalar mesons, for a pseudoscalar and a vector and for a
pseudoscalar and a tensor meson. The
\dsrs\ drive annihilation preferentially 
into specific favoured flavour combinations;
other quantum numbers like spin and angular momentum
of the final-state mesons are seemingly less relevant.

It is not easy to find a microscopic derivation of such a selection rule. 
An attempt was made by Niskanen and Myhrer \cite{Niskanen:1985vi}, who analysed 
quark--antiquark annihilation (necessary for $\NNb\to\rho\pi$) into one or 
two gluons and found, interestingly,  that it is suppressed if the 
$\NNb$ initial state is $\tssz$. A non-perturbative generalisation would be desirable.

\clearpage\markboth{\sl Annihilation dynamics} {\sl Phenomenology}
\setcounter{equation}{0}
\section{Phenomenological analysis}\label{se:phe}
There is an abundant literature on $\NNb$ annihilation. This complex
process has been studied in a variety of models. Most detailed
numerical analyses are now obsolete, to the extent that the
parameters of the models have been tuned to reproduce early and
incomplete sets of data. However, the underlying mechanisms still
deserve to be presented and compared to other possible mechanisms.

In this section, we will expand on the following questions: the pion
multiplicity and the clustering of mesons into resonances; the range
of annihilation; the role of initial and final state interaction; the
importance of symmetries; the probability of producing strange mesons
and the mechanisms of OZI violation; the interpretation of dynamical
selection rules. Ideally, answers to these questions should come from
a full understanding of hadronic interactions which we do not yet have.
\FloatBarrier\subsection{Initial state interaction}\label{phe:sub:ini}
\subsubsection{Overall suppression}
Before considering annihilation mechanisms, it is important to
underline the role of initial-state interactions. If a
specific process $\NNb\to m_1 m_2\ldots$ is calculated in a given model without
accounting for initial-state interactions, its rate is overestimated
by orders of magnitude. In more technical words, a Born-approximation
treatment is unacceptable, while a distorted-wave Born
approximation  can be rather realistic. In any realistic model,
the \NNb\ wave function is dramatically suppressed at short distances
by the cumulative effect of annihilation channels.
\subsubsection{Induced channels}
A serious warning by Green et al.\ (see, e.g., \cite{Green:1988jb}),
is that if initial-state interactions are taken seriously with our
current ideas on nuclear forces, there are ample transition
amplitudes $\NNb\leftrightarrow\N^*\Nb + \mathrm{c.c.}$, or
$\NNb\leftrightarrow\N^*\Nb{}^*$, with at least one baryon or antibaryon
being a spin ($\N^*=\Delta$), orbital or radial excitation of the
nucleon. Similarly strange-meson production can proceed via a
$\overline{\Lambda}\Lambda$ doorway or other hyperon--antihyperon
intermediate states. Hence the quark content of the initial state does
not reduce to $(uud\bar{u}\bar{u}\bar{d})$. This might influence the
conclusions drawn about the hierarchy of various quark diagrams describing
annihilation.

\subsubsection{Selective suppression}
Even for ratios of branching ratios, a pure Born--Oppenheimer treatment of
annihilation might be misleading, as different initial states do not
necessarily experience the same suppression, and in a given partial wave, the
damping of the wave function depends on the momentum range which is
explored (see, e.g., Ref.~\cite{Dover:1991mu}, and discussions later
in this section).

The \NNb\ interaction is investigated in scattering and protonium
experiments, and described in a number of models that combine
long-range meson exchanges and short-range absorption. This subject is
discussed in a previous review article
\cite{Klempt:2002ap}.

As already stressed in Sec.~\ref{kin:sub:imp}, the long-range
potential has a strong spin and isospin dependence, starting with the
one-pion-exchange term which includes a
$\vi{\tau}{1}.\vi{\tau}{2}\,\vi{\sigma}{1}.\vi{\sigma}{2}$ operator
for the central interaction, and something analogous for the tensor
one. In potential models, a dramatic spin--isospin dependence is
induced in the various partial waves contributing to annihilation at
low energy.
\subsubsection{Orbital mixing}
Meson-exchange models also predict some``orbital'' mixing in
natural-parity states of protonium, due to the tensor component of the
\NNb\ interaction.  For instance, the authors of Ref.~\cite{Kercek:1999sc}
 combine the so-called ``A2'' model (to be defined later) with initial-state
protonium wave functions estimated from potential models, and computed
the relative rates for annihilation into two mesons.  Some ratios are
found to be extremely sensitive to details of the $\NNb$ interaction
used to produce the protonium wave-function. In particular,
neglecting the \SLJ3D1\ admixture into the \SLJ3S1\ wave function
sometimes changes the results by an order of magnitude. It was already
noted by Green et al.\ \cite{Green:1985yy,Green:1988jb} that the
\SLJ3P0\ model requires an orbital momentum between the annihilating
quark and antiquark. For an overall S-wave, this is provided by
departures from the harmonic-oscillator behaviour of the wave
function\footnote{For instance, a $J=0$ state of three bosons has pure
$\ell=0$ orbital momentum between any two constituents in the harmonic-oscillator model,
 but this is not true for other type for binding interaction.}.  For a spin $S=1$ system,
an overall $J=1$ state comes either from an S-wave, or, in
presence of tensor forces, from a D-wave.
\subsubsection{Isospin mixing of protonium}
In potential models, the effect of initial-state interaction is
predicted to be particularly important in protonium wave functions, as 
compared to scattering wave functions relevant for annihilation in flight.
This was underlined by Kaufmann and Pilkuhn \cite{Kaufmann:1978vp},
and several other authors. 

In short, a pure $\ppb$ state corresponds to equal weights for 
isospin $I=0$ and $I=1$. However, the charge-exchange potential 
induces transitions from
$\pbp$ to $\nbar\n$, and when the $\pbp$ and   $\nbar\n$ amplitudes 
are recombined into amplitudes of given isospin $I$, one component 
is often much larger than the other one. For instance, $\tpz$ is 
clearly dominated  by its  $I=0$ part. However, this firm
prediction of meson-exchange models is not confirmed by studying the
systematics of branching ratios.  See, for instance, the discussion in
Refs.~\cite{Klempt:1990up,Dover:1991mu,Amsler:1993kg}.  It is 
somewhat of a paradox that the pion-exchange force nicely reproduces the
pattern of fine splitting of P-levels of protonium
\cite{Klempt:2002ap} as well as the hierarchy of hadronic widths (see
Table~\ref{tm:tab:siwidthpp}), but resists experimental checks for
branching ratios.

The explicit calculation of Ref.~\cite{Kercek:1999sc} illustrates the
concerns about isospin mixing in protonium. For instance, the
$\eta\rho^0$ ($I=1$) to $\pi^0\rho^0$ ($I=0$) is calculated to be much
smaller than the experimental value.

In Ref.~\cite{Gutsche:1998fc}, Gutsche et al.\ also discuss this
question of isospin mixing. They estimate the rate for radiative
annihilation $\ppb\to\gamma+X$, with $X$ being $\pi^0$, $\eta$,
$\eta'$, $\rho$ or $\omega$. The transition is sensitive to
interference between the $I=0$ and $I=1$ components of the protonium
wave function, and thus probes the isospin mixing predicted by
potential models.  A rather good agreement is found with the
data. However, the rate for $\gamma\phi$ cannot be reproduced.

\subsubsection{Checking isospin mixing in protonium}
The first attempt to deduce the isospin ratios from experiment was
made in \cite{Klempt:1990up}.  The model assumed that the transition
matrix element for annihilation into two mesons is entirely determined
by the isospins involved.  There are transitions from the $I=0$
initial state to two isoscalars and to two isovectors, and from the
$I=1$ initial state to one isoscalar and an isovector.  The transition
matrix elements are then supposed to be independent of the quantum
numbers of the initial state, apart from a normalisation which could
be different for different initial states.  The results are certainly
model-dependent, and the errors are large.  The isospin-mixing
coefficients are found compatible with the predictions of potential
models, but, due to the large errors, also with the absence of mixing
effects.  The subject was further discussed by Dover et
al.~\cite{Dover:1991mu} and more recently by the Crystal Barrel
collaboration \cite{Abele:2000xt}.

\begin{table}
\caption{\label{phe:tab:is}%
Ratio of isovector to isoscalar fraction of the protonium wave
function for various initial states \protect{$^{2s+1}$L$_J$}.  The
three theoretical values correspond to different \NNb\ potentials, as
compiled in Ref.~\protect\cite{Carbonell:1992wd}.  }
\begin{center}
\renewcommand{\arraystretch}{1.1}
\begin{tabular}{cccccc}
\hline\hline
{Initial}& \multicolumn{3}{c}{Potentials}
&
\multicolumn{2}{c}{Data analysis}\\
{state}& KW & DR1 & DR2 &
{Ref.\cite{Klempt:1990up}}&{Ref.\cite{Abele:2000xt}}\\
\hline\hline
\ssz\                   &  0.68 & 0.68 & 0.8          &
$0.72\,{}^{+0.24}_{-0.18}$  & $0.50\,{}^{+0.48}_{-0.29}$      \\
\tso\                   &  1.22 & 0.95 & 1.26         &
$1.17\,{}^{+0.39}_{-0.28}$  & $1.17\,{}^{+0.30}_{-0.23}$      \\
\tpz\                   &  0.03 & 0.03 & 0.05         &
$1.16\pm 0.34$          & $0.41\,{}^{+0.11}_{-0.09}$      \\
\tpo\                   &  9.4  & 9.7 & 6.5           &
$ 9\pm 5$               &                             \\
\spo\                   &  0.96 & 0.82 & 0.61         &
$0.81\pm 0.51$          &                             \\
\hline\hline
\end{tabular}
\end{center}
\end{table}
\subsubsectionb{Isospin content of $\pbp$ in $\pbd$}
The analysis can be extended to annihilation on deuterium. To our
knowledge, there are no published calculation of the detailed isospin
content of $\pbp$ within $\pbd$. Note that the isospin content
of the neutral $\pbp$ state in $\pbd$ is not necessarily the same as
in protonium, due to the presence of the third hadron.

The transition $\pbard$ into a isovector meson $m_1$ and a isoscalar
meson $m_2$ proceeds via the (squared) isovector component of the
$\rm\pbar N$ subsystem. This component is 1 for $\rm\pbar n$ and
smaller for $\rm\pbar p$. Thus
\begin{equation}\label{phe:eq:pietaomega}
\eqalign{%
\frac{\AF(\pbd\to\pi^0\omega + \n)}{\AF(\pbd\to\pi^-\omega + \p)}
&=   {8.4\pm 0.4\over 12.1\pm 1.4} = 0.69\pm 0.09 \cr
\frac{\AF(\pbd\to\pi^0\eta + \n)}{\AF(\pbd\to\pi^-\eta + \p)}
&=   \phantom{0} {4.9\pm 0.3 \over 8.1\pm 2.0} = 0.61\pm 0.15 }
\end{equation}
The frequencies were obtained by calculating mean values from the
numbers given in Table~\ref{tm:tab:BRexpd}.  Those normalised to the
number of \pbd\ annihilations have been multiplied by 2. The mean
values do not include the results from \cite{Chiba:2000ev}.  The
$\pim\omega$ data from \cite{Bizzarri:1970ta} and
\cite{Bettini:1967bb} are superseded by a reanalysis
\cite{Gaspero:1994tw}.

The reaction \pbp$\to\pi\omega$ is dominated by the \tso\ initial
state; the ratios (\ref{phe:eq:pietaomega}) suggest a mild deviation
from a pure \pbp\ system at annihilation for which ratios $0.5$ would
be expected.  According to Table~\ref{phe:tab:Bpartial}, annihilation
into $\pi\eta$ goes through the \tpz\ isovector component of the \pbp\
system which is calculated to be very small, see
Table~\ref{phe:tab:is}.
\FloatBarrier\subsection{Final state interaction}
When discussing annihilation, one can hardly forget the strong
interaction of the produced mesons, which are likely to rescatter,
form resonances, decay into pions, etc. Even lighter mesons, which
escape faster from the interaction region, might be affected.

Final-state interactions were introduced, e.g., by the Bonn group
\cite{Mull:1994vn} and found to play an important role for the spin
effects associated with $\NNb\to\pi\pi$ or $\mathrm{\overline{K}K}$, and for 
the $\rho\pi$ puzzle. We shall see later in this section that it
is suggested that rescattering could explain the observed deviations from the
OZI rule.

Clearly, if a channel is not or only weakly populated by the main
mechanism of annihilation, it might receive a non-negligible
contribution of rescattering from another final state. In a fictitious
world where $\ppb$ does not couple to $\K\Kb$, the $\pi\pi\to\K\Kb$
reaction would be crucial.

On the other hand, rescattering among channels which are copiously
produced, presumably has little effect on the annihilation frequencies. The loss
towards other channels is compensated for by the feed-back.  More
generally, in a regime of strong production, saturation occurs. If one
adds another mechanism, the production rate grows less than would naively be
expected by adding the squared amplitudes estimated independently.

Consider for instance the following annihilation frequencies for $\omega$ and
$\rho$ production:
\begin{equation}\label{phe:eq:rho-omega-eta}
\frac{\AF(\pbp\to\piz\omega)}{\AF(\pbp \to\rho^0\pi^0)} 
=  0.41 \pm 0.06~,\qquad
\frac{\AF(\pbp\to\rho\eta)}{\AF(\pbp \to\eta\omega)}
= 0.23 \pm 0.06~. 
\end{equation}
There is no obvious dominance of $\rho$ production over $\omega$
production, though the $\rho$ meson is more likely than $\omega$
to be formed in the rescattering of primary mesons.

In short, the final-state interaction is not expected to dominate the
systematics of branching ratios, but should be kept in mind for
channels which are suppressed.
\FloatBarrier\subsection{Pion multiplicity and two-meson doorway scenario}\label{phe:sub:doorw}
An average of five pions are produced in nucleon--antinucleon at rest. 
In a baryon-exchange picture, or in microscopic quark models, it is natural 
to assume that a few meson resonances are primarily produced, 
the observed final states resulting from the decay of these resonances. 
A large, but quantitatively still unknown, fraction of all
annihilation modes proceeds even via two-meson intermediate
states, sometimes called \emph{quasi two-body} annihilation. Insisting on two-body
annihilation as the dominant contribution however does not
account for the large fraction of events leading to high
pion multiplicities. 
\subsubsection{The Vandermeulen model}\label{phe:sub:vm}
Vandermeulen \cite{Vandermeulen:1988hh}
suggested that these high pion multiplicities could come
from an enhanced production of mesons with high mass. He 
observed that meson resonances are preferentially produced 
with nearly the maximal mass which is allowed by phase space. 
The effect can be parametrised by assuming that
annihilation proceeds via two primary mesons and writing the
frequency $\ppb\to m_1+ m_2$ for producing two mesons with
masses $m_1$ and $m_2$ as proportional to
\begin{equation}\label{phe:eq:vandm}
F(q)=q\,\exp\left[-\Rso\sqrt{s-(m_1+m_2)^2}\right]~,
\end{equation}
where $q$ is the momentum of the mesons, as given by
Eq.~(\ref{kin:eq:tbe1}); $s$ is the Mandelstam variable which is $s=4m_\p^2$
for annihilation at rest. 
With a reasonable adjustment at  $\Rso=1.2\;\mathrm{GeV}^{-1}$,
Vandermeulen was able to account for many key features of annihilation
in flight, covering a wide range of antiproton momenta. 
The Crystal Barrel collaboration applied (\ref{phe:eq:vandm}) to annihilation
at rest and derived \cite{Amsler:1993kg}  
$\Rso=0.83$\,GeV$^{-1}$.
The model of Vandermeulen was further developed, e.g., by 
Mundigl et al.\ \cite{Mundigl:1991jp}.

To get more insight into Eq.~(\ref{phe:eq:vandm}), 
we define the annihilation amplitude by $F(q)=q\,|f(q)|^2$. For 
the case of two mesons having the same mass, $f(q)$, thanks to  Eq.~(\ref{kin:eq:tbe2}), reduces to an exponential function, 
\begin{equation}\label{phe:eq:vandm2}
f(q)=\exp\left[-\frac{\Rso}{2} \sqrt{s-(2m_1)^2}\right]=\exp[-q\Rso]~.
\end{equation}
This momentum distribution corresponds to the spatial  distribution 
\begin{equation}\label{phe:eq:vandm3}
S(r) = \frac{1}{\pi}\,\frac{\Rso}{r^2 + \Rso^2}~,
\end{equation}
of a source having a size of $\Rso$, interpreted as the size of the annihilation source. 

Equation~(\ref{phe:eq:vandm}) does not take the centrifugal barrier into
account. The formula can be extended to include a suppression 
of high angular momenta at small momentum $q$
\begin{equation}\label{phe:eq:vdmbw}
F_\ell(q) = qf^2_\ell(q) = q B^2_\ell(q)\,\exp\left[-2q \Rso\right] ~,
\end{equation}
where $f^2_\ell(q)$ is the transition amplitude and the functions $B_\ell$ are given by 
\begin{equation}\label{phe:eq:hippl}
B_0(q) = 1~, \quad
B_1(q) = \sqrt{\frac{2z}{z+1}}~,\quad
B_2(q) = \sqrt{\frac{13z^2}{(z-3)^2+9z^2}}~,\quad  z=(q\Rsi)^2~.
\end{equation}
The decay momenta are measured in units of $1/\Rsi$, where $\Rsi$
corresponds to a strong interaction radius.  In the zero-range
approximation, $\Rsi\to 0$, and the $B_\ell(q)$ become proportional to
$q^{2\ell}$.

Equation~(\ref{phe:eq:vdmbw}) can be read as the product of the two-body
phase-space factor $q$ by the probability of getting the required orbital angular
momentum and linear momentum of the outgoing mesons.

Note that the enhancement of high-mass mesons is qualitatively
equivalent to the pion suppression that was found necessary by Green
and collaborators \cite{Green:1985yy,Green:1988jb}. Schematically,
annihilation produces $\qqb$ pairs in some quantum numbers. The
overlap of such $\qqb$ pair with an actual meson is reduced in the
case of the pion, due to smaller pion size, and its intricate internal
structure.

\subsubsection{An illustration: the pseudoscalar mixing angle from 
\protect$\rm\bf p\bar p$ annihilation.}
Already in 1983, Genz \cite{Genz:1983aj} suggested 
that the ``quark line rule'' could be applied to $\pbarp$  
annihilation into two pseudoscalar mesons and the branching ratios 
could be used to determine the pseudoscalar 
mixing angle. Genz was not able to obtain the right result: the pre-LEAR data
were of low statistical significance and had no redundancy, 
so that dynamical questions  like the influence of phase
space and orbital angular momentum barrier could not be investigated. 
The situation improved once  the Crystal Barrel 
Collaboration determined a large number of branching ratios involving
$\eta$ and $\eta^{\prime}$ mesons \cite{Amsler:1993kg}. 
The Obelix collaboration extended the study to low-energy 
$\nbp$ scattering \cite{Filippi:1999xk}. 

Table \ref{phe:tab:psexp} summarises the results. It gives the initial 
state of protonium which contributes most significantly to the final state,  
the particle against which $\eta^{\prime}$ or $\eta$ recoils, the ratio
of annihilation frequencies, 
\begin{equation}
r_X = \frac{\AF(\NNb\to\eta^{\prime}X)}   
{2^{\delta(X,\eta)}\,\AF(\NNb\to \eta X) }~,
\end{equation}
and the ratio
\begin{equation}
d_X = \frac{\DR(\NNb\to\eta^{\prime}X)}   {2^{\delta(X,\eta)}\,\DR(\NNb\to \eta X) }~,
\end{equation}
of \emph{dynamically corrected} branching ratios, defined as 
\begin{equation}\label{phe:eq:dc}
\DR=\AF/F_\ell(q)~,
\end{equation}
where $F_\ell(q_)$ has the simplified form (\ref{phe:eq:vdmbw}), even
for unequal masses.  The factor $2^{\delta(X,\eta)}=2$ for annihilation
into $\eta\eta$ and $2^{\delta(X,\eta)}=1$ elsewhere accounts for the
Bose symmetry of two identical bosons in final state.

\begin{table}[h!]
\caption{\label{phe:tab:psexp}%
Ratios $r_X$, and ratios $d_X$ corrected with (\ref{phe:eq:vdmbw}),
of annihilation frequencies into $\eta^{\prime}$ and
$\eta$ mesons recoiling against the same particle $X$. The dominant initial
atomic state is given in the first column, the orbital angular momentum $\ell$ 
between the outgoing mesons in the last column.}
\renewcommand{\arraystretch}{1.2}
\begin{center}
\begin{tabular}{clccc}
\hline
\hline
Atomic state &   $X$     & $r_X$ & $d_X$ & $\ell$ \\
\hline
\tpz        &$\pi^0$  & $0.50\pm 0.10$ & $0.53\pm 0.11$ & 0\\
\tpz        &$\pi^+$  & $0.63\pm 0.16$ & $0.65\pm 0.17$ & 0\\
\tpz         &$\eta$   & $0.62\pm 0.07$ & $0.69\pm 0.07$ & 0\\
\tso         &$\omega$ & $0.52\pm 0.07$ & $0.78\pm 0.11$ & 1 \\
\tso         &$\rho$   & $0.42\pm 0.08$ & $0.60\pm 0.11$ & 1 \\
\hline
\hline
\end{tabular}
\end{center}
\renewcommand{\arraystretch}{1.0}
\end{table}

As seen in Table~\ref{phe:tab:Bpartial}, annihilation into two
pseudoscalar mesons proceeds dominantly via the \tpz\ state, and we
shall restrict ourselves to these data.

We further assume 
that the proton and antiproton have no intrinsic $\ssb$ component
and couple, in the notation of Sec.~\ref{mes:sub:psmeson}, 
only to the  $\nnb$ part of the $\eta$ and
$\eta^{\prime}$ wave functions. This is the so-called Zweig or OZI
rule; its validity in $\pbp$ annihilation (where it is also
called Quark Line Rule and abbreviated as QLR) will be discussed later in this section. 
If the $\eta$ and $\eta^{\prime}$ wave functions are
written as  in Sec.~\ref{se:mes} as
\begin{equation}\eqalign{
\ket{\eta\phantom{^\prime}}&=
\cos(\Theta_{\rm PS}-\Theta_{\rm id})\ket{\nnb}-\sin(\Theta_{\rm PS}-\Theta_{\rm id})\ket{\ssb}~,\cr
\ket{\eta^\prime}&=
\sin(\Theta_{\rm PS}-\Theta_{\rm id})\ket{\nnb}+\cos(\Theta_{\rm PS}-\Theta_{\rm id})\ket{\ssb}~,}
\end{equation}
the ratio $d_X$ is given by $d_X=1/\tan^2(\Theta_{\rm PS}-\Theta_{\rm id})$.

Although they do not incorporate any phase-space factor, the
uncorrected ratios $r_X$ are already reasonably consistent. This
indicates that the dynamical corrections are small, i.e., that the
phase-space effects are cancelled out by the Vandermeulen factor. This
requires the size of the annihilation source to be small, compatible
with $\Rso=1/(2m_{\rm p})$, and the interaction radius to be large,
$\Rsi=1.5$\,fm. The systematic errors are are estimated to be about
10\% .

The mean value of the corrected ratios $d_X$ is $0.65\pm 0.07$ with a
$\chi^2 = 3.3$ for 4 degrees of freedom, leading to
\begin{equation}\label{phe:eq:theta}
\Theta_{\rm PS}= -(15.9\pm 1.5)^{\circ}, 
\end{equation}
which is not inconsistent with
other measurements of this quantity. The choice of a larger $\Rso$ or
smaller $\Rsi$ leads to larger mixing angles.  The
importance of the result lies less in the final number but rather in
giving insight into the dynamics of the annihilation process.
\FloatBarrier\subsection{Dynamically corrected branching ratios}\label{phe:sub:dcbr}
The dynamically corrected branching ratios for different two-body
final states, as defined in Eq.~(\ref{phe:eq:dc}) are listed in
Table~\ref{phe:tab:Bpartial}. The corrections are empirical and take
already into account parameters like interaction strengths, wave
function overlap and finite size effects. Hence these ratios should not
be compared to results from a full model calculating annihilation
dynamics. Rather, the rates represent the squares of elementary
transition matrix elements.  The parameters of the Vandermeulen model
were fixed using branching ratios with no open strangeness. A
suppression of these final states due to a penalty for $\ssb$ pair
creation as advocated, e.g., in~\cite{Dover:1992vj} is not yet
accounted for. This aspect will be discussed in
Sec.~\ref{phe:sub:strange}.

The decays from $\tpz$ states into two pseudoscalar mesons are
normalised to $\pi^0\pi^0$, these from the $\tso$ states into vector
plus pseudoscalar to $\rho^0\pi^0$, and from the $\ssz$ states into
tensor plus pseudoscalar to 100\% $\mathrm{a}_2^0\pi^0$.  The
branching ratios \BR\ are taken from Table~\ref{tm:tab:Bpartial} and
Eqs.~(\ref{tm:eq:AN1}).

Table~\ref{phe:tab:Bpartial} provides insight into the hierarchy of
annihilation modes:
\begin{itemize}\itemsep -2pt
\item
The dynamical selection rules, discussed in Sec.~\ref{se:dsr},
are confirmed after the dynamical correction is applied.
The $\rho\pi$ puzzle manifests itself in the smallness
of the branching ratio from isospin $I=1$ initial states
into two isovector mesons ($3^{\rm rd}$ column from bottom) 
compared to those from $I=1$ initial states ($1^{\rm st}$ column). 
The branching ratios called 
$t(I_{\pbp}\to I_1,I_2)=t(1\to 1,1)$ here 
are smaller by about one order of magnitude compared to 
$t(0\to 1,1)$. $I_{1,2}$ are the isospins of the two mesons.

\item
The sum of branching ratios for kaonic decay modes $t(1/2,1/2)$ is of
the same order of magnitude as $t(1\to 1,1)$. In annihilation from
\ppb\ S-states, one isospin component dominates; the \tpz\ initial
states prefers to decay into $\K^+\K^-$.
\begin{table}[!h]
\caption{\label{phe:tab:Bpartial}%
Dynamically corrected branching ratios of selected two-body modes.
For easier comparison, the absolute values
$\DR(\tpz\to\piz\piz )=(133\pm 26)\,10^{-3}$; 
$\DR(\tso\to\piz\rho^0)=(32.2\pm 1.7)\, 10^{-3}$; and 
$\DR(\ssz\to\atmass\piz )=(71\pm 10)\, 10^{-3}$ 
are normalised to $50$ (to account for Bose symmetry), 
$100$ and $100$, respectively. }
\begin{small}
\renewcommand{\arraystretch}{1.1}
\begin{tabular}{c|ccc|ccc|ccc}  \hline\hline
Isospin&\pbp&\hspace*{-2mm}channel\hspace*{-2mm}&DR 
       &\pbp&\hspace*{-2mm}channel\hspace*{-2mm}&DR 
       &\pbp&\hspace*{-2mm}channel\hspace*{-2mm}&DR \\  
\hline
$0\to1+1$&\hspace*{-4mm}\tpz\hspace*{-4mm}&$\pi^0\pi^0$  &\bf 50 &
          \hspace*{-4mm}\tso\hspace*{-4mm}&$\rho^0\pi^0$ &\bf 100&
          \hspace*{-4mm}\ssz\hspace*{-4mm}&$\at^0\pi^0$  &\bf 100 \\
$0\to0+0$&&$\eta\eta$  &\ $12.7\er1.0$
         &&$\omega\eta$ & $115\er6$ 
         &&$\eta \ft$  & $3\er2$ \\
$0\to0+0$&&$\eta\eta^{\prime}$  & $17.5\er1.1$
         &&$\omega\eta^{\prime}$ & $89\er12$
         &&\\
$0\to0+0$&&&
         &&$\phi\eta$ & $0.44\er0.11$
         &&\\
$1\to1+0$&&&
         &&$\pi\omega$& $41\er3$
         &&$\pi \ft$  & $55\er7$ \\    
$1\to1+0$&&$\pi\eta$ & $15.6\er3.2$ 
         &&$\rho\eta$& $26\er3$
         &\\
$1\to1+0$&&$\pi\eta^{\prime}$ & $8.3\er0.5$ 
         &&$\rho\eta^{\prime}$& $15.8\er2.8$
         &&$\pi \ft'$  & $1.38\er0.23$\\  
$1\to1+0$&&&
         &&$\phi\pi$ & $4.3\er0.5$
         &&\\
$1/2+1/2$&&\hspace*{-2mm}$\K^+\K^-$\hspace*{-2mm}& $4.7\er0.4$
&&\hspace*{-2mm}$(\K^*\Kb)_{I=0}\hspace*{-2mm}$ & $0.9\er0.9$
         &&\hspace*{-2mm}$(\K^*_2\Kb)_{I=0}$\hspace*{-2mm}& $9.2\er2.8$ \\
$1/2+1/2$&&\hspace*{-2mm}$\K_{\rm s,l}\K_{\rm s,l}$\hspace*{-2mm}& $<1$
         &&\hspace*{-2mm}$(\K^*\Kb)_{I=1}$\hspace*{-2mm}& $13.7\er2.2$
         &&\hspace*{-2mm}$(\K^*_2\Kb)_{I=1}$\hspace*{-2mm}& $0.8\er0.3$ \\
\hline
$1\to1+1$&\hspace*{-4mm}\tso\hspace*{-4mm}&$\pi^+\pi^-$  & $3.0\er0.2$&
          \hspace*{-4mm}\ssz\hspace*{-4mm}&$\rho^{\pm}\pi^{\mp}$ & $28.0\er4.4$&
          \hspace*{-4mm}\tso\hspace*{-4mm}&$a_2^{\pm}\pi^{\mp}$  & $23.4\er4.6$ \\
$1/2+1/2$&&\hspace*{-2mm}$\K^+\K^-$\hspace*{-2mm}& $1.6\er 0.2$
&&$\hspace*{-2mm}(\K^*\Kb)_{I=0}$\hspace*{-2mm}& $13.6\er 2.2$
&&\hspace*{-2mm}$(\K^*_2\Kb)_{I=0}$\hspace*{-2mm}&$<3$ \\
$1/2+1/2$&&\hspace*{-2mm}$\Ks\Kl\hspace*{-2mm}$  & $1.4\er 0.2$
&&\hspace*{-2mm}$ (\K^*\Kb)_{I=1}$\hspace*{-2mm}&  $2.5\er 1.0$
&&\hspace*{-2mm}$(\K^*_2\Kb)_{I=1}$\hspace*{-2mm}&  $6.5\er 1.8$ \\
\hline\hline
\end{tabular}
\renewcommand{\arraystretch}{1.0}
\end{small}
\end{table}
\item
Annihilation modes $t(0\to 0,0)$ show no systematic behaviour.
For comparison, final states containing an $\eta$ ($\eta^{\prime}$)
meson, should be corrected for their \ssb\ component by multiplying
the DR by $1/(0.65\pm 0.07)$ or  $1/(0.35\pm 0.07)$, respectively.
With this correction the \pbp\ coupling to $\eta\eta$ and 
$\eta\eta^{\prime}$ is about 1/2 of the $\pi\pi$ coupling, while
the coupling to $\omega\eta$ and  $\omega\eta^{\prime}$ is twice
larger than the coupling to $\rho\pi$. The DR for $\eta \ft$ is
very small; of course this transition has a small phase space and
a larger angular momentum barrier (with $\ell=2$). But for the corrections
as suggested here, the DR for $\eta \ft$ remains small.
\item
The annihilation modes $t(1\to 0,1)$ are smaller than 
$t(0\to 1,1)$ by a factor 2 to 4. 
\end{itemize}

\FloatBarrier\subsection{The size of the annihilation source}\label{phe:sub:size}
From \NNb\ scattering data, several ``radii'' were determined
\cite{Klempt:2002ap}. The charge-exchange reaction occurs at typically 
$R_{\rm ce}=2.5$\,fm, the mean strong interaction radius is
$\Rsi=1.5$\,fm and at $R_{\rm an}=1$\,fm, annihilation takes place.
Two further length scales are given by the Compton wave length of the 
pion and proton, $m_\pi^{-1}\sim 1.4\,$fm and 
$m_\p^{-1}\sim 0.2\,$fm, respectively.

\subsubsection{Baryon exchange mechanism}\label{phe:sub:bar-exc}
Baryon exchange was the first mechanism proposed for annihilation, in analogy
with electron exchange in $\mathrm{e}^+\mathrm{e}^-$ annihilation (see
Fig.~\ref{phe:fig:annihee}). This naively  suggests a range of the order of
$1/(2m_\p)\sim 0.1\,$fm, where $m_\p$ is the proton mass. With this sole value 0.1\,fm for any possible range, size or form-factor parameter of a model, 
one would never reproduce the observed ratio of annihilation to elastic \ppb\ cross
sections, nor the smallness of the charge-exchange 
cross-section, nor the occurrence of P-wave annihilation at rest, such as
$\ppb\to\piz\piz$. With large form factors associated with baryon exchange,
annihilation acquires, however, a more realistic spatial extension.
\begin{figure}[!ht]
\includegraphics{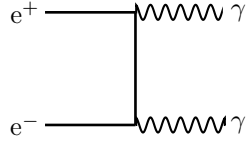}
\caption{Positron--electron annihilation mediated by electron
exchange: the range of the induced absorptive potential is the inverse of twice of the electron mass.
\label{phe:fig:annihee}}
\end{figure}

In principle, baryon exchange  is very appealing, since it
 uses for \NNb\ annihilation  the same baryon--baryon--meson couplings that
enter NN forces.
It is not clear, however, whether the form factors can be extrapolated from the NN
scattering region to the \NNb\ annihilation region. There is also a warning by
Christillin \cite{Christillin:1984up} that the
exchange of a $\Delta$ or other nucleon resonance leads to  a range
and a strength comparable to those of nucleon exchange. Hence there is an ambiguity on
how the series of exchanges should be truncated, before adjusting the
parameters. 
 
However, baryon exchange models have been developed by several authors, in
particular the Bonn--J\"ulich group, see \cite{Mull:1994vn,Mull:1995gz}, and references therein.  
These authors treated annihilation ``adiabatically'', first by 
estimating -- in their approach -- the transition to a few two-meson
channels and mimicking the remaining channels by an empirical
optical potential. They then gradually increased the number of
channels explicitly accounted for. In this framework, one can test
the role of various ingredients such as: the coupling constants in
the meson--nucleon--nucleon vertices, the associated form factors,
the role of final-state interaction, etc. The role of $\Delta$ exchange has also been investigated in this framework \cite{Betz:2002ss}.

\subsubsection{Annihilation range from statistical 
considerations}\label{phe:sub:bar-exc1}

Another estimate of the range for 
annihilation was proposed by Fermi \cite{Fermi:1950jd}. It
is reproduced here in the notation of Amado et al.\
\cite{Amado:1996as}. The phase-space for annihilation into $n$
particles with momenta $\vi{p}{i}$ and energy $E_i$, see
Eq.~(\ref{kin:eq:ps1}), is given by integrating
\begin{equation}
\label{phe:eq:fermi1}
\varrho_n(s,\{\vi{p}{i}\})=\delta^4(\tilde P-\tilde{p}_1-
\cdots\tilde{p}_n) \prod_{i=1}^n 
{\mathrm{d}^3 \vi{p}{i}\over 2E_i}~.
\end{equation}
The integrals with different $n$ having different dimensions, one
should introduce a scale factor $L$ such that the rate for producing
$n$ pions reads
\begin{equation}
\label{theo:eq:fermi2}
R(n)={L^{2n}\over n!}\int \varrho_n(s,\{\vi{p}{i}\})~.
\end{equation}
Adjusting the average multiplicity $\langle n\rangle=\sum n R(n)/\sum
R(n)$ to be 5 gives $L\simeq 1.2\;$fm. This simple reasoning also
reproduced the observed variance $\Delta n\simeq 1$ of the
multiplicity distribution. A more careful calculation accounting for
the $2\pi$ factors of Eq.~(\ref{kin:eq:ps1}) would even increase the
range $L$ \cite{Amado:1996as}. Obviously, the scale factor $L$ is 
related to the size of the fireball from which pions are emitted
and thus to $\Rsi$.

\subsubsection{Quark rearrangement}\label{phe:sub:rearr}
%
%
The quark model explains why a baryon and its antiparticle can
energetically annihilate by simple quark rearrangement, i.e., why the
reaction $({\bar q}{\bar q}{\bar q})+ (qqq)\to
(\bar{q}q)+(\bar{q}q)+(\bar{q}q)$ can occur at rest. A reasonable
phenomenology of both mesons and baryons is, indeed, achieved
\cite{Stanley:1980fe} if it is assumed that the interquark potential
obeys the so-called ``1/2 rule''
\begin{equation}\label{phe:eq:mb1}
V_{\mathrm{qqq}}={1\over 2}\sum_{i<j} V_{\qqb}(r_{ij})~.
\end{equation}
This is obtained by exchanging colour-octets.  A linear confinement
$V=\lambda r$ for mesons is better generalised as a $Y$-shape
interaction for baryons \cite{Dosch:1976gf}
\begin{equation}
\label{phe:eq:mb2 }
V_Y=\lambda \min_J(d_1+d_2+d_3)~,
\end{equation}
where $d_i$ is the distance from the quark $i$ to a junction $J$, whose
location at minimum corresponds to the well-known Fermat--Torricelli
point of elementary triangle geometry.
%
This genuine 3-body interaction is however close to the result for the 1/2 rule, but slightly larger, since \cite{Dosch:1976gf}
\begin{equation}
\label{phe:eq:mb3 }
{1\over2}(r_{12}+r_{23}+r_{31})\le \min_J(d_1+d_2+d_3) \le {1\over \sqrt3} (r_{12}+r_{23}+r_{31})~.
\end{equation}
The variational principle implies that if $V_\mathrm{qqq}\ge \sum
V_{\qqb}(r_{ij})/2$, then $2M(\mathrm{qqq})\ge 3 M(\qqb)$
\cite{Nussinov:1999sx}. Note that quarks and
antiquarks are assumed here to have equal masses. If $m(Q)\gg m(q)$,
then the inequality can be inverted into
$M(\mathrm{qqq})+M(\overline{\mathrm{Q}}
\overline{\mathrm{Q}}\overline{\mathrm{Q}})<
3M(\overline{\mathrm{Q}}\mathrm{q})$ 
\cite{Richard:1992uk}. Very heavy
antibaryons with three units of heavy flavour would not ``annihilate''
on ordinary matter.

In the quark model, the finite size of annihilation is understood from
the composite structure of baryons and mesons. Mesons are produced
according to their ability to make a ``bridge'', i.e., pick up a quark
in N and an antiquark in \Nb\
\cite{Green:1983fd,Ihle:1988mp,Ihle:1987qi}. The baryon size governs
the spatial spread of the final mesons.

This can seen as follows. A typical transition potential is 
\begin{equation}
\label{phe:eq:q1}
\langle\Psi_\mathrm{f}\vert{\cal O}\vert\Psi_\mathrm{f}\rangle~,
\end{equation}
where the operator ${\cal O}$ correspond to various terms of the
interaction Hamiltonian.  One gets a good idea of rearrangement by
estimating the simple overlap integral corresponding to ${\cal
O}=1$. Other matrix elements have similar shape. For describing the
initial state, the individual coordinates $\vec{r}_i$, corresponding
to the labelling of Fig.~\ref{phe:fig:rearr}, can be rearranged into
\begin{figure}[!htbc]
\includegraphics[width=.4\textwidth]{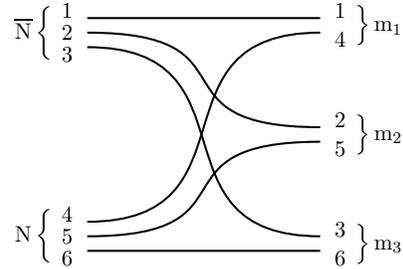}
\caption{Notation for the rearrangement diagram.\label{phe:fig:rearr}}
\end{figure}
\begin{equation}\label{phe:eq:q2}
\eqalignt{%
\vec{r}={\vec{r}_4+\vec{r}_5+\vec{r}_6-\vec{r}_1-\vec{r}_2-\vec{r}_3\over\sqrt6}~,\quad
&\vec{\rho}_{\Nb}={\vec{r}_2-\vec{r}_1\over\sqrt2}~,\qquad
&\vec{\lambda}_{\Nb}={2\vec{r}_3-\vec{r}_1-\vec{r}_1\over\sqrt6}~,\cr
\vec{R}={\vec{r}_1+\vec{r}_2+\vec{r}_3+\vec{r}_4+\vec{r}_5+\vec{r}_6\over\sqrt6}~,\quad
&\vec{\rho}_{\N}={\vec{r}_5-\vec{r}_4\over\sqrt2}~,\qquad
&\vec{\lambda}_{\N}={2\vec{r}_6-\vec{r}_4-\vec{r}_5\over\sqrt6}~,\cr
}
\end{equation}
so that in a harmonic-oscillator model for baryons, the initial wave function
reads, in the centre-of-mass 
\begin{equation}
\label{phe:eq:q3}
\Psi_\mathrm{i}=\left({a\over\pi}\right)^3\exp\left[-{a\over2}
(\vis{\rho}{\Nb}+\vis{\lambda}{\Nb}+\vis{\rho}{\N}+\vis{\lambda}{\N})
\right]\,F(\vec{r})~,
\end{equation}
assuming a simple factorisation of the relative wave function $F$ and the
internal quark or antiquark motion.

Similarly, the final state is described by the internal meson coordinates $\vec{x}_i$ and global meson coordinates $\vec{y}_i$, $i=1,\, 2,\, 3$, 
\begin{equation}
\label{phe:eq:4}
\vec{x}_i={\vec{r}_{i+3}-\vec{r}_i\over\sqrt2}~,\qquad \vec{y}_i={\vec{r}_i+\vec{r}_{i+3}\over\sqrt2}
~,
\end{equation}
and out of the latter, on can built $\vec{R}$ and
\begin{equation}
\label{phe:eq:5}
\vec{\rho}={\vec{y}_2-\vec{y}_1\over\sqrt2}~,\quad\vec{\lambda}={2\vec{y}_3-\vec{y}_1-\vec{y}_2\over\sqrt6}~,
\end{equation}
so that, again assuming a harmonic oscillator and factorisation, the final-state
wave function reads
\begin{equation}
\label{phe:ea:6}
\Psi_\mathrm{f}=\left(b\over\pi\right)^{9/4}
\exp\left[-{b\over2}(\vis{x}{1}+\vis{x}{2}+\vis{x}{3})\right]
\, G(\vec{\rho},\vec{\lambda})~.
\end{equation}
If one keeps an intermediate set of variables
made of $\vec{R}$, $\vec{r}$, $\vec{\rho}$, $\vec{\lambda}$ and
\begin{equation}
\label{phe:eq:6a}
\vec{u}={\vec{x}_2-\vec{x}_1\over\sqrt2}={\vec{\rho}_{\N}-\vec{\rho}_{\Nb}
\over \sqrt2}~,\quad
\vec{v}={2\vec{x}_3-\vec{x}_1-\vec{x}_2\over\sqrt6}={\vec{\lambda}_{\N}-\vec{\lambda}_{\Nb}
\over \sqrt2}~,
\end{equation}
one can integrate over these latter variables and get
\begin{equation}
\label{phe:eq:7}
\langle\Psi_\mathrm{f}\vert\Psi_\mathrm{f}\rangle=
{2a^{9/2}b^{15/4}\over\pi^{9/4}(a+b)}\,
F(\vec{r})\,G(\vec{\rho},\vec{\lambda})\,
\exp\left(-{b\vs{r}\over2}\right)\,
\exp\left(-{a\over2} (\vs{\rho}+\vs{\lambda})\right)~.
\end{equation}
As mentioned earlier, the $\vec{r}$ dependence is governed by the oscillator
parameter $b$ of the meson wave functions, while the spatial
distribution of the final mesons, described by the coordinates
$\vec{\rho}$ and $\vec{\lambda}$, is linked to the baryon size $a$.

This expression (\ref{phe:eq:7}) corresponds to a transition
$\NNb\to3$ mesons. If iterated with its conjugate, it gives the
contribution to the absorptive part of the \NNb\ amplitude, possibly
identified as the driving term of the imaginary part of the optical
potential. One gets an exactly separable potential
\begin{equation}
\label{phe:eq:8}
\Im\mathrm{m}[V]\propto\exp(-b \vs{r}/2) \,\exp(-b\vec{r}{\,}'^{2}/2)~,
\end{equation}
acting between an initial $\NNb$ wave function $F(\vec{r})$ and a
final mesonic wave function $F(\vec{r}{\,}')$, in contrast with the local
character of optical potentials used in current phenomenological
pictures.
%
%
Green et al.\ \cite{Green:1985vu,Green:1988jb} and Ihle et al.\
\cite{Ihle:1988mp} went a little further and studied to what extent
quark rearrangement can describe a large fraction of the observed
annihilation cross-section. The predicted order of magnitude is
reasonable, at least in the framework of simple constituent
models. This means that quark rearrangement is hardly negligible.  It
even opens up the possibility of attempting a first study of the systematics of
branching ratios~\cite{Rubinstein:1966}.  

However, to account for two-body modes, kaon production and the detailed features of 
branching ratios, other quark diagrams have to be included, with some, if not all,
incoming quarks and antiquarks annihilating and some quark--antiquark
pairs being created in the final state out of the released energy.
A phenomenology has been developed, to try to extract from the data the 
relative importance of the various types of diagrams. Different authors have 
reached conflicting conclusions, thus illustrating the difficulties of the art 
of annihilation diagrammatics, which will be presented later in this section.
\subsubsection{Annihilation ranges from the Vandermeulen model}
\label{phe:sub:vm-range}
The values $\Rso=1/(2m_p)\sim 0.1$\,fm for the radius of the
annihilation source, and $\Rsi\sim 1.5$\,fm for the average
interaction radius are suggested by data on annihilation into $\eta$
and $\eta^{\prime}$. They deserve some comments:
\begin{itemize}\itemsep -2pt
\item
A small source cannot provide $\ell>0$ orbital angular
momenta. Without the Blatt--Weisskopf correction, the branching ratios
would scale as $q^3$. The momentum for $\omega\eta^{\prime}$
production is only half of the momentum for $\omega\eta$ production. In
addition, the $\eta^{\prime}$ has a smaller fraction of $\nnb$
quarks than the $\eta$. Hence $r_X$ should be expected to be about
1/10 while experiments give about 1/2. The angular momentum barrier cannot
be large. This is why the $\Rsi$ in Eq.~(\ref{phe:eq:hippl}) must be
large. One could simplify into $\Rsi=R_{\rm an}=1\,$fm but adopting a
larger value for $\Rsi$ gives a slightly better description of the data.
\item
A large source cannot provide large linear momenta, as expressed by
the term $\exp(-2q\Rso)$. Only for small $\Rso$, the momentum
dependence is weak enough to describe simultaneously data on $\ppb\to\pi\eta$ and $\ppb\to\omega\eta^{\prime}$.
\item
$\Rso$ is small compared to the radii used to fit cross sections
\cite{Klempt:2002ap}.  The smallness of the annihilation source
$\Rso\sim 0.1$\,fm underlines the importance of the real part of the
interaction: at $\Rsi\sim 1$\,fm strong interactions lead to a strong attraction which focuses the wave function into a region (of size $\Rso\sim0.1$\,fm) where annihilation takes place, a scenario often underlined by Shapiro \cite{Shapiro:1978wi}.
\end{itemize}
These results illustrate the hot debate about the annihilation
range. Shapiro \cite{shapiro:1996iy} and others insisted that the
annihilation range must be in the order of magnitude of the Compton
length of the annihilating baryons, independently of how the
annihilating objects are constructed from their constituents. On the
other hand, the annihilation range needs to be properly
defined. Nucleons are composite particles, as well as mesons.  As soon
as there is sufficient overlap of the wave functions, {\sl
rearrangement} of quarks can occur and mesons are produced.  Thus
annihilation sets in at large distances \cite{Richard:1988ca}.
\FloatBarrier\subsection{\label{phe:sub:diag}Quark diagrams}
Figure~\ref{phe:fig:diagrams} shows a sample of annihilation diagrams where 
the flavour flow is represented by lines. A2, A3, R2, R3 are 
abbreviations to denote quark diagrams with two or three mesons or 
mesons resonances produced, with or without crossing the lines.
\begin{figure}[!ht]
\includegraphics*[width=.6\textwidth]{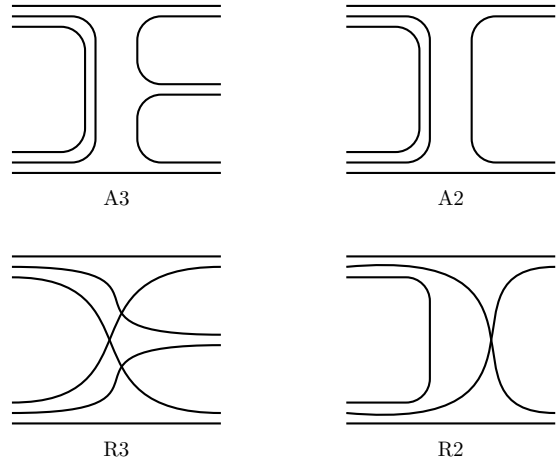}
\caption{\label{phe:fig:diagrams}Annihilation (A2, A3) and rearrangement (R2,R3)
diagrams for $\NNb$ annihilation.}
\end{figure}
These diagrams are not Feynman diagrams, there is no mathematical
prescription as to how to calculate annihilation branching ratios from these
pictures. They have to be supplemented by a model providing initial
and final state wave-functions, and an operator describing pair
creation. So the question arises whether these diagrams can be a
useful guide to annihilation processes.  The main argument in favour
of quark diagrams lies in the \dsrs\ and the observation, at the end
of Sec.~\ref{dsr:sub:disc}, that the \dsrs\ are related to the flavour
flow.

Diagrams with the same topology might be thought to be
equivalent, for instance to those in Fig.~\ref{phe:fig:fsi} 
where either three mesons are created and rescatter into a resonance 
or two mesons are produced and one meson undergoes a
subsequent decay. In actual model calculations, with constituent 
wave functions and empirical creation/annihilation operators, 
the two diagrams are not necessarily equivalent.
\begin{figure}[!ht]
\includegraphics*[width=.8\textwidth]{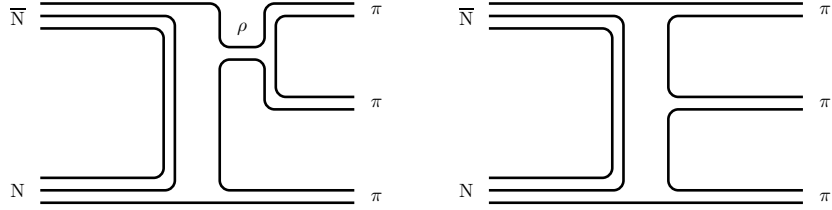}
\caption{Possible mechanisms for $\pi\pi\pi$
production.\label{phe:fig:fsi}}
\end{figure}
\subsubsection{Quark--antiquark creation or annihilation}
Hadron physics requires more than the ``naive quark
model'', an oversimplified approximation to QCD, where the number of
dressed constituents is frozen, as in ordinary quantum mechanics. A
minimal extension has been proposed
\cite{Micu:1969,LeYaouanc:1988fx,Eichten:1978tg,Eichten:1980ms} to
account for the decay of meson and baryon resonances. Usually, a
``\qqbar\ pair-creation operator'' is introduced in an ad-hoc way, its
strength is fitted to reproduce a first resonance width, and then the
model is (rather successfully) applied to predict all other
widths. The calculation of the hadronic widths for, say, $A\to B+C$,
contains an overlap integral involving this creation operator, and the
wave functions of $A$, $B$ and $C$ computed in a specific quark
model. For a review, see the book by Le Yaouanc et
al. \cite{LeYaouanc:1988fx}.

In the physics of decaying resonances, the momenta are rather low, so
it is not too much of a surprise that a single ansatz can account for the
regularities of the observed widths. The current prescription is that
$\qqbar$ is created with vacuum quantum numbers, i.e., in a $\SLJ3P0$
partial wave. This hypothesis is not tested in great detail, since in
a decay $A\to B+C$ , there are not too many possibilities for the
angular momentum between $B$ and $C$ in the final state.
It is thus an audacious enterprise to use the same type of model for describing
annihilation: the momentum of the emitted mesons is much larger; the \qqbar\
creation or annihilation operator is required to work more than once; 
much more freedom is imaginable for the angular momenta, etc.  Quark models 
of annihilation mostly use the $\tpz$ model or a $\tso$ model, where 
the $\qqb$ pair has the quantum number of an isoscalar vector, or a 
combination of both. See, e.g., the discussion by 
Mandrup et al.~\cite{Mandrup:1990be,Mandrup:1991az}, Niskanen and 
Myhrer \cite{Niskanen:1985vi}, Dover and Fishbane \cite{Dover:1986jh}, and 
Maruyama et al. \cite{Maruyama:1988cb}.
\subsubsection{Planar and non-planar diagrams}
A large fraction of $\ppb$ annihilation events, called two-body or
quasi-two-body annihilation, produce two primary mesons. These mesons
may be unstable and decay with a short lifetime. The reaction
$\ppb\to\rho\pi$ with $\rho$ decaying into $\pi\pi$ is a typical
example. There has been much discussion on whether $\rho\pi$
production is better described by the diagram A2 in
Fig.~\ref{phe:fig:diagrams}, without crossing of quark lines, or by
R2, which avoid annihilation and recreation of one quark--antiquark
pair but requires crossing of quark lines.

Besides this specific example, attempts have been made to single out a
few dominant diagrams, leading to definite predictions for the
hierarchy of branching ratios \cite{Genz:1983aj}. Guidance was
sought from two main sources.
\begin{itemize}\itemsep -2pt
\item
The OZI rule, as discussed in Sec.~\ref{mes:sub:Zweig}
 suggests that disconnected
 diagrams are suppressed.  A further step is to assume that only
 diagrams with planar topology (diagrams that can be drawn on a plane
 without intersecting lines) are dominant. 
\item
The suppression of disconnected diagrams is unanimously acknowledged
to become rigorous in the large $N_\mathrm{c}$ limit of QCD, where
$N_\mathrm{c}$ is the number of colours.
However, the question of the suppression of connected but non 
planar diagrams (such as rearrangement)  has been reanalysed by 
Pirner \cite{Pirner:1988mn}. His conclusions do not support 
the claim that planar diagrams dominate.
\end{itemize}

Models assuming dominance of annihilation diagrams over rearrangement
diagrams predict the transition amplitudes $T$ for annihilation into
two isovector and into two isoscalar mesons to be the same.  For
instance, in these models, the rate for $\ppb\to\omega\eta$ is similar
to $\ppb\to\rho\pi$, once some corrections have been applied. One
correction is required for the $\ssb$ component in the $\eta$ wave
function. Another correction is the Vandermeulen factor $F_\ell(q)$
(see Eq.~\ref{phe:eq:vdmbw}).  The dynamically corrected ratios $\DR$
are given in Table~\ref{phe:tab:isiv}, with $\Theta=\Theta_{\rm PS}-
\Theta_{\rm id}+\pi/2$ and $\Theta_{\rm PS}=-15.9^{\circ}$ 
(see Eq.~\ref{phe:eq:theta}).

The ratios in Table \ref{phe:tab:isiv} are largely
incompatible with 1, whilst some have the right order of magnitude.  
With this warning, Table \ref{phe:tab:isiv} could support the 
hypothesis that planar diagrams might in some cases
drive the leading contribution to two-meson annihilation.

\begin{table}[!ht]
\caption{\label{phe:tab:isiv}%
Ratios of dynamically corrected annihilation frequencies for $\ppb$ annihilation into two isoscalar and two isovector mesons.}
\begin{small}
\renewcommand{\arraystretch}{2.0}
\begin{tabular}{rclrcl}
\hline\hline
$\fracd{1}{\cos^2\Theta}\fracd{\DR(\pbp\to\eta\omega)}{\DR(\pbp
\to\rho^0\pi^0)}$ &=&  $1.77 \pm\ 0.10$&
$\fracd{\DR(\pbp \to\rho^0\rho^0)}{\DR(\pbp \to\omega\omega)}$ 
&=& $0.04 \pm\ 0.04$\\
$\cos^4\Theta$ $\fracd{\DR(\pbp \to\piz\piz)}{\DR(\pbp
                   \to\eta\eta)}$ &=& $1.67 \pm\ 0.14 $ &
$\fracd{1}{\cos^2\Theta} 
\fracd{\DR(\pbp \to\eta  \ftmass)}{\DR(\pbp \to\pi^0 \atmass)}$
&$<$& $0.05\quad\ (1\sigma)$\\
\vspace*{-6mm}
&&&&&\\
\hline\hline
\end{tabular} 
\renewcommand{\arraystretch}{1.0}
\end{small}
\end{table}

\subsubsection{Strangeness production} \label{phe:sub:strange}
In models where A2 is small compared to R2, a small fraction of events
contain a pair of kaons. On the other hand, if A2 is the leading mechanism 
of annihilation, and if SU(3) symmetry is approximately valid, a very large number of kaons is expected.
This is a serious problem for models based on A2, since  the overall yield of strange particles is only about $5\%$.

This is why, in models where A2 and other planar diagrams are assumed to
dominate annihilation, it is crucial to introduce an explicit SU(3) breaking, in the form of a
\emph{suppression factor} $\lambda$ for  $\ssb$-creation, compared to $\nnb$. 
A value $\lambda=1$ corresponds to the SU(3) limit. To push the fraction of 
hidden strangeness production close to the  experimental value, models based on planar diagrams need a value as low as $\lambda\sim 0.1$. 

A theoretical foundation for $\ssb$ suppression was given by Dosch and
Gromes \cite{Dosch:1987sf}, who showed that $\ssb$ could be very much
suppressed just above threshold, by a kind of tunnelling effect.  In
the hadronisation following high-energy reactions, the production of
strange quarks is also reduced compared to the production of up and down
quarks. This is why fragmentation models incorporate a strangeness
suppression factor $\lambda$. It increases with energy from values as
low as 0.1 at a few GeV to $\sim 0.3$ at 30 GeV
\cite{Hofmann:1988ui}. $\ppb$ annihilation being a
soft process, a value $\lambda\sim 0.1$ seems not unreasonable. Such
a low value has decisive consequences for the interpretation of
annihilation data. If $\lambda\sim 0.1$ were true, SU(3) symmetry
would be dramatically broken.  The suppression would result in
heavier kaon resonances being more suppressed than lighter kaons, and
also $\mathrm{\overline{K}K}$ dramatically increasing in annihilation
in flight as compared to annihilation at rest.   
Hence there is a possibility of understanding the data by
the dominance of annihilation diagrams supplemented by $\ssb$
suppression.

Fragmentation of hadrons is, however, a rather indirect way to address
the question of $\ssb$ suppression at low energies. Meson decays may
be a better guide. The tensor meson \atmass\ has dominant decay
modes into $\rho\pi$, $\eta\pi$ and into $\K\Kb$. SU(3) relates 
the decays into $\eta\pi$ and into $\K\Kb$. Their
ratio, and the decay branching ratios of all tensor mesons, are fully
compatible with only small 
SU(3) symmetry breaking. The authors of Ref.~\cite{Peters:1995jv} 
fitted 16 decay modes of tensor mesons with 
SU(3) amplitudes allowing
for $\ssb$ suppression. They found $\lambda = 0.8\pm 0.2$: data on
tensor meson decays are compatible with SU(3) and rule out a 
substantial SU(3) symmetry breaking.

\begin{table}[!hbtp]
\renewcommand{\arraystretch}{1.2}
 \caption{\label{phe:tab:jipsi}Selected \protect$\rm J/\psi$ decays probing SU(3)
   symmetry. Data are taken from Ref. \protect\cite{Eidelman:2004wy}.  The
   experimental value is shown, and then corrected for a phase-space
   factor $p^{2\ell+1}$, where $p$ is the momentum, and $\ell$ the
   angular momentum in the final state. The units for
   BR\protect/$p^{2\ell+1}$ are irrelevant. For isospin multiplets,
   we assume normal weights for each charge state or average over the
   data, and display this average, contrary to \protect\cite{Eidelman:2004wy}
   where, e.g., $\rho\pi$ means the sum of the three channels.}
\begin{tabular}{ccccccc}
\hline\hline
Channel&BR ($10^{-3}$)&BR/$p^{2\ell+1}$ &\qquad\qquad &
Channel&BR ($10^{-3}$)&BR/$p^{2\ell+1}$  \\
$\rm p\bar p$  & $2.12\pm0.10$\qquad &1.7 &&
$\rho\pi$&$4.2\ \pm0.5\ $&1.7\\
$\rm n\bar n$&$2.2\ \pm0.4\ $&1.5 &&
$\omega\eta$&$1.58\pm0.16$&  0.6 \\                  
$\Lambda\overline\Lambda$&$1.30\pm0.12$&1.3&&
$\K^*\Kb+{\rm c.c.}$&$2.3\ \pm0.2\ $&0.9\\
$\Sigma\overline\Sigma$&$1.27\pm0.17$&1.3&&
$\pi\pi$&$0.15\pm0.02$&0.04\\
$\Xi\overline\Xi$&$0.9\ \pm0.2\ $&1.1&&
$\K\Kb$&$0.19\pm0.03$&0.06\\
$\Delta\overline\Delta$&$1.10\pm0.29$&1.2&&
\atmass$\rho$\quad &$3.6\ \pm0.7\ $&3.2\\
$\Sigma^*\overline\Sigma{}^*$&$0.52\pm0.07$&0.8&&
$\ftmass\omega$&$4.3\ \pm0.6\ $&3.6\\
                         &&&&
$\K_{2}^*\K$&$3.4\ \pm1.3\ $&2.4\\
\hline\hline
\end{tabular}
\end{table}

The Mark III Collaboration studied J/$\psi$ decays into a vector and a  
pseudoscalar meson in order to find  ''inert" or gluonic components in the  
$\eta$\ and $\eta^{\prime}$\ wave functions. The result was negative. 
One of the (rather numerous) parameters was the suppression of 
$\ssb$ pair creation compared to the creation of $\uub$ or  
$\ddb$ pairs. From their fit a value $\lambda = 0.8$ was deduced
\cite{Coffman:1988ve}.  

We looked for strangeness suppression in J/$\psi$ decays into baryons,
into $\rm p\bar p$, $\rm n\bar n$, $\Lambda\overline\Lambda$,
$\Sigma\overline\Sigma$, $\Xi\overline\Xi$, $\Delta\overline\Delta$,
and $\Sigma^*\overline\Sigma{}^*$. After correcting for phase space
(i.e., after division by the respective decay momenta) the squared
invariant couplings are similar in size, but scale with $0.75^{n_s}$
where $n_s$ is the number of $\ssb$ pairs created. The creation of a
second $s\bar s$ pair in $\Xi\overline\Xi$ is certainly a
non-perturbative process but still governed by $\lambda\sim 0.75$!

In short, the physics of tensor mesons and charmonium decay does not
support $\ssb$ suppression, and hence calls, in the case of $\NNb$ annihilation,
for an important role of non-planar, rearrangement diagrams.

Perhaps strangeness production in $\NNb$ cannot be explained in a simple 
uniform manner and should instead be examined for each type of final state.
Table~\ref{phe:tab:kproda} compares (dynamically
corrected) branching ratios for annihilation into two strange
mesons with those for annihilation into two isovector mesons.
The data are grouped into processes in which the two vector
mesons come from isoscalar ($D$-coupling) and isovector 
($F$-coupling) protonium states. $D$ or $F$ coupling belong to 
standard SU(3) notation \cite{Eidelman:2004wy}.

The ratios with $D$- and $F$-coupling differ remarkably  
while being internally compatible.
There is an astonishing consistency between
these ratios when processes of the same SU(3) structure
are compared. Again, the flavour flow is  
responsible for gross features 
of two-body annihilation processes.

The ratios with $F$-coupling suggest that A2 might be
the relevant quark line diagram, possibly with a `penalty
factor' $(\sim 0.5)$ for producing a $s\bar s$ pair. 
Then the rearrangement diagram R2 must contribute strongly
to the production of two isovector mesons from isoscalar
initial states. 
To conclude, it seems that the data on strangeness production support a
scenario where rearrangement diagrams are important.

The problem of strangeness production is not restricted to the
question of the relative importance of A2 vs.\ R2 diagrams.  Another
mechanism has been proposed by Ellis et al.~\cite{Ellis:1995ww}, where
strange quarks and antiquarks are extracted from the nucleon or
antinucleon sea, leading to specific signatures, for instance for the
spin effects in $\ppb\to\overline{\Lambda}\Lambda$, and for the
violation of the OZI rule (on which more in the next subsection).
Holinde et al.~\cite{Mull:1995sd} suggested a possible role of
initial-state interaction, via $\NNb\to\overline{Y}Y$, to describe the
$\ppb\to\phi\phi$ annihilation in flight. The mechanism can also be
applied to a violation of the OZI rule, since the hyperons $Y$ and
$\overline{Y}$ can produce a $\phi$ together with light mesons by
simple rearrangement.

\begin{table}[h]
\caption{\label{phe:tab:kproda}
Dynamically corrected branching ratios for annihilation into strange mesons
and into two isovector mesons 
in the final state for different SU(3) flavour couplings.
A ``$ + \mathrm{c.c.}$'' is implied for every final 
state with a $\K{}^*$ or $\K_{2}^*$.}
\begin{small}
\renewcommand{\arraystretch}{2}
\begin{tabular}{ccccccc}
\hline\hline
\multicolumn{3}{c}{\hspace*{-3mm}$D$-coupling}& \qquad&\multicolumn{3}{c}{$F$-coupling}\\
\hline
$\fracd{\DR\left(\tpz\to \K\Kb\right)} {\DR\left(\stpz\to\pi\pi\right)}$ 
& \hspace*{-3mm}=\hspace*{-2mm} 
& $0.047 \pm\ 0.004$
&\qquad 
&\hspace*{-3mm}$\fracd{\DR\left(\tso\to \K\Kb\right)}
{\DR\left(\ttso\to\pi\pi\right)}$ 
& \hspace*{-3mm}=\hspace*{-3mm} & $1.00\pm\ 0.20$\\
\hspace*{-3mm}$\fracd{\DR\left(\tso\to \K^*\Kb\right)}
{\DR\left(\stso\to\rho\pi\right)}$ & \hspace*{-3mm}=\hspace*{-2mm} &  $0.049 \pm\ 0.008$&&
\hspace*{-3mm}$\fracd{\DR\left(\ssz\to \K^*\Kb\right)}
{\DR\left(\tssz\to\rho\pi\right)}$ & \hspace*{-3mm}=\hspace*{-3mm} &$0.58 \pm\ 0.13$ \\ 
\hspace*{-3mm}$\fracd{\DR\left(\ssz\to \K\K{}^*_2(1430)\right)} 
{\DR\left(\stso\to\atmass\pi\right)}$ & \hspace*{-3mm}=\hspace*{-2mm} &  $0.033 \pm\ 0.010$ &&
\hspace*{-3mm}$\fracd{\DR\left(\tso\to \K\K{}^*_2(1430) \right)}
{\DR\left(\ttso\to\atmass\pi\right)}$ & \hspace*{-3mm}=\hspace*{-3mm} &$0.34 \pm\ 0.14$ \\ 
\vspace*{-6mm}
&&&&&&\\
\hline\hline
\end{tabular}\end{small}
\renewcommand{\arraystretch}{1.0}
\end{table}
\FloatBarrier\subsection{Violation of the OZI rule}\label{phe:sub:OZI}
The quark-line rule, or OZI rule, has already been introduced for mesons,
see sec.~\ref{mes:sub:Zweig}.  To first order, the rule forbids
production of $\phi$ mesons from initial systems like \pbp\ with up
and down quarks only.  The vector meson mixing angle,
$\Theta_{V}=39^{\circ}$ for the quadratic GMO mass formula, allows for
the small $\phi\to\pi^+\pi^-\pi^0$ decay rate, and also $\phi$
production from \pbp\ is permitted with an expected ratio
\begin{equation}
d_X({\rm vector, expected}) = 
\tan^2(\Theta_{V}-\Theta_{\rm id}) = 0.004~,
\end{equation}
where $d_X({\rm vector})$ is defined as the ratio of the dynamically
corrected annihilation frequencies
\begin{equation}
\label{phe:eq:ozi}
d_X({\rm vector}) = {\DR(\pbp\to X\phi)\over\DR(\pbp\to X\omega)}~,
\end{equation}
while the  $r_X(\rm vector)$ are the corresponding ratios
without dynamical corrections. The ratios $r_X$ and $d_X$
are defined analogously for the tensor mesons
$\ft(1270)$ and $\ft(1525)$.

Different analyses of bubble chamber data gave 
results in the range from 0.07 to 0.23 for the ratio of annihilation 
frequencies $\AF(\pbar N\to\pi\phi)$ to 
$\AF(\pbar N\to\pi\omega)$ \cite{Dover:1989zs}.
Based on a selection of results, a mean value
$0.085\er0.030$ was given in \cite{Reifenroether:1991ik}.
Obviously $\phi$ production is much
stronger than the anticipated ratio  $0.004$ suggests.

At LEAR, $\phi$ production was studied by the 
Asterix~\cite{Reifenroether:1991ik}, 
Obelix~\cite{Ableev:1994uy,Alberico:1998uj,Alberico:1998fr,Filippi:1999ym}, 
and Crystal Barrel collaborations~\cite{Amsler:1993xd,Amsler:1995up}. 
In Table~\ref{phe:tab:ozi} we give ratios  $r_X({\rm nonet})$ and $d_X({\rm nonet})$ 
for $X=\pi^0$ and $X=\eta$ from the two initial states 
\tso\ and \spo .
\begin{table}[htb]
\caption{Ratios of $\phi/\omega$ and $\mathrm{f}_2(1525)/\mathrm{f}_2(1270)$
production. Given are the \pbp\ initial state, the recoil meson,  
the ratio without and with
dynamical corrections, and the ratio expected from the nonet
mixing angle.  In the last column, the isospin component is
given which gives the dominant contribution to the corresponding
final states with kaons.  
}
\label{phe:tab:ozi}
\begin{small}
\renewcommand{\arraystretch}{1.1}
\begin{tabular}{cccccc}  \hline\hline
\pbp\ & $X$     & $r_X({\rm vector})$  & $d_X({\rm vector})$  &  From GMO& $\K^*\Kb$  \\
\ttso\ & $\pi^0$ & 0.099\er0.012  & 0.105\er0.013  &  0.004 & $I=1$\\
\stso\ & $\eta$  & 0.0035\er0.0009& 0.0038\er0.0010&  0.004 & $I=1$\\
\tspo\ & $\pi^0$ & 0.005\er0.009  & 0.005\er0.009  &  0.004 & $I=0$\\
\sspo\ & $\eta$  & 0.10\er0.07    & 0.12\er0.09    & 0.004  & $I=0$\\
\hline
\pbp\ & $X$     & $r_X({\rm tensor})$  & $d_X({\rm tensor})$  & From GMO & $\K^{*}_2\Kb$  \\
\tssz\ & $\pi^0$ & 0.020\er0.004  & 0.025\er0.005 & 0.012  & $I=0$\\
\hline
\pbp\ & $X$    & $r_X({\rm vector})$  & $d_X({\rm vector})$ & From GMO& $\K^{*}\Kb^{*}$ \\
\tssz\ & $\rho$   & 0.018\er0.007 & 0.030\er0.010  & 0.004  & both\\
\sssz\ & $\omega$ &0.0095\er0.0035& 0.017\er0.006  & 0.004  & both\\
\ssz\ & $\gamma$ & 0.25\er0.09   & 0.026\er0.09   & 0.004  & both\\
\hline\hline
\end{tabular}
\renewcommand{\arraystretch}{1.0}
\end{small}\end{table}

The ratios vary over a wide range: the \pbp\ annihilation frequencies
from the \tso\ state to $\phi\eta$, from the \spo\ state
to $\phi\pi$, and the frequency to $\mathrm{f}_2(1525)\pi$ 
are all of the order of magnitude expected 
from the meson nonet mixing angles. A few processes give a moderately
large $\phi$ production rate like $\phi\rho$ and $\phi\omega$. 
For some reactions however, the OZI violation is really large, the 
$\phi/\omega$ ratio being about $10\%$ or larger.

Three interpretations of this large excess of $\phi$ production
compared to $\omega$ production have been pursued.
Figure~\ref{phe:fig:phi} sketches the three scenarios.
\begin{figure}[tbh]
\includegraphics[width=.90\textwidth]{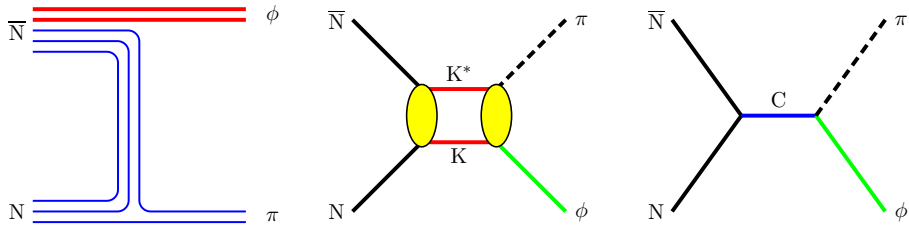}
\caption{\label{phe:fig:phi}
Diagrams which could contribute to $\phi$ production: as shake-off of
hidden \ssb\ pairs in the nucleon wave function, via rescattering of
Kaons from secondary decays, and from formation of four-quark exotic
states.}
\end{figure}

Dover and Fishbane~\cite{Dover:1989zs} link the excess in $\phi$
production to the production of four-quark exotics with hidden strangeness
($qs\bar q\bar s$), e.g., to the tail of the $C(1480)$
meson~\cite{Bityukov:1987yd}\footnote{\footnotesize The C(1480) was
observed as $\phi\pi$ resonance and interpreted as $(ns\bar n\bar s)$
resonance. However, it was never confirmed.}.

The nucleon wave function is known to contain an
\ssb\ component. This is evident from deep inelastic 
scattering~\cite{Barone:1999yv} 
or from the so-called $\pi$-nucleon $\sigma$ term 
$\sigma_{\rm\pi N}$~\cite{Procura:2003ig}. There is no guarantee however
that virtual \qqb\ pairs may be shaken off in an OZI rule violating
diagram. If this is assumed, the sharp selectivity of the
processes leading to large OZI rule violating effects can potentially
be understood~\cite{Ellis:1995ww,Ellis:1999er} as originating from
the \ssb\ component being negatively polarized.

The PSI group studied OZI rule violating effects in \NNb\ annihilation
in a series of
papers~\cite{Locher:1994cc,Gortchakov:1996im,vonRotz:1999ef}. See,
also, \cite{Buzatu:1995xn,Zou:1996fg,Buzatu:1996ff}.  The large
$\pi\phi$ annihilation frequencies were interpreted by rescattering
into $\phi$ mesons of $\K\Kb$ pairs from $\K\Kb^*+\K^*\Kb$ annihilation
and or $\rho\pi$ rescattering from $\rho^{+}\rho^{-}$.  The large rate
for $\gamma\phi$ channel was understood as the effect of $\rho\phi$ and
$\omega\phi$ production, and vector meson dominance.  The study was
extended in~\cite{Markushin:1998bv} to include $\phi\pi^+\pi^-$
production in \ppb\ annihilation at rest and in flight.

We notice that the ratios $r_X$ and $d_X$ are large when the $\phi$ and 
$\omega$ are produced from the isospin component which gives a large 
contribution to kaonic final states. Rescattering of, e.g., $\K\Kb$ in
the $\K^*\Kb$ final state is certainly proportional to the frequency
with which $\K^*\Kb$ with the correct isospin is produced. Indeed, the OZI
violation is found to be correlated with the $\K^*\Kb$ production strength of
the required isospin. This observation supports the rescattering 
interpretation of the strong OZI rule violation. As emphasized 
in~\cite{Ellis:1999er}, there are more processes in which OZI rule
violating effects can be studied and possibly linked to a
hidden \ssb\ component in the \NNb\ wave function. 

Decays of the J/$\psi$ can help to elucidate the problem further. A
sample is given in Table \ref{phe:tab:Jpsi}.  The J/$\psi$ wave
function has little hidden strangeness, and any $\pi\phi$ production
must be due to rescattering. In reversing the argument, the absence of
$\pi\phi$ in J/$\psi$ decays could possibly be interpreted as evidence
that the large OZI rule violating in \NNb\ annihilation must come from
the nucleon wave function.

\begin{table}[!ht]
\caption{\label{phe:tab:Jpsi} Selected decays modes of J$/\psi$}
\renewcommand{\arraystretch}{1.2}
\begin{tabular}{ccccc}
\hline\hline
J/$\psi\to$ & $\eta\omega$  & $\pi\omega$  & $\pi\phi$  &  $\rm K^*K$  \\
Rate $(10^{-4})$            & $15.8\pm 1.6$   & 
$4.2\pm 0.6$ & $<$0.068     & $92\pm  6$ \\
\hline\hline
\end{tabular}
\renewcommand{\arraystretch}{1.0}
\end{table}
From the $\eta\omega /\pi\omega$ ratio, isospin violation can be
estimated to occur at the 14\% level. The $\K^*\Kb$ final state was
not analysed to identify isospin breaking effects; however a
$\K^*\K\,(I=1)$ contribution of more than $10^{-3}$ can be
expected. Rescattering may then lead to $\pi\phi$ production at a
detectable level.

The problem was studied in~\cite{Anisovich:1996xr} using a dispersion
theoretical approach. The interpretation of large $\phi\pi$ frequency
in \pbp\ annihilation as an effect of rescattering was confirmed. The low
$\phi\pi$ production in J/$\psi$ decays is due to its larger phase
space, and is compatible with the rescattering mechanism  for OZI rule
violating effects.
\FloatBarrier\subsection{Flavour flow and flavour symmetry}\label{pheno:sub:SU3}
\label{phe:sub:SU3}
Support for the use of quark-line diagrams can be
found by comparing branching ratios for annihilation into
specific two-body final states.
In Table~\ref{phe:tab:rat10}, we compare dynamically corrected
annihilation frequencies DR, using Eq.~(\ref{phe:eq:vdmbw}). 
The  numerators correspond to final states with
one isovector and one isoscalar meson, and the denominators to 
annihilation into two isovector or two isoscalar mesons. 

\begin{table}[h!]
\caption{\label{phe:tab:rat10}%
Ratios of dynamically corrected annihilation frequencies DR into an isovector
and an isoscalar meson compared to production of two isoscalar or
two isovector mesons. Bose symmetry is taken into account. 
} 
\begin{small}
\renewcommand{\arraystretch}{2.2}
\begin{tabular}{rr}
\hline\hline
$\fracd{1}{2\cos^2\Theta}  \fracd{\DR(\pbp\to\pi^0\eta)}{\DR(\pbp\to\pi^0\pi^0)}                       
= 0.24\pm 0.08$  &
$\fracd{1}{2\sin^2\Theta}  \fracd{\DR(\pbp\to\pi^0\eta^{\prime})}{\DR(\pbp\to\pi^0\pi^0)}        
= 0.24\pm 0.7\phantom{0}$ \\
$\fracd{\cos^2\Theta}{2}   \fracd{\DR(\pbp\to\pi^0\eta)}{\DR(\pbp\to\eta\eta)}                  
= 0.40\pm 0.10$  &
$\fracd{\cos^4\Theta}{2\sin^2\Theta}  \fracd{\DR(\pbp\to\pi^0\eta^{\prime})}{\DR(\pbp\to\eta\eta)}        
= 0.48\pm 0.14$  \\
$\fracd{\sin^2\Theta}{2}  \fracd{\DR(\pbp\to\pi^0\eta)}{\DR(\pbp\to\eta\eta^{\prime})}                       
= 0.31\pm 0.10$   &
\qquad\quad$\cos^2\Theta\,  \fracd{\DR(\pbp\to\pi^0\eta^{\prime})}{\DR(\pbp\to\eta\eta^{\prime})}
= 0.39\pm 0.06$ \\
$\fracd{1}{\cos^2\Theta}   \fracd{\DR(\pbp\to\rho^0\eta)}{\DR(\pbp\to\rho^0\pi^0)}          
=  0.40\pm 0.11$   &
$\fracd{1}{\sin^2\Theta}   \fracd{\DR(\pbp\to\rho^0\eta^{\prime})}{\DR(\pbp\to\rho^0\pi^0)}  
=  0.45\pm 0.13$  \\
$\fracd{\DR(\pbp\to\rho^0\eta)}{\DR(\pbp\to\omega\eta)}  
=  0.23\pm 0.06$   &
$\fracd{\cos^2\Theta}{\sin^2\Theta}   \fracd{\DR(\pbp\to\rho^0\eta^{\prime})}{\DR(\pbp\to\omega\eta)}  
=  0.26\pm 0.10$  \\
$\fracd{\DR(\pbp\to\rho^0\eta^{\prime})}{\DR(\pbp\to\omega\eta^{\prime})}  
=  0.18\pm 0.06$  &
$\fracd{\DR(\pbp\to\rho^0\eta^{\prime})} {\DR(\pbp\to\omega\eta^{\prime})}   
=  0.18\pm 0.06$  \\
$\fracd{\DR(\pbp\to\omega\pi^0)}{\DR(\pbp\to\rho^0\pi^0)}
=  0.41\pm 0.06$   & 
$\cos^2\Theta \, \fracd{\DR(\pbp\to\omega\pi^0)}{\DR(\pbp\to\omega\eta)}
= 0.23\pm 0.06$   \\ 
$\sin^2\Theta\, \fracd{\DR(\pbp\to\omega\pi^0)}{\DR(\pbp\to\omega\eta^{\prime})}
=  0.18\pm 0.07$ &
$\fracd{\DR(\pbp\to\ftmass\pi^0)}{\DR(\pbp\to\atmass\pi^0)}
=  0.53\pm 0.09$  \\ 
\vspace*{-6mm}
&\\
\hline\hline
\end{tabular}
\end{small}
\renewcommand{\arraystretch}{1.0}
\end{table}
The ratios require some corrections. In the process $\ppb\to \pi\eta$,
the initial $\ppb$ system couples only to the $\nnb$ component and
not to the $\ssb$ component.  The OZI rule reduces the amplitude for
this process by $\cos\Theta$, and the branching ratio by
$\cos^2\Theta$. Correspondingly, $\ppb\to \pi\eta^{\prime}$ is reduced
by $\sin^2\Theta$, where $\Theta$ is defined in
Eq.(\ref{meson:eq:psmix}). We use $\cos^2\Theta =0.65\pm0.07$ and  
$\sin^2\Theta = 0.35\pm0.07$.

The table reveals a surprise: most results are about
compatible with each other, except for those containing 
the DR for $\omega\eta^{\prime}$ (which are too large).
The mean value of those ratios gives $0.29\pm 0.03$.
with a $\chi^2/N_F = 2$. 
There is no a priori reason why these ratios should be similar.
The two final states $\pi^0\eta$ and $\pi^0\pi^0$ are
produced from the \tpz\ initial state, $\rho\eta$
and $\omega\eta$ from \tso , \ftmass$\pi^0$ 
and \atmass$\pi^0$ from \ssz . 
Table~\ref{phe:tab:rat10} includes
final states with two light mesons like $\pi$ and $\eta$,
or with two massive mesons like $\omega$ and $\rho$;
the ratio is formed using pseudoscalar, vector and tensor
mesons in arbitrary combinations.
In some cases, broad mesons are only in the numerator, in
others only in the denominator. Obviously, the flavour content
and the flavour coupling are the decisive ingredients. 
Table~\ref{phe:tab:rat10} supports the conjecture made 
in Sec.~\ref{dsr:sub:disc} that the flavour flow has a decisive
impact on annihilation dynamics.

In Table~\ref{phe:tab:rat10}, the two annihilation modes  
$\ppb\to\rho^0\rho^0$  
and $\ppb\to\eta$\ftmass\  are excluded. The frequency for annihilation into 
$\ppb\to\rho^0\rho^0$ is compatible with zero, and there
is no known reason for this effect. The reaction
$\ppb\to\eta\ftmass$ needs $\ell=2$ and is very close
to threshold; the measured frequency is compatible with zero
but is also expected to be small. 
\subsubsectionb{SU(3): quark-line rule, $s$-channel resonances and baryon
exchange}
As we have seen \pbp\ annihilation can be discussed in rather different
languages. Quark models describe annihilation in terms of planar and
non-planar diagrams (often called annihilation and rearrangement
diagrams). On the other hand, \pbp\ annihilation may prefer to proceed
via a few $s$-channel resonances (e.g., by mixing between the \pbp\
system and $(\bar{q}\bar{q}qq)$ states close in mass). In this case, a
description in terms of $s$-channel amplitudes may be more
appropriate. Or, alternatively, baryons and mesons might be the
relevant degrees of freedom, and \pbp\ annihilation could be most
efficiently described by baryon exchange
amplitudes. Figure~\ref{phe:fig:coupl} visualises the different
approaches.
\begin{figure}[htb]
\begin{tabular}{c@{\hspace*{.5cm}}c@{\hspace*{.5cm}}c}
\hspace*{-.2cm}\includegraphics[width=0.31\textwidth]{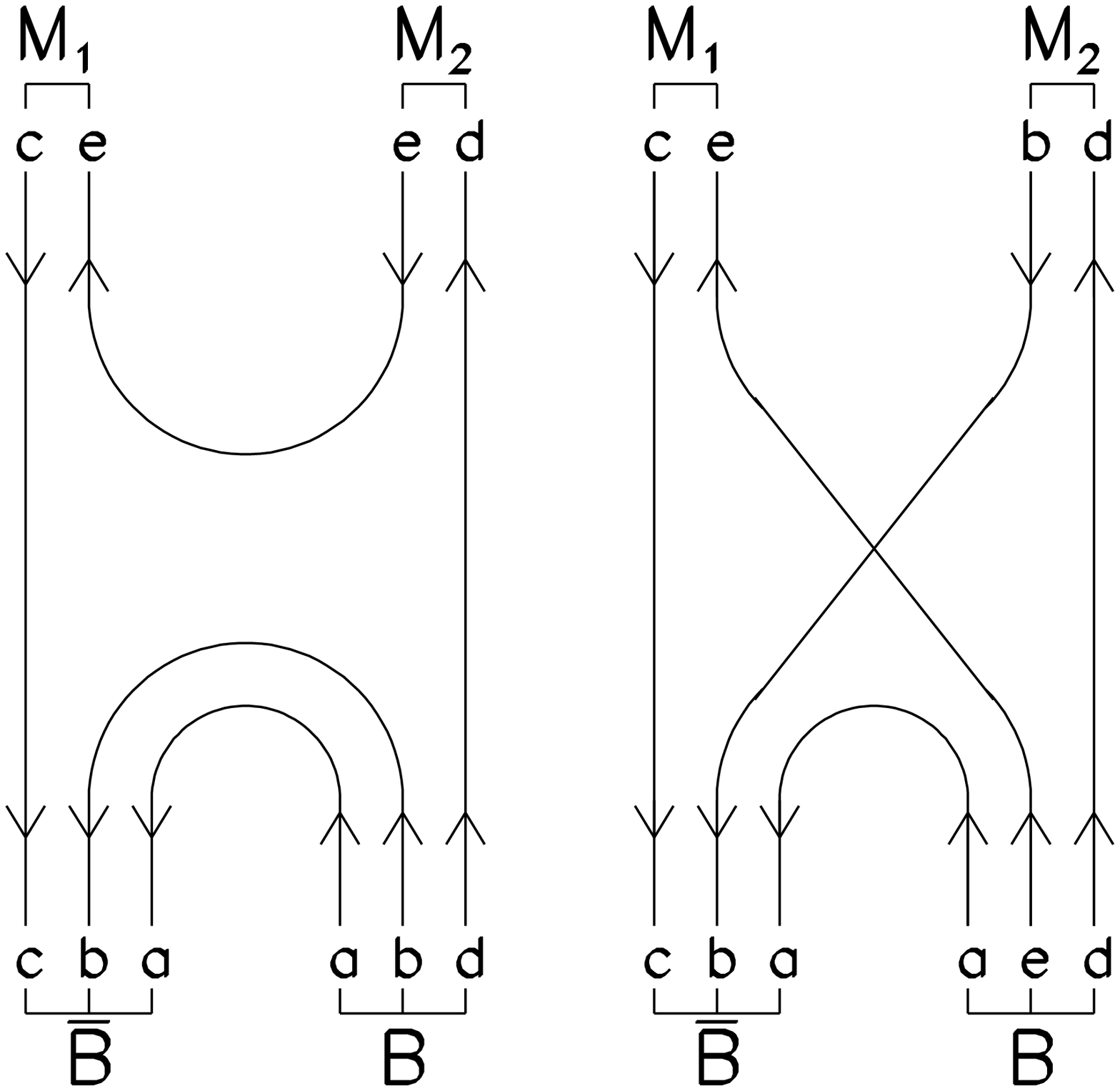}&
\includegraphics[width=0.31\textwidth]{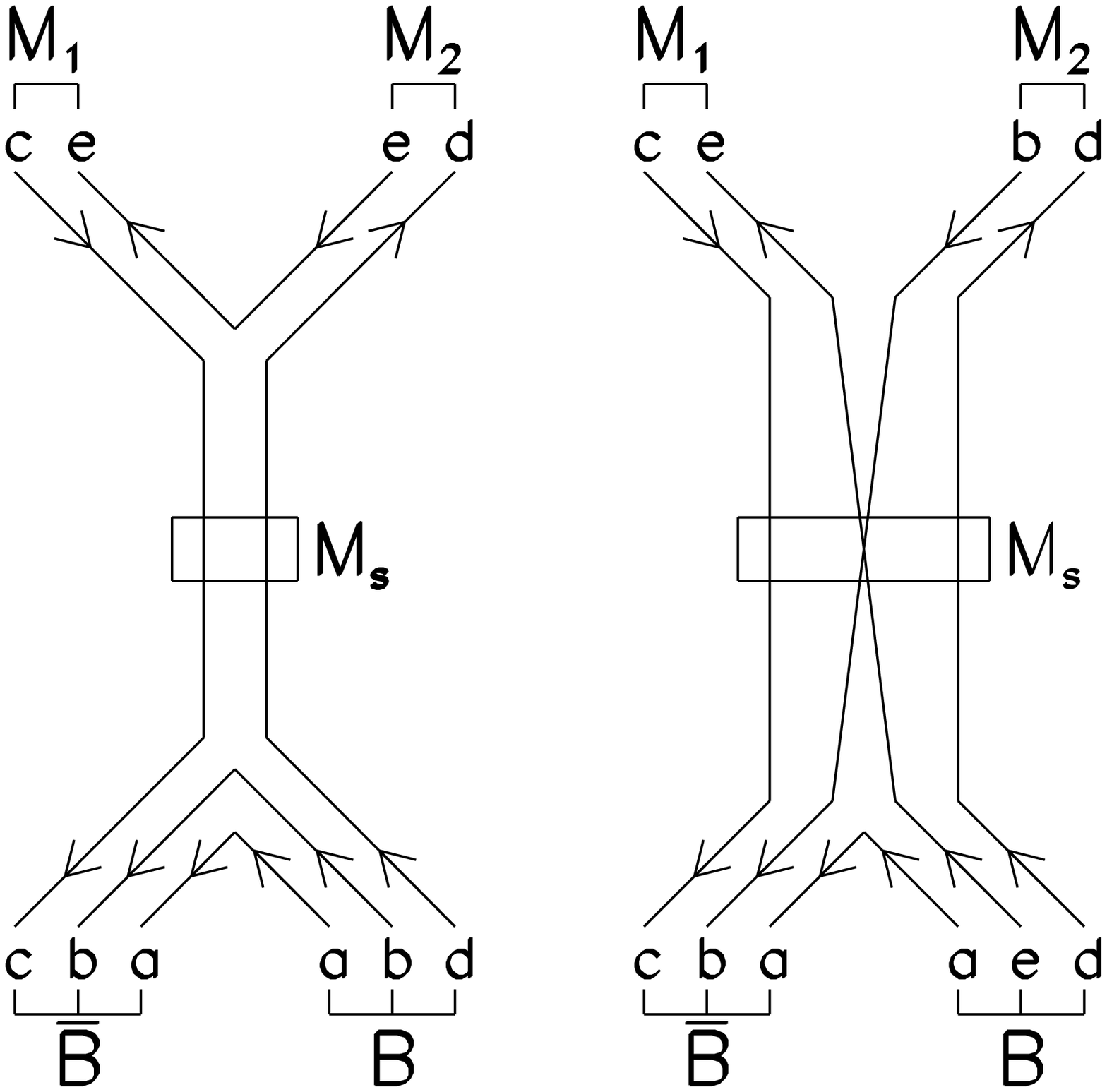}&
\includegraphics[width=0.31\textwidth]{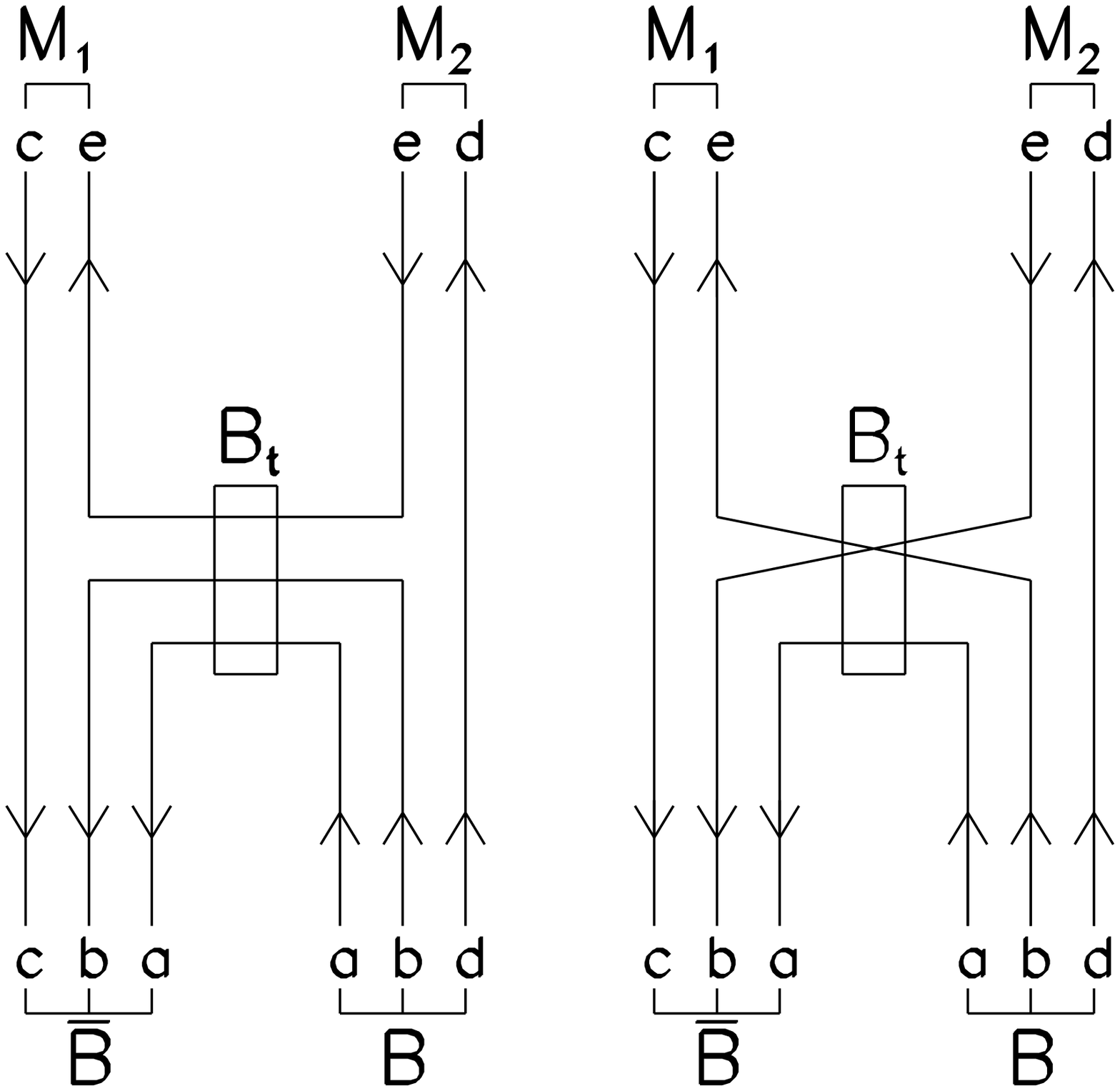}\\
\end{tabular}
\caption{\label{phe:fig:coupl}
Annihilation of protons and antiprotons in different
coupling schemes. From left to right: the quark line coupling
depicting the flavour flux, the $s$-channel coupling with meson
formation in the $s$-channel, and baryon
exchange coupling where baryons are exchanged in the $t$-channel.
}
\end{figure}
The question to be addressed is which scheme is best suited to
incorporate the most important aspects of annihilation, and in
particular, whether {\em dynamical selection rules} find a natural
interpretation in one of the three coupling schemes. The three
descriptions are related by unitary matrices which were developed in
\cite{Klempt:1996ws}.

Figure \ref{phe:fig:coupl} shows for each coupling scheme two basic
diagrams. Their meaning is illustrated using the quark line coupling
scheme.  The two outgoing mesons can be symmetric or antisymmetric
with respect to their exchange leading to a doubling of diagrams.
In any diagram, one antiquark and one quark reach the final
state as  `spectators'. The spectators can be either 
$(u,\bar{u})$ or $(d,\bar{d})$, again leading to a doubling of 
diagrams\footnote{More precisely, the spectator either belongs or 
not to an antisymmetric $(u,d)$ pair.}. Hence 8
diagrams are needed to describe the annihilation process. These are
shown in figure~\ref{phe:fig:qlr}. In case of two mesons in the final
state belonging to the same multiplet (e.g. for annihilation into two
pseudoscalar mesons), the generalised Pauli principle requires R$^+_3$
to vanish.

\begin{figure}[!ht]
\hspace{-0.8cm}
\begin{tabular}{c@{\hspace*{.5cm}} c@{\hspace*{.5cm}}c}
\hspace*{.4cm}\includegraphics[width=0.31\textwidth]{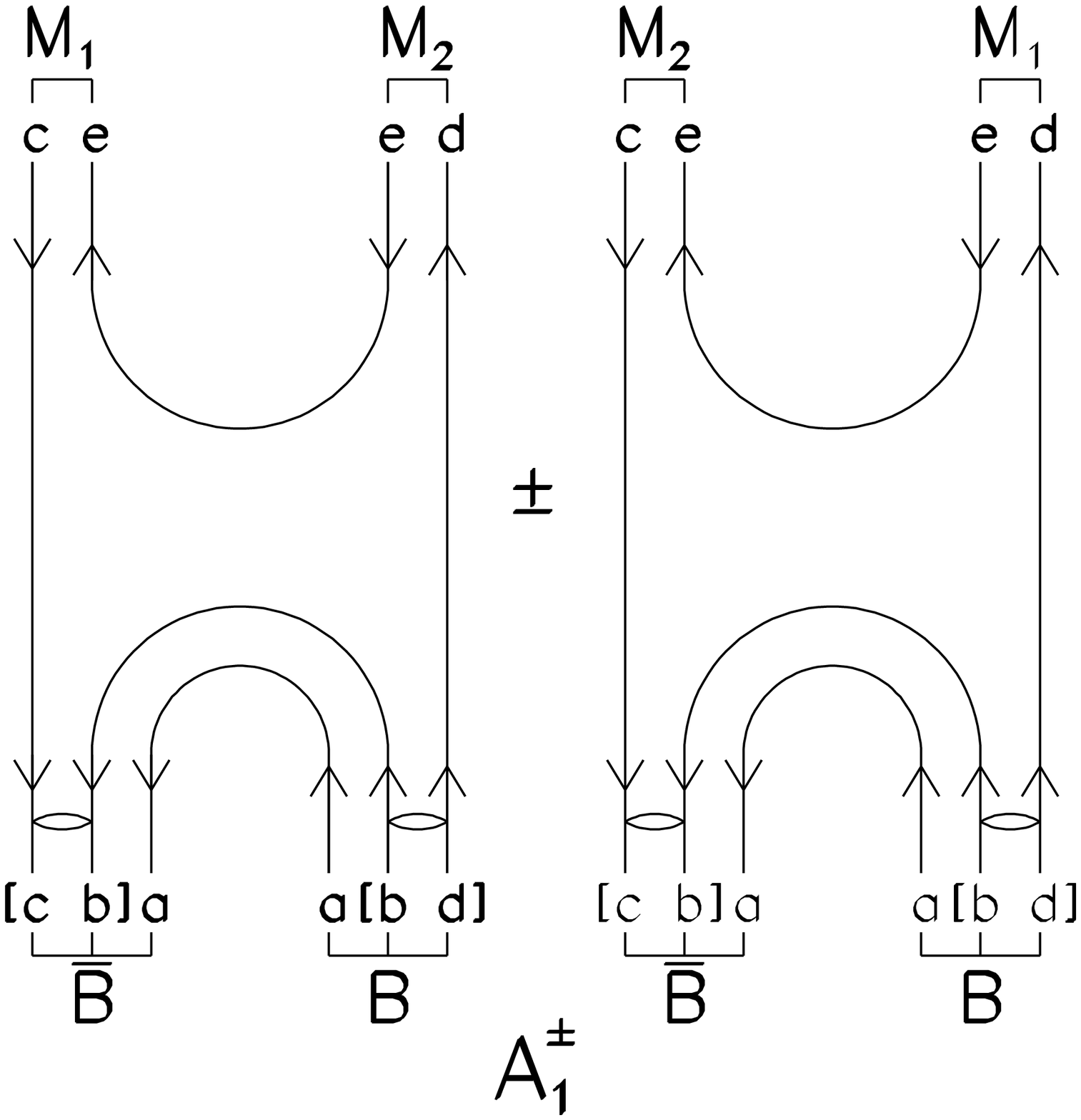}&
\includegraphics[width=0.31\textwidth]{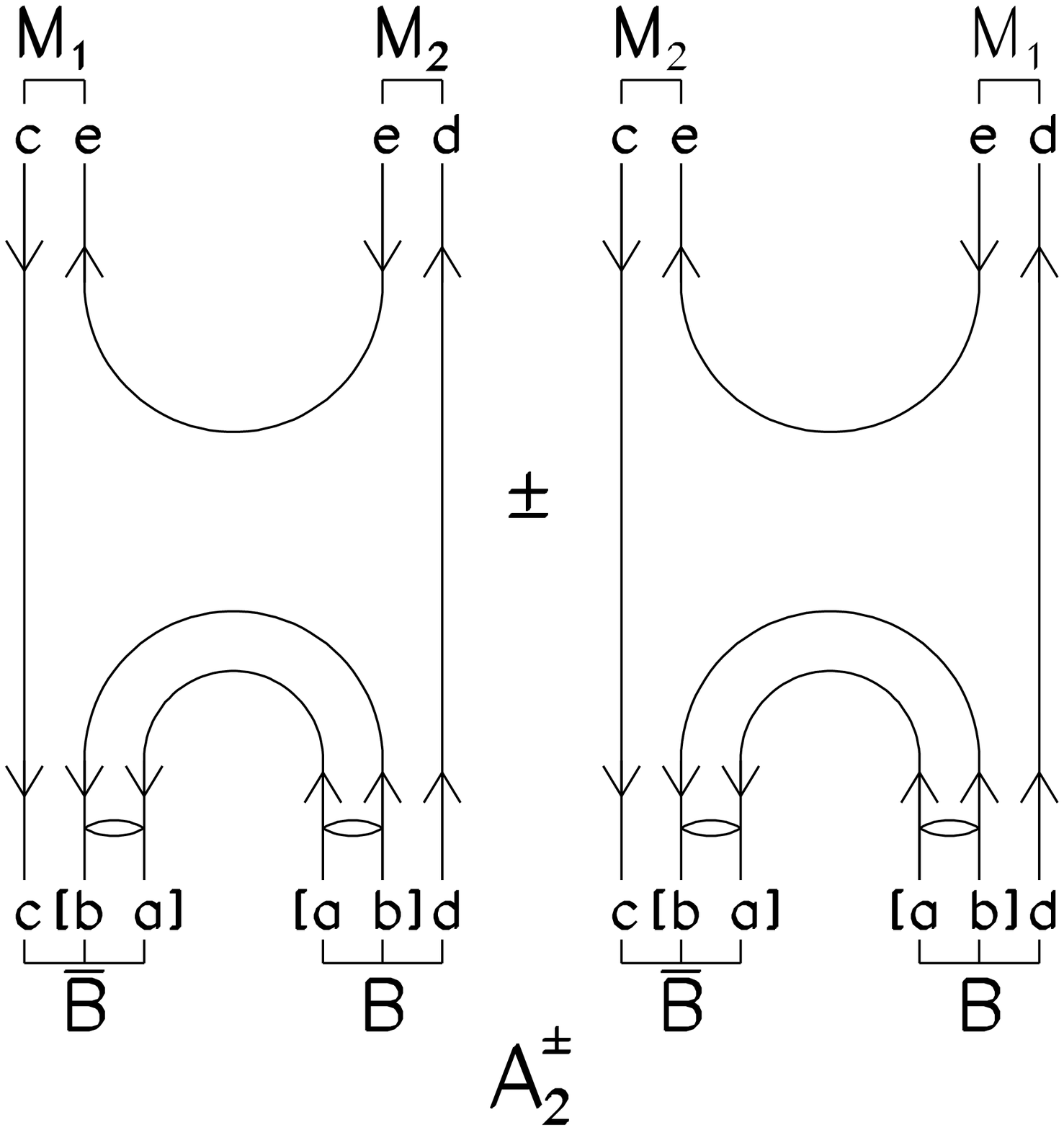}&
\includegraphics[width=0.31\textwidth]{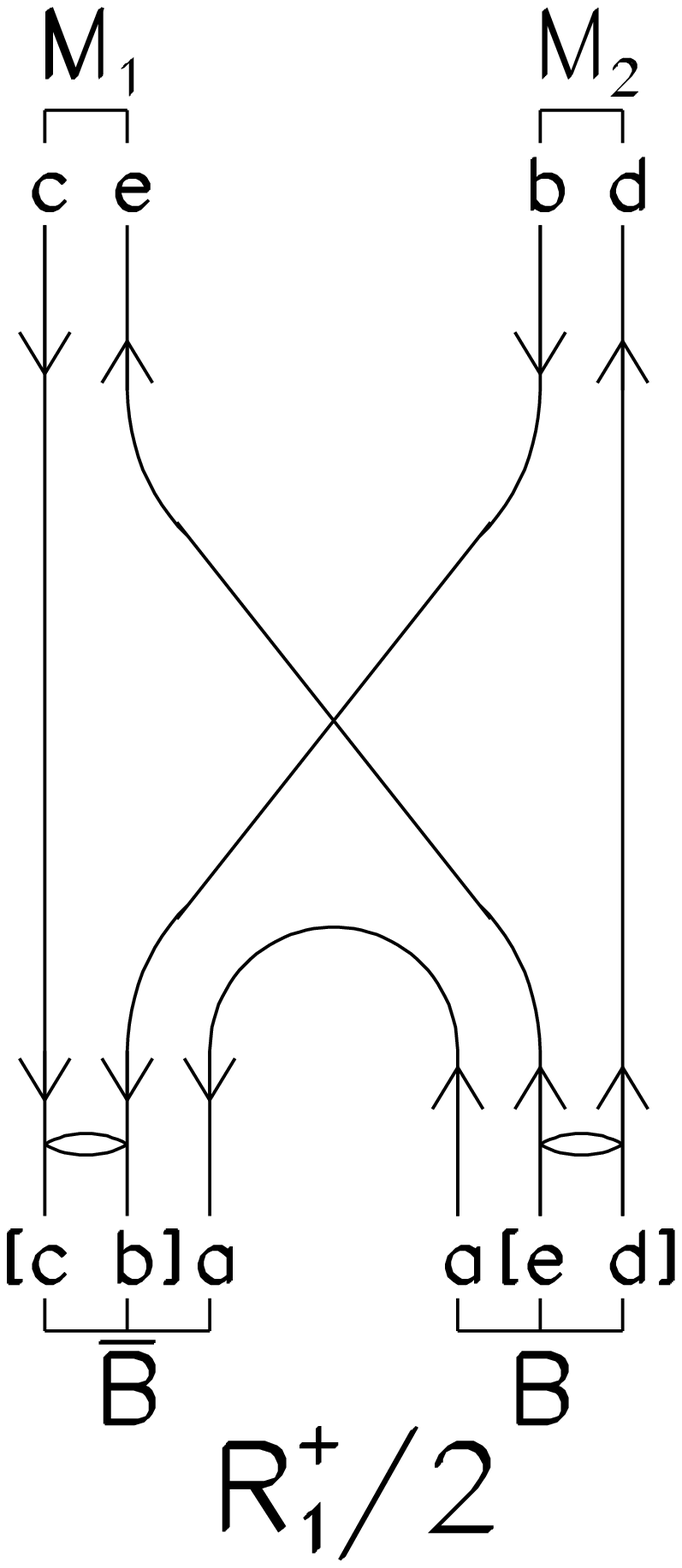}\\[.2cm]
\hspace*{.4cm}\includegraphics[width=0.31\textwidth]{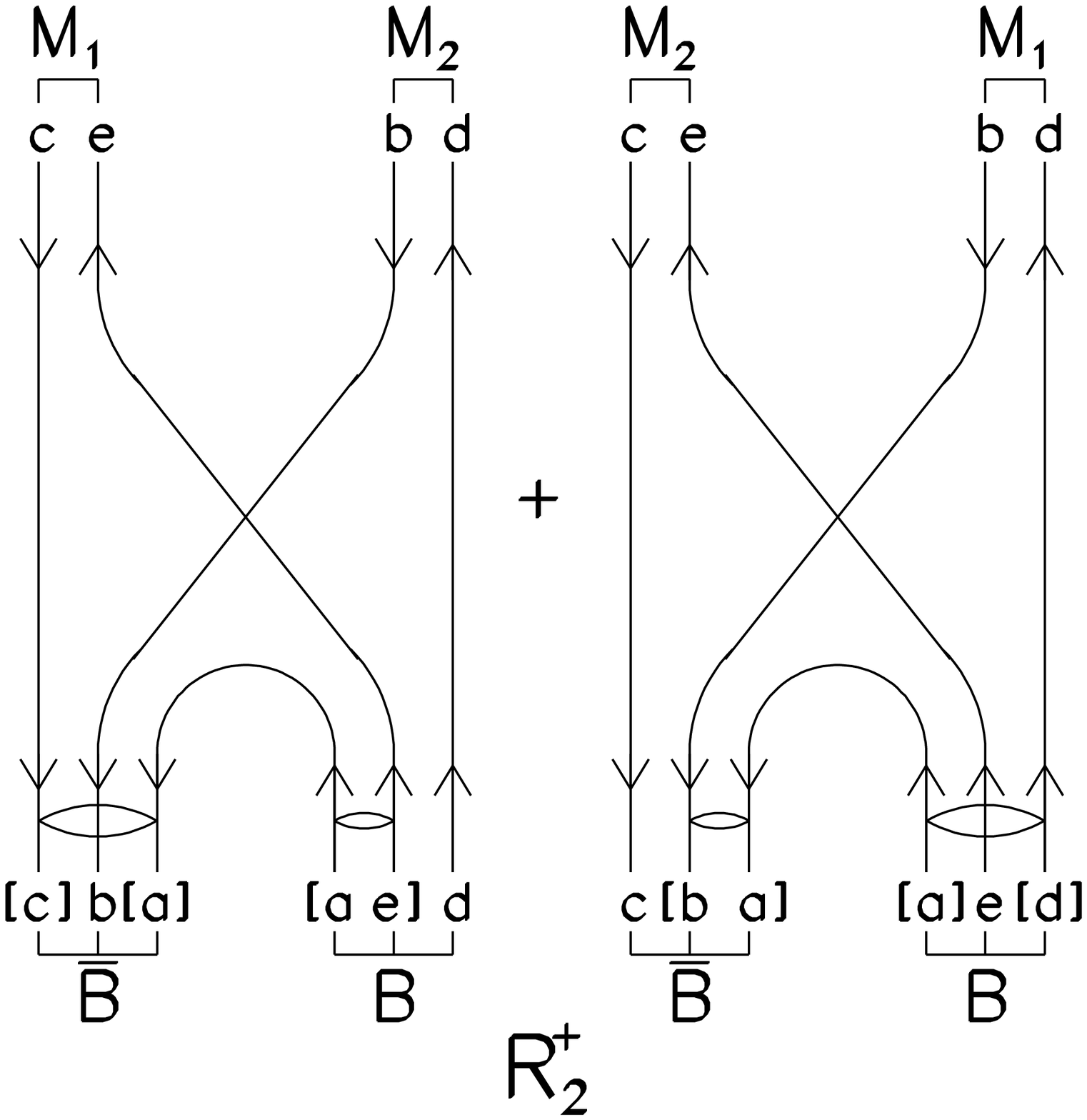}&
\includegraphics[width=0.31\textwidth]{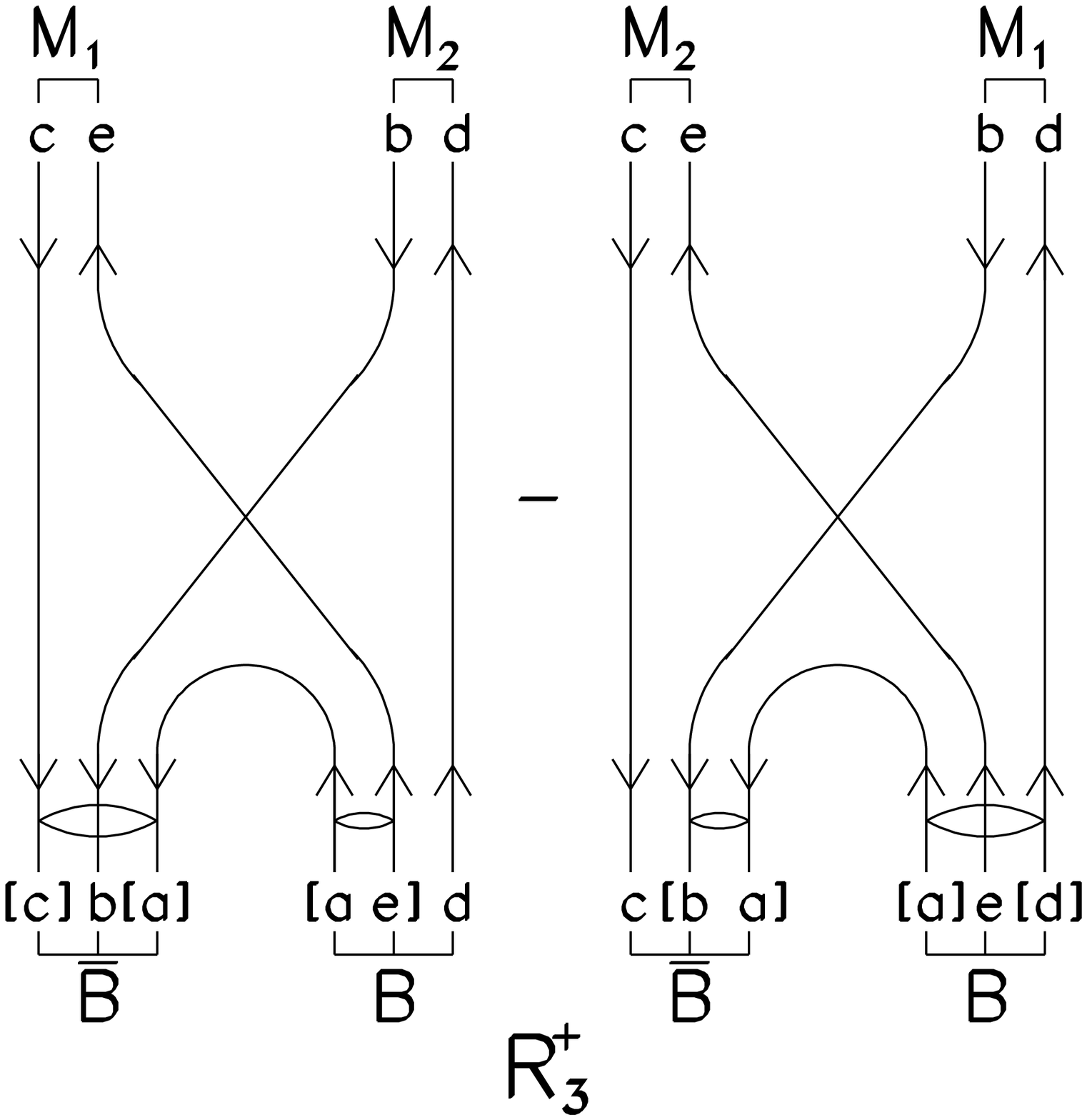}&
\includegraphics[width=0.31\textwidth]{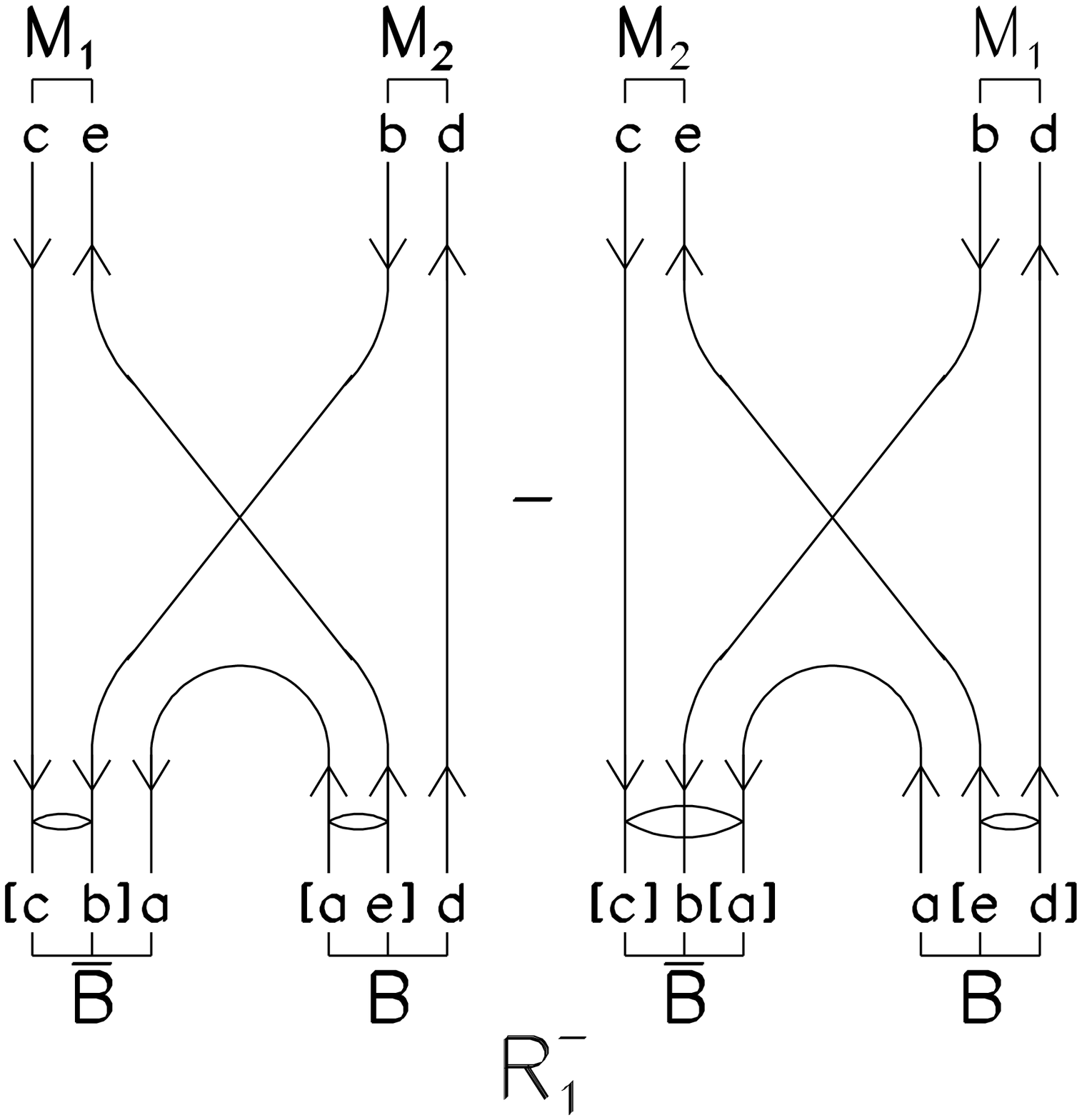}\\
\end{tabular}
\caption{\label{phe:fig:qlr}
Explicit representation of quark-line amplitudes of all
independent couplings in the quark-line scheme. Pairs of
(anti-)quarks antisymmetric with respect to their exchange
are connected by small loops.
}
\end{figure}

For a quantitative analysis we use the dynamically corrected
branching ratios of Table~\ref{phe:tab:Bpartial}. These should be related to the squared SU(3) transition matrix elements. In the fits
we use a strangeness suppression factor $\lambda=0.7$ to $0.8$. The
results do not depend critically on this assumption. A SU(3) error of
20\% is introduced. The latter is added quadratically to the
experimental errors of Table~\ref{phe:tab:Bpartial}.

All three coupling schemes allow us to fit the data
with identical $\chi^2$. There are 10 measured branching ratios for annihilation into two pseudoscalar mesons and 11 for annihilation
into a vector and a pseudoscalar meson, 7 amplitudes in the former
case and 8 in the latter one. Since the pseudoscalar mixing angle
has been fixed from these data (Eq.~\ref{phe:eq:theta}), the 
number of degrees of freedom is 2 in both cases. These two
sets of branching ratios are fitted with $\chi^2=1.3$  For annihilation into
a pseudoscalar and a tensor meson, there are only 8 data points and 8
amplitudes; the data are reproduced with $\chi^2=0$. 
The three coupling schemes give identical descriptions of the
data. Even within one coupling scheme, different solutions exist
with identical or similar $\chi^2$. Hence care has to be taken in
interpreting the results.
\subsubsection{The quark coupling scheme}
Figure~\ref{phe:fig:qlr} shows the decomposition of the quark coupling
scheme. There are 8 amplitudes which may be fit to the values given in Table~\ref{phe:tab:qcs},
\begin{table}[!ht]
\caption{\label{phe:tab:qcs}Best-fit values for the amplitudes in the quark coupling scheme.}
\renewcommand{\arraystretch}{1.2}
\begin{tabular}{rrrrrrrrr}
\hline\hline
\ & \ $\rm A^+_1$ \ & \  $\rm A^+_2$ \ & \  $\rm R^+_1$ \ & \  $\rm R^+_2$ \ & \  $\rm R^+_3$ \ & \
$\rm A^-_1$ \ & \  $\rm A^-_2$ \ & \  $\rm R^-_1$\\
\hline PS--PS& 1.6 & $-2.1$ & 9.7  & $-0.3$ &   -  & 1.4 & $-1.5$ & 1.2 \\
PS--V & 2.3 & $-1.9$ & 10.9 & 1.2  & $-3.9$ & 1.1 &$-1.8$ &$ -1.6$ \\
PS--T & 1.3 & $-1.8$ & 8.9  & $-2.8$ & $-1.0$ & 0.8 & $-1.5$ & $-1.8$ \\
\hline\hline
\end{tabular}
\renewcommand{\arraystretch}{1.0}
\end{table}
where a positive sign has been chosen for the amplitudes $\rm
A^{\pm}_1$. The largest contribution is given by $\rm R^+_1$. Before
discussing the meaning and significance of this result, a comment is
made concerning the amplitudes $\rm A^{\pm}_1$ and $\rm
A^{\pm}_2$. They have opposite signs and are about equal in
amplitude. The approximate relations $\rm A^{\pm}_1 + A^{\pm}_2\simeq
0$ hold for all five annihilation processes from initial S-states. For
annihilation into two pseudoscalar meson, 
the data are also approximately compatible with this relation with an
alternative solution where $\rm\vert A^{+}_1\vert \gg \vert
A^{+}_2\vert $. Setting $\rm A^{\pm}_1 + A^{\pm}_2 = 0$  substantially helps
further discussion: annihilation into two isovector
mesons (first line in Table~\ref{phe:tab:Bpartial}) is given by the SU(3)
matrix element $\rm\vert 2A^{+}_1 + 2A^{+}_2 + R^+_1\vert^2$.  $\rm
R^+_1$ can be chosen positive or negative; for $\rm A^{+}_1 +
A^{+}_2 \neq 0$ two values are found; both are large but differ in
the precise number. Only for $\rm A^{+}_1 + A^{+}_2 = 0$, does $\rm
R^+_1$ have a unique value. As long as the interest is in a qualitative
understanding, it suffices to state that rearrangement diagrams give a
substantial contribution to annihilation and should not be
neglected. This statement is compatible with the findings presented
above (see Sec.~\ref{phe:sub:rearr}).

Among the rearrangement diagrams, $\rm R^+_1$ plays the dominant role.
This result has to be interpreted with some precaution. Annihilation
into $\rho\pi$ from the \tso\ initial state is driven by
$\rm\frac{1}{2}R^+_1$, annihilation into $\omega\eta$ by
$\rm\frac{1}{2}R^+_1+ R^+_2$. Obviously, there is a small and a large
negative value of $\rm R^+_2$ which can satisfy the relation. So we
can conclude only that the data are compatible with  $\rm R^+_2$ and
$\rm R^+_3$ both being small but   $\rm R^+_2$ could be large.
The relation $\rm A^{\pm}_1 + A^{\pm}_2 \sim 0$ reproduces the
dynamical selection rules governing strangeness production. The
$\rho\pi$ puzzle and the generalisation to $\pi\pi$ and \atmass\
is related to the dominance of $\rm R^+_1$.
\subsubsection{The $s$-channel coupling scheme}

The 8 SU(3) amplitudes can be decomposed into
$s$-channel couplings denoted as
$s_1 , s_{8_{ss}}, s_{8_{sa}}, s_{8_{as}}$, $s_{8_{aa}},
s_{10}, s_{\bar{10}}, s_{27} $.
The coupling scheme classifies the SU(3) structure of the intermediate
state in the $s$-channel. The couplings of the \pbp\ system and of the
two mesons to an octet intermediate state ($\bar{q}q$)
or ($\bar{q}\bar{q}qq$) can be symmetric or antisymmetric; this
flexibility leads to four amplitudes. The intermediate state could also
be flavour singlet, decuplet, \{$\bar{10}$\} or a \{27\}-plet.

As stated above, the fit with all amplitudes leads to identical
results. But the $s$-channel coupling scheme leads to a different
interpretation. The four annihilation diagrams go through a $\bar{q}q$
intermediate state; obviously no $s$-channel decuplet, \{$\bar{10}$\}
or  \{27\}-plet can contribute to these diagrams. More interesting is
the observation that the $s$-channel decuplet, \{$\bar{10}$\} or 
\{27\}-plet amplitudes can all be set to zero without a significant
deterioration of the fit.  This fact is due to the dominance of
$R^+_1$ which can be decomposed into $s_1 , s_{8_{ss}}, s_{8_{sa}},
s_{8_{as}}, s_{8_{aa}}$ only.  If \pbp\ annihilation dynamics is
interpreted in the $s$-channel coupling scheme, there are no
``exotic'' components even though four-quark $\bar{q}\bar{q}qq$
``crypto-exotic'' components do play a very significant role. The
(generalised) $\rho\pi$ puzzle is thus interpreted as absence of
exotic states in the $s$ channel of \pbp\ annihilation. The strange
pattern in strangeness production finds however no straightforward
interpretation in the $s$--channel coupling scheme.

\subsubsectionb{The $t$-$u$-channel coupling scheme}

The 8 amplitudes of the $t$-$u$-channel coupling scheme are given by
$B_1 , B_{8_{ss}}, B_{8_{sa}}, B_{8_{as}}$, $ B_{8_{aa}}$, $B_{10},
B_{\bar{10}}, B_{27}$ where $B_{8_{ss}}$ e.g. stands for an octet
baryon in the $t$- or $u$-channel with symmetric N--N--meson coupling to
both mesons. These 8 amplitudes fit data again with a $\chi^2$ as
given above; suspiciously, singlet baryon exchange plays the most
important role.

In the baryon-exchange picture, contributions due to
\{$\bar{10}$\} and \{27\}-plet exchanges could be expected
to be small; further, the four N--N--PS and N--N--V couplings
(with different symmetries ss, as, sa, ss) should be related by
one $F/D$ ratio for the N--N--PS coupling and one
$F/D$ ratio for the N--N--V coupling. Predictions for these ratios are
model dependent. SU(6) predicts $F/D= 2/3$ \cite{Gursey:1964fk}
while an analysis of hyperon decays yields
$F/D=0.575\pm 0.016$ \cite{Song:1996mi}. The N--N--V coupling
is less well established and considered as free parameter here.

The data are completely incompatible with these constraints. Relaxing
the $F/D$ ratio yields unreasonably small $F/D$  and still a bad fit.
Also, there is no link of the amplitudes with the dynamical selection
rules. In this simple form baryon exchange does not provide additional
insight. Likely, the exchange of excited baryons would be needed
to achieve a better understanding of annihilation dynamics using meson
and baryons as fundamental actors. But this would be in conflict with
the spirit of this study in which simple interpretations of the branching
ratios are searched for.

Details of the fit method and results based on older data
can be found in a thesis at Mainz~\cite{walter_diss}. The results
are very similar to the ones obtained here.

\subsubsection{Discussion}
A similar range of energy release is involved in $\pbp$ annihilation
and in ${\rm J}/\psi$ decays, and a comparison between these two
processes is instructive.  Data on annihilation into a vector and a
pseudoscalar meson are collected in Table~\ref{dsr:tab:jpbar}. The
branching ratios for $\rm J/\psi$ are taken
from~\cite{Eidelman:2004wy}, for \pbp\ annihilation from
Table~\ref{tm:tab:Bpartial} in Sec.~\ref{tm:sub:BRanalpbp}. We use
only $\pbp$ annihilation from the $\tso$ initial state since its
$\jpc$ quantum numbers are identical to those of the $\rm J/\psi$.
\begin{table}[h!]
\caption{\label{dsr:tab:jpbar}
Annihilation fractions for $\pbp$ annihilation and  branching rations for ${\rm J}/\psi$ decays
into vector and pseudoscalar mesons.
}
\renewcommand{\arraystretch}{1.2}
\begin{tabular}{cccccccc}
\hline\hline
               & $\rho^0\pi^0$   & $\omega\eta$       & $\omega\eta^{\prime}$ 
               & $\phi\eta$      & $\phi\eta^{\prime}$& $\K^{*+}\K^-$ 
               & $\K^{*0}\Kb{}^0$ \\
${\rm J}/\psi$ & $100$             &$ 37.3\pm3.8$    &$3.9\pm0.6$
               & $15.4\pm1.7$       &$7.8\pm0.9$     &$59.1\er4.7$
               & $49.6\pm4.7$      \\
$\pbp$         & $100$             &$109.0\pm5.2$    &$57.1\pm7.6$
               & $0.38\pm0.10$      &x                &$7.0\pm1.2$
               &$7.0\pm1.2$        \\
\hline\hline
\end{tabular}
\renewcommand{\arraystretch}{1.0}
\end{table}
The branching ratio for $\pbp\to\rho^0\pi^0$ from the
\tso\ state is $21.2\times 10^{-3}$, whilst from  
${\rm J}/\psi$ decay it is about 5 times smaller, $4.23\times 10^{-3}$.
This is probably due to the larger phase-space available for 
${\rm J}/\psi$ which gives access to more final states. To facilitate
the comparison, the branching ratios for $\rho^0\pi^0$ are normalised to 100. 

Production of strange mesons is much larger in ${\rm J}/\psi$ decays
than in $\pbp$ annihilation. The rate in ${\rm J}/\psi$ is of the
magnitude expected by simple SU(3) considerations. It is anomalously
low in $\ppb$.  This reduction has often been
interpreted as being due to the dramatic suppression of $\ssb$ pair creation in low
energy physics. In Sec.~\ref{phe:sub:strange}, the reduction was suggested
to arise from large contributions from rearrangement processes in which
no new quarks need to be created. These processes are, of course, absent
in ${\rm J}/\psi$ decays where all quarks in the final state have to originate
from the vacuum. This fact enhances $\phi$ production in 
${\rm J}/\psi$ decays when it recoils against the $\eta$ or 
$\eta^{\prime}$. In $\pbp$ annihilation, there is no or little
$\ssb$ in the wave function, and $\phi$ production is small.

In ${\rm J}/\psi$ decay, the branching ratios decrease when the mass of
the produced mesons increases, by the ordinary phase-space effect.  A
very interesting feature of $\ppb$ annihilation is that this reduction
is not observed. In annihilation dynamics, production of large masses
is preferred over production of high momenta, as noted by
Vandermeulen~\cite{Vandermeulen:1988hh}, and discussed in
Sec.~\ref{phe:sub:vm}. In atomic physics, a similar effect is observed
in the Auger effect, with low momenta highly preferred. This is
understood by a better overlap of the wave function, and this
overcomes  phase-space considerations. Perhaps a derivation of the 
Vandermeulen effect is to be sought in the quark wave function of 
higher mass mesons, as compared to those of light mesons.

Annihilation into two vector mesons was excluded in the SU(3) analysis.
In Sec.~\ref{dsr:sub:ksks} it was shown that the daughter mesons from
the two vector mesons can interact before leaving the strong 
interaction region. This may be the reason that the dynamical
selection rules are not observed in this case. In particular
the strong $\omega\omega$ channel, in absence of a strong $\rho^0\rho^0$
counterpart, requires a large contribution of decuplet or 27--plet 
four--quark configurations in the intermediate state. It is interesting
to note that in the $\gamma\gamma\to\rho\rho$ channel, isotensor 
interactions are required above the $\rho\rho$ 
threshold~\cite{Brandelik:1980jv}
and this may be the reason why `exotic' exchanges are realistic. 
The ratio of $\gamma\gamma\to\rho^0\rho^0$~\cite{Achard:2004us} and 
$\gamma\gamma\to\rho^+\rho^-$~\cite{Achard:2004ux} measured
at LEP by L3 support~\cite{Sarantsev04} the previous findings 
of the TASSO  collaboration.

\clearpage\markboth{\sl Annihilation dynamics} {\sl Conclusions}
\setcounter{equation}{0}
\section{Conclusions and outlook}
\label{se:conc}
In this review we have discussed how a nucleon and an antinucleon
annihilate to create mesons in the final state.  In $\NNb$
annihilation, part of the incoming mass, 2 GeV for annihilation at
rest, is transformed into pure energy that hadronises into mesons. The
fraction presumably increases from a vanishing value for pure
rearrangement to reach 1 or 2 GeV for processes involving annihilation
of quark--antiquark pairs in the initial state.  The experiments
performed at LEAR have boosted our knowledge about mesons formed in
the annihilation process; inclusive and exclusive final states are
often known with good accuracy.  The multiplicity distribution of
charged and neutral mesons and their momentum distribution finds an
interpretation in a thermodynamic picture. A fireball with 120\,MeV
temperature and size $1/m_{\pi}$ annihilates into pions. The number of
charged and neutral pions also follow a statistical distribution.
This simple picture misses, however, the importance of meson
resonances and their production in two-body annihilation modes.
These reactions are very important for an understanding of the
annihilation process.

A systematic approach to annihilation dynamics requires 
annihilation frequencies to be measured for a large number 
of reactions. We now have `complete' information on frequencies
for annihilation into two pseudoscalar mesons, one pseudoscalar
and one vector or tensor meson, and into two vector mesons; the
information is complete in the sense that the frequencies of
all kinematically allowed annihilation modes are known. 
Unfortunately, this is only true for the $\pbp$ initial state; data
on $\pbn$ annihilation are still obscured by the spectator
proton in $\pbd$ annihilation, and data on low-energy $\nbp$
annihilation into exclusive final states are still scarce.

A given two-body annihilation channel can be produced from 
different atomic states or partial waves of the \pbp\ system.
The assignment of a fraction of an annihilation frequency to
a specific partial wave requires further experimental input. 
This could come from experiments using polarized antiprotons and 
protons; such experiments were not carried out at LEAR. The initial atomic 
states can be restricted when events are tagged by observation of
a coincident X-ray from the $\pbp$ atomic  cascade. In some
cases a series of frequency measurement at different target
pressures is sufficient to constrain the quantum numbers of the 
initial state from which annihilation occurred. 

Using these techniques, annihilation frequencies were 
recalculated to yield branching ratios for $\pbp$ annihilation
from a specified initial atomic state to two-meson final states. 
These are the numbers which should be compared to models
of $\NNb$ annihilation. 

Strong interaction physics, which is the domain of Quantum
ChromoDynamics, has evolved considerably in recent years.  However,
precise predictions dealing with $\NNb$ annihilation remain out of
reach. Models need to be developed to identify the relevant degrees of
freedom and the effective forces and symmetries. In an attempt to
identify the leading mechanisms in annihilation dynamics, the
experimental branching ratios were corrected dynamically using a model
proposed by Vandermeulen and extended here to account for centrifugal
barrier effects. The corrections parametrise the finite size of the
$\pbp$ source, and the preference for annihilation into mesons with a
high mass.  The parameters of the model were determined by the
requirement that the pseudoscalar mixing angle be reproduced from
branching ratios for annihilation into $\eta$ and $\eta^{\prime}$
mesons.

The dynamically corrected annihilation branching ratios 
provide a surprise: there are annihilation 
modes which seem to be suppressed in comparison with other modes
even though they are  perfectly legitimate and compatible with all
known selection rules. Here we quote Carl Dover~\cite{Dover:1992vj}:

{\it
The search for signatures of quark-gluon dynamics in \NNb\
annihilation is somewhat analogous to the search of the phase
transition from a hadron gas to a quark-gluon plasma in relativistic
ion transitions. The signal must be isolated from a background of
statistical processes characteristic of a system with many degrees of
freedom. For the \NNb\ system, an important role is played by
quasi-two-body intermediate states, or ''doorway states'', which
display directly the selection rules arising from baryon exchange or
quark dynamics. ... These {\em dynamical selection rules} provide key 
signatures of the annihilation mechanism. 
}

It has been shown that the dynamical selection rules can be grouped into two
classes of observations. The first selection rule is observed in
$\pbp$ annihilation into two isovector mesons ($\pi\pi , \rho\pi$ and
$\atmass\pi$). These frequencies are large from initial states which
allow annihilation into the two neutral (non-strange) mesons
(i.e., into $\piz\piz$ etc), and they are small from initial states
forbidding two neutral mesons by charge conjugation. The selection
rule is known, at least for annihilation into a pseudoscalar and a
vector meson, as the $\rho\pi$ puzzle. It has found different dynamical
interpretations which however missed the generalisation to $\pi\pi$
and $\atmass\pi$. The analysis of the flavour flow in quark line
diagrams linked the first class of selection rules to the suppression
of decuplet or 27-plet four-quark intermediate states.  According to
this interpretation, exotic (or non-octet) four-quark states could
have a large mass. Annihilation modes requiring these
intermediate states are then suppressed.  Non-exotic four-quark
configurations are not only allowed as intermediate states but provide
even the leading contribution to annihilation dynamics. The large
branching ratios for $\rho\pi$, $\omega\eta$, $\atmass\pi$ and other
meson pairs from isoscalar initial states are driven by this mechanism.
This observation does not claim the existence of four-quark resonances
in this mass range but states only that four-quark configurations can
be formed at the $\NNb$ mass if, and only if, they carry flavour-octet
quantum numbers.

The second selection rule is found in annihilation into two strange
mesons. At least in annihilation from S-wave orbitals one isospin in
the initial $\NNb$ state makes a dominant contribution.  In an
analysis of the flavour flow, the selection rule can be traced to a
symmetry property between quark line annihilation diagrams.  Likely,
the symmetry pattern reflects the symmetry properties of the quark
pair annihilation/creation operator. It is often advocated that $\qqb$
pairs couple to the gluon fields with vacuum quantum numbers; this is
the basis of the $\tpz$ model.  The annihilation and recreation of a
$\qqb$ pair with zero total angular momentum may also be governed by
instanton-induced forces. However, the origin of the symmetry
responsible for the second dynamical selection rule is so far
unexplored.

There have been attempts to apply the modern concepts of
strong-interaction physics to $\NNb$ annihilation, at an
exploratory level; however the first results obtained are interesting 
\cite{Zahed:1987us,Halasz:2000ye}.

Lattice simulations, no longer restricted to the quenched
approximation, have made dramatic progress. However, systems of two
interacting hadrons are at the edge of current possibilities, as
demonstrated by the somewhat contradictory results obtained by
different groups on the pentaquark. For a review, see, e.g.,
\cite{Sasaki:2004vz}.

The method of effective Lagrangians, which has a wide range of
applications, is particularly suited for low-energy strong
interactions, a domain where QCD can hardly be applied
directly. Instead, a Lagrangian having the appropriate symmetries, in
particular chiral symmetry, can describe  hadronic processes with
only a few parameters. The field was at
first restricted to mesons, and evaluating the $\pi\pi$ scattering
length was a typical challenge. The method is now applied to
meson--baryon interaction and to nuclear forces. Recent reviews 
can be found in Refs.~\cite{Hong:2001qh,Gasser:2003cg,Meissner:2004yy}.

QCD itself is sometimes considered in a variant with a large number
$N_{\rm c}$ of colour degrees of freedom, instead of the actual $N_{\rm
c}=3$. It has been shown that the $N_{\rm c}\to\infty$ limit provides
a simpler picture, in which just a few diagrams contribute, the other
being suppressed by powers of $N_{\rm c}$. For instance, many features
of the charmed baryons are understood from considerations based on
$1/N_{\rm c}$. It has also been emphasised
\cite{Manohar:1984ys,Cohen:2003yi} that a well-understood large
$N_{\rm c}$ and a well-controlled chiral theory gives compatible
results.  Unfortunately, it is still difficult to apply large-$N_{\rm
c}$ methods to $\NNb$ annihilation. There are debates about the
hierarchy of diagrams and the selection rules suggested by the
$N_c\to\infty$ view of annihilation.

It is fair to say that the aim of most annihilation experiments was to
study the spectrum of mesons and discover new meson resonances, 
rather than being solely aimed at the
study of strong interaction dynamics in annihilation. Instead,
annihilation dynamics was a side product of the experiments. The main
objective was meson spectroscopy and the search for new forms of
hadronic matter. These were quasinuclear and four-quark states when
LEAR was started.  Later, the fashion changed to glueballs or hybrids.

The study of the meson spectrum has been rather successful, both in early
bubble-chamber experiments and in recent LEAR experiments. A review on
light-meson physics, and in particular the implications of LEAR
results, is found, e.g.,  in~\cite{Amsler:2004ps}.  For mesons that were
already known, the role of LEAR experiments was twofold: their
existence was confirmed and their structure and decay properties were
studied. But also a large number of mesons was discovered in $\NNb$
annihilation.  LEAR has provided evidence for several new meson
resonances, and in some sectors such as the scalar sector
$\jpc=0^{++}$, there may be too many mesons to be accommodated by
$\qqb$ states, even when radial excitations are included. Some of the
best experts on the physics of mesons have also proposed a multiquark
interpretation of the excess mesons\footnote{We benefited from many discussions with Lucien Montanet on this subject.}.

When this review was started, the authors envisaged some apologetic
words to mention baryonium as the main motivation for building the
LEAR facility. See Sec.~\ref{se:intro} and 
Ref.~\cite{Klempt:2002ap}. Ironically, this review was finished at the end of
2004, and almost every day there is a paper published suggesting that
one of the new mesons, such as X$(3872)$, D$_s(2632)$, etc., or new
exotic baryons such as $\theta^+(1540)$ are of multiquark or
hadron--hadron nature. See, e.g., \cite{Jaffe:2004wv,Nicolescu:2004in,Maiani:2004vq,Barnes:2004ay}.

Throughout this review, a comparison has sometimes been attempted between $\pbp$
annihilation and J$/\psi$ decay. The similarities and differences
certainly deserve to be studied more closely. The study of $\NNb$ annihilation does
not aim at remaining an isolated field. Topics such as the topology of
quark diagrams, the rate of hidden strangeness, the production of
high-mass resonances occur in several processes; in particular the
decay of particles containing heavy quarks. It is hoped that an
unified picture of hadronisation will emerge when analysing the results
collected at future antiproton facilities and heavy quark factories.

\clearpage\markboth{Annihilation dynamics} {\sl Acknowledgments}
\clearpage\setcounter{equation}{0}
\section*{Acknowledgments}
We would like to thank our collaborators on the topics covered by this
review, in particular the members of the Asterix and Crystal Barrel
experiments, J.~Carbonell, G.~Ihle, J.G.~K\"orner, and H.J.~Pirner.
We have benefited in recent or past discussions with the broad and critical
expertise of C.B.~Dover, P.~Fishbane, A.M. Green,
T.~Kalogeropoulos, J.A.~Niskanen, I.S. Shapiro, J.~Vandermeulen, and many other
colleagues. This review would not have been possible without several
friendly and animated discussions with the members of the Obelix
collaboration.

CJB acknowledges support from the Rutherford Appleton Laboratory and the British Particle Physics and Astronomy Research Council.
 
JMR is grateful to the Humboldt Foundation for its generous support and to the members of HISKP, Bonn, for their hospitality.

\clearpage\markboth{Annihilation dynamics} {\sl Bibliography}
\end{document}